\documentstyle[12pt,epsfig,fleqn]{report}
\def\Journal#1#2#3#4{{#1}{\bf #2}, #3 (#4)}

\def\NIMA{{Nucl. Instrum. Methods}~{\bf A}}
\def\NPA{{Nucl. Phys.}~{\bf A}}
\def\NPB{{Nucl. Phys.}~{\bf B}}
\def\PLB{{Phys. Lett.}~{\bf B}}

\def\PRL{Phys. Rev. Lett.\ }
\def\PRD{{Phys. Rev.}~{\bf D}}
\def\PRC{{Phys. Rev.}~{\bf C}}
\def\ZPC{{Z. Phys.}~{\bf C}}

\begin{document}
\newcommand{\bcc}{\begin{center}}
\newcommand{\ecc}{\end{center}}
\newcommand{\bitm}{\begin{itemize}}
\newcommand{\eitm}{\end{itemize}}
\newcommand{\bdes}{\begin{description}}
\newcommand{\edes}{\end{description}}
\newcommand{\benn}{\begin{enumerate}}
\newcommand{\eenn}{\end{enumerate}}
\newcommand{\bary}{\begin{array}}
\newcommand{\eary}{\end{array}}
\newcommand\be{\begin{equation}}
\newcommand\ee{\end{equation}}
\newcommand\bea{\begin{eqnarray}}
\newcommand\eea{\end{eqnarray}}
\newcommand{\ese}{\end{subequations}}
\newcommand{\bse}{\begin{subequations}}
\newcommand{\beq}{\begin{eqalignno}}
\newcommand{\eeq}{\end{eqalignno}}
\newcommand{\bfig}{\begin{figure}}
\newcommand{\efig}{\end{figure}}

\pagestyle{plain}

\thispagestyle{empty}
\bcc
{\Large {\bf Study of Initial and Final State Effects in Ultrarelativistic Heavy Ion Collisions Using Hadronic Probes}}
\\
\vspace*{2.5cm}
A Dissertation Presented\\
\vspace*{0.5cm}
by\\
\vspace*{0.5cm}
Anuj Kumar Purwar\\
\vspace*{0.5cm}
to\\
\vspace*{0.5cm}
The Graduate School\\
\vspace*{0.5cm}
in Partial Fulfillment of the Requirements\\
\vspace*{0.5cm}
for the Degree of\\
\vspace*{0.5cm}
Doctor of Philosophy\\
\vspace*{0.5cm}
in\\
\vspace*{0.5cm}
Physics\\
\vspace*{0.5cm}
Stony Brook University\\
\vspace*{1.0cm}
December 2004
\ecc

\pagebreak

\renewcommand{\thepage}{\roman{page}}

\thispagestyle{plain}
\bcc
State University of New York\\
at Stony Brook\\
The Graduate School\\
\vspace*{0.5cm}
\rule{8cm}{0.2pt}\\
Anuj Kumar Purwar\\
\vspace*{1.0cm}
We, the dissertation committee for the above candidate for the  
Doctor of Philosophy degree, hereby recommend acceptance of this 
dissertation.\\
\vspace*{1cm}
\rule{8cm}{0.2pt}\\
Thomas K. Hemmick (Advisor)\\
Professor, Department of Physics \& Astronomy\\
\vspace*{1cm}
\rule{8cm}{0.2pt}\\
Gerald E. Brown (Chair)\\
Professor, Department of Physics \& Astronomy\\
\vspace*{1cm}
\rule{8cm}{0.2pt}\\
John Hobbs\\
Professor, Department of Physics \& Astronomy\\
\vspace*{1cm}
\rule{8cm}{0.2pt}\\
 Edward J. O'Brien \\
Physicist, Physics Department, Brookhaven National Lab \\
\vspace*{1cm}
This dissertation is accepted by the Graduate School\\
\vspace*{1cm}
\flushright\rule{8cm}{0.2pt}\\
Dean of the Graduate School   \\
\ecc

\pagebreak

\bcc
{\Large {\bf Abstract of the Dissertation}}\\
\vspace*{0.5cm}
{\Large {\bf Study of Initial and Final State Effects in Ultrarelativistic Heavy Ion Collisions Using Hadronic Probes}}\\
\vspace*{0.5cm}
by\\
\vspace*{0.5cm}
Anuj Kumar Purwar\\
\vspace*{0.5cm}
Doctor of Philosophy\\
\vspace*{0.5cm}
in\\
\vspace*{0.5cm}
Physics\\
\vspace*{0.5cm}
Stony Brook University\\
\vspace*{0.5cm}
2004
\ecc

It has been theorized that if heavy nuclei (e.g. Au, Pb) are collided at sufficiently high energies, we might be to recreate the conditions that existed in the universe a few microseconds after the Big Bang: a phase transition into a new state of matter in which quarks and gluons are deconfined  (Quark-Gluon Plasma or QGP). However, we never directly get to see the QGP because as the matter cools it recombines (hadronizes) into ordinary subatomic particles. In this dissertation we attempt to shed some light on:
\benn
\item Properties of the final state of produced matter in Au+Au collisions at $\sqrt{s_{NN}}=200$ GeV. As the hot, dense system of particles from the collision zone cools and expands, light nuclei like deuterons and anti-deuterons can be formed, with a probability proportional to the product of the phase space densities of its constituent nucleons. Thus, invariant yield of deuterons, compared to the protons and neutrons from which they coalesce, provides information about the size of the emitting system and its space-time evolution. 

The transverse momentum spectra of $d$ and $\bar{d}$ in the
range $1.1<p_T<4.3$~GeV/$c$ were measured at mid-rapidity and were found to 
be less steeply falling than proton (and antiproton) spectra. 
A coalescence analysis comparing the deuteron
and antideuteron spectra with that of proton and antiproton was performed and
the extracted coalescence parameter $B_2$ was found to increase with $p_T$, 
indicating an expanding source. 

\item The initial conditions can be probed by looking at the nuclear modification factor $R_{cp}$ from particle production in forward and backward directions in a ``control'' experiment using d+Au collisions at $\sqrt{s_{NN}}=200$ GeV. This can allow us to distinguish between effects that could potentially be due to deconfinement, versus effects of cold nuclear matter.

We found hadron $R_{cp}$ to be suppressed at forward rapidities
(d going direction). This is qualitatively consistent with 
shadowing/saturation type effects in the small-$x$ region being probed at 
forward rapidities. $R_{cp}$ was enhanced at backward rapidities (Au going 
region).
\eenn

\pagebreak

\tableofcontents



\listoffigures

\listoftables

\chapter*{Acknowledgements\markboth{Acknowledgements}{Acknowledgements}}
First, of all I would like to acknowledge my advisor Tom Hemmick for
his guidance and support through the ups and the downs of Ph.D research.
His infectious enthusiasm for physics was very inspiring. 
The group at Stony Brook is like a large family: always there for you. I thank
Barbara Jacak, Axel Drees, Ralf Averbeck and Vlad Panteuv. At every step of 
the way I got help from other students and postdocs: Jane Burward-Hoy, Sergey 
Butsyk, Felix Matathias, Federica Messer, Julia Velkovska, Mike Reuter and 
Sasha Milov. I also received a lot of assistance from other people in the 
PHENIX collaboration including Joakim Nystrand, Rickard du Rietz, Chun Zhang,
Jamie Nagle, Ming X. Liu and Youngil Kwon. I specially thank Pat Peiliker, 
Diane 
Siegel, Pam Burris and Socoro Delqualgio for the administrative support. I 
also thank Rich Hutter for the hardware support. Victor Weisskopf once said:

   {\it There are three kinds of physicists, as we know, namely the machine 
builders, the experimental physicists, and the theoretical physicists. If we 
compare those three classes, we find that the machine builders are the most 
important ones, because if they were not there, we could not get to this 
small-scale region. If we compare this with the discovery of America, then, I 
would say, the machine builders correspond to the captains and ship builders 
who really developed the techniques at that time. The experimentalists were 
those fellows on the ships that sailed to the other side of the world and then
jumped upon the new islands and just wrote down what they saw. The 
theoretical physicists are those fellows who stayed back in Madrid and told 
Columbus that he was going to land in India.}

	This dissertation and the corresponding research would not have been
possible the work done by the people of the Collider Accelarator department,
as well as people who worked hard during the initial construction of PHENIX,
allowing me to reap the fruit of their labor.

Special thanks goes to my parents and brother and sister, who 
nurtured my interest in science, when I was younger. I also thank Rohini
Godbole, who guided me through my M.S. thesis in Indian Institute of Science,
at Bangalore.

This acknowledgement would not be complete without thanking the authors of 
popular science books like Nigel Calder (Einsteins Universe), Rudolf 
Kippenhahn (100 Billion Suns), Heinz W. Pagels (The Cosmic Code), Richard P. 
Feynman and Matthew Sands (Surely You Are Joking Mr. Feynman) and Martin 
Gardener (This Ambidextrous Universe), who led me down this career. This
dissertation is dedicated to them.

\pagebreak

\renewcommand{\thepage}{\arabic{page}}

\setcounter{page}{1}

\pagestyle{headings}

\chapter{Introduction}
A few microseconds~\cite{microbb} after the Big Bang, the universe consisted 
of a hot and dense plasma of deconfined quarks and gluons. As the 
universe expanded and cooled, this deconfined plasma coalesced into protons 
and neutrons (hadronization), followed by the formation of bound nuclei 
(nucleosynthesis). Finally atoms and molecules were formed after a few 
thousand years. A sketch of this timeline is shown in Figure~\ref{fig:bbphase}.

\begin{figure}[h!]
\bcc
\includegraphics[width=0.75\linewidth]{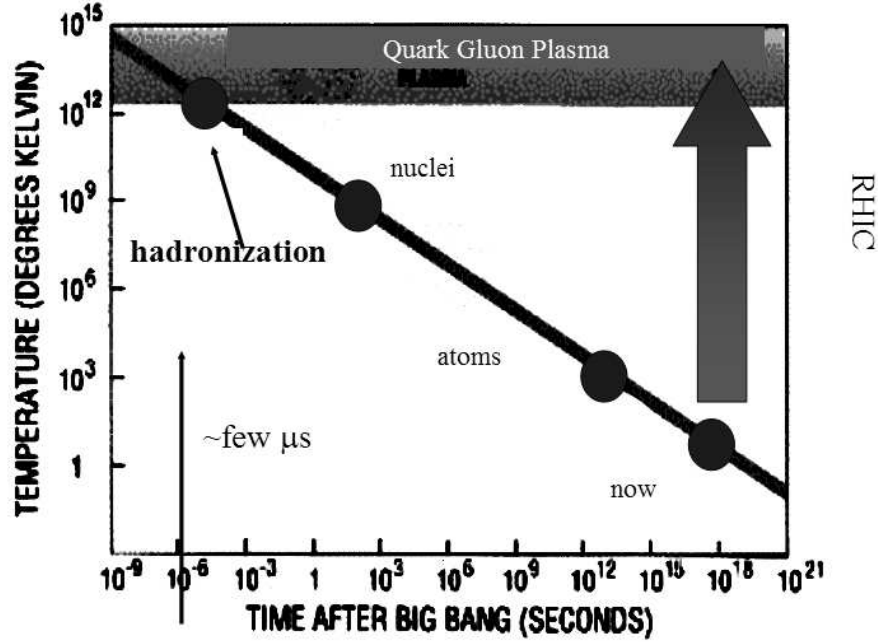}  
\ecc
\caption{Big Bang timeline and Quark-Gluon-Plasma (QGP).}
\label{fig:bbphase}
\end{figure}

\section{Ultrarelatvistic Heavy Ion Collisions and QGP}
	It has been theorised that if heavy nuclei are smashed together at
very high energies (Ultrarelativistic heavy ion collisions) by means
of particle accelerators, we might be to recreate the conditions that
existed in the universe in that early epoch, in the lab. At sufficiently high 
energies, it is expected that the kinetic energy of the colliding nuclei gets 
converted into heat, leading to a phase transition into a new state of 
matter: the Quark-Gluon Plasma (QGP), in which quarks and gluons are 
deconfined. Quantum Chromodynamics (QCD), the theory of the Strong Interaction 
predicts~\cite{qcdprediction} that at a well determined temperature
($T_c \simeq 150 - 180$ MeV for zero net baryon density\footnote{Total baryon
number equal to zero or in other words the amount of matter and anti-matter is
approximately equal.}) ordinary hadronic matter undergoes a phase transition
from color singlet hadrons to a deconfined medium consisting of colored quarks
and gluons. Lattice QCD calculations predict that the energy density at this 
transition point: $\epsilon (T_c) \simeq 0.7-1.0$ GeV/fm$^3$, almost 10 times
the density of nuclear matter. A phase diagram of nuclear matter in equilibrium
is shown in Figure~\ref{fig:nucl_phase}.

\begin{figure}[h!]
\bcc
\includegraphics[width=1.0\linewidth]{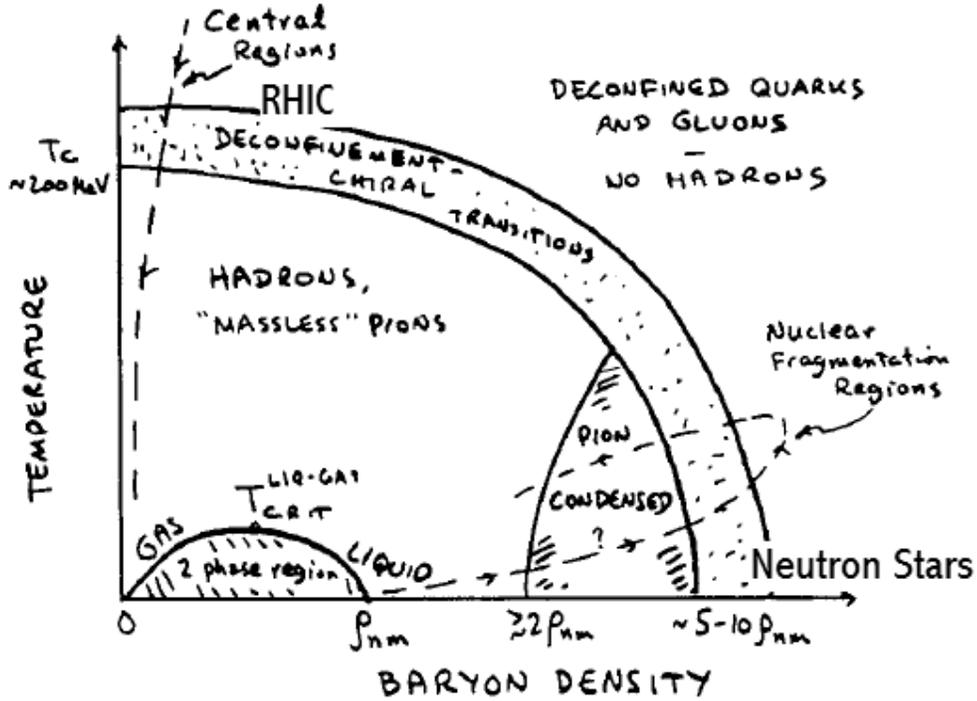}  
\ecc
\caption{A phase diagram of nuclear matter (nuclear density along $x$-axis
and temperature along $y$-axis) shows how deconfinement can occur extreme
conditions of temperature (Big Bang, RHIC) and density (dense stellar matter 
like neutron stars). This figure is taken from 1983 NSAC Long Range 
Plan~\cite{nsac}.}
\label{fig:nucl_phase}
\end{figure}

The holy grail of ultrarelativistic heavy ion collisions is the discovery
and characterisation of the Quark-Gluon Plasma (QGP). Discovering the QGP
is not an easy task, because we never see the bare quarks and gluons. Even
if QGP is produced in an experiment, it subsequently hadronizes into the usual
menagerie of hadrons. So we never get to directly see the QGP, and can only 
hope to infer it's existence from indirect means. Some of the traditional
QGP signatures are briefly outlined below:
\benn
\item {\bf Dilepton production:} A quark and an anti-quark can interact via 
a virtual
photon $\gamma^*$ to produce a lepton and an anti-lepton $l^+l^-$ (often
called dilepton). Since the leptons interact only via electromagnetic means,
they usually reach the detectors with no interactions, after production. As a
result dilepton momentum distribution contains information about the 
thermodynamical state of the medium (For reviews see~\cite{ruuskanen}).
\item {\bf Thermal Radiation:} Similar to dilepton production, a photon and 
a gluon
can be produced via $q + \bar{q} \rightarrow \gamma + g$. Since the 
electromagnetic interaction isn't very strong, the produced photon usually 
passes to the detectors without any interactions after production. And just 
like dileptons, the momentum distribution of photons can yield valuable 
information about the momentum distributions of the quarks and gluons that 
make up the plasma, giving us a window into it's thermodynamical properties
(for a review see~\cite{lichard}).
\item {\bf Strangeness Enhancement:} Production of strange quarks requires
a larger amount of energy compared to ordinary u and d quarks. The high energy
densities in QGP are conducive for $s\bar{s}$ production, leading to an
enhancement in the number of strange particles as compared to the strangeness
production in p+p collisions~\cite{strangeness}.
\item {\bf $J/\psi$ suppression:} In a Quark-Gluon-Plasma (QGP), color 
screening due
the presence of free quarks and gluons (similar to Debye screening seen in 
QED), the $J/\psi$ particle --- a bound state of charm and anti-charm quarks 
$c\bar{c}$ --- can dissociate. This leads to a suppression of $J/\psi$
production, a classic signature first predicted by Matsui and Satz~\cite{satz}.
\item {\bf HBT:} The Hanbury-Brown-Twiss effect --- first used to measure the 
diameter of a star~\cite{hbt_original} --- is also used to in high energy 
nuclear experiments, by measuring the space-time(or energy-momentum) 
correlation of identical particles emitted from an extended source. In 
ultrarelativistic heavy ion collisions, an HBT measurement can yield 
information about size and the matter distribution of the source.
\item {\bf Jet suppression:} In nucleon collisions, energetic partons 
(jets) can be produced via hard scatterings. In presence of deconfined matter,
they interact strongly, leading to energy loss $\simeq$ GeV/fm, mostly due to 
gluon bremmstrahlung processes. This results in a decrease in the yield of 
high energy particles or jet suppression~\cite{gyulassy_jetsup}.
\eenn

Discovery of QGP is beyond the scope of this dissertation, instead we shall 
have to satisfy ourselves by studying the behavior of nuclear matter at 
extreme conditions of temperature and density, via ultrarelativistic heavy 
ion collisions and trying to shed some light on a) properties of the final 
state of produced matter, and b) the initial conditions that led to this. As 
a result this dissertation will have two main thrusts:

\benn
\item Exploration of the final state effects of the produced
matter from Au+Au collisions at $\sqrt{s_{NN}} = 200$ GeV, by studying the
production of the simplest nuclei: deuterons and anti-deuterons.  

\item Study the effect of cold nuclear matter in d+Au collisions at 
$\sqrt{s_{NN}} = 200$ GeV, by looking at particle production in forward
and backward directions.  
\eenn

\section{Deuterons and anti-deuterons as probes of final state effects}
Previous measurements indicate that high particle 
multiplicities~\cite{mult1,phobosback} and large $\bar{p}/p$ ratios prevail 
at RHIC, which is expected for a nearly net baryon free 
region~\cite{ppg006,phobospbar,starpbar}. As the hot, dense system of 
particles cools, it expands and the mean free path increases until the 
particles cease interacting (``freezeout''). At this point, light nuclei like 
deuterons and antideuterons ($d$ and $\bar{d}$) can be formed, with a 
probability proportional to the product of the phase space densities of it's
constituent nucleons~\cite{csernai,mekjian}. Thus, invariant yield of 
deuterons, compared to the protons~\cite{ppg009,ppg026} from which they 
coalesce, provides information about the size of the emitting system and its 
space-time evolution. We use the PHENIX Time-Of-Flight (TOF) detector along 
with
the central arm tracking chambers: Drift Chamber (DC) and Pad Chambers (PC3)
to detect deuterons. We measure momentum and time of flight and use it to
obtain the mass, which is used for particle identification (PID). Corrections
are then applied to account for limited acceptance, detector efficiencies and
myraid other minutae that are the bane of experimentalists all over the world.
We eventually obtain the corrected invariant yields $Ed^3N/dp^3$ and look
really really hard at them, paying special attention to the shapes of the
spectra and compare them with proton yields to glean some information about the
spacetime evolution of the collision zone.

\section{Particle multiplicities at forward and backward rapidities 
as probes of initial state effects}

Particle multiplicities have yielded some of the most interesting insights at
RHIC (Relativistic Heavy Ion Collider). Data from the Au + Au collisions at 
$\sqrt{s_{NN}} = 200$ GeV at mid-rapidity indicated a suppression~\cite{phenix_130supre,phenix_130cent,star_130supre,phenix_200pi} of particle 
yields as compared to the expectation from naive scaling of p+p collisions. 
This was consistent either with a) jet suppression i.e, suppression of high 
$p_T$ particles due to energy loss in the deconfined medium or b) due to the 
depletion of low-$x$\footnote{$x$ is the momentum fraction carried by the
parton.} partons due to gluon saturation processes as predicted
by the Color-Glass Condensate (CGC) hypothesis. In order to figure out whether
this suppression in Au+Au collisions was due to final state effects (the 
Quark-Gluon-Plasma (QGP)) or due to initial state effects (gluon saturation 
effects), a control experiment was done by colliding deuteron and gold nuclei
at the same energy. The Run 3 data with d + Au collisions at 
$\sqrt{s_{NN}} = 200$ GeV, showed an enhancement at mid-rapidity~\cite{phenix_nosup,star_nosup}. Similar 
effects have been seen at lower energies and go by the name of Cronin effect 
and are usually attributed to multiple scattering of partons in the initial 
state. Obviously this was inconsistent with the CGC (gluon saturation) 
hypothesis, which predicted a suppression in particle 
multiplicities~\cite{cgc_klm,gluonsat1,gluonsat2}
for d+Au collisions. However, all hope wasn't lost: the scale at which the
gluon saturation effects occur, provided an escape hatch and it turns out that
although particle multiplicities are not suppressed at mid-rapidity, if we
look at forward rapidity (in the deuteron going direction) we can explore
the low-$x$ region of the Au nucleus. And depending upon the saturation 
scale, we might be able to see suppression. In the second half of this 
dissertation, we seek to measure charged hadron multiplicities at 
forward rapidity (approximate pseudorapidity range $1.2<\eta<2.0$) using the 
PHENIX Muon Arms (which b.t.w weren't supposed to detect hadrons). By looking
at the particle multiplicities scaled with those at peripheral collisions,
which is the lazy man's way of getting around the need to use p+p data, at 
forward and backward rapidities and their variation with centrality (or
impact parameter) we hope to shed some light on the issue of initial 
conditions. 

\section{Some jargon}
The field of the relativistic heavy ion physics is saturated with jargon, a 
minefield for the uninitiated. Before we embark on the rest of this 
dissertation, here is a brief description of some of the commonly used terms:
\bitm
\item{\bf Center of mass energy:} a.k.a. $\sqrt{s}$, this is the Lorentz
invariant quantity:
\be
s = (p_1+p_2)_\mu(p_1+p_2)^\mu
\label{eq:s_def1}
\ee 
For nuclei with energy $E_i$ and 3-momentum $\mathbf{p_i}$, it reduces to:

\be
\sqrt{s} = \sqrt{m_1^2+2E_1E_2-2\mathbf{p_1\cdot p_2}+\mathbf{m_2^2}}
\label{eq:s_def2}
\ee
For instance at RHIC (Run 2 and 3), the center-of-mass energy per nucleon is 
$\sqrt{s_{NN}}=200$ GeV.

\begin{figure}[h!]
\bcc
\includegraphics[width=0.5\linewidth]{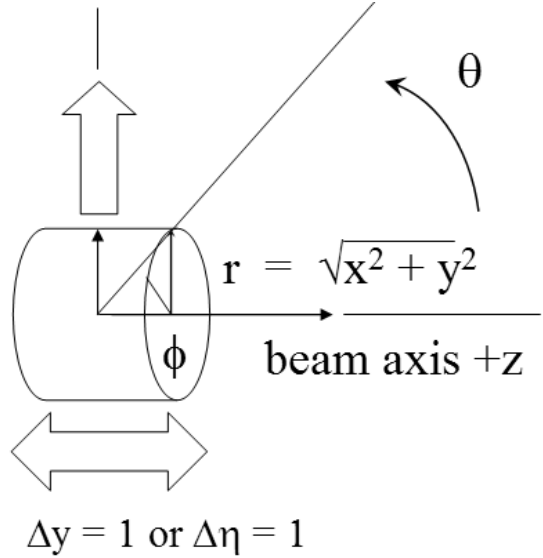}  
\ecc
\caption{Beam axis, transverse momentum $p_T$ and rapidity $y$.}
\label{fig:beam_axis}
\end{figure}

\item {\bf Tranverse momentum $p_T$:} this is simply the projection of a 
particle's momentum perpendicular to the collision axis: $z$ (see 
Figure~\ref{fig:beam_axis}).
\be
p_T = p\sin\theta
\label{eq:pt_def}
\ee
where $\theta$ is the polar angle along the $z$-axis. A common variable derived
from this is the transverse energy (or mass) $m_T=\sqrt{p_T^2+m_0^2}$.
\item {\bf Rapidity $y$:} this defines the longitudinal motion scale for a 
particle of mass $m_0$ moving along $z$-axis (see Figure~\ref{fig:beam_axis}):
\be
y = \frac{1}{2}\log \left( \frac{E+p_z}{E-p_z} \right)
\label{eq:y_def}
\ee
Since there is cylindrical symmetry around the collision axis, this allows us 
to describe the 4-momentum of particle in terms of its 
transverse momentum $p_T$, rapidity $y$ and the transverse energy $m_T$ as:
\be
p^{\mu}=(m_T \cosh y, p_T \cos \phi_0, m_T\sinh y)
\label{eq:4momentum}
\ee

\item {\bf Pseudorapidity $\eta$:} derived from rapidity (Eq.~\ref{eq:y_def}),
this variable is used when the particle in question is unidentified i.e., $m_0$
is not known:
\be
\eta = -\log \left( \tan\frac{\theta}{2} \right)
\label{eq:eta_def}
\ee
Where $\theta$ is the angle w.r.t. the beam axis. $\eta$ is often used to 
describe geometrical acceptances of detectors.
\item {\bf Invariant yield:} the
invariant differential cross section of a particle is the probability of 
obtaining $d^3N$ particles in the phase space volume $dp^3/E$ in a given number
of events $N_{event}$:

\be
\frac{1}{N_{event}}E\frac{d^3N}{dp^3} = \frac{d^3N}{N_{event}p_Tdp_Tdy} 
\label{eq:yield_def}
\ee

In cylindrical coordinates $dp^3=dp_xdp_ydp_z$ reduces to 
$p_Tdp_Td\phi m_t\cosh ydy$. Due to azimuthal symmetry we get a factor of 
$1/2\pi$, resulting in the form:
\be
\frac{1}{N_{event}}E\frac{d^2N}{dp^2}=\frac{d^2N}{2\pi N_{event}p_Tdp_Tdy}
\label{eq:ptspectra}
\ee

Using $dN/p_Tdp_T = dN/m_Tdm_T$, we get our final form:
\be
\frac{1}{N_{event}}E\frac{d^2N}{dp^2}=\frac{d^2N}{2\pi N_{event}m_Tdm_Tdy}
\label{eq:mtspectra}
\ee
\item {\bf Centrality:} when the two nuclei collide, there can be range of
impact parameters. Events with a small impact parameter are known as central
events whereas events with a large impact parameter are called peripheral (see
Figure~\ref{fig:impact_par}), and the variation in impact parameters is 
called centrality.

\begin{figure}[h!]
\bcc
\includegraphics[width=0.5\linewidth]{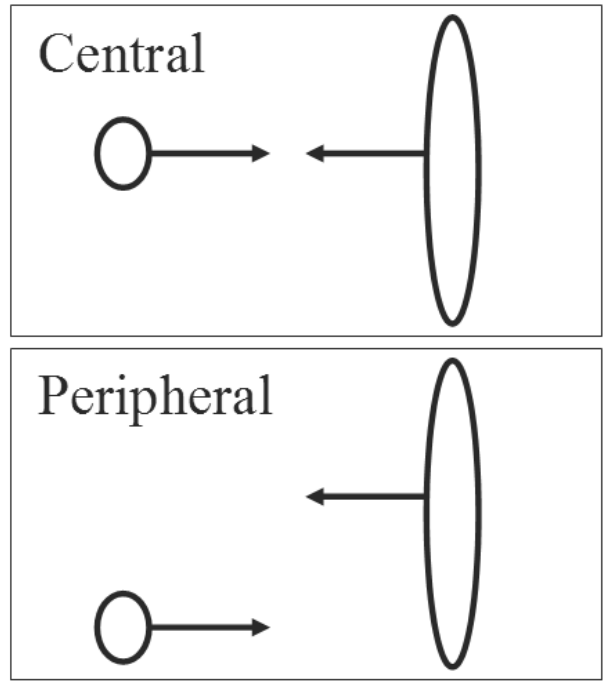}  
\caption{Centrality is related to impact parameter: large impact parameter 
events are called peripheral and small impact parameter events are called 
central.}
\label{fig:impact_par}
\ecc
\end{figure}

\item {\bf Minimum Bias:} this is the collection of events containing all 
possible ranges of impact parameters. This is important so that our data does
not have any bias due to events that might be triggered by specific 
signals e.g. presence of a high $p_T$ particle.
\eitm

\section{Organization of thesis}
This work is organised as follows: in Chapter 2 we describe the experimental 
setup at PHENIX. Our measurements of deuterons and anti-deuterons are 
described in Chapters 3 and 4. Chapters 5 discusses the background for the
nuclear modification factor, while Chapters 6 and 7 are devoted to the 
measurement of particle multiplicities at forward (and backward) rapidities. 
Finally Chapter 8 summarizes all our results. Bon Voyage!

\chapter{Experimental Facilities and Setup}
In this chapter we shall give an overview of the Relativistic Heavy Ion 
Collider (RHIC) and the PHENIX detector~\cite{phenixnim} alongwith the 
subsystems that were used for deuteron/anti-deuteron measurement. 

\section{RHIC facility}
In order to have any hope of discovering the Quark-Gluon-Plasma, 
experimentalists have tried to collide heavy nuclei at the highest possible
energies obtainable subject to the usual constraints of technology and funding.
Most of the past experiments have been fixed target experiments, in which a 
beam of a given species, e.g. proton (p) or lead (Pb) at CERN Super Proton 
Synchrotron (SPS), is incident on a fixed target of the appropriate 
material e.g. Pb at SPS. Since the lab frame is not same as the beam frame,
for a given beam energy, the actual center-of-mass energy is lesser as compared
to a colliding beam accelerator like RHIC. Typical center-of-mass energies 
at the Alternating Gradient Synchrotron (AGS) at BNL were in the range 
2.5 -- 4.5 GeV, and at SPS typical $\sqrt{s}$ = 17 GeV (for Pb+Pb).

RHIC consists of 2 counter-circulating rings  capable of accelerating any 
nucleus on any other, with a top energy (each beam) of 100 GeV/nucleon Au+Au 
and 250 GeV polarized p+p. The tunnel is 3.8 km in circumference and contains 
powerful superconducting dipole magnets to guide the beams at these energies.

\begin{figure}[h!]
\includegraphics[width=1.0\linewidth]{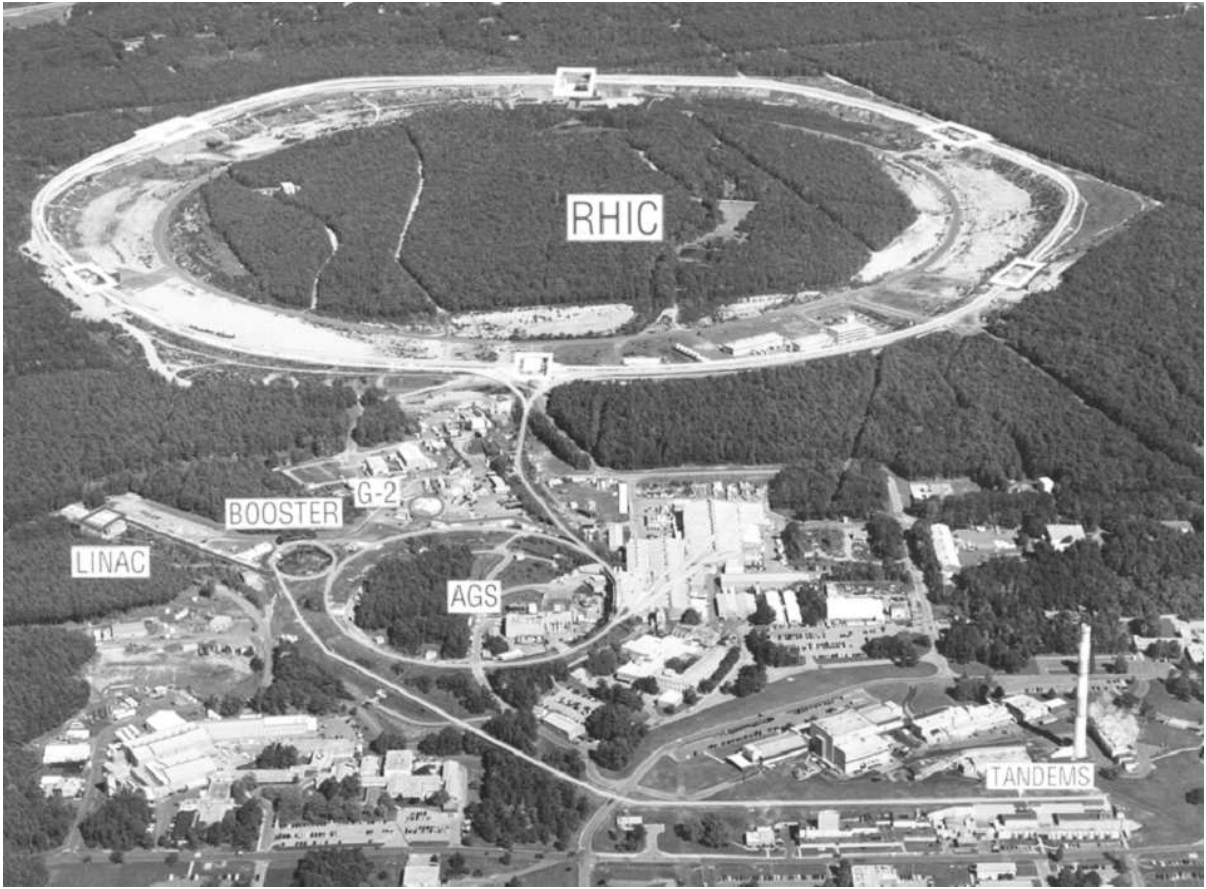}  
\caption{A picture of the Relativistic Heavy Ion Collider (RHIC) complex at
Brookhaven National Lab (BNL) in Long Island, NY. The RHIC ring with associated
systems: Tandems and AGS booster can also be seen.}
\label{fig:rhic_pic}
\end{figure}

Before the high energy heavy ion collisions can occur, the ions undergo 
several steps:
\benn 
\item After removing some of the electrons from the atom, the Tandem Van de 
Graaff uses static electricity to accelerate the resulting ions into the the
Tandem-to-Booster line (TTB). For p+p collisions, the Linear Accelerator 
(Linac) is used to provide energetic protons ($\approx$ 200 MeV).
\item  The Booster synchrotron is a compact, powerful circular accelerator that
propels the ions closer to the speed of light ($\approx$ 37\%) and feeds them 
into the Alternating Gradient Synchrotron (AGS).
\item The AGS further accelerates the ions to 99.7\% of the speed of light 
and injects them into the AGS-To-RHIC (ATR) transfer line, where a switching
magnet directs the ion bunches to either the clockwise RHIC ring or the 
anti-clockwise RHIC ring.
\item Once in the RHIC rings, the ions are accelerated by radio waves
(RF) to $\gamma$ = 70 or equivalently 99.995\% the speed of light. Finally they
are collided at the six interaction points where the four experiments reside:
BRAHMS, PHENIX, PHOBOS and STAR. A typical central Au+Au event as taken in the
PHENIX detector is shown in Figure~\ref{fig:phenix_event}.
\eenn

\begin{figure}[h!]
\bcc
\includegraphics[width=0.75\linewidth]{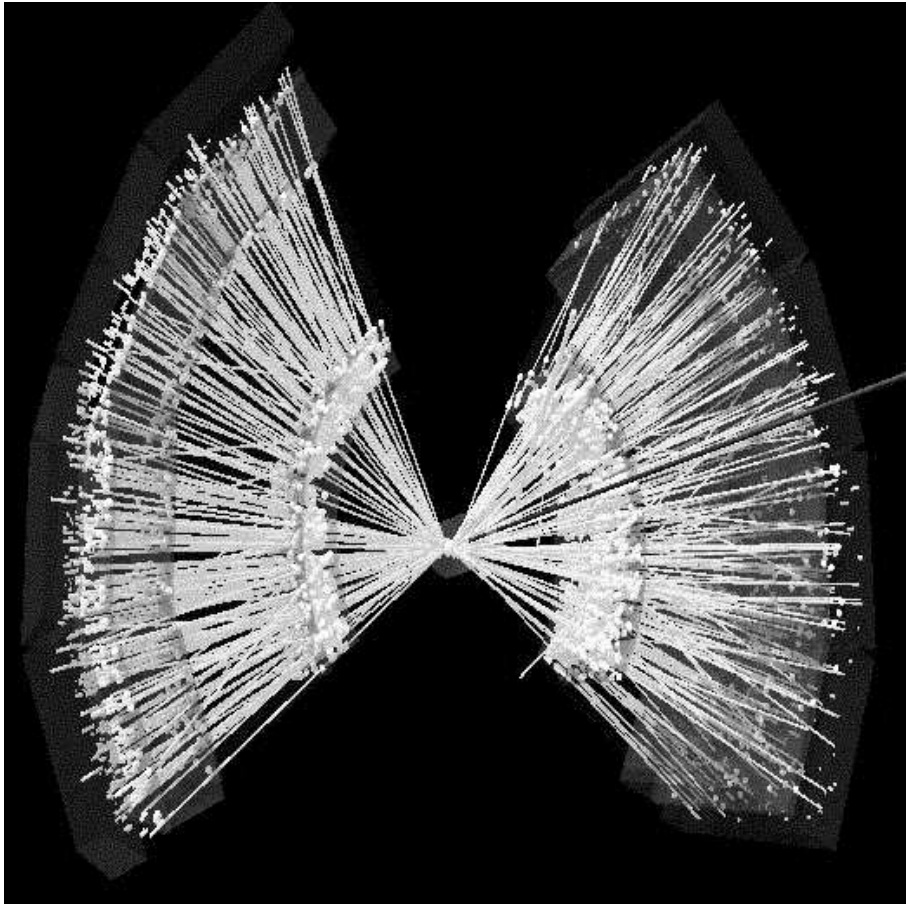}  
\ecc
\caption{View of a central Au+Au event in PHENIX as taken in Run 2 (2002).}
\label{fig:phenix_event}
\end{figure}

\section{PHENIX detector overview}
PHENIX~\cite{phenixnim} ({\bf P}ioneering {\bf H}igh {\bf E}nergy {\bf N}uclear
 {\bf I}nteraction e{\bf X}periment) is a versatile detector designed to study 
a maximal set of observables including the production of leptons, photons and 
hadrons over a wide momentum range. It is capable of taking events at a high 
rate and do selective triggering for rare processes. A detailed overview of 
PHENIX and its subsystems is given in Reference~\cite{phenixnim}. 

PHENIX consists of four
spectrometers: Central Arms (East and West) at midrapidity ($|\eta|<0.35$) and
the Muon Arms (North and South) at forward and backward rapidities. The 
detector layout in front and side view can be seen in 
Figure~\ref{fig:phenix_both}). The information from the PHENIX Beam-Beam 
Counters (BBC) and Zero-Degree Calorimeters (ZDC) is used for triggering 
and event selection. The BBCs are {\v C}erenkov-counters surrounding the 
beam pipe in the pseudorapidity interval $3.0<|\eta|<3.9$, and provide the 
start timing signal. The ZDCs are hadronic calorimeters 18~m downstream of the 
interaction region and detect spectator neutrons in a narrow forward cone. 

\begin{figure}[h!]
\bcc
\includegraphics[width=0.75\linewidth]{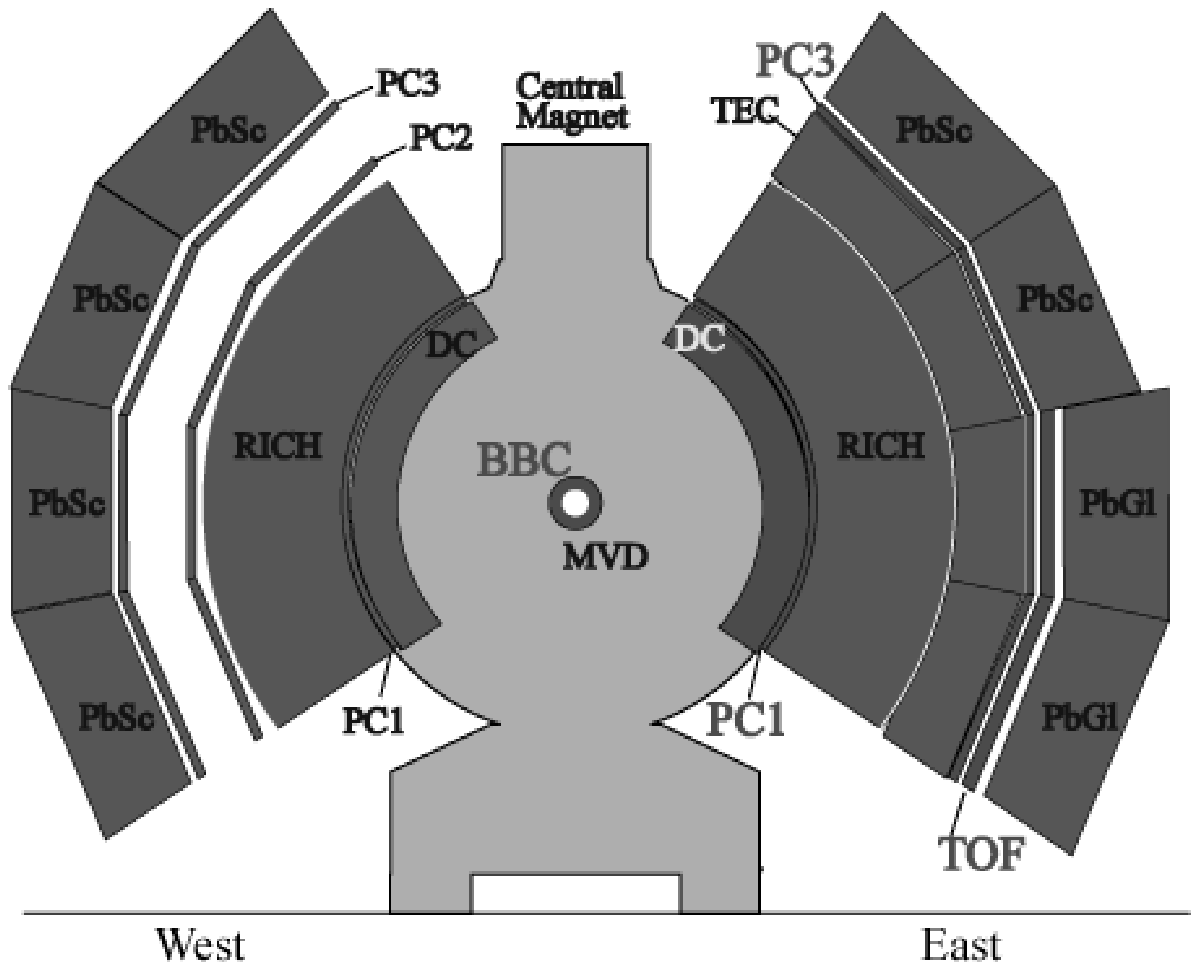}  
\includegraphics[width=0.75\linewidth]{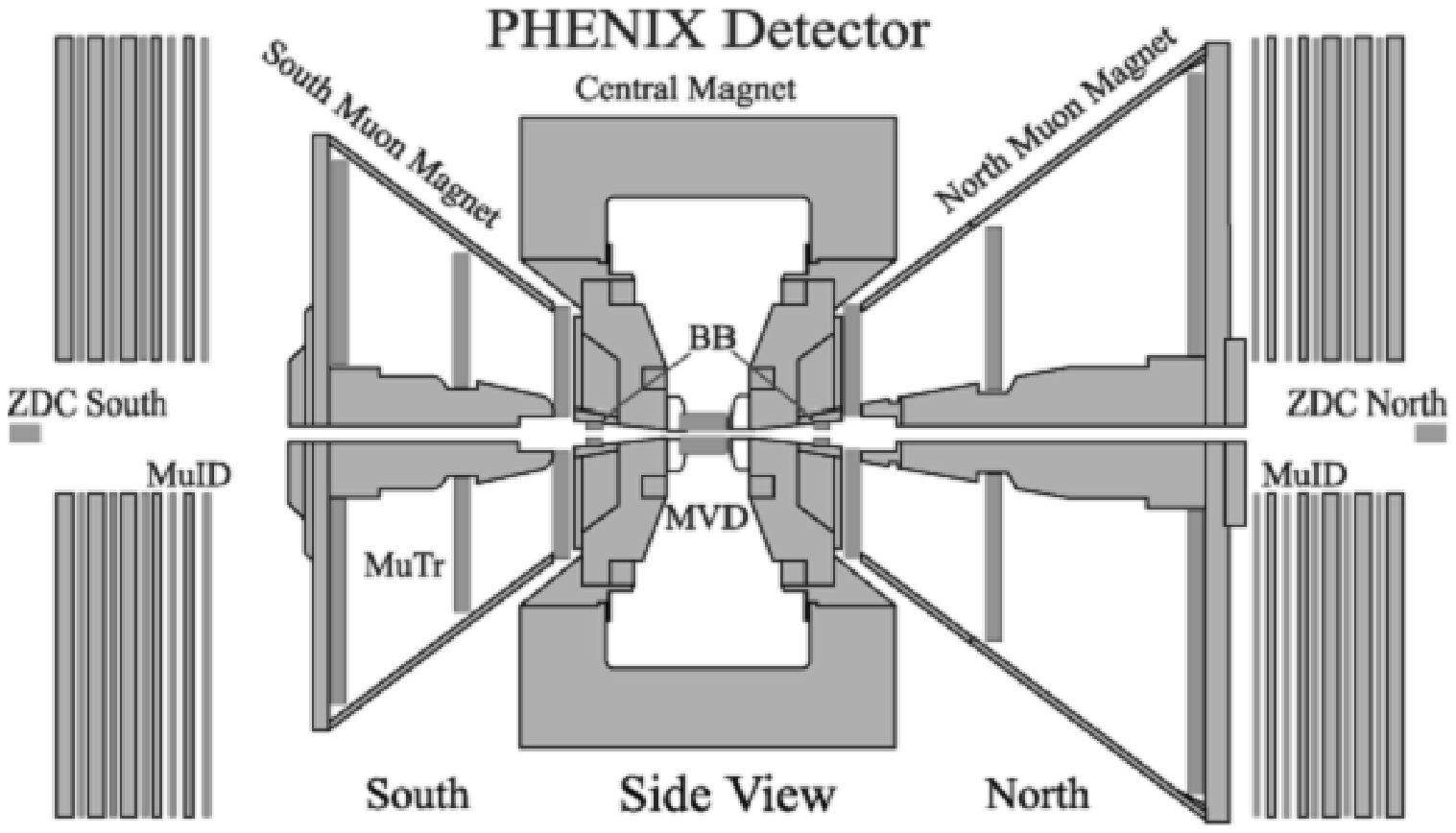}  
\ecc
\caption{Top panel shows Front view of PHENIX (beam going into the page) 
showing the Central Arm with the Time-Of-Flight (TOF) detector as in Run 2 
(2002). The bottom panel shows the PHENIX Muon Arms at forward and backward
rapidities as set up in Run 3 (2003).}
\label{fig:phenix_both}
\end{figure}

\clearpage
\section{Central Arms}
The Central Arm Detectors are placed radially around the beam axis extending
from 2.5 m to 5 m. See top panel in Figure~\ref{fig:phenix_both} for a 
schematic drawing the PHENIX Central Arms. They contain the following 
subsystems:
\benn
\item {\bf Central Magnets (CM)} create an axial field around the interaction vertex 
with a field integral $\approx$ 0.78 T.m. perpendicular to the beam axis, 
with a uniformity of 2 parts in $10^3$. This field bends the tracks into the
detector acceptance and helps the tracking detectors in momentum determination.
\item {\bf Charged tracking chambers:} there are two Drift Chambers (DC), three
Pad Chambers (PC) and one Time Expansion Chamber (TEC). The DC determines $p_T$
by measuring the charged particle trajectories in $r-\phi$ plane, as they
curve in the axial magnetic field produced by the Central Magnets. The PCs
aid measurement of longitudinal momentum and get 3D hits for pattern
recognition by providing spatial resolution ($\simeq$ few mm) along $r-\phi$ 
and $z$ directions. The TEC uses differential energy loss of a traversing 
particle to improve $e, \pi$ seperation and can help in track reconstruction
using drift times of ionization products in a gas mixture in a manner 
similar to DC.
\item {\bf Ring Imaging Cherenkov Detector (RICH)} (one in each arm) uses 
photomultiplier tubes (PMTs) to detect electrons via their characteristic 
Cherenkov emissions. Since electrons have a low mass, they emit Cherenkov 
light at lower momenta as compared to other ``contaminants'' like pions.
\item {\bf Time-Of-Flight system (TOF)} gives accurate measurement of the time of
flight of a particle from vertex, aiding in particle identification (PID).
\item {\bf Electromagnetic Calorimeters} measure energy deposition from e.m. showers
, using two methods: Lead Glass (PbGl) and Lead Scintillator (PbSc). They are
unique in being able to detect $\gamma$s (photons) and $\pi^0$s.
\eenn

In our first analysis ($d$, $\bar{d}$ measurement) we primarily used the Drift
Chamber and the Time-Of-Flight subsystems. These will be discussed in more
detail in subsequent sections. The Muon Arms --- used in the second analysis 
(hadron multiplicities in forward and backward rapidities) will be discussed
in Chapter 5.

\subsection{Drift Chambers} 

\begin{figure}[h!]
\bcc
\includegraphics[width=0.5\linewidth]{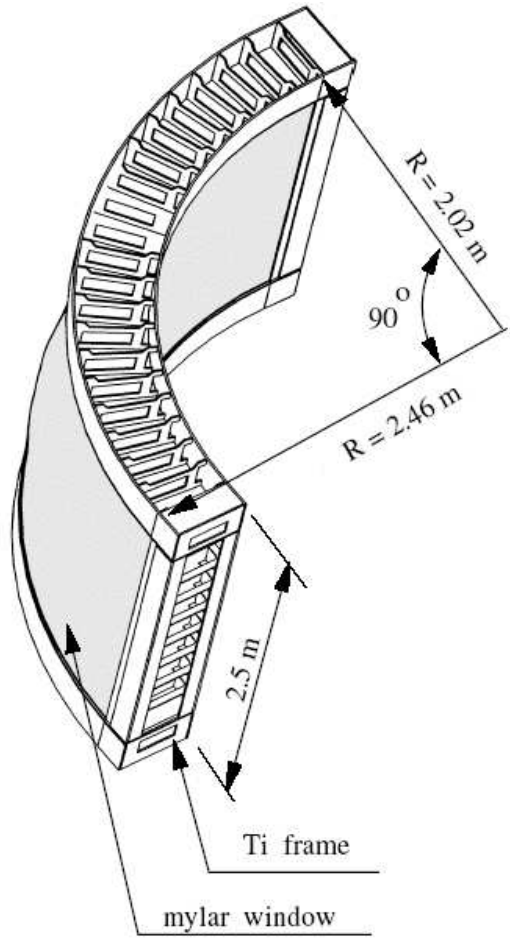}  
\ecc
\caption{Geometry of the DC, showing the titanium frame and 20 keystones.}
\label{fig:dc_geom}
\end{figure}

\begin{figure}[h!]
\bcc
\includegraphics[width=1.0\linewidth]{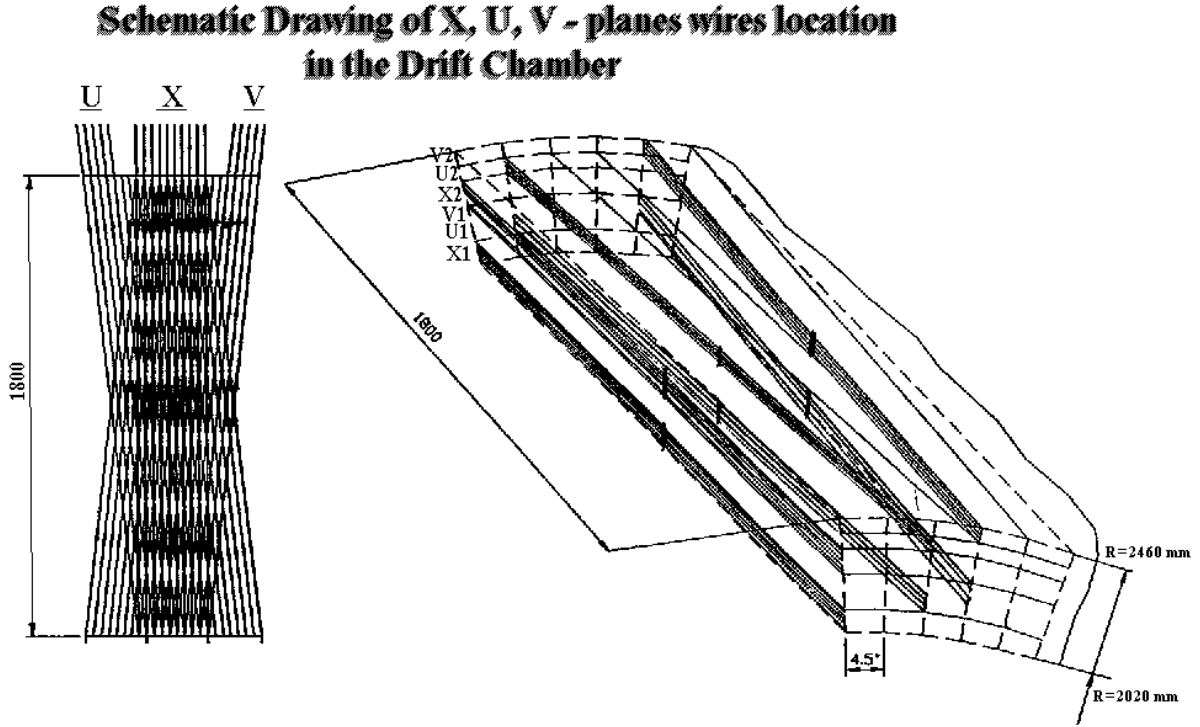}  
\ecc
\caption{Arrangement of the X, U and V wires in the DC.}
\label{fig:uv_section}
\end{figure}

The PHENIX Drift Chambers are wire chambers and reconstruct the trajectory of
a particle by using the time difference of the primary ionization (when the
particle first passes through the detector) and the time the charge signal
arrives on the sense wire.  They are cylindrically shaped and located 
radially 2 to 2.4 m from the $z$-axis, and 2 m transversally along the beam 
direction, and each have an azimuthal acceptane of 90$^o$. They consist of 
two independent gas volumes in East and West Arms, enclosed by 5 mil Al mylar 
windows in a cylindrical titanium frame. Each chamber is subdivided into 20 
equal sectors (called keystone) of 4.5$^o$ in $\phi$ each of which contains
6 wire modules stacked radially: X1, U1, V1, X2, U2 and V2. 
Figure~\ref{fig:dc_geom} shows the geometry of the DC frame. Each module has
4 sense (anode) planes and 4 cathode planes forming cells wth 2-2.5 cm drift
space in $\phi$ direction. The X wires run along $z$-axis and can only
reconstruct $x-y$ information, whereas the U and V layers are tilted 
$\approx \pm 6^o$ to the $z$-axis and are called stereo layers as they can
be used to obtain $z$ information. A schematic diagram showing the relative
arrangement of the U, V and X layers in DC is shown in 
Figure~\ref{fig:uv_section}. In addition to the cathode and the sense 
(anode) wires, the DC has potential and gate wires to shape the field and
remove front back ambiguity. In total the DC has 6500 anode wires and about
13, 000 readout channels. A 50/50 mixture of argon and ethane is used as the
working gas in the chambers. An angular deflection $\alpha$ in the magnetic 
field alongwith the measured hits in the DC are used to reconstruct the 
momentum of the particle. The DC was designed with a single wire resolution 
of 150 $\mu$m and single wire efficiency greater than 99\% for good tracking 
efficiency for high multiplicities at RHIC.
\clearpage
\subsection{Time-Of-Flight}

\begin{figure}[h!]
\bcc
\includegraphics[width=0.5\linewidth]{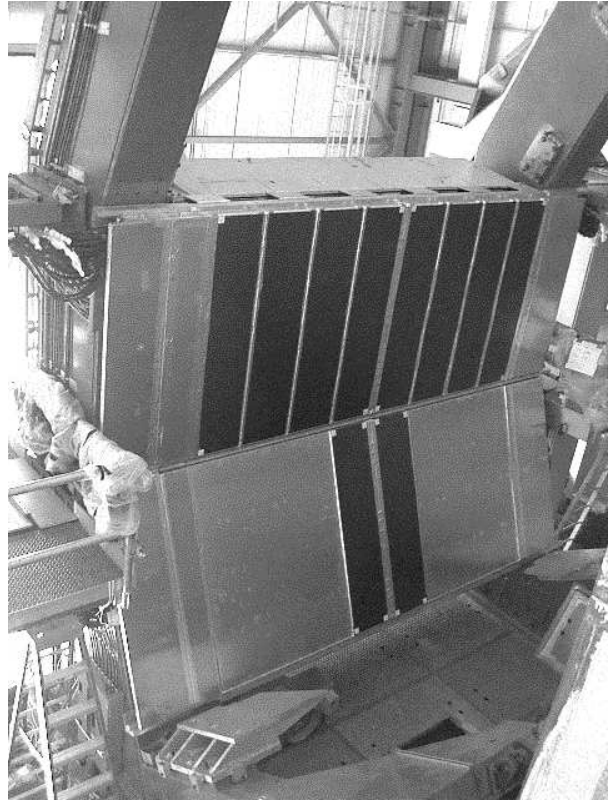}  
\ecc
\caption{The TOF with 8 panels in the upper sector and 2 panels in
the lower sector, as mounted in the East Arm.}
\label{fig:tof_pic}
\end{figure}

\begin{figure}[h!]
\bcc
\includegraphics[width=0.5\linewidth]{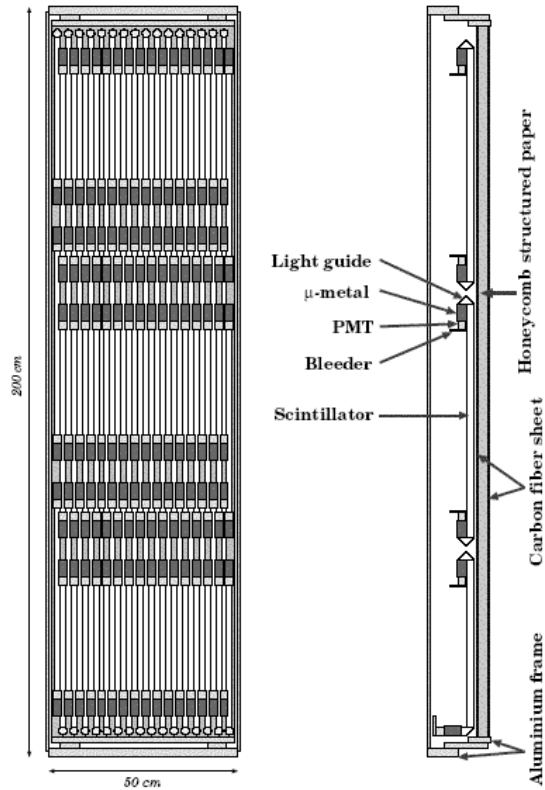}  
\ecc
\caption{Diagram of a single TOF panel showing the plastic scintillator with
PMTs at each end with corresponding light guides and supports.}
\label{fig:tof_panel}
\end{figure}

The Time-Of-Flight subsystem, used as a particle identification device for
hadrons, is present only in the East Arm and covers $\pi/4$ in azimuth.  
A picture of the TOF detector mounted in the East Arm is shown in 
Figure~\ref{fig:tof_pic}. It is located 5.0 m away from the vertex and
consists of 1000 elements of plastic scintillation counters with 
photomultiplier tube (PMT) readouts. It consists of two sectors: top sector has
8 panels and the lower sector has 2 panels. Each panel has 960 scintillator
counters along $r-\phi$ direction with 1920 PMTs readouts collectively called 
slats. A schematic diagram showing a TOF panel with PMTs is shown in 
Figure~\ref{fig:tof_panel}. Each slat provides time and longitudinal position 
information of the
particles that hit the slat. The timing information, alongwith the momentum
from the DC enables us to determine the mass (see Eq.~\ref{eq:m2}), thus 
giving us a powerful method of particle identification. More details on this
are given in the next chapter.

\section{Centrality determination}
The event centrality is measured in PHENIX using the Zero Degree Calorimeters 
(ZDC) and the Beam Beam Counters (BBC) in conjunction. ZDCs are small 
hadronic calorimeters positioned 18 m upstream and downstream of the 
interaction point and detect the energy deposited by the spectator neutrons 
during the collisions. ZDCs are so postioned so that the spectator neutrons, 
which do not get bent by the magnets, hit them directly allowing them to be 
used as event triggers in each RHIC experiment. The BBCs are positioned 
radially around the $z$-axis at 1.44 m from the interaction point and measure
charged particle multiplicities in $3.0<\eta<3.9$. The correlation between the
BBC charge sum and the ZDC energy deposit enables us to determine event 
centrality, because the more peripheral the collision (large impact parameter)
greater is number of spectator neutrons in the ZDC, whereas the more central
the collision (small impact parameter) the greater is charged particle 
multiplicity in the BBC and correspondingly fewer spectator neutrons make it into the ZDC. Figure~\ref{fig:zdc_bbc} shows a scatter plot of the BBC vs ZDC
response.

\begin{figure}[h!]
\bcc
\includegraphics[width=0.75\linewidth]{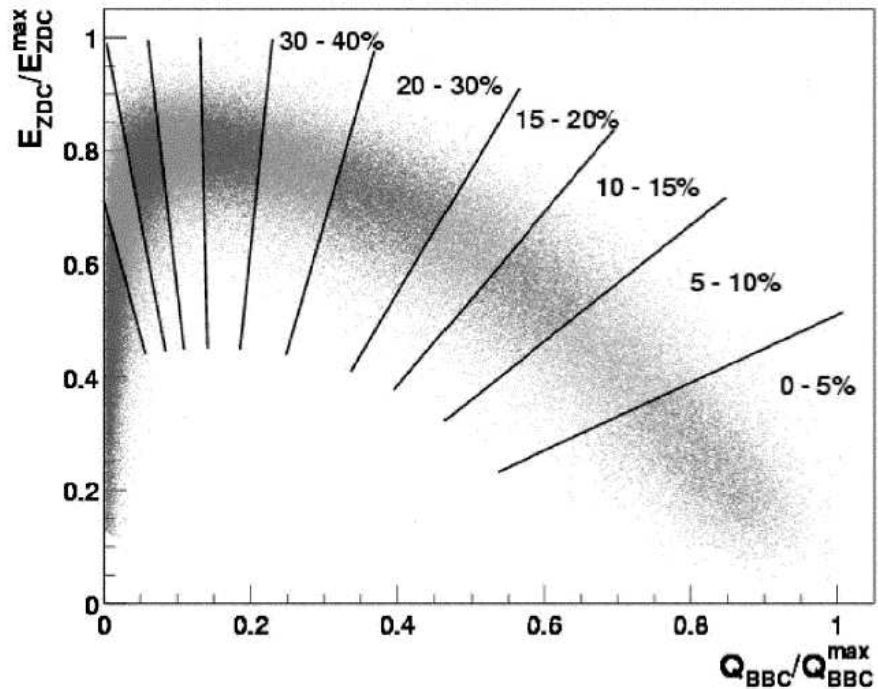}  
\ecc
\caption{BBC vs ZDC scatter plot shows how centrality is determined in 
PHENIX.}
\label{fig:zdc_bbc}
\end{figure}

\section{Event reconstruction and data stream}
At RHIC, the Au beams cross at a frequency of 9.4 MHz, which is the master 
clock frequency for all PHENIX subsystems. The ZDC and BBCs are used for
triggering and setting the start time for an event. Whenever the vertex or the
trigger subsystems are triggered (Level 1 triggers) data is sent from each 
of the detector subsystems by optical fiber in a raw digitized format. These
Level 1 triggers allow data taking even at high collision rates by selecting
interesting events e.g. an event in which a high $p_T$ electron was detected
in the RICH. After passing the quality checks from the Online Monitoring 
Systems, the data is stored on tape. This raw data is now ready for
calibration and reconstruction. After it is reconstructed, it saved in an 
easily digestible format: the nanoDSTs(nDSTs). nDSTs come in various flavors 
depending upon the analysis and the detector subsystem being used. For example,
data from the PHENIX Central Arms is saved in Central Track Nanodsts (CNTs), 
which we use for the first analysis. Further details on the PHENIX data 
acquisition and reconstruction can be found in Reference~\cite{phenixnim}.

\chapter{Data reduction for $d, \bar{d}$ measurement}

 In this Chapter, we discuss how deuterons and
anti-deuterons are identified and seperated based on mass squared 
distributions. We obtain raw momentum distributions which are then corrected 
for detector acceptances, efficiency, occupancy effects. There is also a 
section on sources of systematical uncertainties.

\section{Event and track selection cuts}

In the Year-2 of running (2001-2002), events are selected in PHENIX using the 
Beam Beam Counters (BBCs), which measure the event vertex position along $z$ 
and also set the start time for other detectors including the Drift Chamber 
(DC). In conjunction with the Zero Degree Calorimeters (ZDCs), as described 
in the previous Chapter, BBCs are also used to determine the event centrality.
For this analysis, we used Minimum Bias (MB) events, which are essentially 
minimally triggered events in which the BBC and the ZDC fired and are as 
unbiased as can be made. Event vertex was restricted to $|z|<35$ cm of the 
collision vertex, primarily for reasons to do with detector acceptances.

In addition we selected runs based on quality (QA), by excluding runs in 
which there were known subsystem problems e.g. too many dead channels in the 
DC, or timing problems in the Time Of Flight hodoscopes (TOFs). Bad runs were 
also flagged by looking at quantities like the average particle multiplicities
and average momentum. In total we analysed 21.6 million minimum bias events, 
after excluding events rejected by our global cuts. The minimum bias cross 
section corresponds to $92.2^{+2.5}_{-3}$\% of the total inelastic Au+Au 
cross section (6.8 b)~\cite{ppgmb}. In addition, tracks were selected to
optimize signal and reduce background, by using the following cuts:
\bitm
\item DC track quality=31 or 63. These quality bit numbers give detail 
regarding matching the hits from the Drift Chamber UV stereo layers (in 
addition to the X1, X2 layers) to the Pad Chamber 1 (PC1) hits. Quality 63
is the best and indicates a unique PC1 and a unique UV hit, whereas quality
31 means an ambiguous PC1 hit and a best choice UV hit. 
\item $|\sigma_{\phi}(TOF, PC3)|< 2.5$ and $|\sigma_z(TOF, PC3)|< 2.5$
detector matching cuts for TOF and PC3. We look
at the residual between the track projection from the vertex and actual hit,
both along azimuthal $\phi$ direction and the $z$ direction. After normalizing
these to account for change with momentum, we have a set of residuals in 
$\sigma$. 
\item Momentum $p>0.5$ GeV. Low momentum particles aren't well reconstructed 
due to acceptance of charged particles as they bend too much, as well as 
energy loss effects.
\item TOF $E_{loss} > 0.0014\beta^{-5/3}$ GeV. The energy deposited in the
TOF scintillator: $E_{TOF}$ is plotted in a scatter plot vs $\beta$ in 
Fig.~\ref{fig:elosscut}. Using a form motivated by the Bethe-Bloch 
formula~\cite{bethe-bloch}, we can exclude tracks that did not deposit the 
minimum ionizing particle energy, thus allowing us to reject background.
\eitm

\begin{figure}[h]
\bcc
\includegraphics[width=1.0\linewidth]{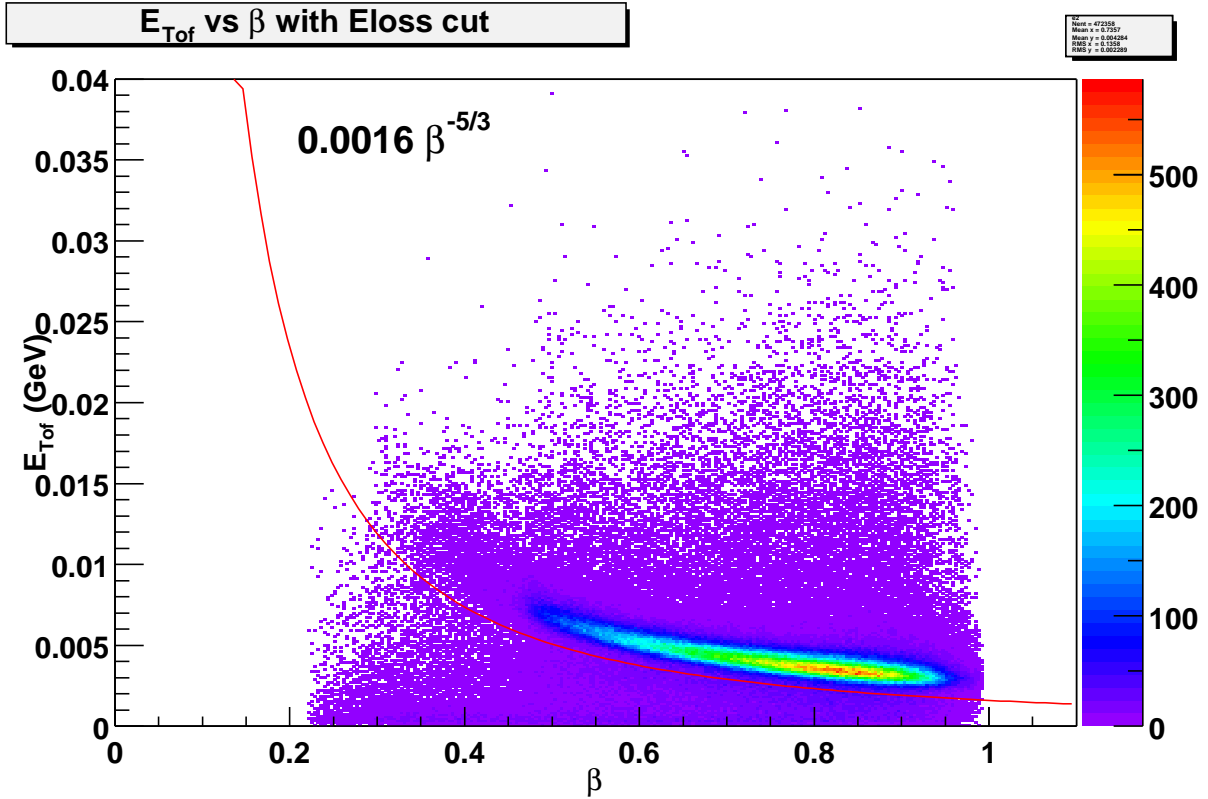}  
\ecc
\caption{$\beta$ dependent energy loss cut for TOF to reduce background.}
\label{fig:elosscut}
\end{figure}

In addition to above cuts we excluded the low gain sector E0 of TOF (this 
can lead to a $z$ dependence of the mass widths) by requiring TOF Slat $<767$.

\section{Detector resolutions}

Particle identification was done by measuring the momentum $p$ using the Drift
Chamber and the time-of-flight $t$ using the TOF counter. Using the standard
relativistic relationship between mass and momentum:
\be
p = m\frac{\beta}{\sqrt{1-\beta^2}}
\ee
we can obtain an expression for $m^2$ using momentum $p$ (from DC), the 
time-of-flight $t$ (from TOF), the pathlength $d$ traversed by the particle 
from the collision vertex to the TOF detector:
\be
m^2 = p^2(\frac{t^2c^2}{d^2} - 1)
\label{eq:m2}
\ee

\begin{figure}[h]
\bcc
\includegraphics[width=1.0\linewidth]{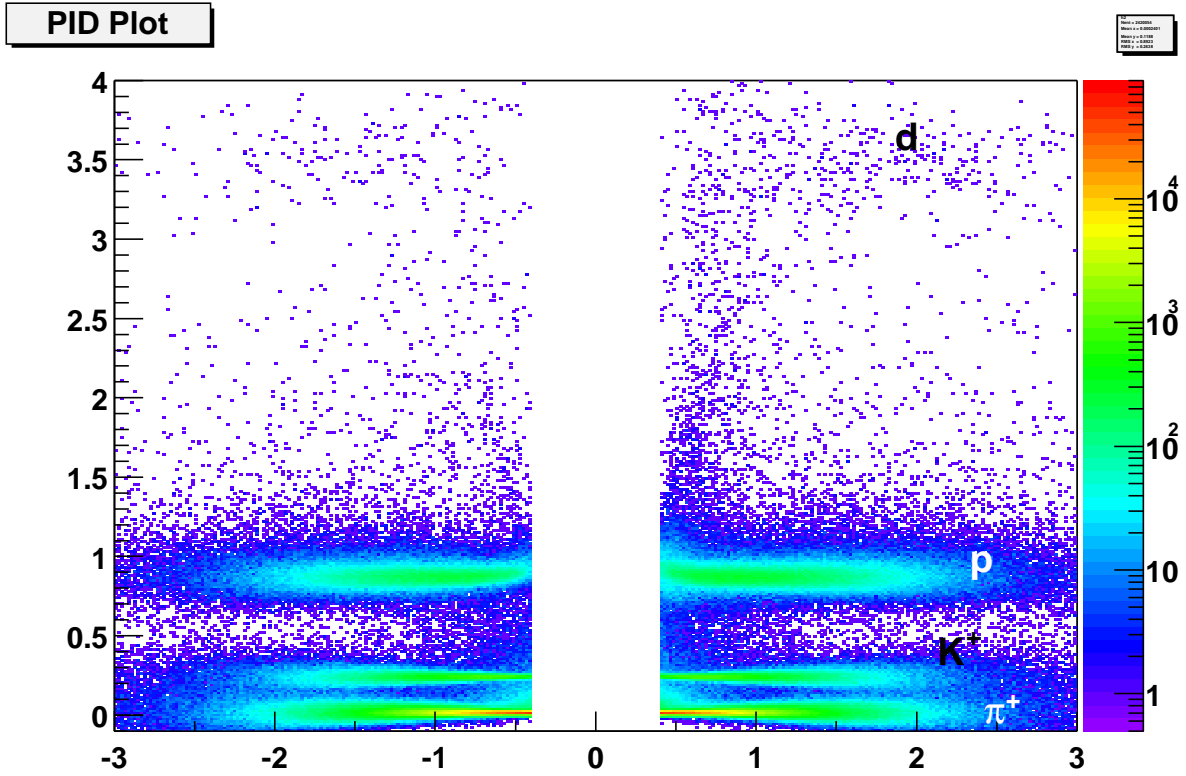}  
\ecc
\caption{PID scatter plot with signed momentum (negatives on left and 
positives on right) along $x$-axis and $m^2$ (calculated using 
Eq.~\ref{eq:m2}) along $y$-axis.}
\label{fig:pidplot}
\end{figure}
A typical PID scatter plot with $p$ along $x$-axis and $m^2$ along the $y$-axis
is shown in Fig.~\ref{fig:pidplot}. Particles are on right side (along positive
 $p$) and anti-particles are on the left. We can clearly see bands 
of different particle species like pions, kaons and protons. A faint deuteron
band can also be seen.

Our ability to discriminate between different particle species depends on the 
resolutions of detectors being
used. For example, better DC resolution enables us to go to higher values of
tranverse momentum $p_T$ by better signal to noise ratio. Similarly improved
TOF resolution will give a better seperation between particle species. 
One way to estimate the detector resolutions for DC and TOF is by measuring 
the width of the $m^2$ bands of particles. Since the DC determines momentum
by measuring the angle by which a track bends in the magnetic field, it's
momentum resolution depends upon it's intrinsic angular resolution 
$\sigma_\alpha$. In addition other factors like the multiple scattering of a 
charged particle as it travels up to the drift chamber due to the intervening 
matter (MVD cladding, air, DC mylar window, He-bag mylar window etc.) also play
a role at different ranges. The momentum (in GeV) determined by DC is related 
to the angle of bending $\alpha$ (in mrad) by:
\be
p = \frac{87}{\alpha}
\label{eq:p87}
\ee
Where, 87 mrad GeV is simply the field integral: 
\be
K_1 = \int_{0.3/R_{DC}}lBdl
\label{eq:bdl}
\ee
 This gives us the momentum resolution as:
\be
\left(\frac{\delta p}{p}\right)^2 = \left(\frac{\sigma_\alpha}{K_1}p\right)^2 
+  \left(\frac{\sigma_{ms}}{K_1\beta}\right)^2
\label{eq:momres}
\ee
where $\sigma_\alpha$ is the angular resolution of the DC, $\sigma_{ms}$ is 
the multiple scattering term and $\sigma{tof}$ is the timing resolution of the
TOF.
Using above equation (Eq.~\ref{eq:momres} we can derive a relation between
the width of the $m^2$ bands and the DC angular resolution ($\sigma_\alpha$), 
multiple scattering ($\sigma_{ms}$) and TOF resolution ($\sigma_{tof}$):
\be
\sigma_{m^2}^2 = \frac{\sigma_\alpha^2}{K_1^2}\left(4m^4p^2\right)+ 
\frac{\sigma_{ms}^2}{K_1^2}\left[4m^4\left(1+\frac{m^2}{p^2}\right)\right]+
\frac{\sigma_{tof}^2c^2}{L^2}\left[4p^2\left(m^2+p^2\right)\right]
\label{eq:m2width}
\ee

Hence by looking at the width of the $m^2$ distributions of various particle
species we can estimate various resolution terms. In order to determine the 
contributions due to multiple scattering and 
angular resolution, we calculated the $m^2$ using CNTs from certain selected
runs 29531, 29999, 30015 and 30069 from Run 2 Au+Au (about 2.5 million events).
After making some standard cuts as before we used the measured $p$ and $t$
to make histograms of the mass squared distribution. A $m^2$ histogram in a
given momentum range is shown in Fig.~\ref{fig:m2plot}. We can see sharp
pion and kaon peaks around the expected Particle Data Group~\cite{pdg} 
values, followed by a broader
proton peak. And finally there is a little deuteron peak on the extreme right.
As a first order estimate for particle identification, we made simple straight
line cuts by assuming that all particles in the $m^2$ range [-0.15, 0.15] are 
pions, all particles in [0.15, 0.35] range are kaons and all particles in the
[0.5, 1.25] range are protons. We fitted gaussians and 
double gaussians to the resulting $m^2$ histograms to each particle species. 
Some of these fits, for the momentum range $1.0 < p< 1.2$ GeV are shown for 
positive particles in Fig.~\ref{fig:pion_sigma1_3} for pions,
in Fig.~\ref{fig:kaon_sigma1_3} for kaons and Fig.~\ref{fig:proton_sigma1_3} 
for protons. Systematic uncertainties in fits were obtained 
from the difference between the gaussian and the digaussian fits. Thus, we 
obtained the mean and the sigma of the $m^2$ bands for different momentum bins.

\begin{figure}[h]
\begin{center}
\epsfig{file = 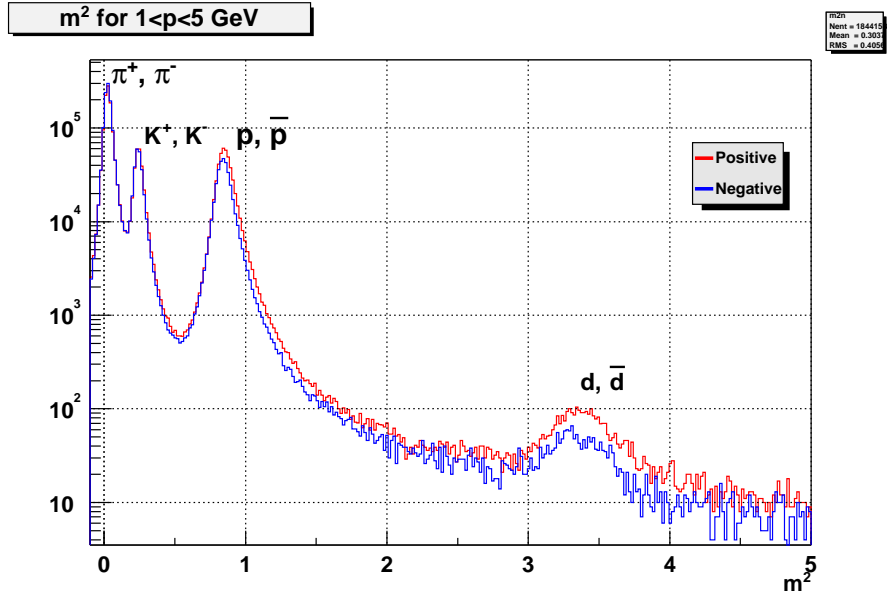, width = 12cm, clip = }
\end{center}
\caption{$m^2$ histogram showing peaks for different (anti)particle species.}
\label{fig:m2plot}
\end{figure}

\begin{figure}[h]
\centerline{\epsfig{file = 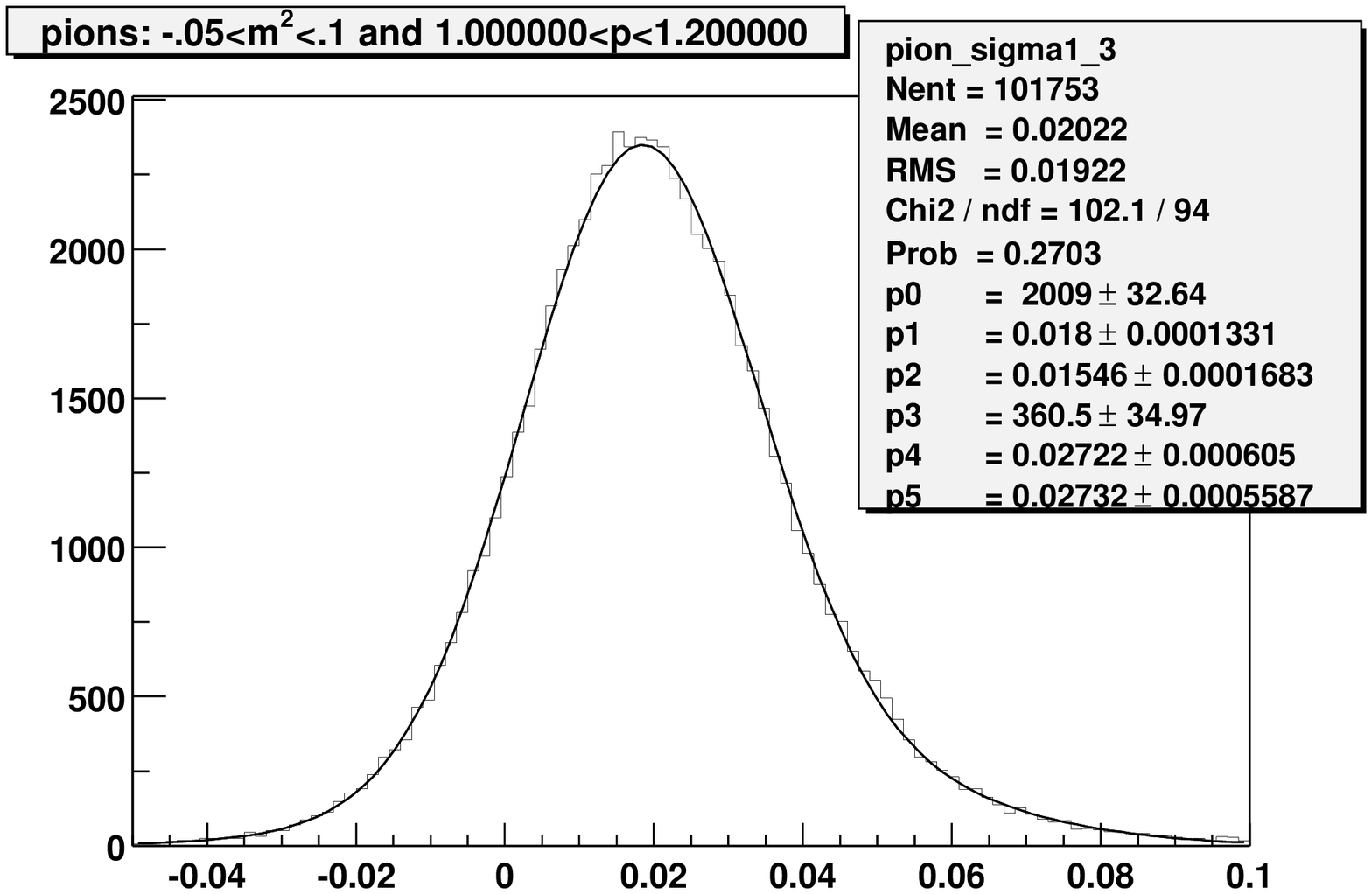, width = 12cm, clip = }}
\caption{Digaussian fits to $m^2$ histograms for pions in the momentum range 
$1.0 < p< 1.2$ GeV}
\label{fig:pion_sigma1_3}
\end{figure}

\begin{figure}[h]
\centerline{\epsfig{file = 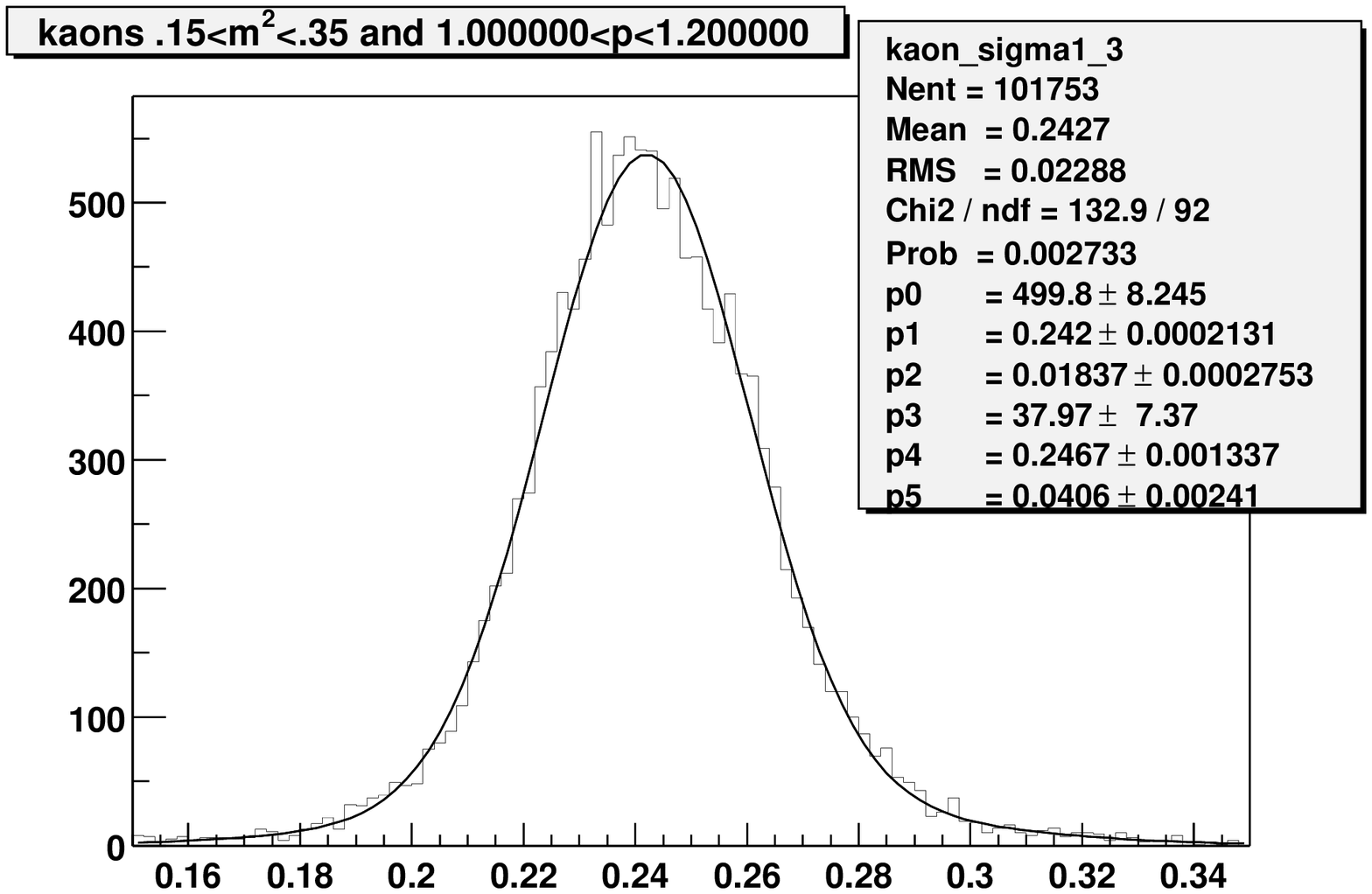, width = 12cm, clip = }}
\caption{Digaussian fits to $m^2$ histograms for kaons in the momentum range
 $1.0 < p< 1.2$ GeV}
\label{fig:kaon_sigma1_3}
\end{figure}

\begin{figure}[h]
\centerline{\epsfig{file = 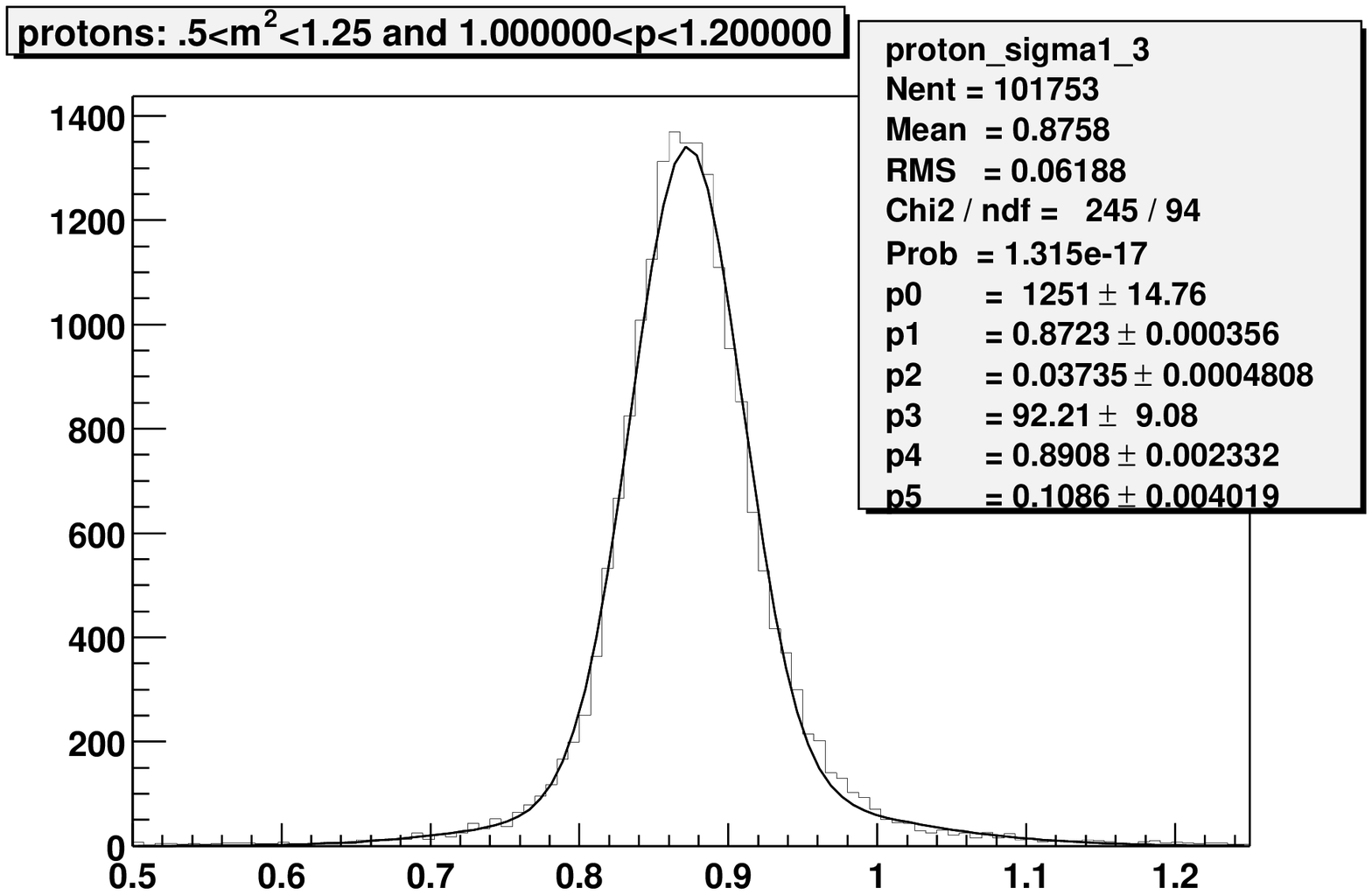, width = 12cm, clip = }}
\caption{Digaussian fits to $m^2$ histograms for protons in the momentum 
range $1.0 < p< 1.2$ GeV}
\label{fig:proton_sigma1_3}
\end{figure}

\begin{figure}[h]
\centerline{\epsfig{file = 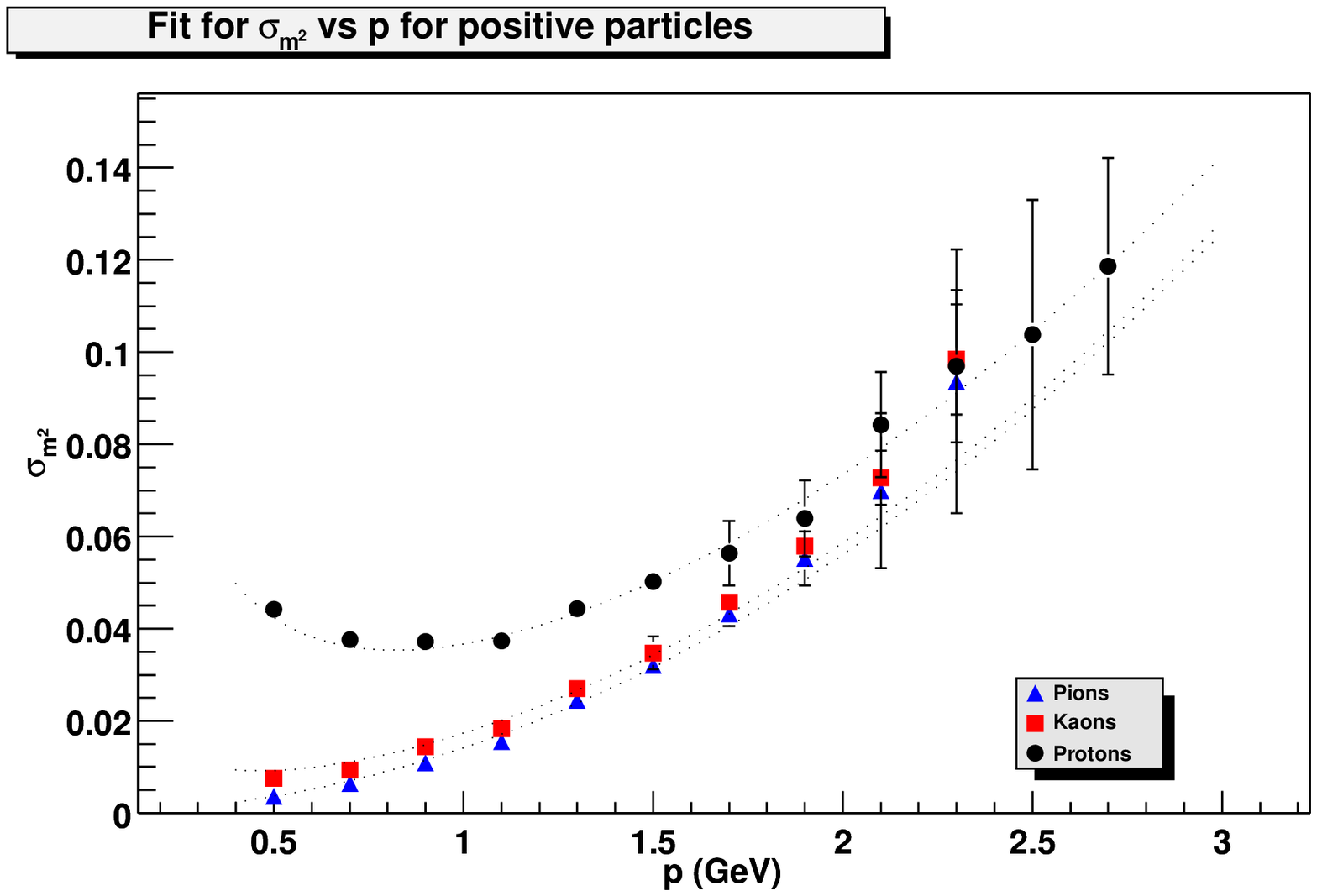, width = 12cm, clip = }}
\caption{Fit for parameters for positive particles 
(see Table~\ref{tab:regular}).}
\label{fig:posfitplot}
\end{figure}

We made a simultaneous three parameter fit to pions, kaons and protons with 
$\sigma_\alpha$, $\sigma_{ms}$ and  $\sigma_{tof}$ as the parameters, 
first for positive particles and then for negative particles. The fit for 
$m^2$ bands is shown in Fig~\ref{fig:posfitplot} (for positive particles).
From these fits we obtained $\sigma_\alpha$ = 0.83 mrad, $ \sigma_{ms}$ = 
0.92 mrad/GeV, and $\sigma_{tof}$ = 120 ps. Values of the fit parameters for 
Au+Au data (Run 2) are tabulated in Table~\ref{tab:regular}.
\begin{table}[h]
\caption{Momentum resolution parameters for Au+Au data (Run 2).}
\begin{center}
\begin{tabular}[]{|l|l|l|l|}
\hline
 Type& $\sigma_\alpha$ (mrad) &$\sigma_{ms}$ (mrad GeV) & $\sigma_{tof}$ (ps)\\
\hline
 Positives & $0.83\pm 0.09$ & $0.92 \pm 0.03$  & $120 \pm 2$ \\
\hline
 Negatives & $0.89\pm 0.07$ & $0.75 \pm 0.02$  & $119 \pm 2$ \\
\hline
\end{tabular}
\label{tab:regular}
\end{center}
\end{table}

The contribution to the $m^2$ width of protons from each of these terms is 
shown in Fig.~\ref{fig:contribpr}. Multiple scattering $\sigma_{ms}$ is
important at low momenta and for more massive particles like protons and 
deuterons as it varies primarily as a function of speed $\beta$. This term 
alongwith acceptance effects is the main reason why our deuteron and 
anti-deuteron spectra are limited at the lower end of momentum around 
1.1 GeV/c. At high values of momenta, our $m^2$ width is limited by the timing
resolution $\sigma_{tof}$. Obviously the better our resolution, narrower are
the $m^2$ bands and better is our particle identification (PID). 

\begin{figure}[h!]
\centerline{\epsfig{file = 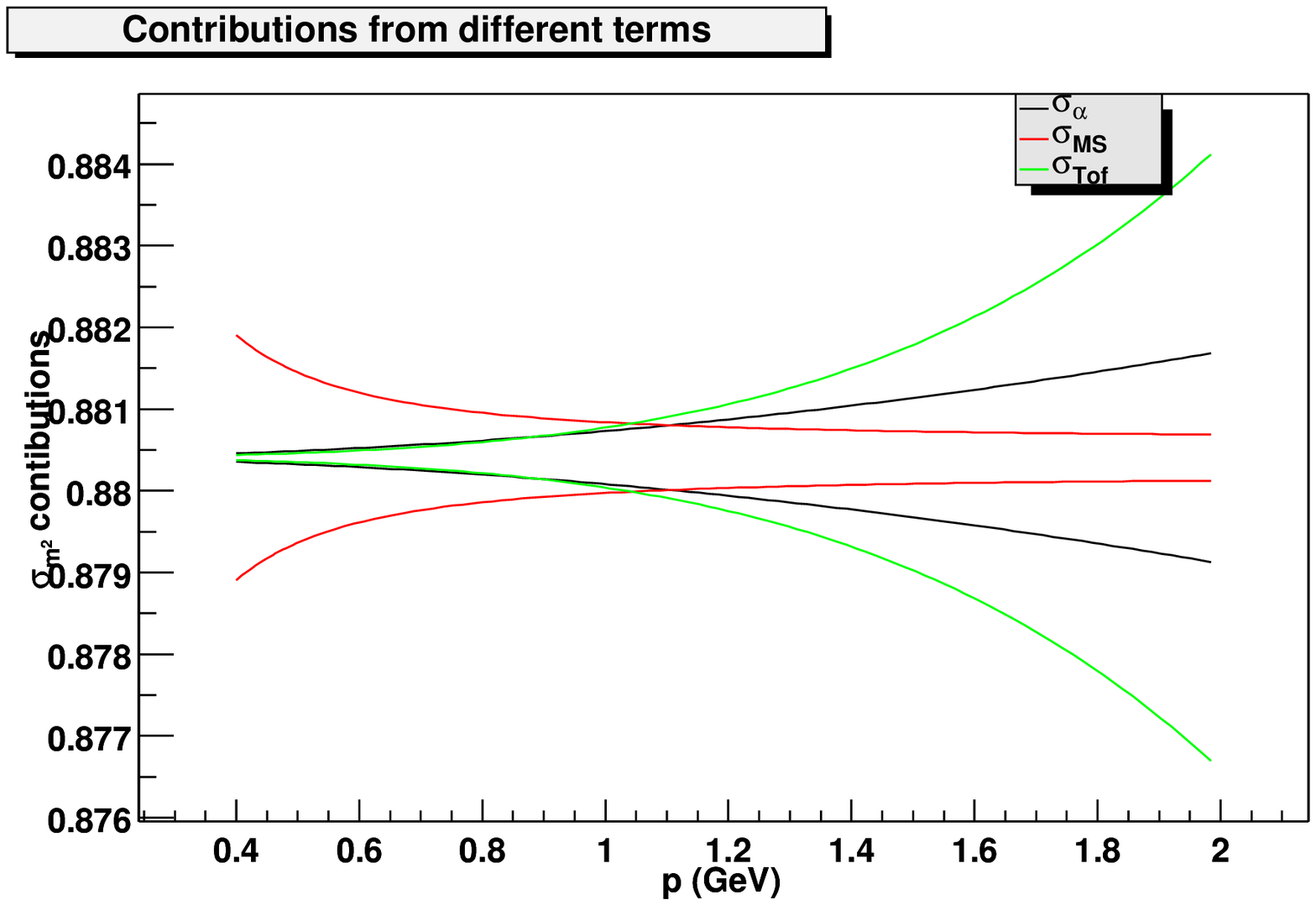, width = 12cm, clip = }}
\caption{Contribution to the PID cuts for protons from different terms}
\label{fig:contribpr}
\end{figure}

We also put in new calibrations for drift velocity $v_d$, by looking at the
variation of proton mass width as a function of change in drift
velocity $\Delta v_d$, (see Fig.~\ref{fig:vd}) and taking the
value of $v_d$ which leads to the narrowest mass width for protons.
If we assume an angular resolution of 0.83 mrad, then at low momenta  
momentum resolution is limited by multiple scattering to 
$\delta p/p \approx 0.7$\% and at high momenta it is limited by the angular 
resolution of the DC to $\delta p/p^2 \approx 1.0 \pm 0.1$ \%. This is often
represented as: $\delta p/p \approx 0.7\% \oplus 1\% p$ GeV/$c$.
A plot of the momentum resolution as a function of $p_T$ is shown in 
Fig~\ref{fig:momres}. We also
checked that the momentum resolution does not vary as a function of centrality.

\begin{figure}[h]
\centerline{\epsfig{file = 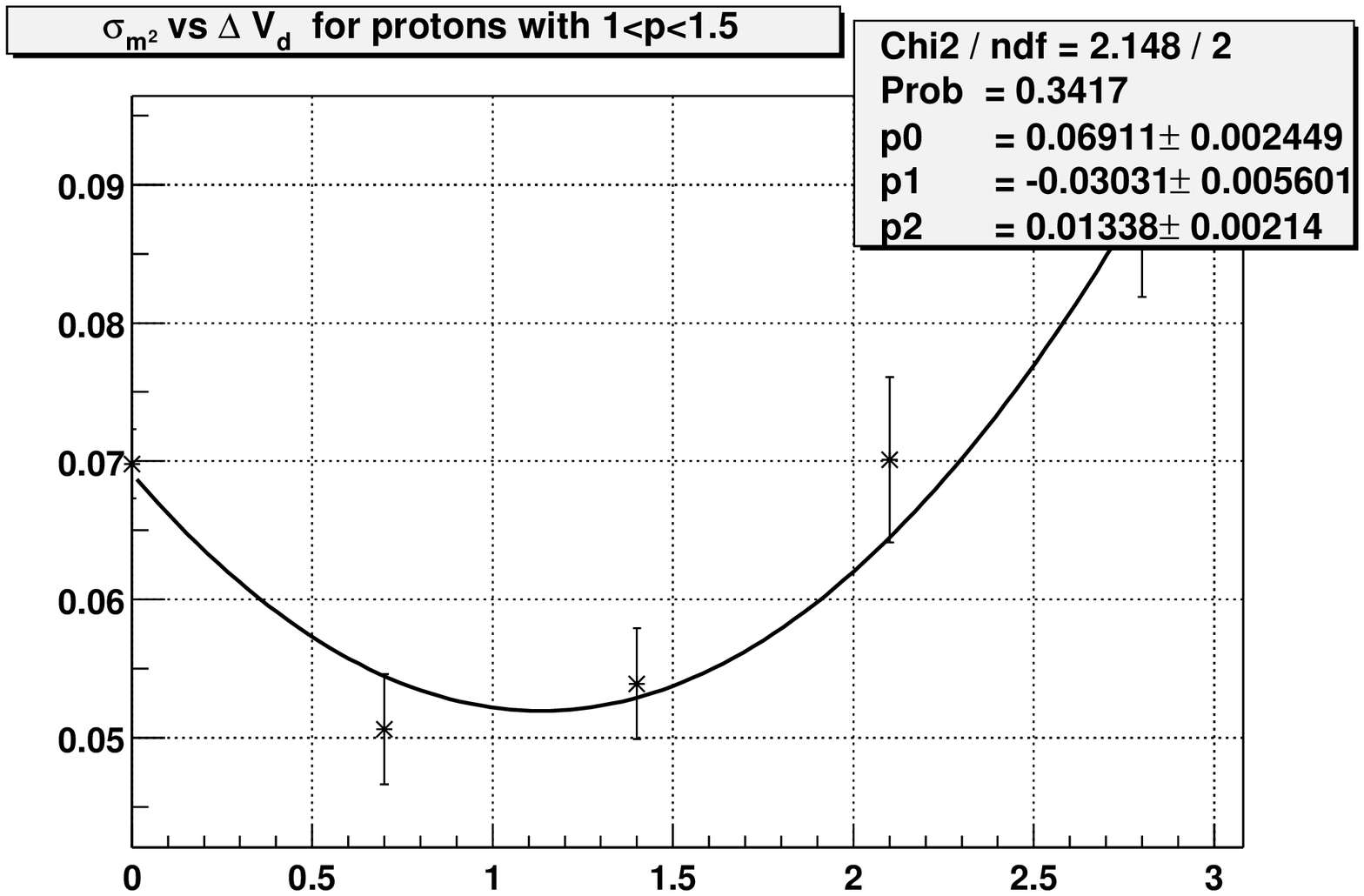, width = 12cm, clip = }}
\caption{Width of proton $m^2$ vs change in drift velocity $v_d$.}
\label{fig:vd}
\end{figure}

A comparision of the $m^2$ values of particles with those from the Particle 
Data Book~\cite{pdg} indicated a 4\% discrepancy. And these values changed 
with momentum. This is commonly seen if there is an offset in the momentum
scale as well as offset in TOF timing. A correction of about 2\% $\pm$ 0.7\%
for the momentum scale was applied to remove these offsets.

\begin{figure}[h]
\centerline{\epsfig{file = 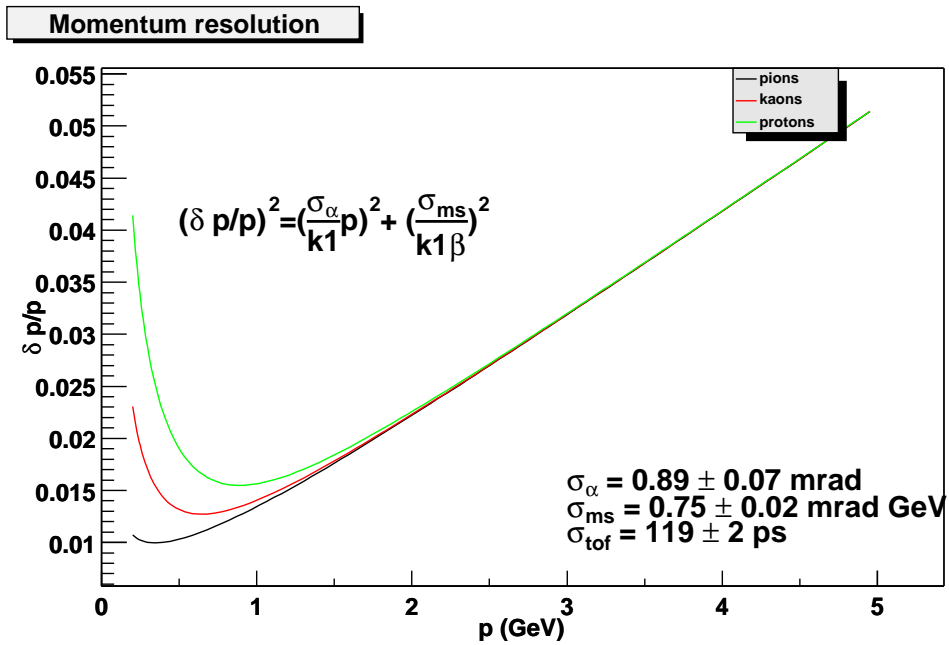, width = 14cm, clip = }}
\caption{Momentum resolution}
\label{fig:momres}
\end{figure}

\clearpage

\section{Signal Extraction}

Finally after all these cross checks on resolutions we are now ready to start
extract (anti)deuteron signal from our $m^2$ histograms. As mentioned 
before, in Fig.~\ref{fig:m2plot}, we can see sharp pion and kaon peaks around
followed by a broader proton peak. And finally there is a little deuteron 
peak on the extreme right. Our simplistic PID technique of using just straight
line cuts by taking all particles in a given $m^2$ range as a particular 
particle species (e.g., pions in $m^2$ range of [-0.15, 0.15]) is not good 
enough. As we can see there is background under the $d$, $\bar{d}$ peak from
various sources like mismatched momenta and tail from proton peaks. A simple
straight line cut will include too much background, so we fit a gaussian with 
a background function (either $e^{-x}$ or $1/x$) and extract the number of 
deuterons under the gaussian.

We made several $m^2$ histograms around the expected mass squared value of 
deuterons in different momentum bins  of 400 MeV/c widths (except last bin, which is 800 MeV/c wide to increase statistics) from 1.1 to 4.3 GeV/c. For the minimum bias data, the fits for deuterons are shown in Fig.~\ref{fig:ppg0201_0_92_final} and for anti-deuterons in Fig.~\ref{fig:ppg0202_0_92_final}. This was repeated for two other centrality classes --- 0-20\% and  20-92\% --- in Figs.
~\ref{fig:ppg0201_0_20_final}, ~\ref{fig:ppg0201_20_92_final},
~\ref{fig:ppg0202_0_20_final} and ~\ref{fig:ppg0202_20_92_final}.

Using the gaussian fits, we can calculate the raw count of deuterons by:
\be
N_d = \sqrt{2\pi}A_d\sigma_d
\label{eq:deutraw}
\ee
where $A_d$ is the amplitude and $\sigma_d$ is the sigma of the gaussian. We
used $N_d$ as a fit parameter instead of amplitude, by inverting the above
relation to obtain $A_d$ in terms of $N_d$. In this case the area under 
the gaussian is written as:
\be
\frac{N_d}{\sqrt{2\pi}\sigma}\exp\left({-\frac{x^2}{2\sigma^2}}\right)
\label{eq:deutgaussian}
\ee
The width of the gaussian was restricted within a narrow range using our 
knowledge of the momentum resolution parameters as determined 
before~\ref{eq:m2width}.  A plot showing the variation of mean and the sigma
of the peak obtained by this method is shown in Fig.~\ref{fig:dsigmapt}.

\begin{figure}[h]
\bcc
\includegraphics[width=1.0\linewidth]{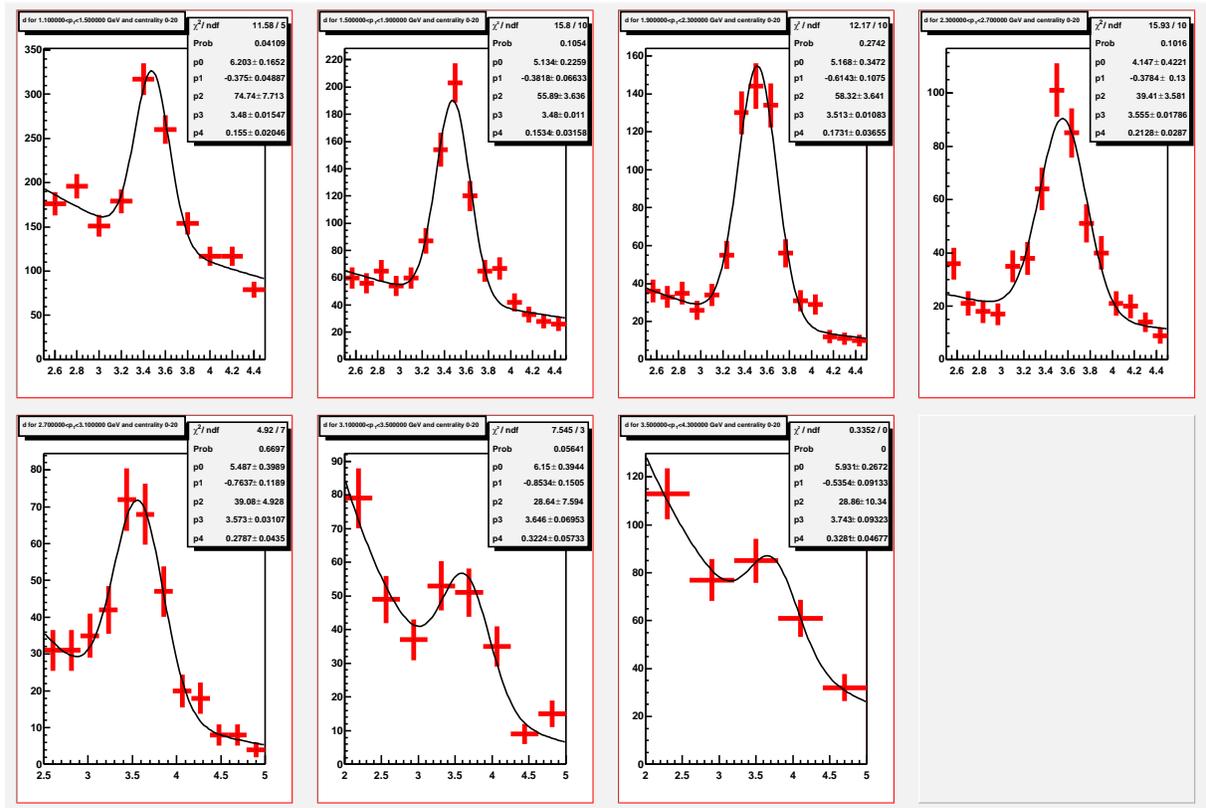}  
\ecc
\caption{Gaussian fit with $e^{-x}$ background for d in 0-20\% centrality for
$p_T$ ranges 1.1-1.5, 1.5-1.9, 1.9-2.3, 2.3-2.7, 2.7-3.1, 3.1-3.5, 3.5-4.3 
GeV/c (starting from top left).}
\label{fig:ppg0201_0_20_final}
\end{figure}

\begin{figure}[h]
\begin{center}
\includegraphics[width=1.0\linewidth]{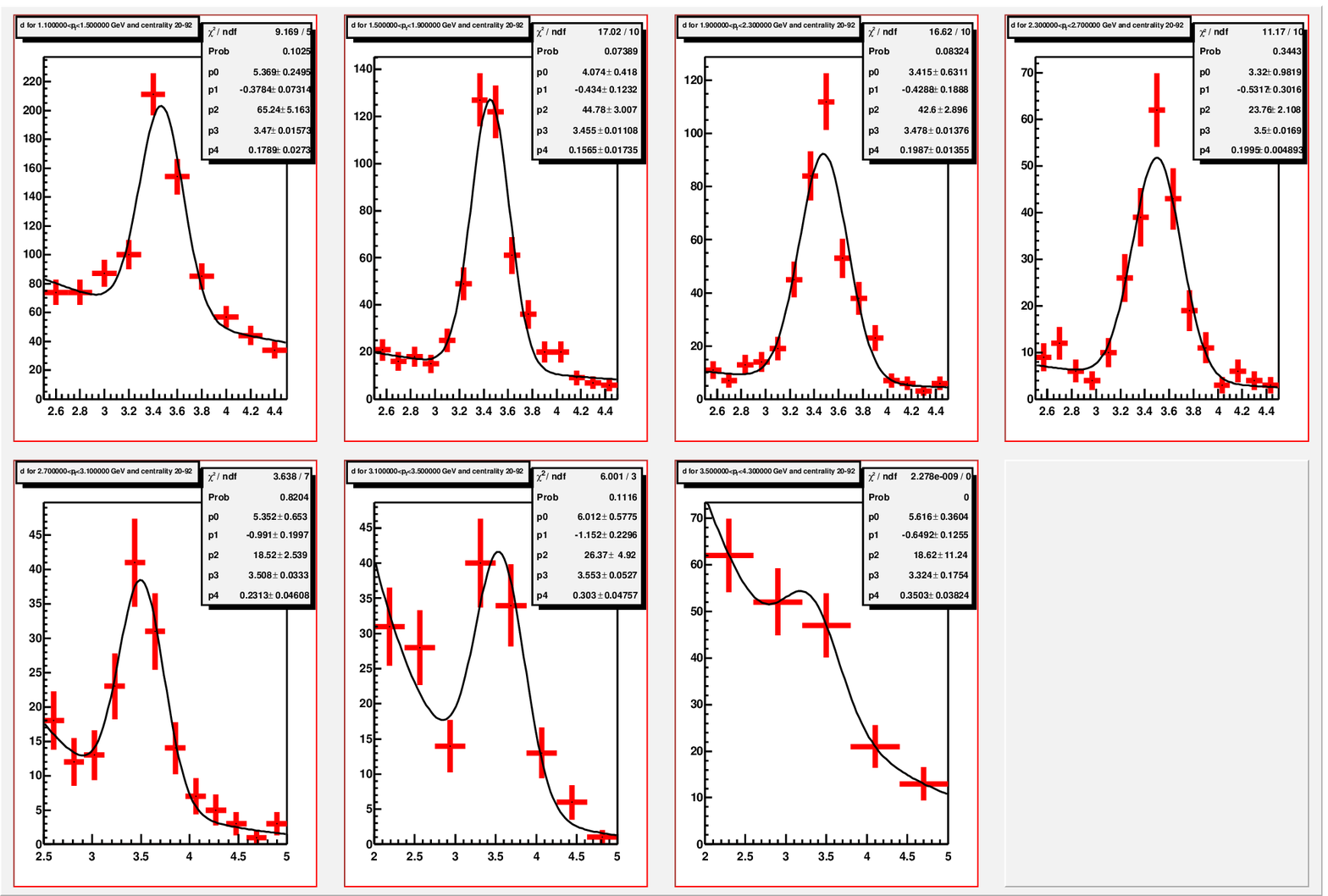}  
\end{center}
\caption{Gaussian fit with $e^{-x}$ background for d in 20-92\% centrality for
$p_T$ ranges 1.1-1.5, 1.5-1.9, 1.9-2.3, 2.3-2.7, 2.7-3.1, 3.1-3.5, 3.5-4.3 
GeV/c (starting from top left).}
\label{fig:ppg0201_20_92_final}
\end{figure}

\begin{figure}[h]
\begin{center}
\includegraphics[width=1.0\linewidth]{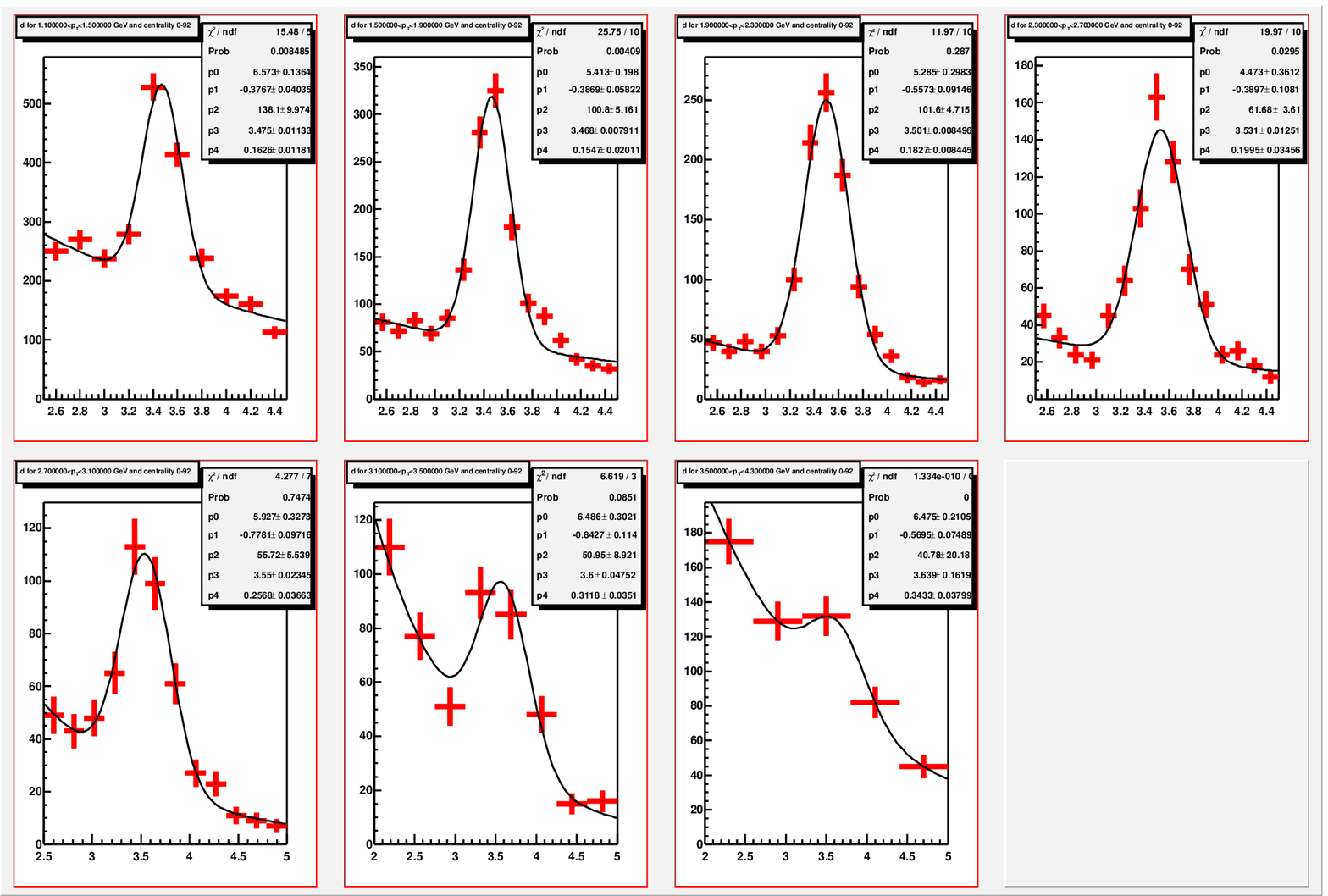}  
\end{center}
\caption{Gaussian fit with $e^{-x}$ background for d in 0-92\% centrality for
$p_T$ ranges 1.1-1.5, 1.5-1.9, 1.9-2.3, 2.3-2.7, 2.7-3.1, 3.1-3.5, 3.5-4.3 
GeV/c (starting from top left).}
\label{fig:ppg0201_0_92_final}
\end{figure}

\begin{figure}
\begin{center}
\includegraphics[width=1.0\linewidth]{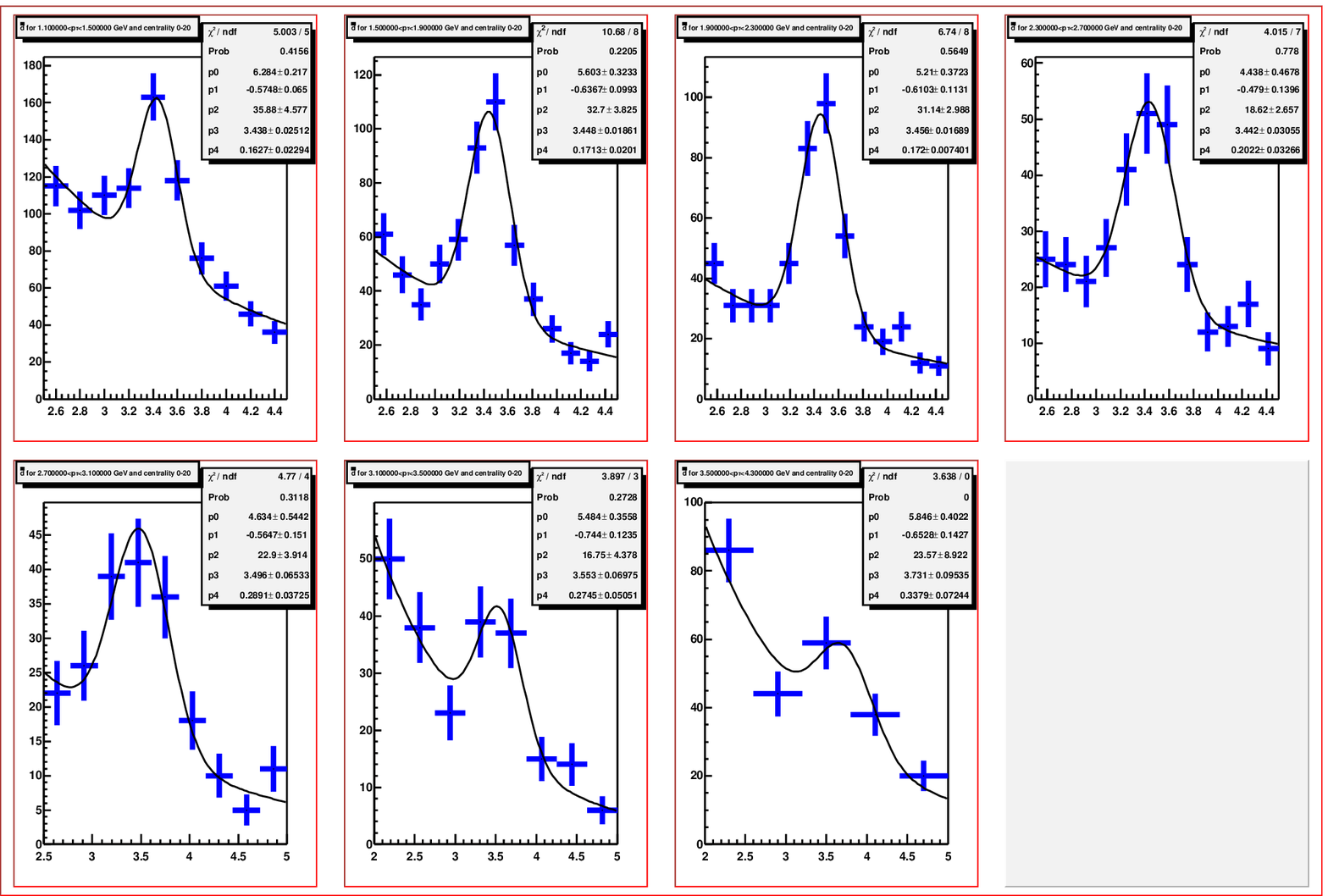}  
\end{center}
\caption{Gaussian fit with $e^{-x}$ background for dbar in 0-20\% centrality 
for $p_T$ ranges 1.1-1.5, 1.5-1.9, 1.9-2.3, 2.3-2.7, 2.7-3.1, 3.1-3.5, 3.5-4.3
GeV/c (starting from top left).}
\label{fig:ppg0202_0_20_final}
\end{figure}

\begin{figure}
\begin{center}
\includegraphics[width=1.0\linewidth]{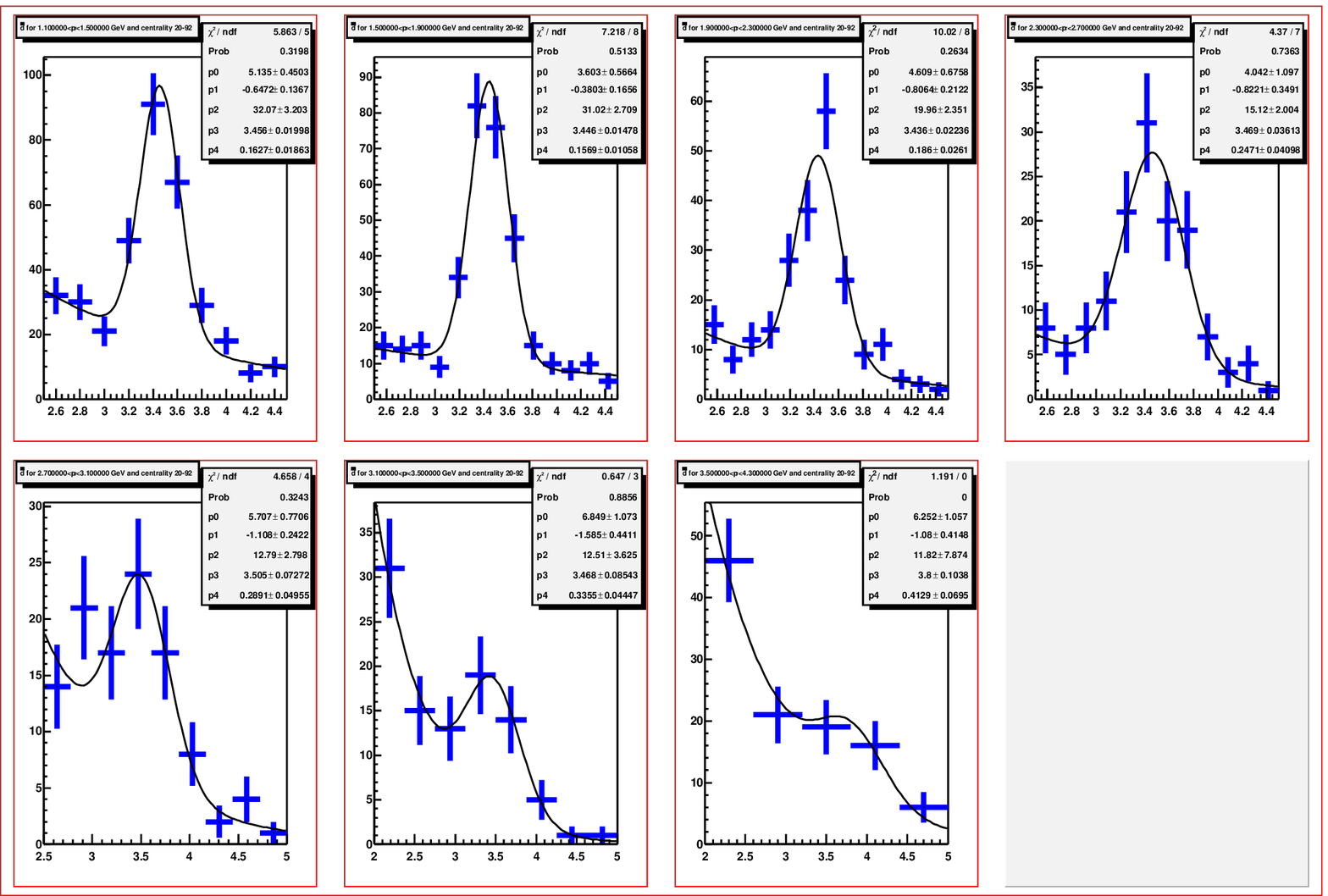}  
\end{center}
\caption{Gaussian fit with $e^{-x}$ background for dbar in 20-92\% centrality 
for $p_T$ ranges 1.1-1.5, 1.5-1.9, 1.9-2.3, 2.3-2.7, 2.7-3.1, 3.1-3.5, 
3.5-4.3 GeV/c (starting from top left).}
\label{fig:ppg0202_20_92_final}
\end{figure}

\begin{figure}
\begin{center}
\includegraphics[width=1.0\linewidth]{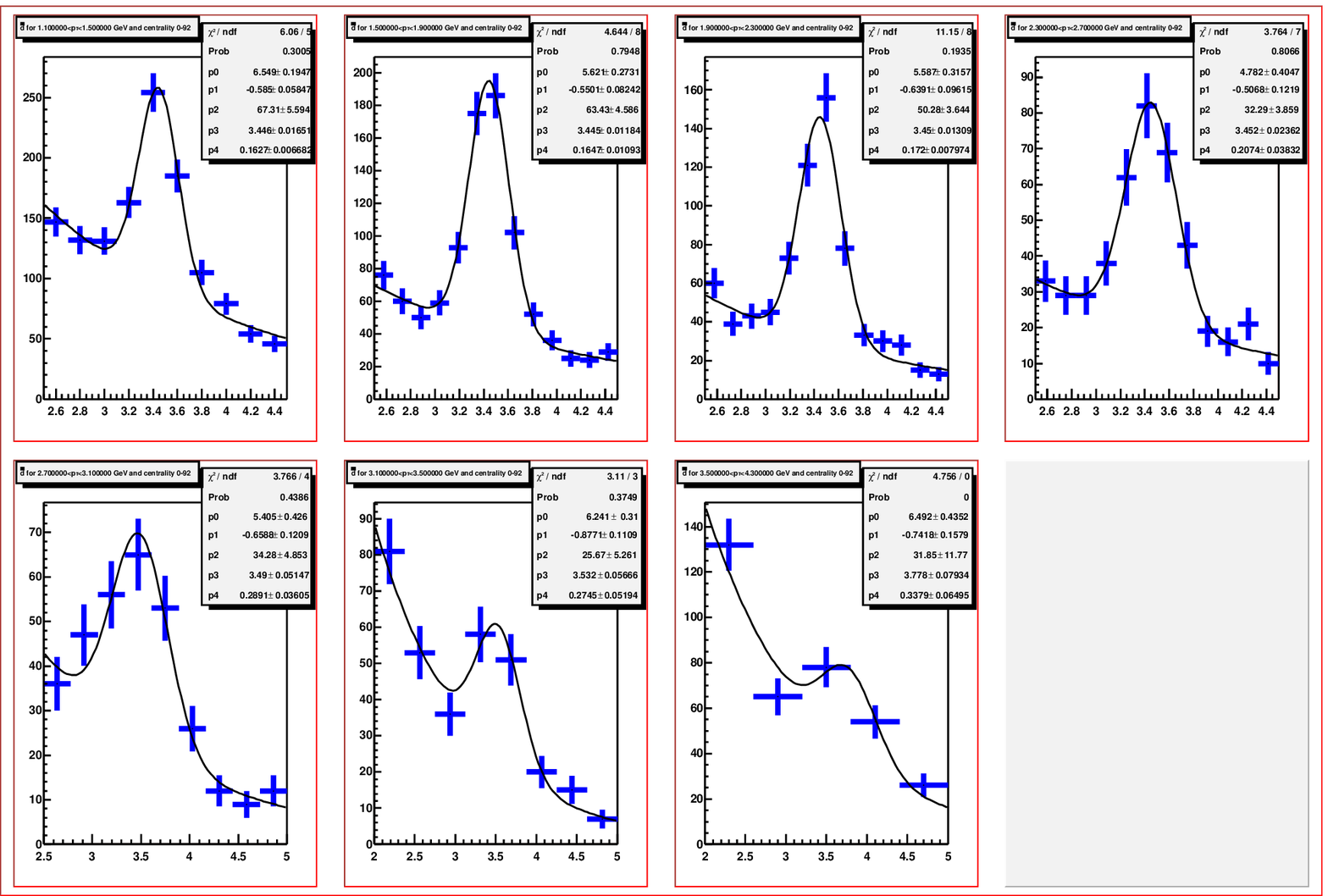}  
\end{center}
\caption{Gaussian fit with $e^{-x}$ background for dbar in 0-92\% centrality 
for $p_T$ ranges 1.1-1.5, 1.5-1.9, 1.9-2.3, 2.3-2.7, 2.7-3.1, 3.1-3.5, 3.5-4.3
GeV/c (starting from top left).}
\label{fig:ppg0202_0_92_final}
\end{figure}
\clearpage

The raw spectra deuteron (anti-deuteron) spectra are shown
in Fig ~\ref{fig:dyield_raw_final} and the raw counts (for 21.6 M events) as 
extracted from the fitting procedure are listed in Table~\ref{tab:rawyields}.
We notice that the raw spectra fall off at low $p_T$. This is primarily 
because of the heavy mass of deuterons, which leads to a smaller acceptance. 
A plot showing the PHENIX acceptance for different particle species as a 
function of momentum and rapidity is shown in Fig.~\ref{fig:phe_acc}.
\begin{figure}[h]
\begin{center}
\epsfig{file = 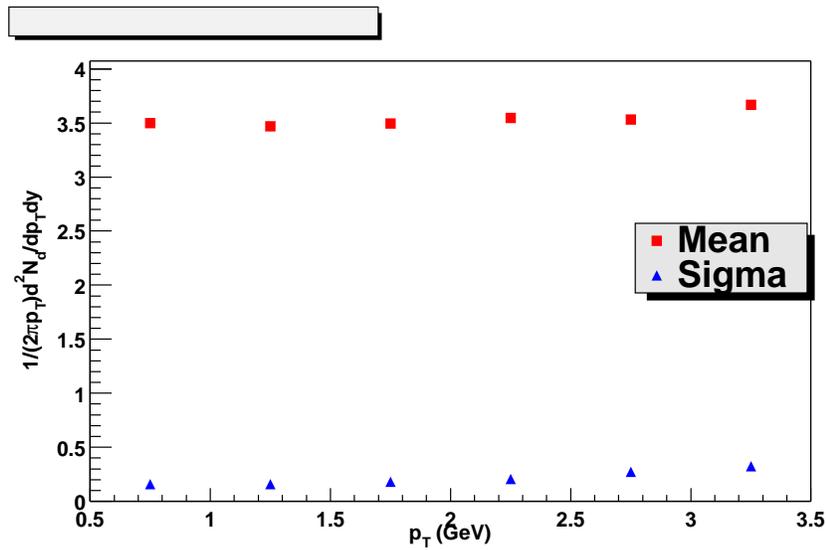, width = 12cm, clip = }
\end{center}
\caption{Variation of the mean and sigma of $m^2$ centroid for anti-deuterons}
\label{fig:dsigmapt}
\end{figure}

\begin{figure}[h]
\begin{center}
\epsfig{file = 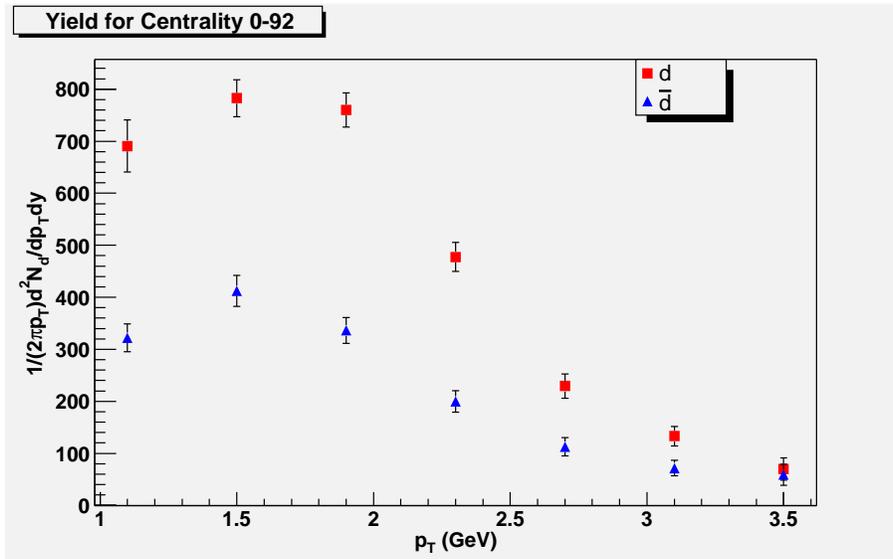, width = 12cm, clip = }
\end{center}
\caption{Raw spectra for deuterons and anti-deuterons (min. bias)}
\label{fig:dyield_raw_final}
\end{figure}

\begin{table}[h]
\caption{Raw Yields vs $p_T$ (mid-point) for different centralities} 
\bcc
\begin{tabular}[]{|l|l|l|l|}
\hline
Centrality &$p_T$ GeV& Raw counts for $d$ &  Raw counts for $\bar{d}$ \\
\hline
	&1.3	&	690.712	$\pm$	49.868	&	322.114	$\pm$	26.9392	\\
	&1.7	&	782.783	$\pm$	35.6892	&	412.272	$\pm$	29.811	\\
	&2.1	&	760.001	$\pm$	32.999	&	336.306	$\pm$	24.556	\\
Min. Bias&2.5	&	477.309	$\pm$	27.9426	&	199.812	$\pm$	20.3159	\\
	&2.9	&	229.359	$\pm$	23.4378	&	112.669	$\pm$	17.5777	\\
	&3.3	&	133.208	$\pm$	18.8056	&	71.6043	$\pm$	14.913	\\
	&3.9	&	69.7416	$\pm$	21.3508	&	58.9473	$\pm$	20.2146	\\
\hline
	&1.3	&	378.291	$\pm$	31.4306	& 171.522 $\pm$ 21.9933		\\
	&1.7	&	443.66	$\pm$	28.3072	&	210.57 $\pm$ 20.5607		\\
	&2.1	&	449.276	$\pm$	26.1366	&	207.874$\pm$  20.1791		\\
0-20\%	&2.5	&	308.455	$\pm$	23.2434	&	116.245 $\pm$ 16.7561		\\
	&2.9	&	145.113	$\pm$	19.6931	&	75.9248 $\pm$ 14.242		\\
	&3.3	&	72.1734	$\pm$	14.8395	&	46.3732 $\pm$ 12.3767		\\
	&3.9	&	50.032	$\pm$	18.0059	&	40.9745 $\pm$ 15.8305		\\
\hline
	&1.3	&	314.905	$\pm$	24.9601	&	153.976	$\pm$	15.5005	\\
	&1.7	&	338.779	$\pm$	21.5489	&	203.035	$\pm$	16.7537	\\
	&2.1	&	313.051	$\pm$	20.1975	&	129.831	$\pm$	13.8321	\\
20-92\%	&2.5	&	177.778	$\pm$	16.5595	&	76.1233	$\pm$	11.838	\\
	&2.9	&	87.5999	$\pm$	13.9707	&	39.7836	$\pm$	10.2824	\\
	&3.3	&	70.3182	$\pm$	13.1204	&	31.136	$\pm$	8.79026	\\
	&3.9	&	31.368	$\pm$	13.1671	&	17.3637	$\pm$	11.5297	\\
\hline
\end{tabular}
\ecc
\label{tab:rawyields}

\end{table}

\begin{figure}[h]
\begin{center}
\epsfig{file = 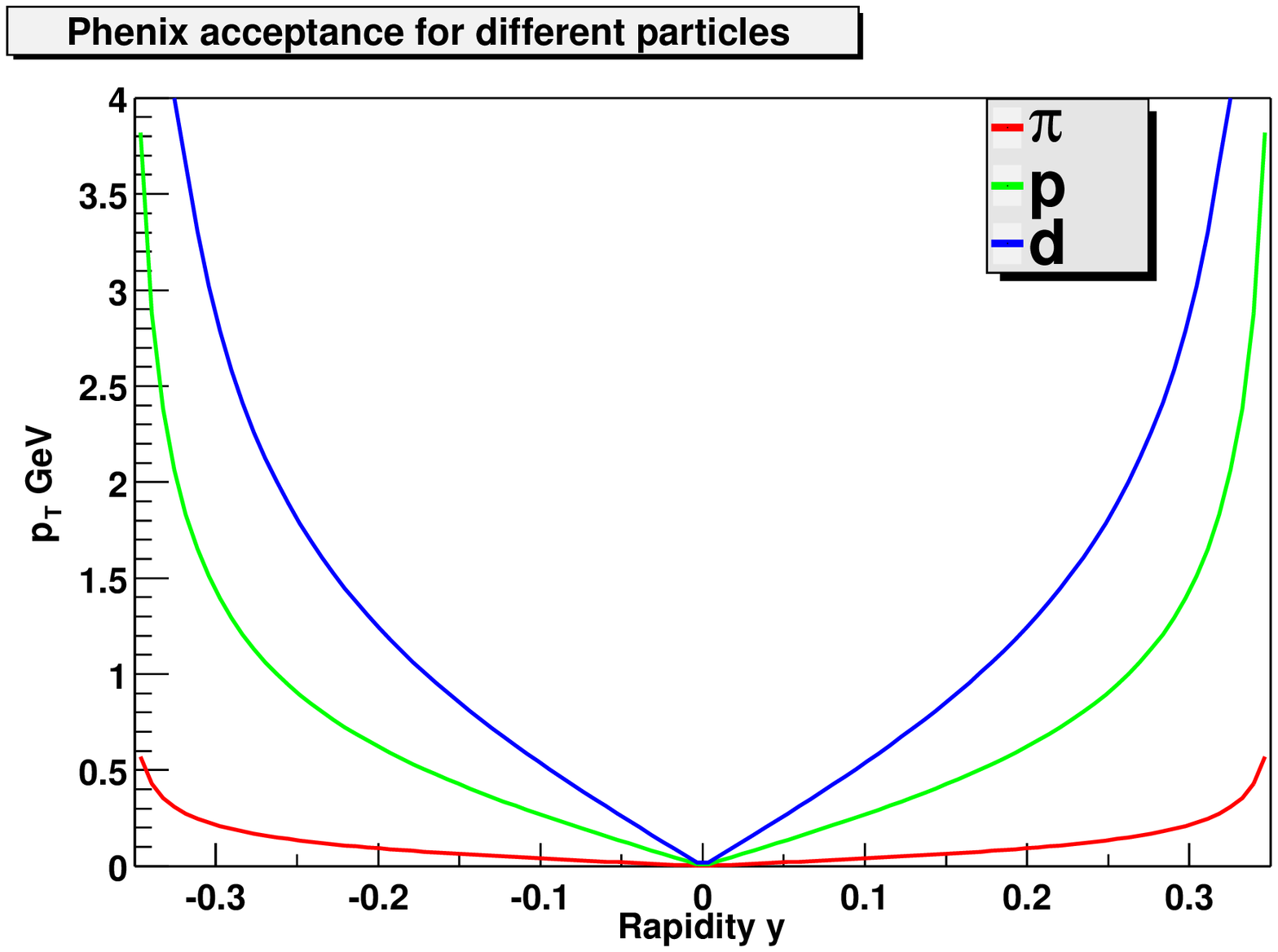, width = 12cm, clip = }
\end{center}
\caption{Phenix acceptance for deuterons and anti-deuterons with other 
particles ($\pi, K, p$).}
\label{fig:phe_acc}
\end{figure}

\clearpage

\section{Corrections}

Before we can obtain our final spectra from the raw particle counts, we need 
to correct for factors like:
\bitm
\item {\bf Detector acceptance:} limited PHENIX acceptance means our raw counts need
to be corrected up.
\item {\bf Detector efficiency:} Detectors can be fickle, sometimes certain
channels have to be switched off, at other times low voltages can lead to lower
detection capabilities. This has to be corrected for.
\item  {\bf Track reconstruction:} the algorithms that do event reconstruction and
determine tracks from hits aren't always 100\% efficient. This reconstruction
efficiency can also depend on event centrality. E.g. there are more tracks in 
central events, leading to a greater probability of error and misidentification
of a track.
\item {\bf Hadronic annihilation factors:} the GEANT package~\cite{geant}, which is used to 
simulate how a particle tracks through a given detector does not implement
hadronic interactions for clusters like deuterons. In addition, anti-deuterons
also have a probability to annihilate before they are fully tracked and 
identified causing us to undercount them. This has be to be corrected.
\item {\bf Detector occupancy effects:} detector and track reconstruction 
efficiencies are often affected by high multiplicity events. E.g. before a 
slower particle like deuteron hits a detector channel, a faster particle might
have hit and so the slower particle might not register. This effect is greater
for heavier and slower particles like deuterons. Since this varies with 
particle multiplicities, it leads to a centrality dependent correction.
\item {\bf Momentum bins:} since particle counts are
not flat over a given momentum bin putting the data points at the bin center 
would be incorrect. E.g. in the $p_T$ range 1.1 to 1.5 GeV/c the raw counts 
increase, as a result the midpoint of the $p_T$ bin is too low. This can be 
rectified by taking the mean $p_T$ of each bin using the following
expression:
\be
p_{Bin} = \frac{\int_{p_1}^{p_2} p_T f(p_T)dp_T}{\int_{p_1}^{p_2} f(p_T)dp_T}
\label{eq:binmid}
\ee
where $f(p_T)$ is a function (gaussian) used to fit the raw 
yields~\ref{fig:dyield_raw_final}. Due to shape of the raw spectra, a 
polynomial or a gaussian give similar results.
\eitm
\subsection{Single particle efficiency}

In order to correct our raw spectra due to limitations of detector acceptance, 
efficiencies and track reconstruction, we use Monte Carlo (MC) simulation
techniques. Single particle tracks are generated and then reconstructed using
PHENIX simulation package.
One million deuteron (anti-deuteron) events were generated using the
package EXODUS over full azimuthal coverage and rapidity $|y| < 0.6$. The 
output files in OSCAR format were run through the PHENIX simulation package 
PISA. These packages try to reflect the detector characteristics as accurately 
as possible. The single particle tracks thus generated were reconstructed to 
obtain the correction factors by taking the 
ratio of the reconstructed $p_T$ distributions with the input EXODUS $p_T$ 
distributions (restricted to $|y| < 0.5$ in order to get corrections for unit
rapidity interval). This gives us the acceptance, 
efficiency of reconstruction, matching cuts and so on.

\be
\frac{dN_{gen}/dp_T}{dN_{reco}/dp_T} = \epsilon_{acc} \times \epsilon_{eff}
\ee

where

\be
\epsilon_{eff} = \epsilon_{track}\times \epsilon_{match} \times \epsilon_{PID}\times \epsilon_{active\ area}
\ee

In order to account for the finite momentum resolution effects the input 
$p_T$ distribution was weighted with the weight function:

\be
W(p_T) = \frac{e^{-m_t/T_{eff}}}{f(p_T)}
\ee

where 
\be
f(p_T) = 1 + 10e^{-2p_t}
\ee
is the EXODUS low $p_T$ enhanced distribution. The input distribution was 
enhanced at low $p_T$ to increase our statistics since we lose low momentum
particles due to acceptance effects. Since the inverse slope
parameter $T_{eff}$ (used for the fit to spectra) is known to be different 
for different centralities, we
changed it's value to check if this has any effect on the MC for different
centralities. We found no change in the correction function for different
centralities.

\begin{figure}[h!]
\begin{center}
\epsfig{file = 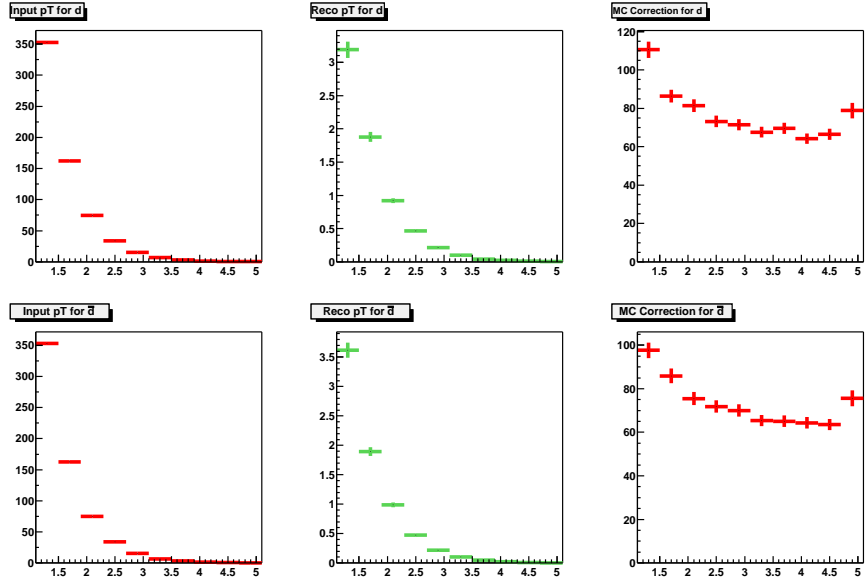, width = 12cm, clip = }
\end{center}
\caption{Top left panel shows input EXODUS distribution, middle panel 
shows the reconstructed output distribution and right panel shows the 
single particle MC
correction as a function of $p_T$ for deuterons (top panels) and 
anti-deuterons (bottom panels) for min. bias data.}
\label{fig:ppg020mccor_final}
\end{figure}

We also need to check that our MC simulations match each detector's
characteristics as accurately as possible. To cross check this, we look at
track and hit distributions for each of our track cuts (like matching cuts) 
both in MC and data. Some comparisions for acceptances between data and MC 
are shown in in Appendix A. We also need to ensure that our matching cuts (or 
track residuals): $\sigma_{\phi}(TOF, PC3)$ and $\sigma_z(TOF, PC3)$ are well
matched between data and MC. Matching cuts kept too wide lead to too 
much background, whereas if they are kept too narrow then we lose signal. If
the matching cuts are not well matched between data and MC then the corrections
can have errors. In order to avoid this, we plot the matching cuts 
$\sigma_{\phi}(TOF, PC3)$ and $\sigma_z(TOF, PC3)$ against variables 
like $z$, centrality, $p_T$ and so on and look for differences between data
and MC. These plots are also available in Appendix A.

This finally gives us a set of
bin by bin corrections for deuterons (anti-deuterons) as shown in 
Fig.~\ref{fig:ppg020mccor_final} for minbias data. 
The values of the bin-by-bin corrections are tabulated in 
Table~\ref{tab:mcfinal}. This is then fit to a polynomial
and used to interpolate to get corrections (see 
Fig.~\ref{fig:ppg020_mcfit}). This interpolation is necessary 
even though we have generated bin-by-bin corrections, because values for raw 
counts are being plotted at mean $p_T$ of each bin, instead of the midpoint of 
the bin.

\begin{table}[h!]
\caption{Single particle Monte Carlo corrections}
\bcc
 \begin{tabular}[ ]{|l|l|l|}
\hline
$p_T$ GeV & Deuteron & Anti-deuteron \\
\hline
1.3	&	110.513	$\pm$	4.2204	&	97.615	$\pm$	3.50381	\\
1.7	&	86.3182	$\pm$	3.3729	&	85.8968	$\pm$	3.35731	\\
2.1	&	81.4226	$\pm$	3.34997	&	75.51	$\pm$	2.98608	\\
2.5	&	73.1532	$\pm$	2.98991	&	71.8395	$\pm$	2.89169	\\
2.9	&	71.4168	$\pm$	2.92981	&	70.0016	$\pm$	2.81757	\\
3.3	&	67.5995	$\pm$	2.78592	&	65.4321	$\pm$	2.59155	\\
3.7	&	69.5334	$\pm$	2.95599	&	65.0881	$\pm$	2.64015	\\
4.1	&	64.1365	$\pm$	2.69051	&	64.347	$\pm$	2.613	\\
4.5	&	66.524	$\pm$	2.92311	&	63.553	$\pm$	2.62167	\\
\hline
\end{tabular}
\ecc
\label{tab:mcfinal}
\end{table}

\begin{figure}[h!]
\begin{center}
\epsfig{file = 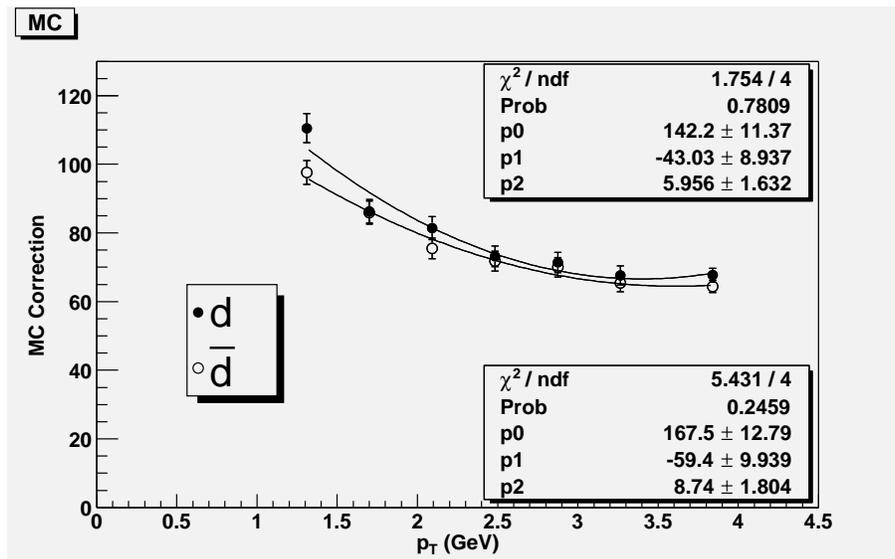, width = 12cm, clip = }
\end{center}
\caption{Final single particle Monte Carlo (MC) corrections with fits for interpolation}
\label{fig:ppg020_mcfit}
\end{figure}

\clearpage

\subsection{Loss of tracks due to occupancy effects}
In high multiplicity events, tracks can be lost. Sometimes, the track 
reconstruction  algorithms can misidentify tracks in the presence of a large 
number of hits. At other times the track of slower particle like a deuteron
might not register at a detector, because it has already been hit by a fast 
particle like electron or pion. This effect is greater for heavier particles
as they are slower. It also depends on the event centrality because more tracks
can get misidentified in high multiplicity events. In order to correct for 
this, we embed a simulated single particle MC track in a real event. We then 
see if this simulated track gets reconstructed. We generated 
these corrections using code developed by Jiangyong Jia~\cite{jiaembed}. The 
embedding correction as a function of $p_T$ for different centralities is 
shown in Figs.~\ref{fig:embdcor_0_92_final}, ~\ref{fig:embdcor_0_20_final}, 
~\ref{fig:embdcor_20_92_final} for the min. bias, 0-20\% centrality and 
20-92\% centrality data respectively. The embedding correction is flat with
$p_T$ within errors. The final values are calculated by intergrating over the
entire $p_T$ range and are tabulated in Table~\ref{tab:embed}.

\begin{figure}[h!]
\begin{center}
\epsfig{file = 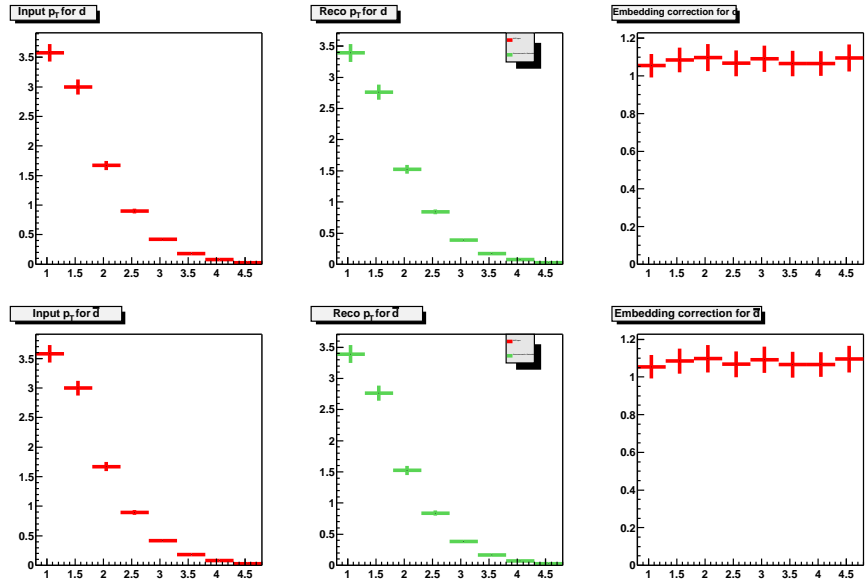, width = 12cm, clip = }
\end{center}
\caption{Top left panel shows input distribution, middle panel shows the 
output distribution (after embedding) and right panel shows the embedding 
correction as a function of $p_T$ for deuterons (top panels) and 
anti-deuterons (bottom panels) for min. bias data. }
\label{fig:embdcor_0_92_final}
\end{figure}

\begin{figure}[h!]
\begin{center}
\epsfig{file = 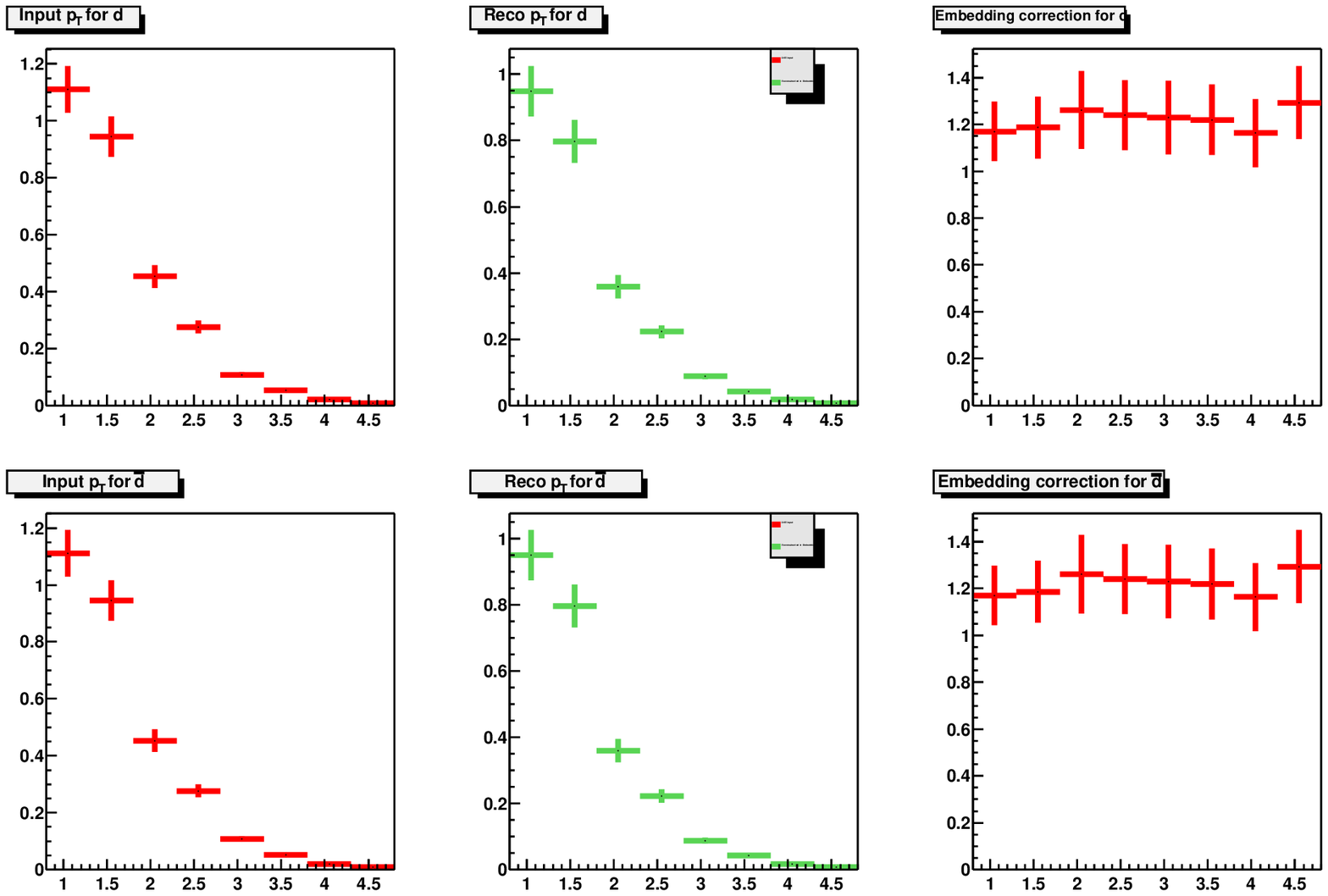, width = 12cm, clip = }
\end{center}
\caption{Top left panel shows input distribution, middle panel shows the 
output distribution (after embedding) and right panel shows the embedding 
correction as a function of $p_T$ for deuterons (top panels) and 
anti-deuterons (bottom panels) for 0-20\% centrality data. }
\label{fig:embdcor_0_20_final}
\end{figure}

\begin{figure}[h!]
\begin{center}
\epsfig{file = 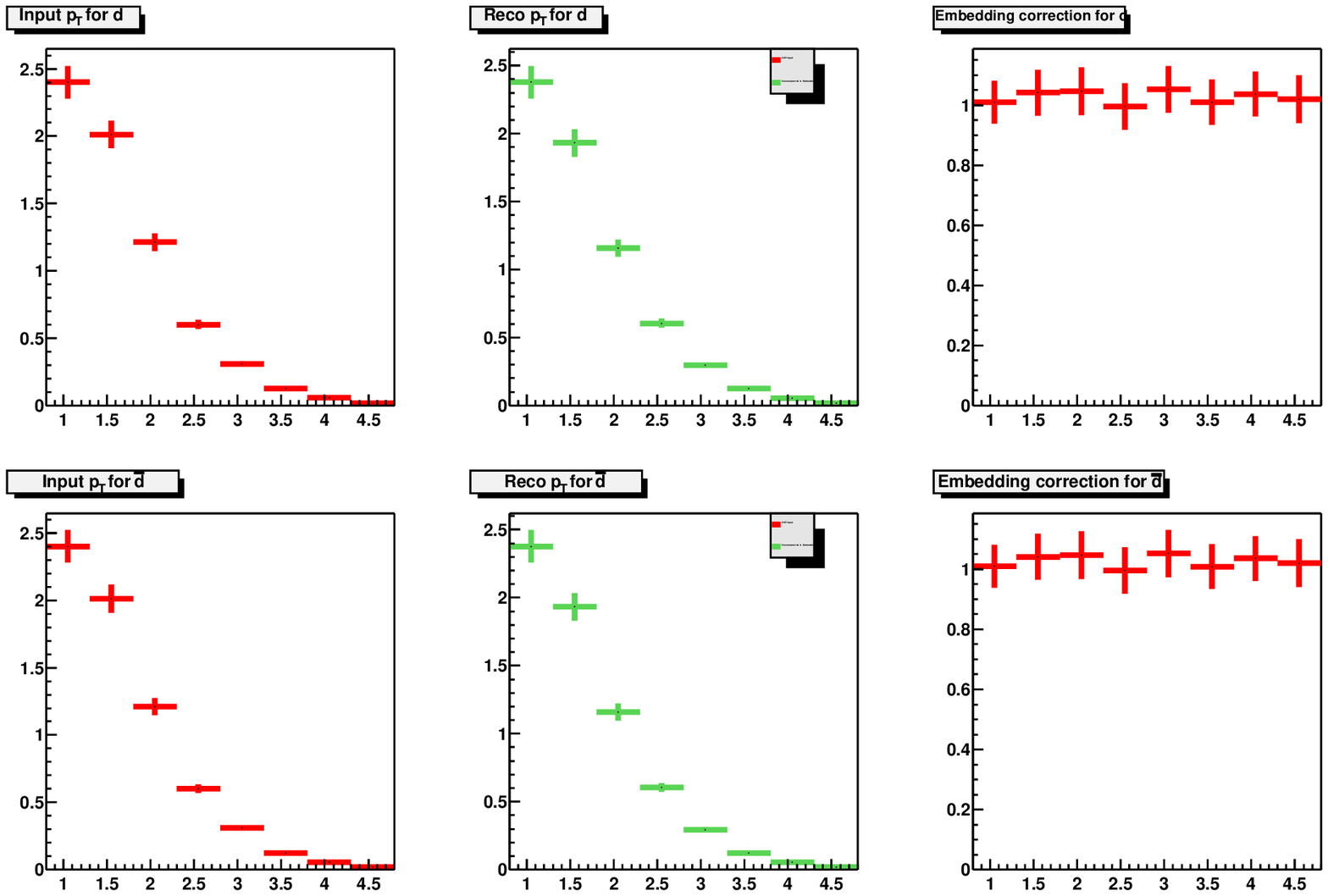, width = 12cm, clip = }
\end{center}
\caption{Top left panel shows input distribution, middle panel shows the 
output distribution (after embedding) and right panel shows the embedding 
correction as a function of $p_T$ for deuterons (top panels) and 
anti-deuterons (bottom panels) for 20-92\% centrality data. }
\label{fig:embdcor_20_92_final}
\end{figure}

\begin{table}[h!]
\caption{Embedding correction}
\bcc
\begin{tabular}[ ]{|l|l|}
\hline
Centrality 		& Correction factor \\
\hline
Min Bias: 0-92\% 	& 1.0691 $\pm$ 0.0274\\

Most Central: 0-20\%  	& 1.1979 $\pm$ 0.0711\\

Mid-Central 20-92\% 	& 1.0224 $\pm$ 0.0310\\
\hline
\end{tabular}
\ecc
\label{tab:embed}
\end{table}

The (anti-) proton spectra used for comparision were also corrected for 
feed-down from $\Lambda$ and $\bar{\Lambda}$ decays by using a simulation 
tuned to reproduce the particle composition: $\Lambda/p$ and 
$\bar{\Lambda}/\bar{p}$ measured by PHENIX at 130 GeV~\cite{lambda130}.
The systematic error in proton yields from the feed 
down corrections is estimated at 6 \%.

\subsection{Hadronic absorption of $d$/$\overline{d}$}

Since the hadronic interactions of nuclei 
are not treated by GEANT, a correction needs to be applied to account for the 
hadronic absorption of $d$ and $\bar{d}$ (including annihilation). The $d$- 
and $\bar{d}$-nucleus cross sections are calculated from parameterizations
 of the nucleon and anti-nucleon cross sections~\cite{Moiseev}: 
\begin{equation}
\sigma_{d/\bar{d},A} =  [\sqrt{\sigma_{N/\bar{N},A}} +  \Delta_d ]^{2}
\label{eq:dbarcs}
\end{equation}

The limited data 
available on deuteron induced interactions~\cite{Jaros,Abdurak} indicate that 
the term $\Delta_d$ is independent of the nuclear mass number $A$ and that 
$\Delta_d = 3.51\pm 0.25$~mb$^{1/2}$. The hadronic absorption varies 
only slightly over the applicable $p_T$ range and is $\approx$ 10\% for $d$ 
and $\approx$ 15\% for $\bar{d}$. A plot of the (anti)deuteron
survival correction is shown in Fig.~\ref{fig:anncor}. Special thanks goes to
Joakim Nystrand for determination of this correction. The background 
contribution from deuterons 
knocked out due to the interaction of the beam with the beam pipe is estimated
using simulation and found to be negligible in the momentum range of our 
measurement. 

\begin{figure}[h!]
\begin{center}
\epsfig{file = 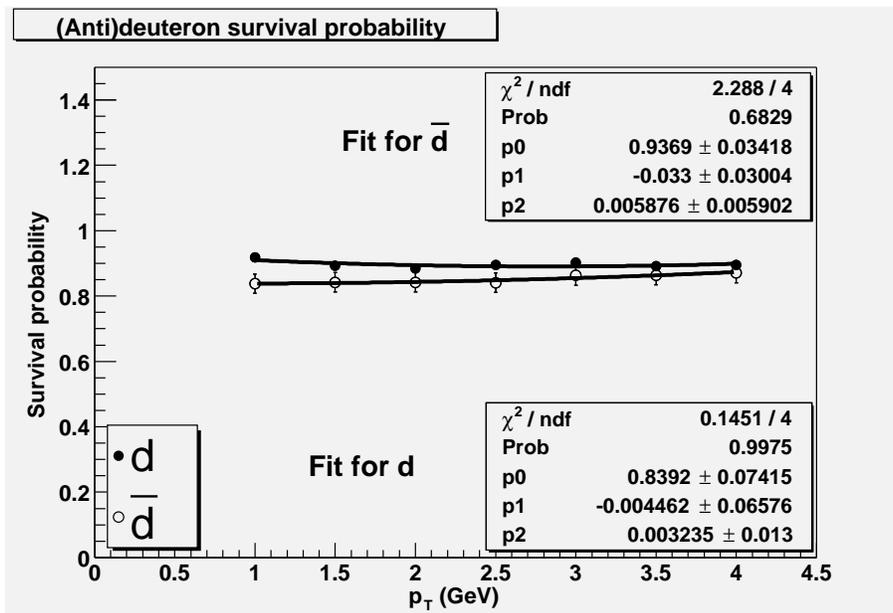, width = 12cm, clip = }
\end{center}
\caption{Correction for (anti-)deuteron survival vs $p_T$~\cite{lundwork}.}
\label{fig:anncor}
\end{figure}

The total error is dominated by the systematic errors in the correction 
factors, and we estimate these to be 1.5\% (3.5\%) for deuterons 
(anti-deuterons).

\chapter{$d, \bar{d}$ yields and implications}

The hadrons produced in the collision zone carry information about the 
nature of the collision, as well its size and composition. In particular, 
the $p_T$ behaviour of the spectra can yield information about the dynamics
of the collision, while the particle yields and abundances can help us to 
determine the chemical composition.
\section{Spectra}
\begin{figure}[h!]
\includegraphics[width=1.0\linewidth]{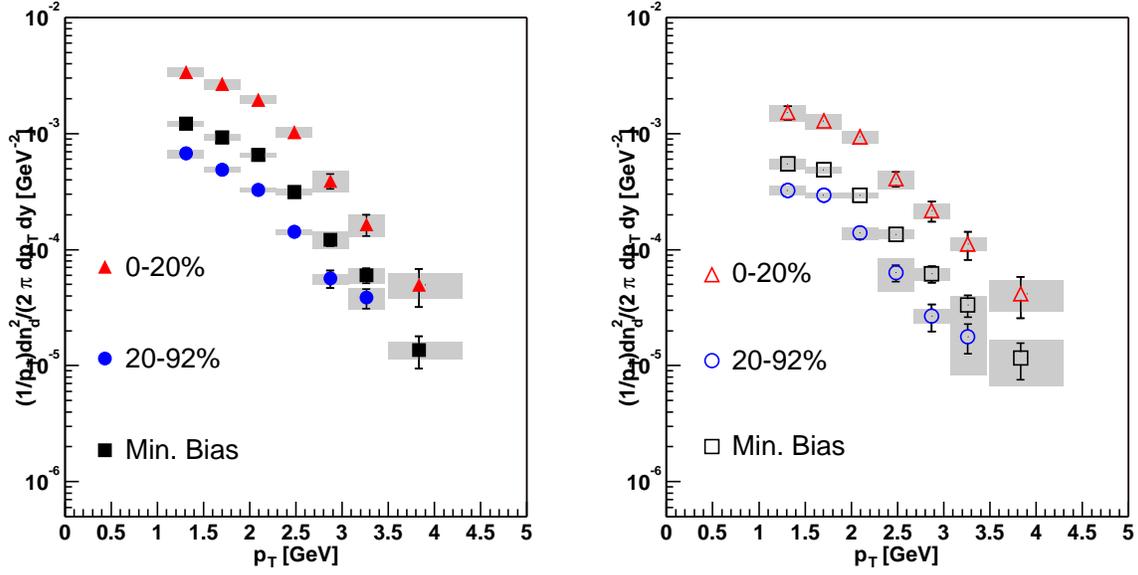}  
\caption{Corrected $\bar{d}, d$ yields vs $p_T$ for different centralities.}
\label{fig:ppg020_fig2}
\end{figure}

\begin{figure}[h!]
\epsfig{file = 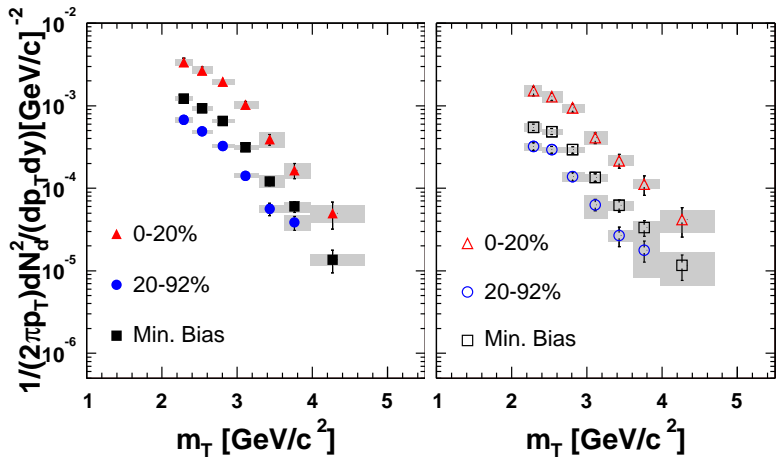, width = 12cm, clip = }
\caption{Corrected $\bar{d}, d$ yields vs $m_T$ for different centralities.}
\label{fig:ppg020_mtspectra}
\end{figure}

After applying the corrections described previously, we obtain the 
(anti-)deuteron spectra in the momentum range $1.1<p_T<4.3$ GeV/c  for two 
centrality classes: 0-20\% (most central), 20-92\% (mid-central) and for 
minimum bias events. The spectra are shown in 
Figure~\ref{fig:ppg020_mtspectra}. The $x$-axis has the transverse 
mass $m_T = \sqrt{m^2+p_T^2}$, while the $y$-axis has the invariant yield. 
We immediately
notice that the $d$, $\bar{d}$ spectra have a shoulder arm shape and do not
fall in a straight line, but show curvature in the region of lower $m_T$. 
This is indicative of hydrodynamic
flow, which pushes heavier particles to higher $p_T$ as a result of 
interactions. Another consequence of is that deuteron spectra are flatter 
compared to the proton spectra. Final 
corrected invariant yields are given in Table~\ref{tab:coryields}.

\begin{table}[h!]
\caption{Corrected Yields at the mean of each $p_T$ bin for different 
centralities} 
\bcc
\begin{tabular}[ ]{|l|l|l|l|}
\hline
Centrality &$p_T$ [GeV]& $Ed^3N/dp^3$ (deuterons) &$Ed^3N/dp^3$ (anti-deuterons) \\
\hline
	&	1.31193	&	0.00338179	$\pm$	0.000364594	&	0.00151508	$\pm$	0.000203537	\\
	&	1.70166	&	0.00268111	$\pm$	0.000254032	&	0.0012878	$\pm$	0.000135466	\\
	&	2.09137	&	0.00196259	$\pm$	0.000139802	&	0.00093872	$\pm$	0.000103305	\\
0-20\% 	&	2.48116	&	0.00103001	$\pm$	8.88E-05	&	0.000407605	$\pm$	6.10E-05	\\
	&	2.87113	&	0.00039078	$\pm$	5.72E-05	&	0.000216166	$\pm$	4.22E-05	\\
	&	3.26136	&	0.000165438	$\pm$	3.48E-05	&	0.00011187	$\pm$	3.02E-05	\\
	&	3.83075	&	4.99E-05	$\pm$	1.80E-05	&	4.18E-05	$\pm$	1.62E-05	\\
\hline
	&	1.31193	&	0.000671812	$\pm$	7.05E-05	&	0.000321646	$\pm$	3.49E-05	\\
	&	1.70166	&	0.000488571	$\pm$	4.62E-05	&	0.000293652	$\pm$	2.68E-05	\\
	&	2.09137	&	0.000326346	$\pm$	2.50E-05	&	0.000138651	$\pm$	1.64E-05	\\
20-90\%	&	2.48116	&	0.000141669	$\pm$	1.45E-05	&	6.31E-05	$\pm$	1.01E-05	\\
	&	2.87113	&	5.63E-05	$\pm$	9.50E-06	&	2.68E-05	$\pm$	7.07E-06	\\
	&	3.26136	&	3.85E-05	$\pm$	7.37E-06	&	1.78E-05	$\pm$	5.06E-06	\\
	&	3.83075	&	7.46E-06	$\pm$	3.14E-06	&	4.19E-06	$\pm$	2.78E-06	\\
\hline
	&	1.31193	&	0.00121155	$\pm$	0.000120745	&	0.000548677	$\pm$	5.09E-05	\\
	&	1.70166	&	0.000928171	$\pm$	7.76E-05	&	0.000486214	$\pm$	4.00E-05	\\
	&	2.09137	&	0.000651407	$\pm$	3.89E-05	&	0.000292861	$\pm$	2.62E-05	\\
Min. Bias.	&	2.48116	&	0.000312731	$\pm$	2.25E-05	&	0.000135107	$\pm$	1.48E-05	\\
	&	2.87113	&	0.000121189	$\pm$	1.41E-05	&	6.19E-05	$\pm$	1.02E-05	\\
	&	3.26136	&	5.99E-05	$\pm$	8.85E-06	&	3.33E-05	$\pm$	7.06E-06	\\
	&	3.83075	&	1.36E-05	$\pm$	4.20E-06	&	1.16E-05	$\pm$	3.99E-06	\\
\hline
\end{tabular}
\ecc
\label{tab:coryields}
\end{table}

\clearpage
\section{Systematical uncertainties}

Our systematic uncertainties fall in two categories:
\benn

\item Errors that vary point to point as a function of $p_T$. Most errors fall
in this category and include detector matching in both $\phi$ and $z$, energy
losse cut in TOF, momentum scale, PID error 
etc. We calculate the $p_T$ dependent errors, by varying our cuts to 
generate spectra and then looking at the difference between the new yields and
the final yield. The combined point to point value of these systematic 
errors is listed in Table~\ref{tab:ppg020sys}. 

\item Errors that are constant as a function of $p_T$, for
example embedding, absolute normalisation, annihilation/hadronic interaction 
correction, feeddown correction (for proton yields). These systematic 
uncertainties 
are tabulated in Table~\ref{tab:normsys} and are explained below:
\eenn

\begin{table}[h!]
\caption{$p_T$ dependent absolute systematic errors from different sources
(matching, eloss and PID) are added in quadrature}
\bcc
\begin{tabular}[ ]{|l|l|l|l|}
\hline
Centrality   &$p_T$ & deuterons & anti-deuterons \\
\hline

	&	1.31193	&	0.000179042	&	0.000165652	\\
	&	1.70166	&	0.000189072	&	0.000134881	\\
	&	2.09137	&	8.09E-05	&	4.04E-05	\\
0-20\%	&	2.48116	&	7.30E-05	&	5.31E-05	\\
	&	2.87113	&	7.55E-05	&	2.31E-05	\\
	&	3.26136	&	3.47E-05	&	1.09E-05	\\
	&	3.83075	&	1.14E-05	&	1.23E-05	\\
\hline
	&	1.31193	&	5.06E-05	&	2.67E-05	\\
	&	1.70166	&	1.80E-05	&	1.11E-05	\\
	&	2.09137	&	9.92E-06	&	1.54E-05	\\
20-90\%	&	2.48116	&	7.20E-06	&	1.97E-05	\\
	&	2.87113	&	5.61E-06	&	3.74E-06	\\
	&	3.26136	&	7.97E-06	&	9.45E-06	\\
	&	3.83075	&	3.05E-06	&	3.35E-06	\\
\hline
	&	1.31193	&	5.91E-05	&	5.30E-05	\\
	&	1.70166	&	5.14E-05	&	2.36E-05	\\
	&	2.09137	&	2.30E-05	&	1.02E-05	\\
Min.Bias&	2.48116	&	1.69E-05	&	1.12E-05	\\
	&	2.87113	&	2.08E-05	&	7.57E-06	\\
	&	3.26136	&	8.73E-06	&	6.01E-06	\\
	&	3.83075	&	2.26E-06	&	5.00E-06	\\
\hline
\end{tabular}
\ecc
\label{tab:ppg020sys}
\end{table}

\subsection{$p_T$ dependent systematic uncertainties}

Sources of the systematic uncertainties of type I, which vary with $p_T$ are
briefly described below:

\benn
\item {\bf Matching Systematics}:
As mentioned in the previous chapter, in order to reduce our background,
we made detector matching cuts, by tracking the particle from the collision
vertex to the detector and looking at the difference between the expected hit
in the detector to the actual hit in both $\phi$ and $z$ directions. We looked
at the mean and sigma of our detector matching variables $\sigma_{\phi}(TOF)$, 
$\sigma_{z}(TOF)$, $\sigma_{\phi}(PC3)$ and $\sigma_{z}(PC3)$ vs various 
variables of interest like $z$, $p_T$ and centrality. These plots are shown 
in Appendix A. Some variation of these quantities is seen upto a variation of 
0.5 $\sigma$. In order to determine our systematic uncertainty we generate new
increase the matching cut from 2.5 $\sigma$ to 3 $\sigma$ (in both data and MC)
and take the ratio. See Figure~\ref{fig:sysplot_match}.

\begin{figure}
\begin{center}
\epsfig{file = 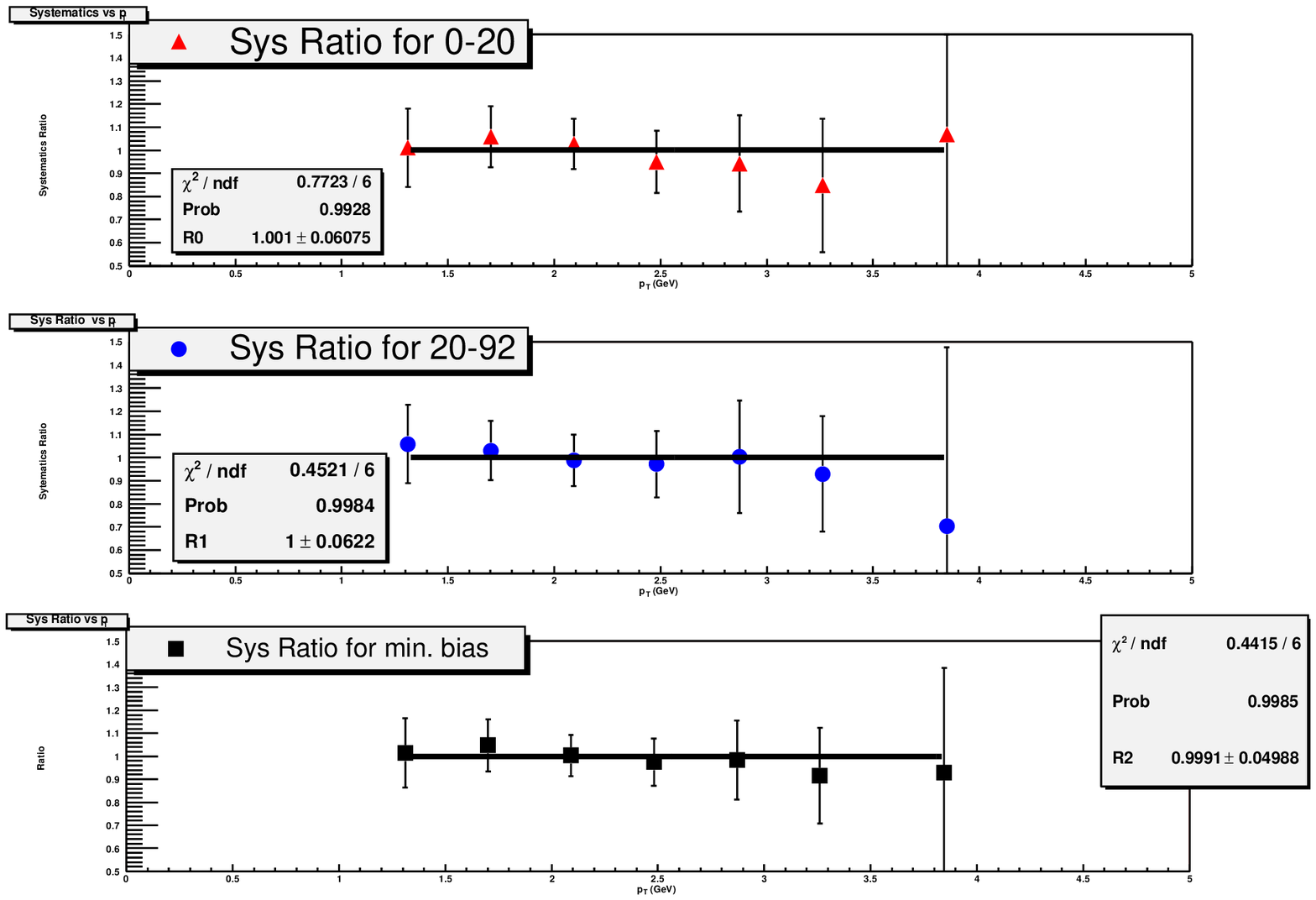, width = 12cm, clip = }
\end{center}
\caption{Systematic error estimate for matching cut}
\label{fig:sysplot_match}
\end{figure}

\item {\bf TOF $E_{loss}$ Systematics}: 
We shifted the $E_{loss}$ cut in the TOF scintillator from 
$E_{loss} > 0.0014\beta^{-5/3}$ GeV to
$E_{loss} > 0.0016\beta^{-5/3}$ GeV (see Fig.~\ref{fig:sysplot_etof}) to obtain
the systematic uncertainty. 

\begin{figure}[h!]
\begin{center}
\epsfig{file = 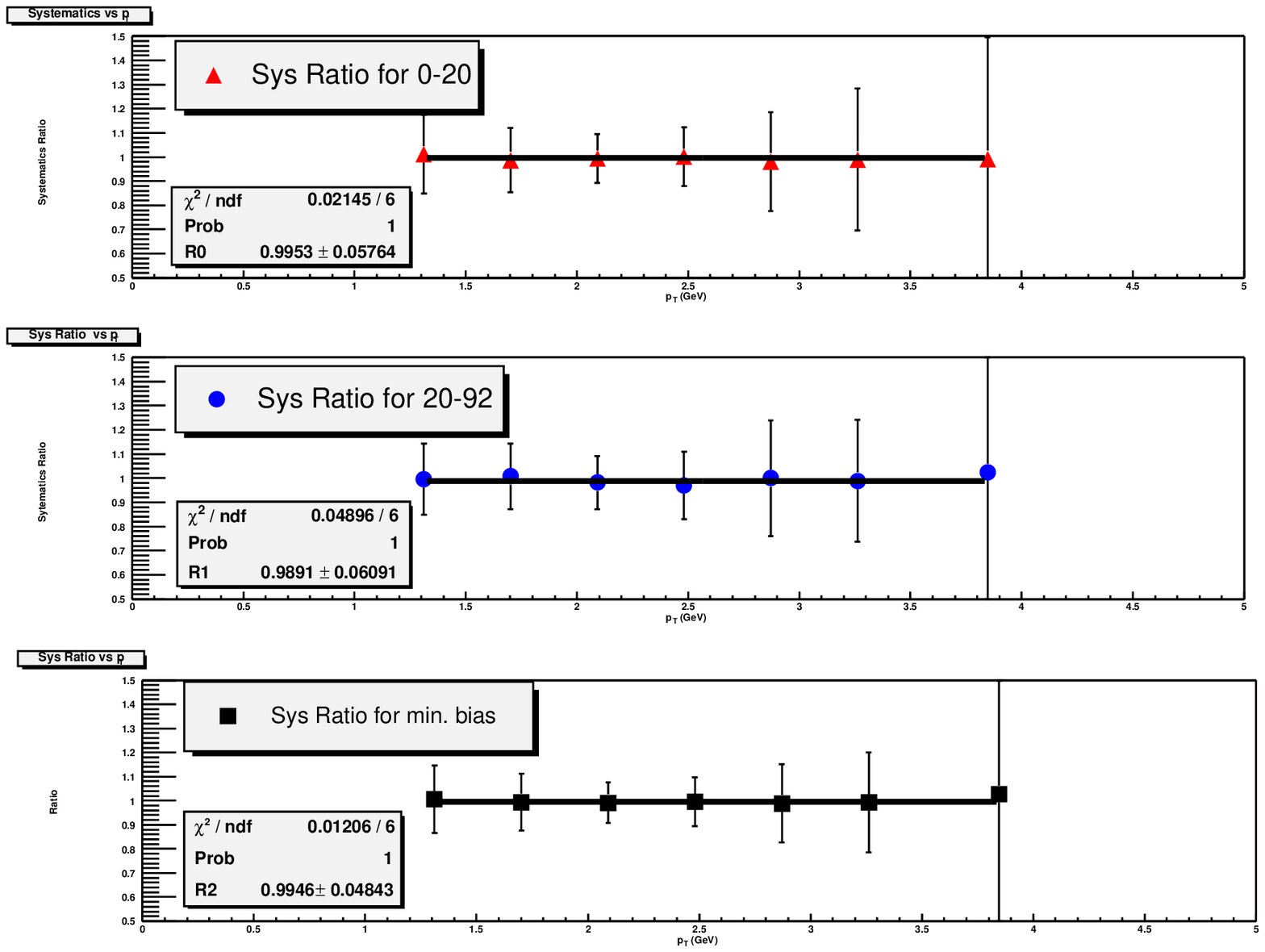, width = 12cm, clip = }
\end{center}
\caption{Systematic error estimate for TOF $E_{loss}$ cut}
\label{fig:sysplot_etof}
\end{figure}

\item {\bf PID Systematics}: 

In order to estimate the uncertainty in our particle identification (PID) we 
used three different methods:

\benn
\item The systematic error due to fitting is estimated by comparing the
yields from two different functional forms ($1/x$ and $e^{-x} $for the 
background (see Fig.~\ref{fig:sysplot_byx}). 
\item The binning of the $m^2$ histograms is changed to see how it affects 
the fits (see Fig.~\ref{fig:sysplot_bin}). 
\item The momentum resolution parameters are changed in the fitting routine, 
to see how it affects the width of $m^2$ histograms and hence the fits 
(see Fig.~\ref{fig:sysplot_respar}). 
\eenn

\begin{figure}[h!]
\begin{center}
\epsfig{file = 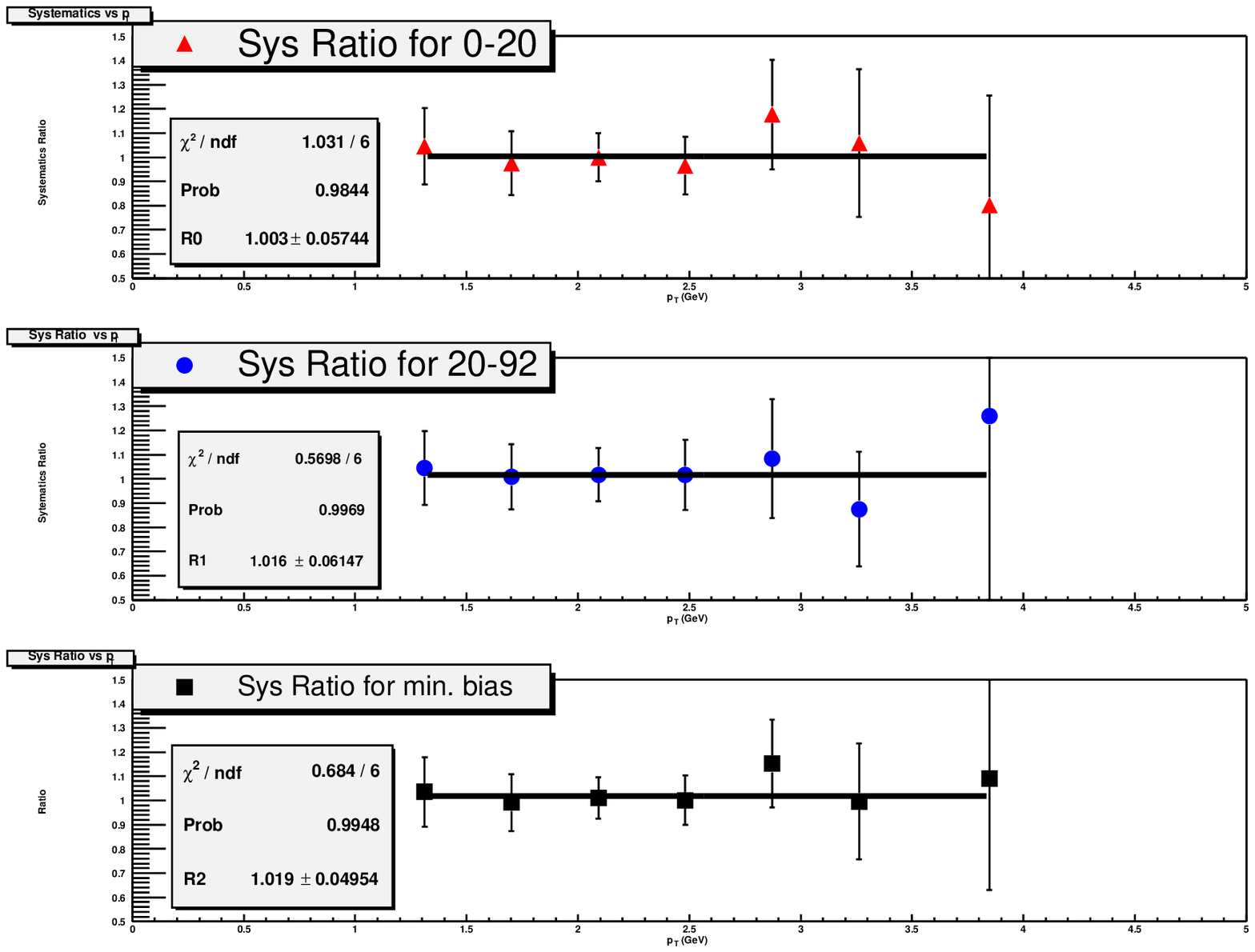, width = 12cm, clip = }
\end{center}
\caption{Systematic uncertainty for PID (binning)}
\label{fig:sysplot_bin}
\end{figure}

\begin{figure}[h!]
\begin{center}
\epsfig{file = 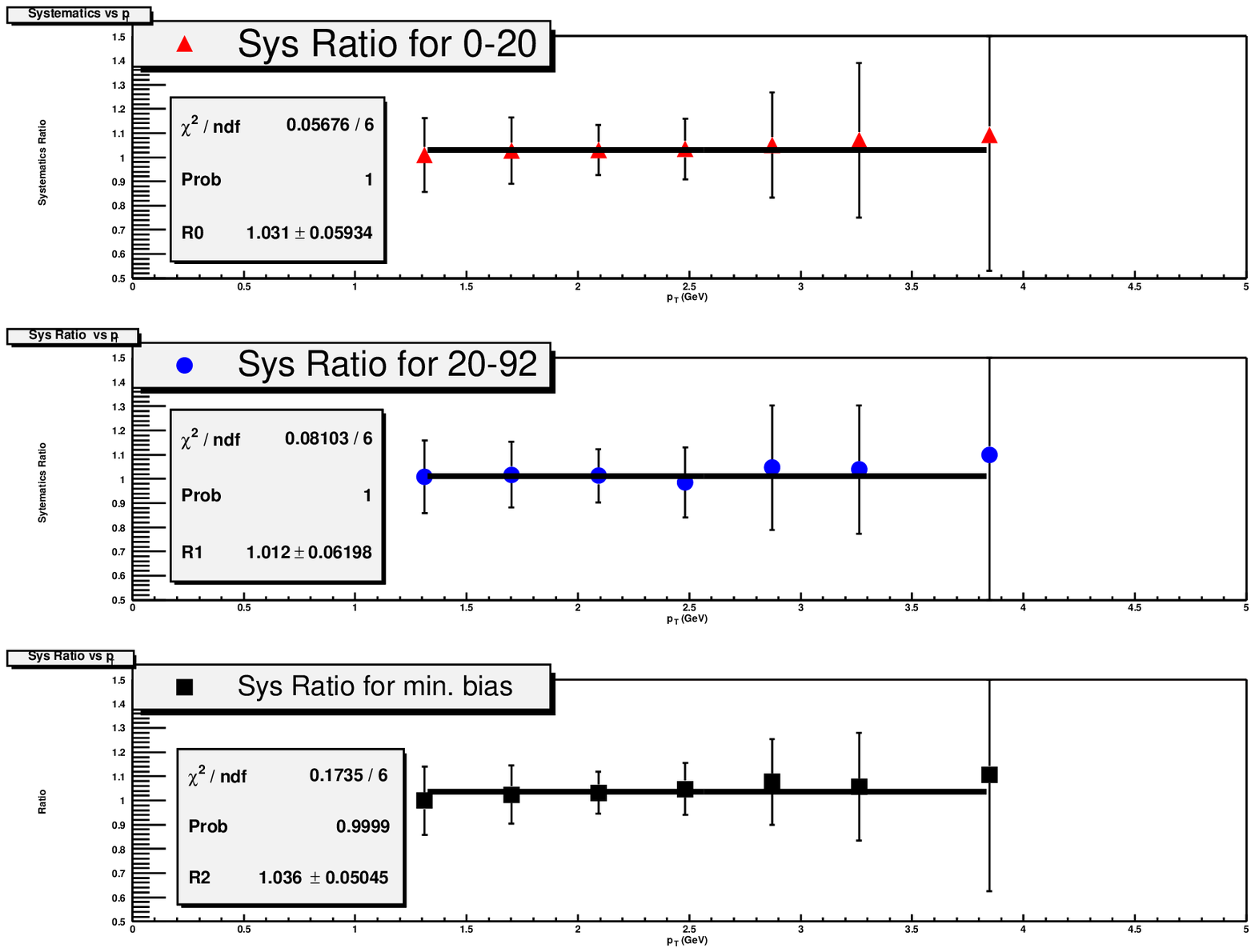, width = 12cm, clip = }
\end{center}
\caption{Systematic uncertainty for PID (varying momentum resolution
parameters)}
\label{fig:sysplot_respar}
\end{figure}

\begin{figure}[h!]
\begin{center}
\epsfig{file = 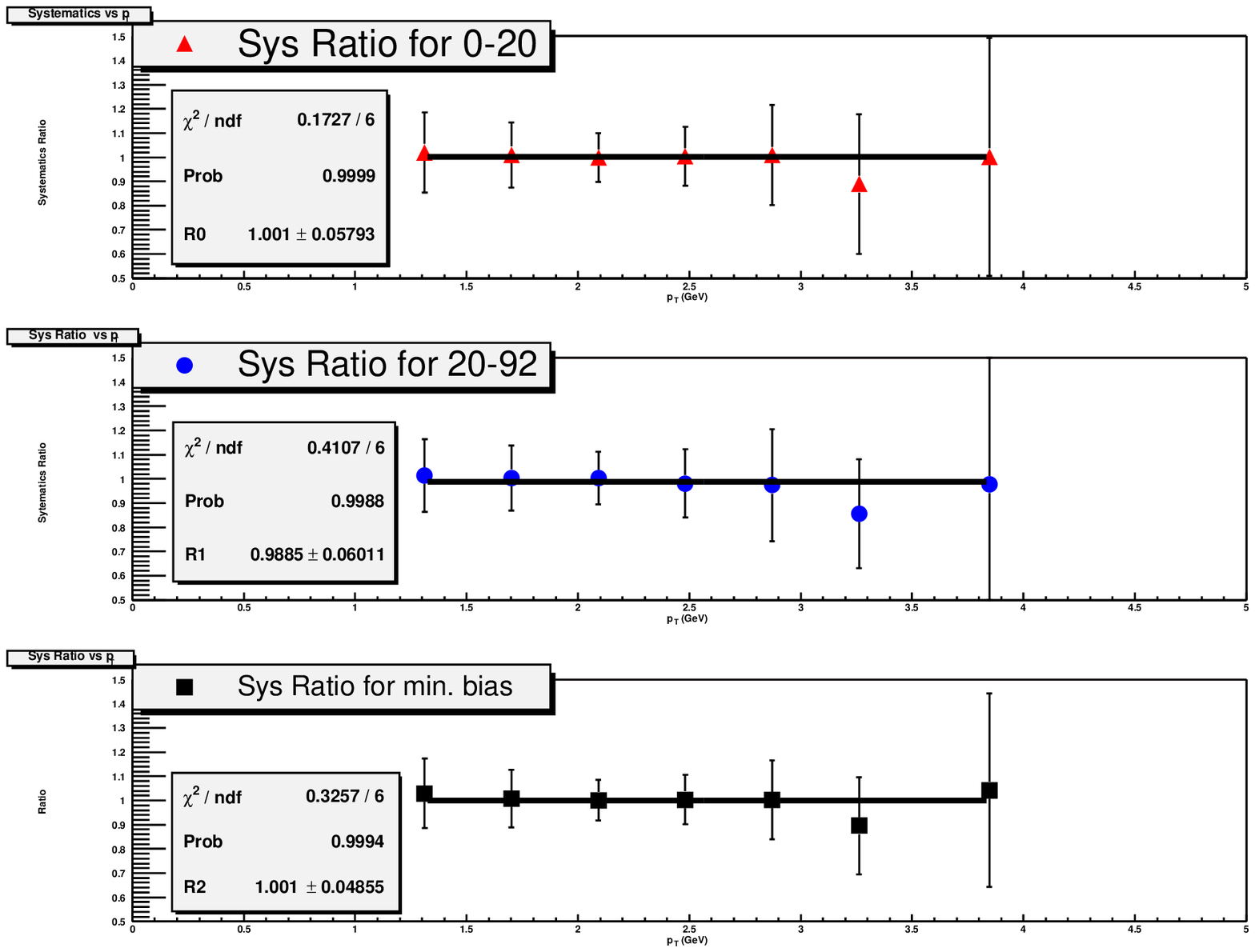, width = 12cm, clip = }
\end{center}
\caption{Systematic uncertainty for PID (using $1/x$ function for 
background instead of $e^{-x}$)}
\label{fig:sysplot_byx}
\end{figure}

\item {\bf Momentum Scale Systematics}: 
The error in our estimate for the momentum scale was found to be 0.71\%. This 
can lead to a systematic uncertainty in our yields that varies as a function 
of $p_T$ reaching $\approx$ 3\% at 4.0 GeV (see Fig.~\ref{fig:sysplot_scale}). 
This was generated by doing a simple Monte Carlo, by generating yields 
assuming the momentum scale to vary by $\pm$ 0.71\%.

\eenn

\begin{figure}[h!]
\begin{center}
\epsfig{file = 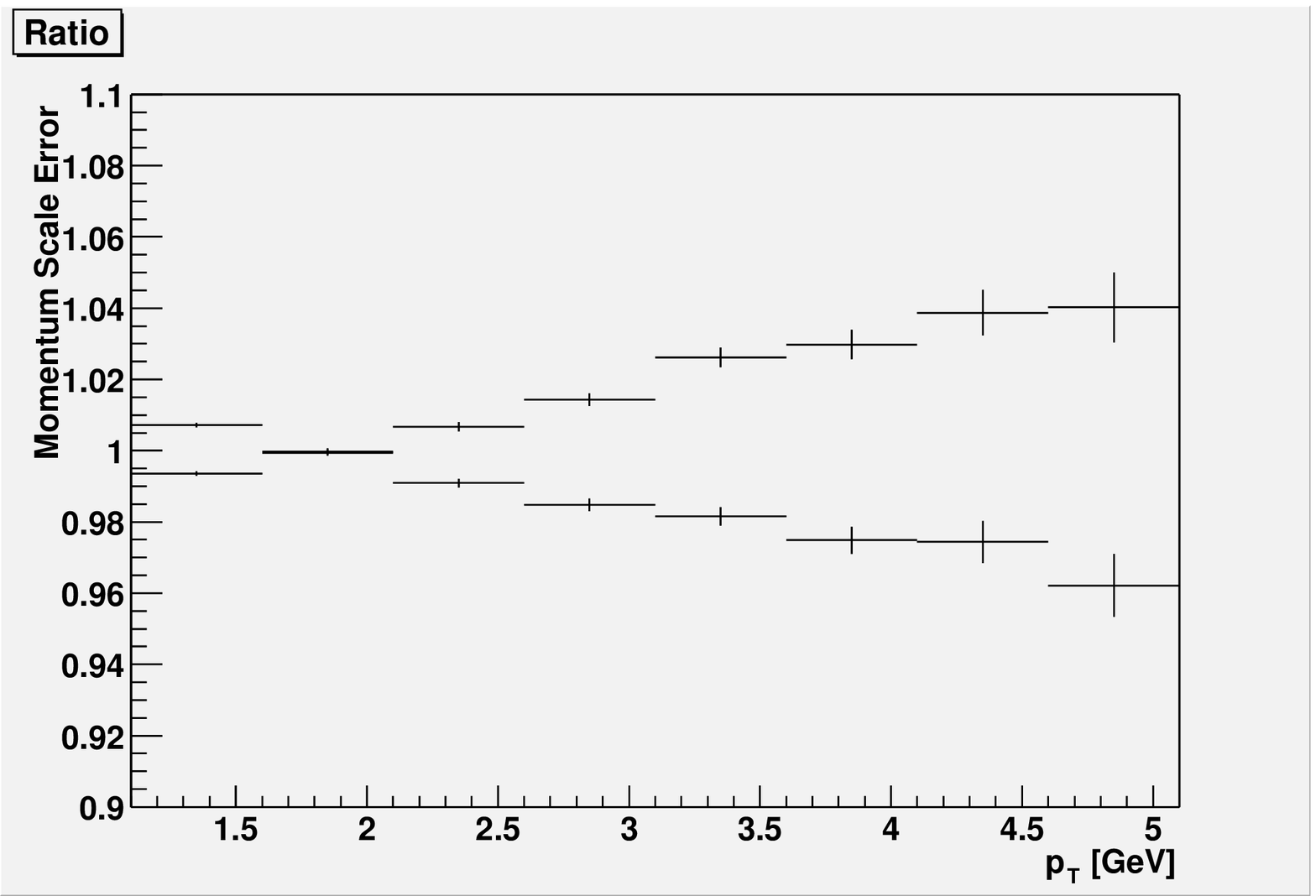, width = 12cm, clip = }
\end{center}
\caption{Systematic uncertainity due to momentum scale}
\label{fig:sysplot_scale}
\end{figure}

\subsection{$p_T$ independent systematic uncertainties}

Sources of the systematic uncertainties of type II, which are independent of
$p_T$ are briefly described below:

\begin{table}[h]
\caption{$p_T$ independent systematic uncertainties from different sources}
 \begin{tabular}[ ]{|l|l|l|l|}
\hline
Centrality 		& Annihilation $d(\bar{d})$& Embedding & Absolute Normalisation\\
\hline
Min Bias: 0-92\% 	&1.5\% (3.5\%)		& 2.74\%	&2.5\%\\

Most Central: 0-20\%  	&1.5\% (3.5\%)		& 7.11\%	&2.5\%\\

Mid-Central 20-92\% 	&1.5\% (3.5\%)		& 3.1\%		&2.5\%\\
\hline
\end{tabular}
\label{tab:normsys}
\end{table}

\benn
\item {\bf Annihilation correction systematics}: 
Systematic uncertainty due to survival (annihilation) correction 
for deuterons is 1.5\% and for anti-deuterons is 3.5\%.

\item {\bf Embedding correction systematics}: 
As mentioned in the previous chapter, in high multiplicity events, we can lose
tracks due to less efficient track reconstruction as well multiple hits in a 
given detector element. This was corrected for by embedding a simulated event
in a real event and running the reconstruction software on it. The uncertainty
in this correction is tabulated in Table~\ref{tab:embed}.

\item {\bf Vertex Dependence}: We varied the BBC $z$-vertex from 35 cm to 
30 cm and found no change in yields.

\item {\bf Absolute normalisation systematics}: 
Small variations in efficiencies of detectors from run to run, can lead to an
uncertainty in our absolute normalisation. To determine this we look at the 
average number of tracks in the TOF (Fig.~\ref{fig:ntofsys}), 
Drift Chamber(Fig.~\ref{fig:ndchsys}) and PC3 (Fig.~\ref{fig:npc3sys})
from run to run. The variance of this number gives our systematic uncertainty
in absolute normalisation ($\approx$ 2.5\%).

\begin{figure}[h!]
\begin{center}
\epsfig{file = 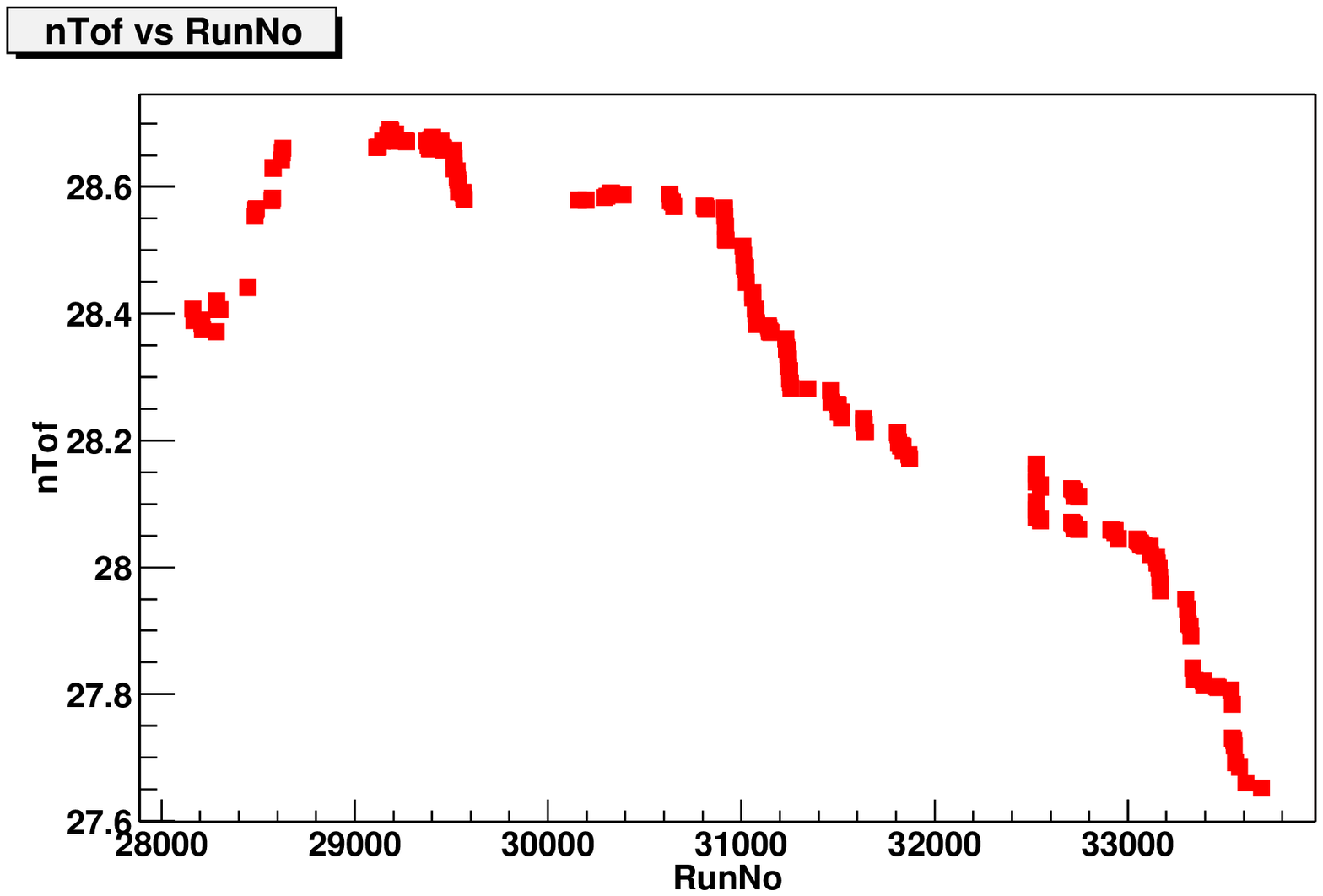, width = 12cm, clip = }
\end{center}
\caption{Run by run variation in average number of TOF tracks}
\label{fig:ntofsys}
\end{figure}

\begin{figure}[h!]
\begin{center}
\epsfig{file = 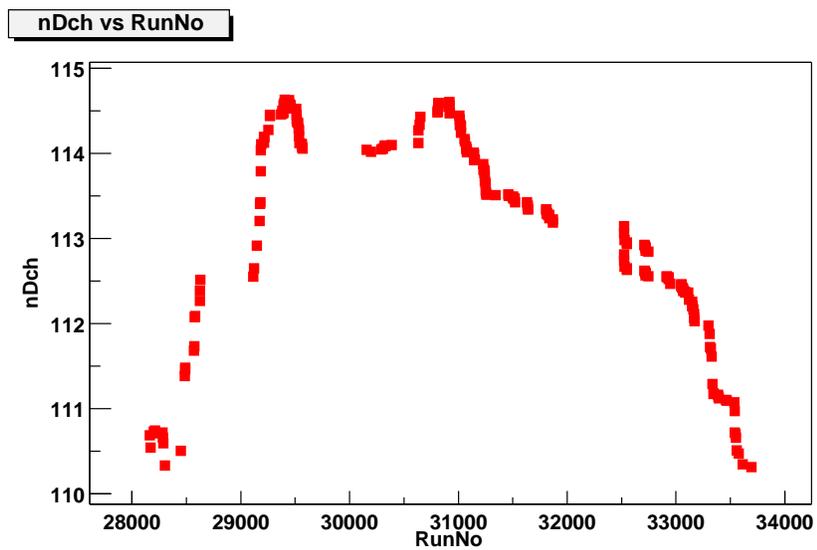, width = 12cm, clip = }
\end{center}
\caption{Run by run variation in average number of Drift Chamber tracks}
\label{fig:ndchsys}
\end{figure}

\begin{figure}[h!]
\begin{center}
\epsfig{file = 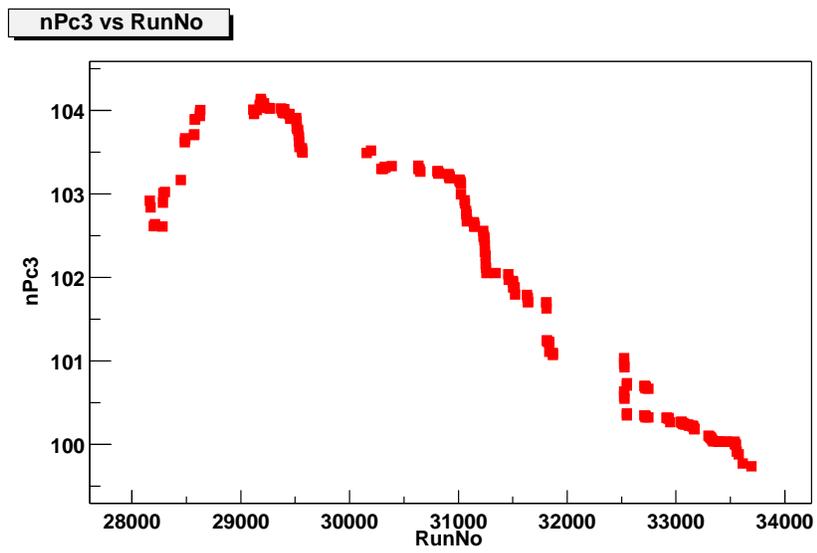, width = 12cm, clip = }
\end{center}
\caption{Run by run variation in average number of PC3 tracks}
\label{fig:npc3sys}
\end{figure}

\eenn

\clearpage
\section{$\bar{d}/d$ ratios and implications for $\bar{n}/n$ ratio}

The $\bar{d}/d$ ratios vs $p_T$ are shown in 
Figures~\ref{fig:ppg020_ratio2}, ~\ref{fig:ppg020_ratio0} and 
\ref{fig:ppg020_ratio1} for different centralities, and the values are listed
in Table\ref{tab:ratios}. The $\bar{d}/d$ 
ratio does not change as we go from one centrality to the other. We find 
that $\bar{d}/d =$ 0.47 $\pm$ 0.03 for minbias data. This is consistent 
with the square of the $\bar{p}/p = 0.73\pm 0.01$~\cite{ppg026}, within the 
statistical and the systematic errors, as expected if (anti-)deuterons are 
formed by coalescence of comoving (anti-)nucleons. 

\begin{figure}[h!]
\begin{center}
\epsfig{file = 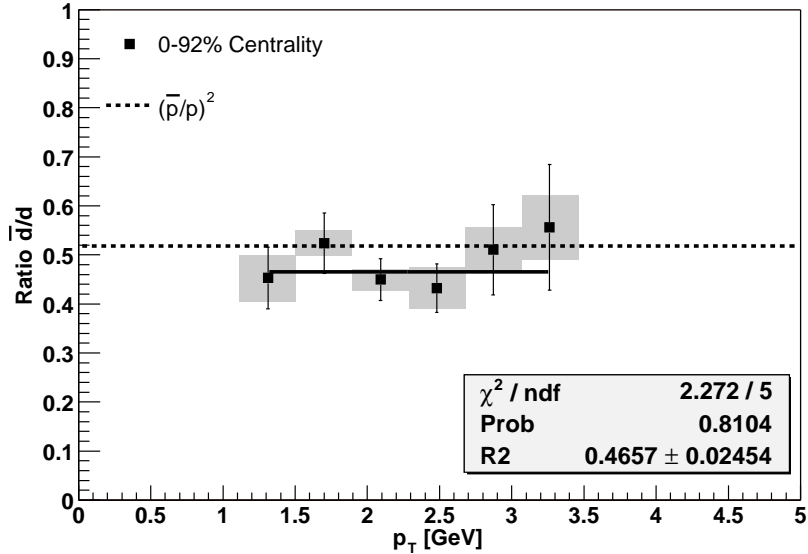, width = 12cm, clip = }
\end{center}
\caption{$\bar{d}/d$ vs $p_T$ for min. bias (grey bars depict systematic 
uncertainties)}
\label{fig:ppg020_ratio2}
\end{figure}

\begin{figure}[h!]
\begin{center}
\epsfig{file = 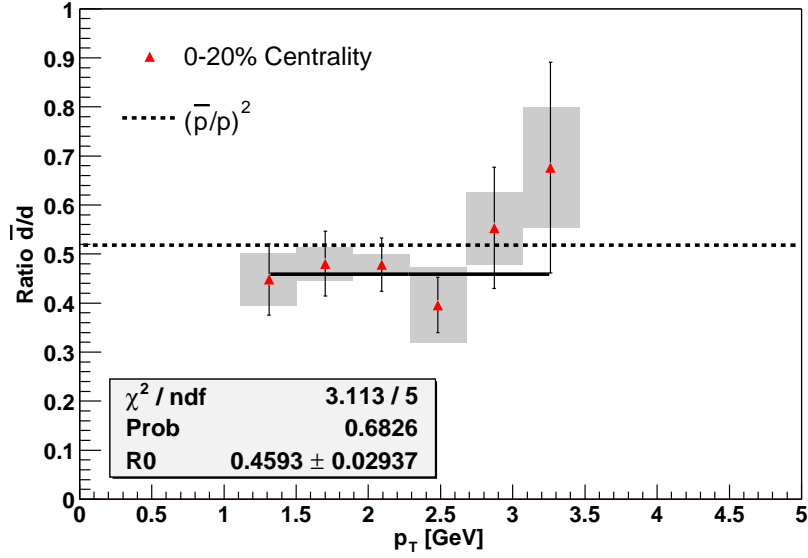, width = 12cm, clip = }
\end{center}
\caption{$\bar{d}/d$ vs $p_T$ for 0-20\% most central (grey bars depict 
systematic uncertainties)}
\label{fig:ppg020_ratio0}
\end{figure}

\begin{figure}[h!]
\begin{center}
\epsfig{file = 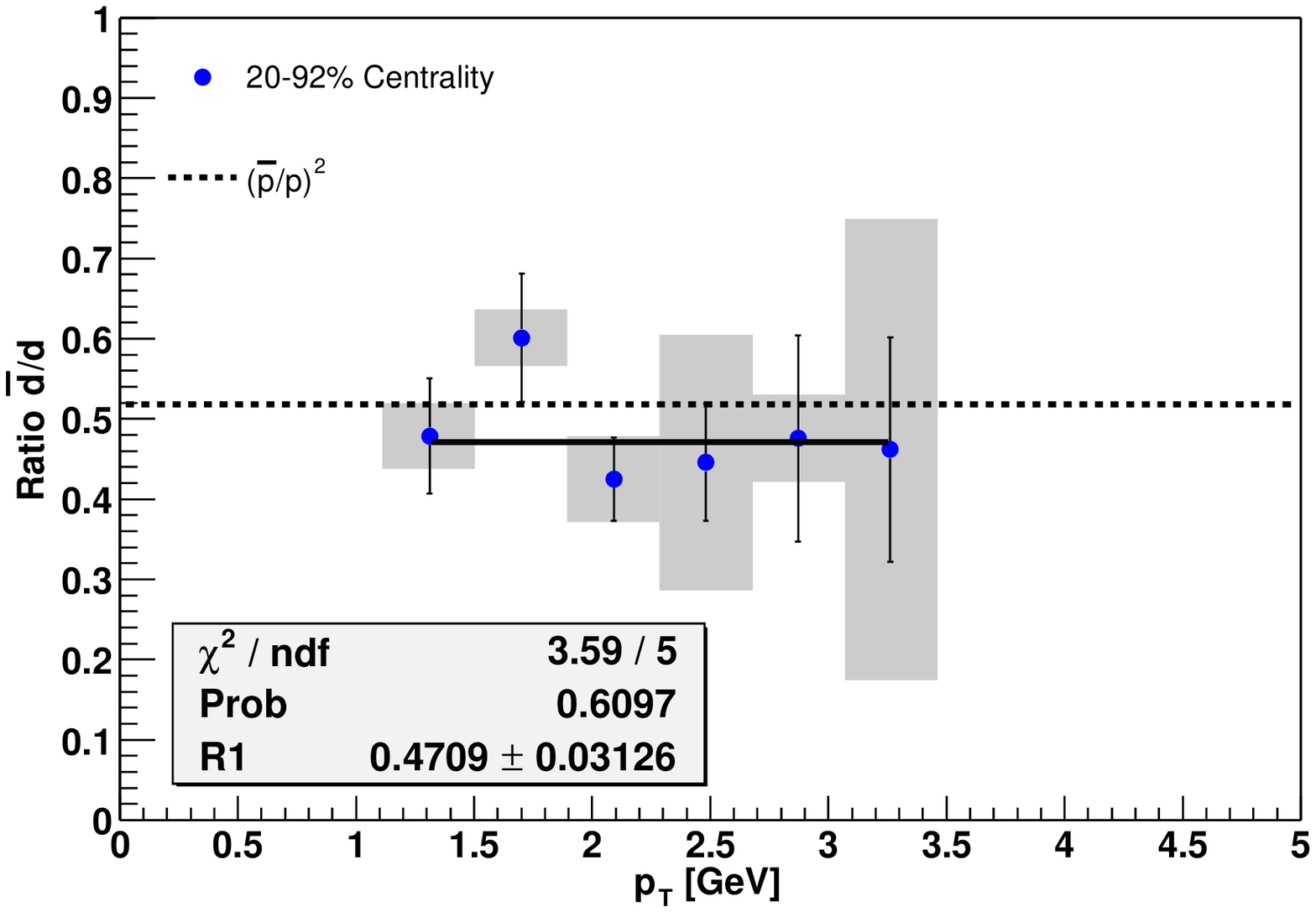, width = 12cm, clip = }
\end{center}
\caption{$\bar{d}/d$ vs $p_T$ for 20-92\% mid-central (grey bars depict 
systematic uncertainties)}
\label{fig:ppg020_ratio1}
\end{figure}

The systematic uncertainties were calculated by making same cuts as for the 
spectra as outlined in previous section, and then taking the $\bar{d}/d$ 
ratio. For details see Figures~\ref{fig:ratiosys_bin},
~\ref{fig:ratiosys_respar},~\ref{fig:ratiosys_byx},~\ref{fig:ratiosys_etof} 
and~\ref{fig:ratiosys_match}.

\begin{figure}[h!]
\begin{center}
\epsfig{file = 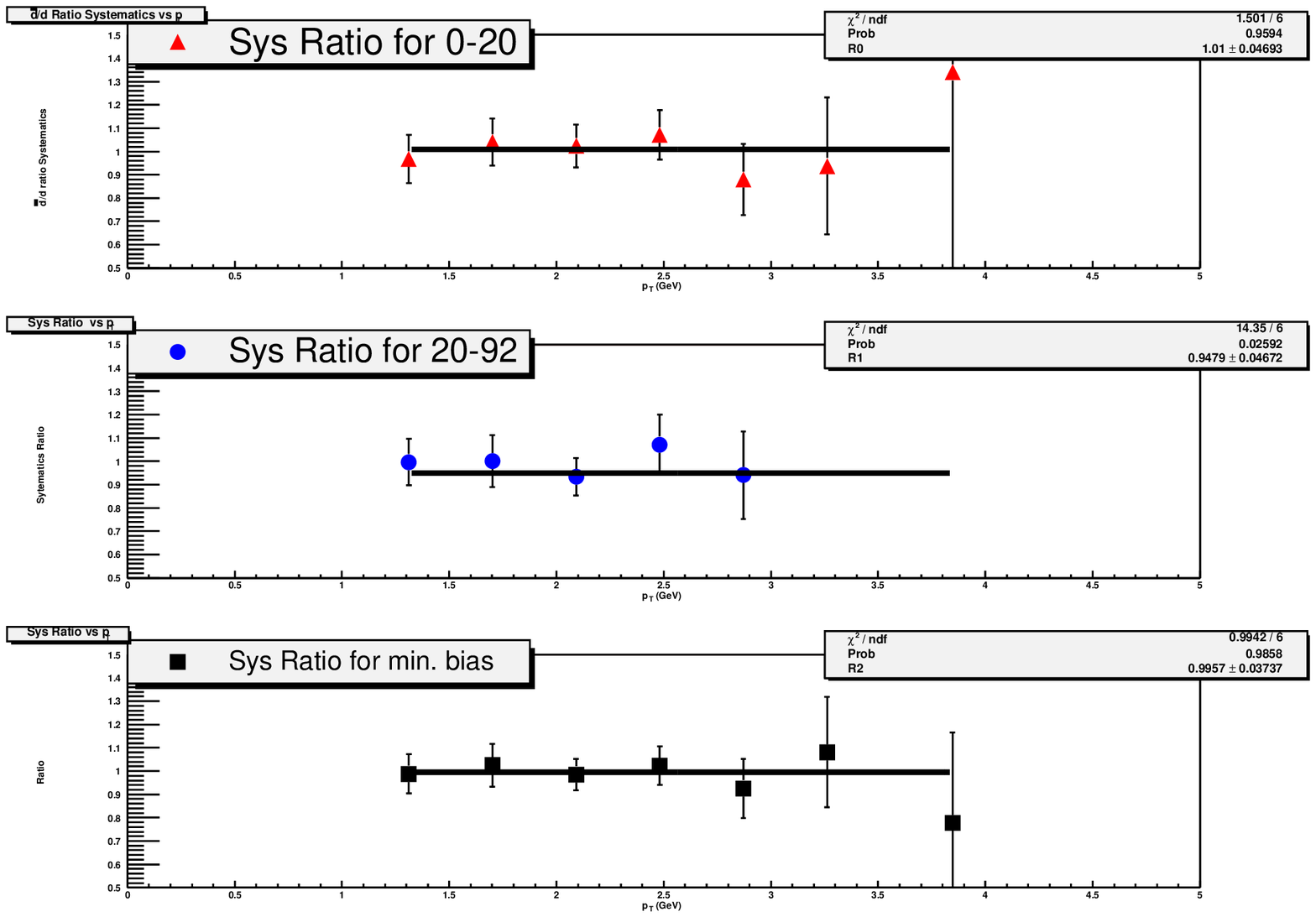, width = 12cm, clip = }
\end{center}
\caption{$\bar{d}/d$ ratio systematic uncertainty for PID (binning)}
\label{fig:ratiosys_bin}
\end{figure}

\begin{figure}[h!]
\begin{center}
\epsfig{file = 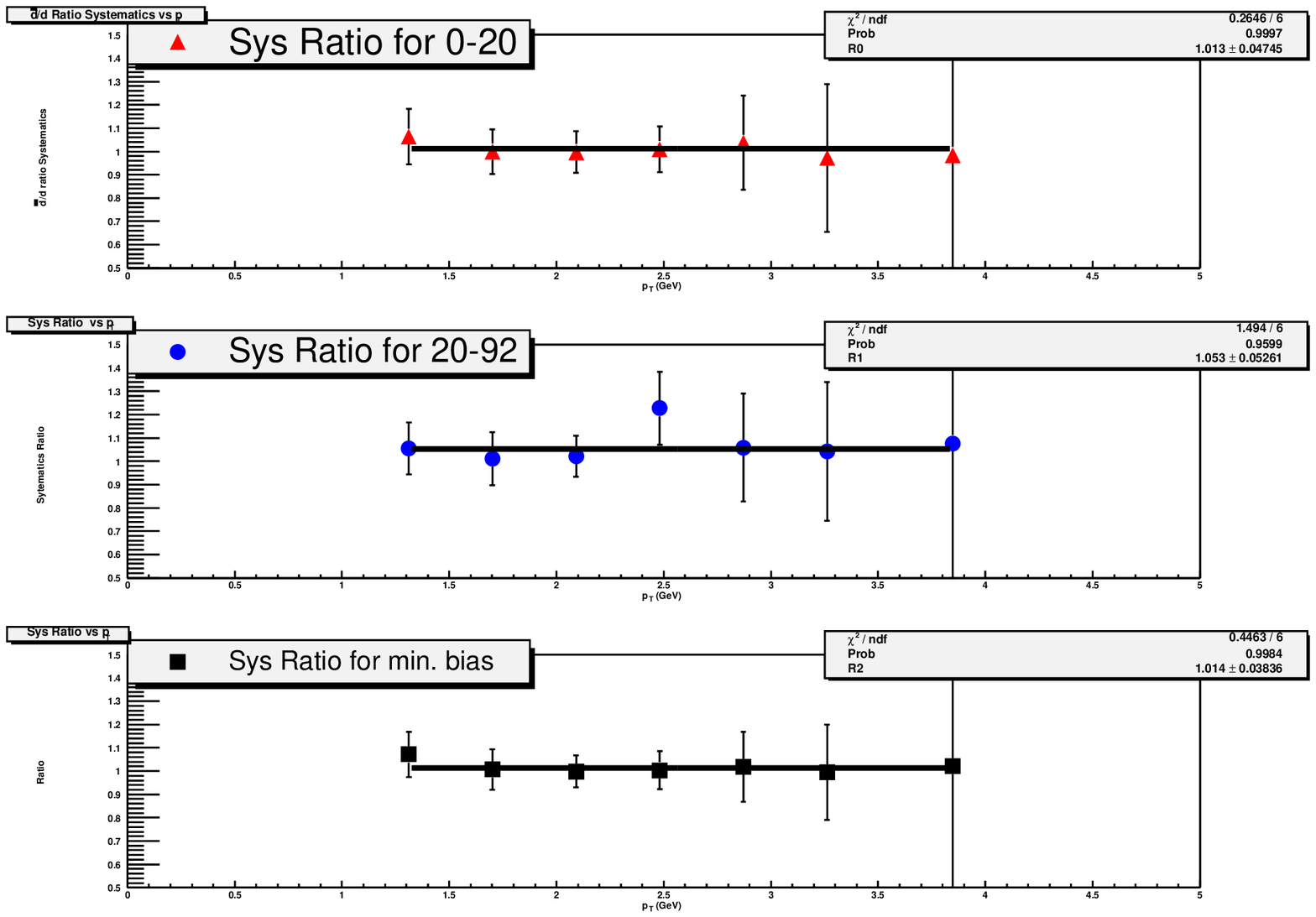, width = 12cm, clip = }
\end{center}
\caption{$\bar{d}/d$ ratio systematic uncertainty for PID 
(varying momentum resolution parameters)}
\label{fig:ratiosys_respar}
\end{figure}

\begin{figure}[h!]
\begin{center}
\epsfig{file = 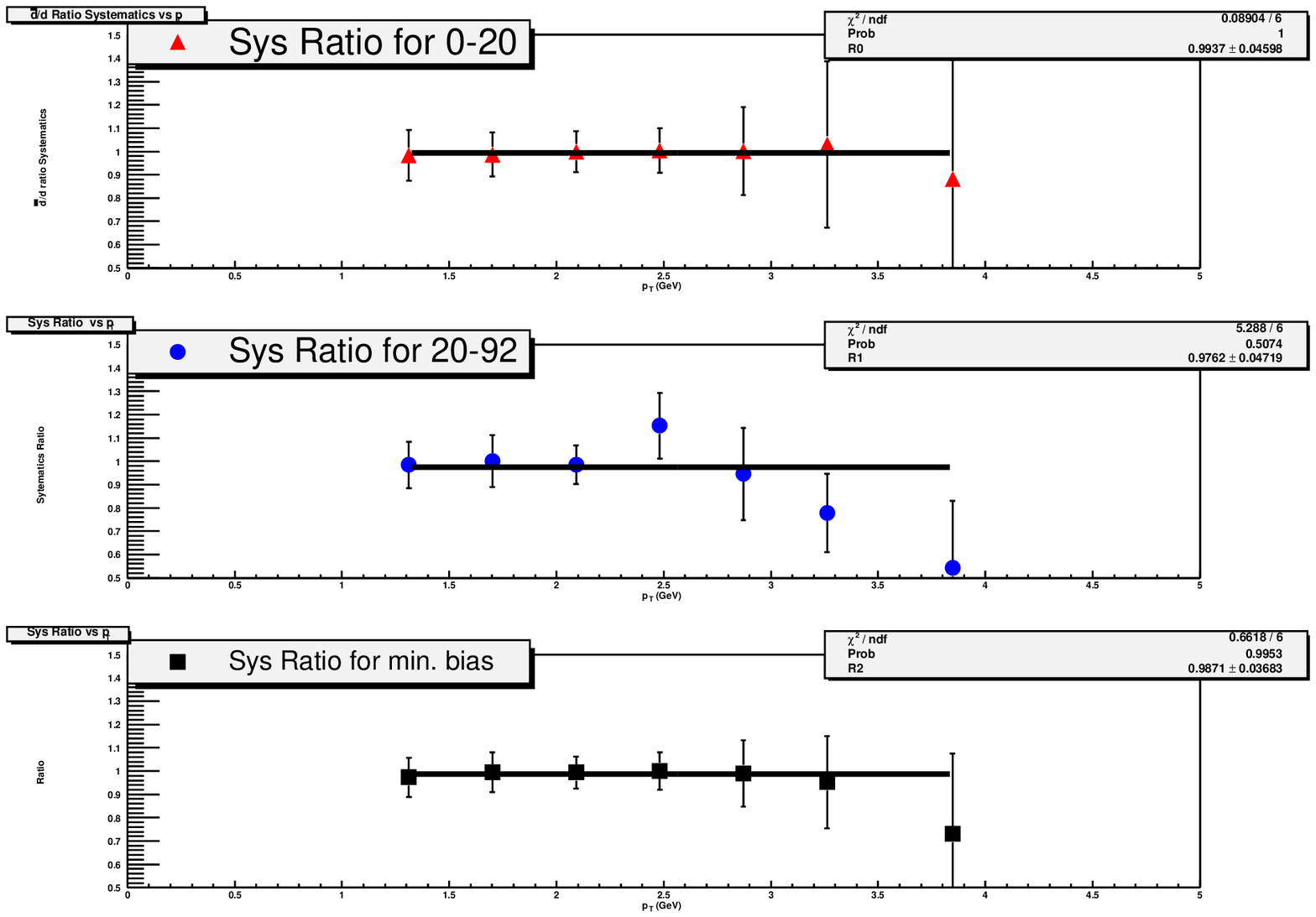, width = 12cm, clip = }
\end{center}
\caption{$\bar{d}/d$ ratio systematic uncertainty for PID 
(using $1/x$ function for background instead of $e^{-x}$)}
\label{fig:ratiosys_byx}
\end{figure}

\begin{figure}[h!]
\begin{center}
\epsfig{file = 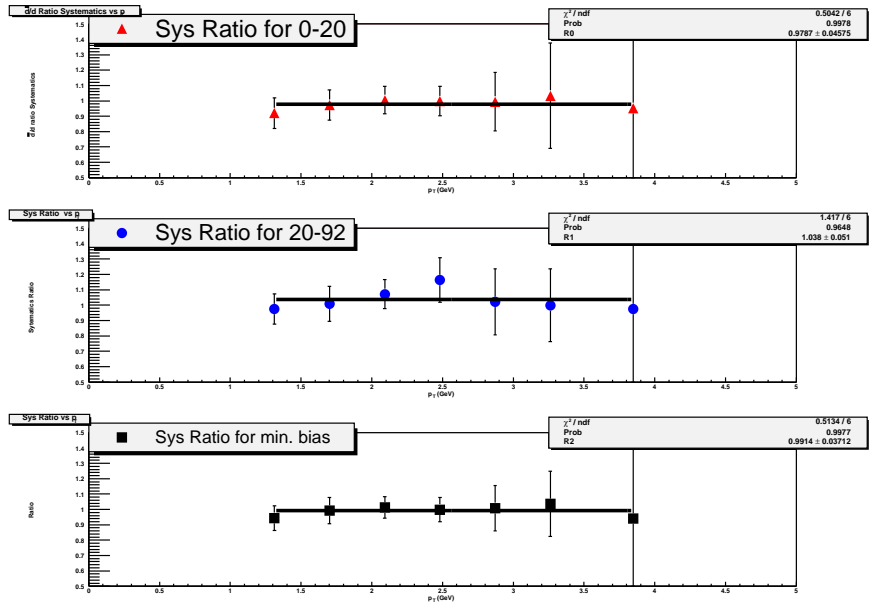, width = 12cm, clip = }
\end{center}
\caption{$\bar{d}/d$ ratio systematic uncertainty due to Etof cut.}
\label{fig:ratiosys_etof}
\end{figure}

\begin{figure}[h!]
\begin{center}
\epsfig{file = 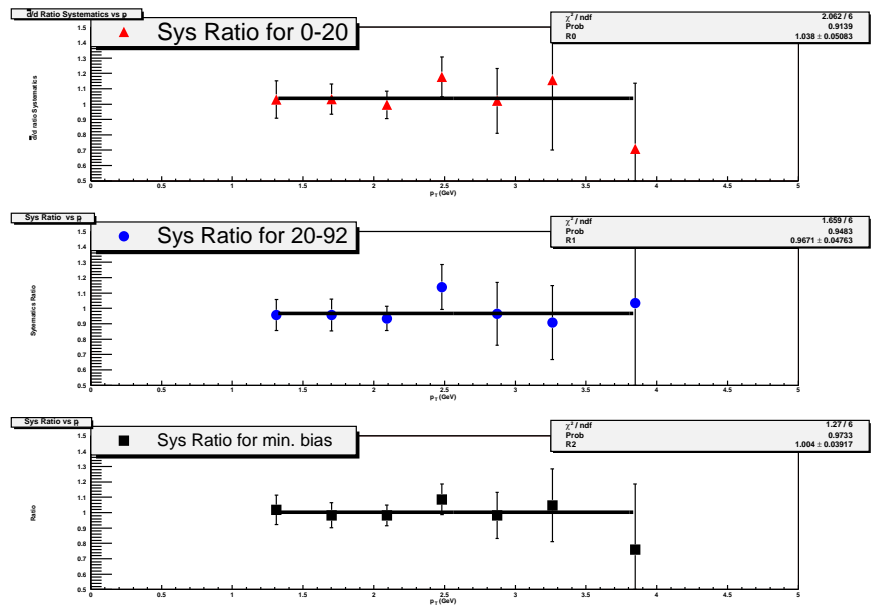, width = 12cm, clip = }
\end{center}
\caption{$\bar{d}/d$ ratio systematic uncertainty due to matching cuts.}
\label{fig:ratiosys_match}
\end{figure}

\begin{table}[h!]
\caption{Values of $\bar{d}/d$ ratio  at the mean of each $p_T$ bin for 
different centralities} 
\bcc
\begin{tabular}[ ]{|l|l|l|l|l|}
\hline
Centrality &$p_T$ [GeV]& $\bar{d}/d$ Ratio &Stat. Errors &Sys. Errors \\
\hline
	&	1.31193	&	0.448011	&	0.0723876	&	0.0501816	\\
	&	1.70166	&	0.480323	&	0.0661106	&	0.0284824	\\
	&	2.09137	&	0.478307	&	0.0543533	&	0.0118666	\\
0-20\% 	&	2.48116	&	0.395729	&	0.0564264	&	0.0759096	\\
	&	2.87113	&	0.553165	&	0.123666	&	0.0709565	\\
	&	3.26136	&	0.676205	&	0.214811	&	0.120164	\\
	&	3.83075	&	0.837902	&	0.435682	&	0.389956	\\
\hline				
	&	1.31193	&	0.478774	&	0.0716046	&	0.0366907	\\
	&	1.70166	&	0.601043	&	0.0797217	&	0.0267981	\\
	&	2.09137	&	0.424859	&	0.0518895	&	0.0511881	\\
20-90\%	&	2.48116	&	0.44557		&	0.0729878	&	0.158388	\\
	&	2.87113	&	0.475816	&	0.128523	&	0.0509976	\\
	&	3.26136	&	0.461791	&	0.139587	&	0.286625	\\
	&	3.83075	&	0.561234	&	0.379337	&	0.448365	\\
\hline								
	&	1.31193	&	0.452872	&	0.0627106	&	0.0441257	\\
	&	1.70166	&	0.523841	&	0.0616634	&	0.0168406	\\
	&	2.09137	&	0.449582	&	0.0424838	&	0.012414	\\
Min. Bias&	2.48116	&	0.432023	&	0.0493521	&	0.0386692	\\
	&	2.87113	&	0.510429	&	0.0921352	&	0.0406558	\\
	&	3.26136	&	0.555987	&	0.128219	&	0.0620898	\\
	&	3.83075	&	0.849898	&	0.380574	&	0.365119	\\
\hline

\end{tabular}
\ecc
\label{tab:ratios}
\end{table}

The ratio $\bar{n}/n$ can be estimated from the data based on the 
thermal chemical model. Assuming thermal and chemical equilibrium, 
the chemical fugacities are determined from the particle/anti-particle 
ratios~\cite{heinz_prc99}:
\begin{equation}
\frac{E_A(d^3N_A/d^3p_A)}{E_{\bar{A}}(d^3N_{\bar{A}}/d^3p_{\bar{A}})} = 
\exp\left(\frac{2\mu_A}{T}\right) = \lambda_A^2
\label{eq:fugacity}
\end{equation}

Using the ratio $p/\bar{p}$, the extracted proton fugacity 
is $\lambda_p = \exp(\mu_p/T)$ = 1.17 $\pm$ 0.01.
Similarly, using the $d/\bar{d}$ ratio, the extracted deuteron fugacity is 
$\lambda_d = \exp[(\mu_p+\mu_n)/T]$ = 1.46 $\pm$ 0.05.
From this, the neutron fugacity can be estimated to be
$\lambda_n = \exp(\mu_n/T)$ = 1.25 $\pm$ 0.04, which results in 
$\bar{n}/n$ = 0.64 $\pm$ 0.04. These estimates, along with equality of the
coalescence parameter $B_2$ (to be discussed in detail in later sections)
for $d$ and $\bar{d}$ indicate that, within errors, $\mu_n \geq \mu_p$. This 
is expected since the entrance Au+Au channel has larger 
net neutron density than net proton density. This effect is also seen in the
kaon ratio: $K^-/K^+ = 0.933 \pm 0.007$~\cite{ppg026}, which is slightly less 
than unity. Most particle ratios compare well with the thermal model 
predictions~\cite{braun_thermal}.

\clearpage
\section{$T_{eff}$, $dN/dy$ and $<p_T>$}

The $p_T$ shapes of the particle spectra an be quantified by looking at the
spectral slopes. It has been observed at lower beam energies, that particle
invariant yields often exhibit an exponential slope in the transverse mass 
$m_T$. This is parameterized in terms of the inverse slope parameter 
$T_{eff}$ as follows:
\be
\frac{d^2N}{2\pi N_{evt}m_tdm_tdy} = Ae^{-m_t/T_{eff}}
\label{eq:teff}
\ee

We fitted the above function to the (anti-)deuteron spectra in the range
$1.1<p_T<3.5$ GeV/$c$ for different centralities. The fits can be seen 
Figures~\ref{fig:ppg020_mt2},~\ref{fig:ppg020_mt0}, and ~\ref{fig:ppg020_mt1}. 
Deuterons are depicted by red squares, whereas anti-deuterons are depicted by 
blue triangles and the line overlaid on the spectra is the fit. 
The spectra have a shoulder arm shape characteristic of hydrodynamic
flow, which pushes heavier particles to higher $p_T$ as a result of 
interactions. The inverse slopes ($T_{eff}$) of the spectra are
tabulated in Table~\ref{tab:teff}. The deuteron inverse slopes of
$T_{eff}$ = 500--520 MeV are considerably higher than the $T_{eff}$ =
300--350 MeV observed for protons~\cite{ppg009,ppg026}. However, we also
observe that our spectra curve downwards at low $p_T$ and are not very well 
described by a simple $m_T$. This is indicates that source in
the collision zone is not static, but expanding. A more sophisticated
treatment incorporating hydrodynamical flow effects was developed by 
R. Scheibl and U. Heinz ~\cite{heinz_prc99}.

\begin{figure}[h!]
\begin{center}
\epsfig{file = 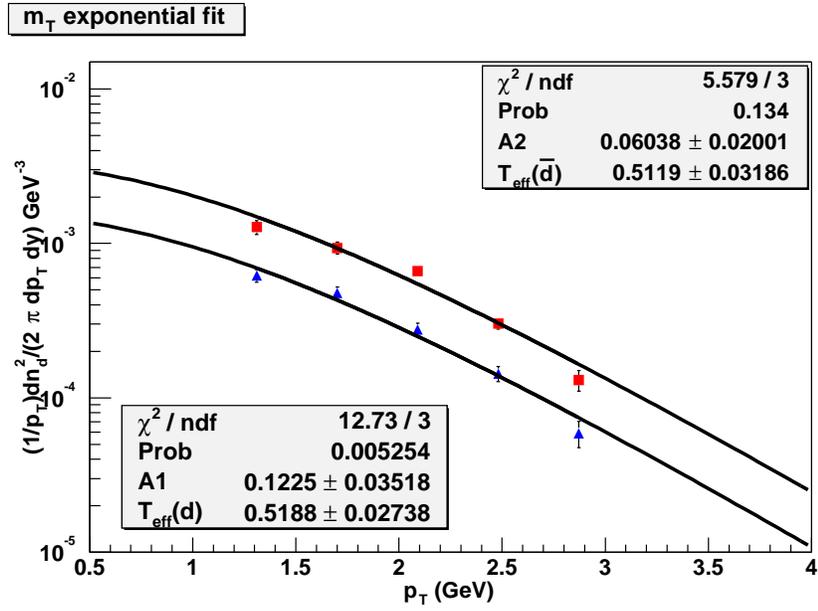, width = 12cm, clip = }
\end{center}
\caption{$T_{eff}$ fits using a $m_T$ exponential for min. bias data.}
\label{fig:ppg020_mt2}
\end{figure}

\begin{figure}[h!]
\begin{center}
\epsfig{file = 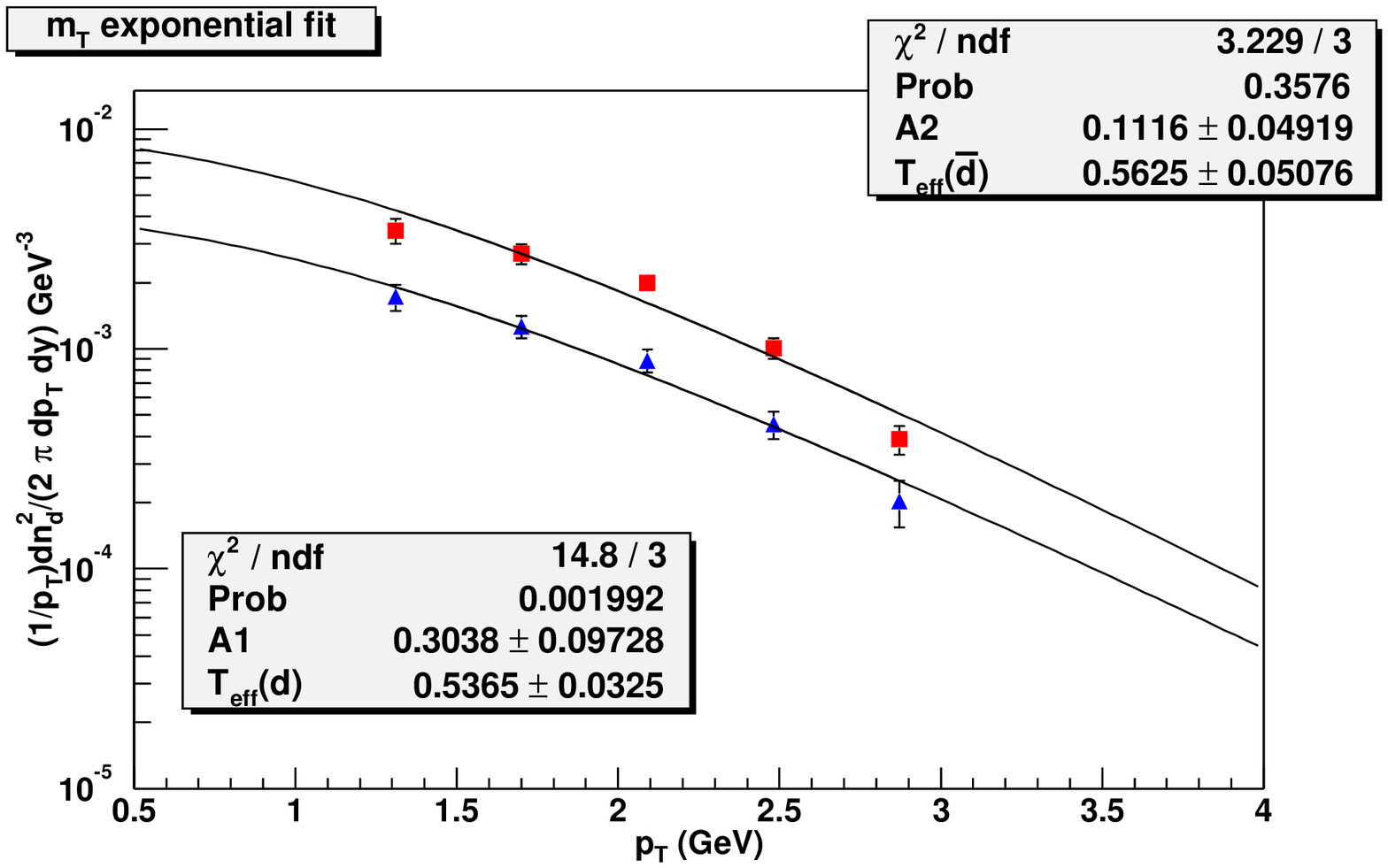, width = 12cm, clip = }
\end{center}
\caption{$T_{eff}$ fits using a $m_T$ exponential for 0-20\% centrality.}
\label{fig:ppg020_mt0}
\end{figure}

\begin{figure}[h!]
\begin{center}
\epsfig{file = 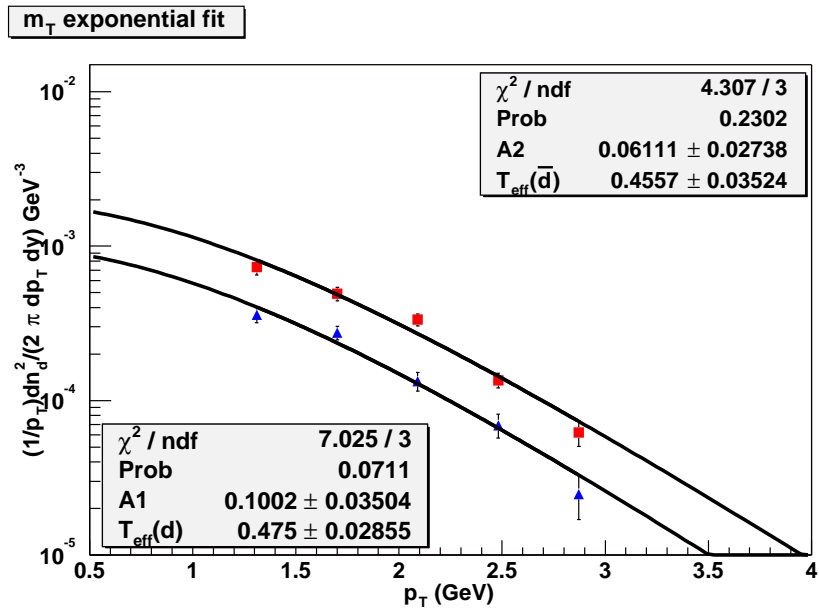, width = 12cm, clip = }
\end{center}
\caption{$T_{eff}$ fits using a $m_T$ exponential for 20-92\% centrality.}
\label{fig:ppg020_mt1}
\end{figure}

\clearpage

The invariant yields $dN/dy$ and the average transverse momenta 
$\langle p_T \rangle$ are obtained by summing the data over $p_T$ and using 
a functional form to extrapolate to low $m_T$ regions where we have no data. 
$dN/dy$ was calculated as follows:
\be
\frac{dN_d}{dy} = \int_0^\infty 2\pi p_T f(p_T)dp_T
\ee
where $f(p_T)$ is the function that gives us the yield as a function
of $p_T$:
\be
\frac{d^2N_d}{2\pi dp_T dy} = f(p_T)
\ee

In order to minimise our errors, we subdivided this integral into
three regions:
\benn
\item $p_T$ in the range 0 -- 1.1 GeV (the beginning of our experimental 
data), where we used an extrapolated function. For the extrapolation function
we used a Boltzmann distribution of the form:

\be
\frac{d^2n}{2\pi m_tdm_tdy} = Am_Te^{-m_t/T_{eff}}
\label{eq:boltzfit}
\ee
for the extrapolation fit as it gives a slightly better 
$\chi^2/n.d.f.=4.8/3$ vs. $\chi^2/n.d.f.=5.6/3$ for $m_T$ exponential. The 
fits are shown in Figures ~\ref{fig:ppg020_boltz2},
~\ref{fig:ppg020_boltz0},~\ref{fig:ppg020_boltz1}).

\item $p_T$ in the range 1.1 -- 4.3 GeV, for which we numerically integrated 
our data.
\item $p_T$ in the range 4.3 -- $\infty$ GeV, where we again extrapolated.
\eenn

\begin{figure}[h!]
\begin{center}
\epsfig{file = 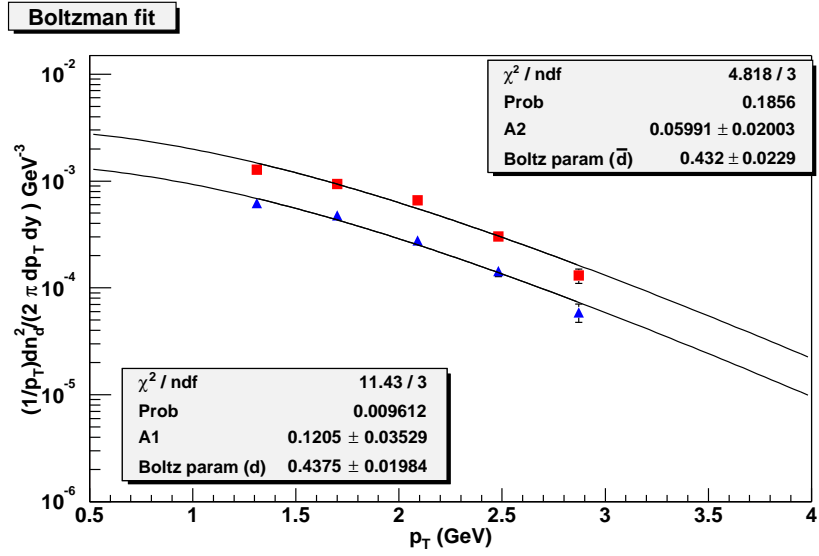, width = 12cm, clip = }
\end{center}
\caption{Fits using a Boltzmann Distribution for min. bias data.}
\label{fig:ppg020_boltz2}
\end{figure}

\begin{figure}[h!]
\begin{center}
\epsfig{file = 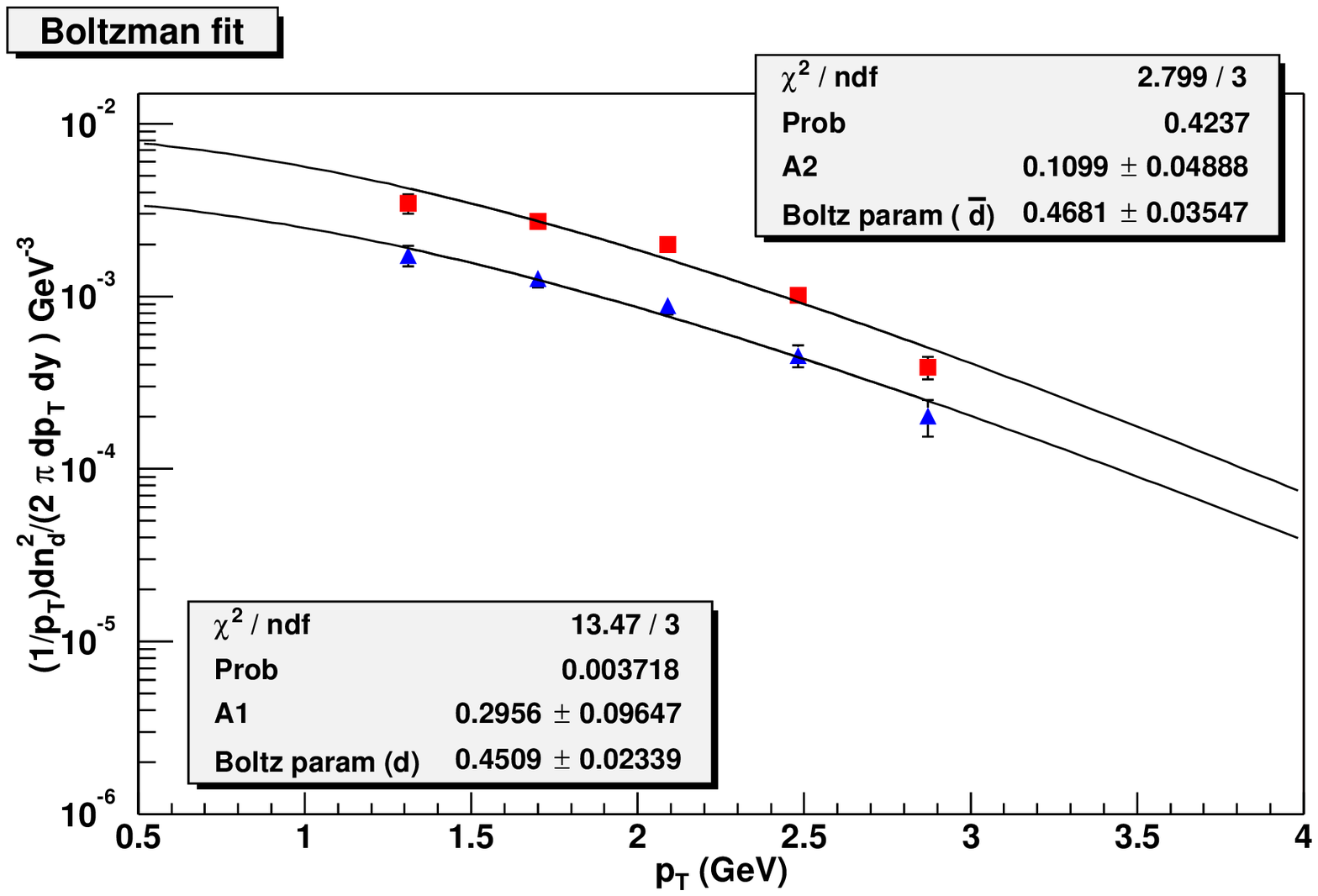, width = 12cm, clip = }
\end{center}
\caption{Fits using a Boltzmann Distribution for 0-20\% centrality.}
\label{fig:ppg020_boltz0}
\end{figure}

\begin{figure}[h!]
\begin{center}
\epsfig{file = 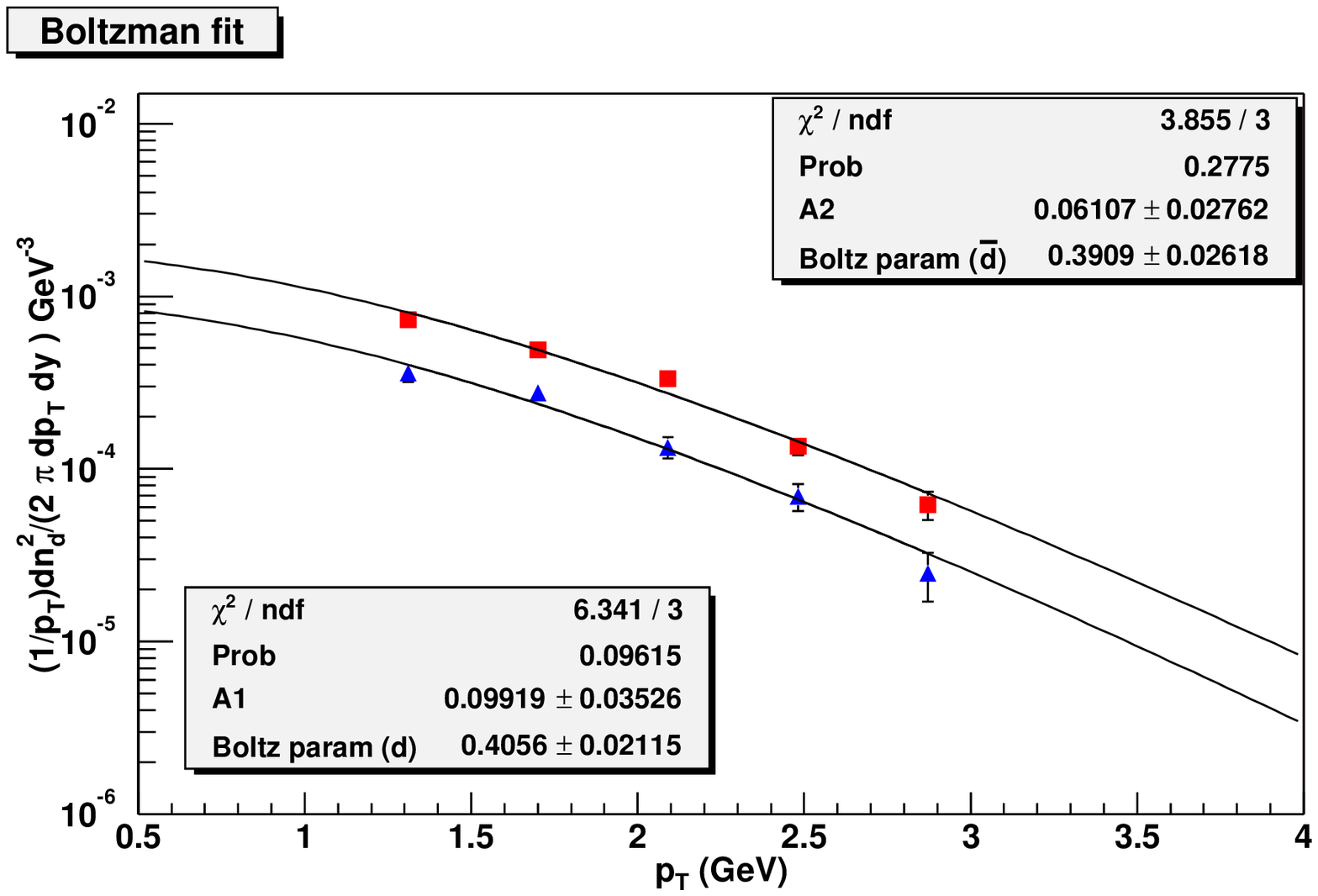, width = 12cm, clip = }
\end{center}
\caption{Fits using a Boltzmann Distribution for 20-92\% centrality.}
\label{fig:ppg020_boltz1}
\end{figure}

\begin{table}
\caption{\label{tab:teff} 
The inverse slope parameter $T_{eff}$ obtained from a $m_T$
exponential fit to the spectra along with multiplicity $dN/dy$ and mean 
transverse momentum $\langle p_T \rangle$ obtained from a Boltzman 
distribution for different centralities:}
\begin{tabular}{lll}
\hline
\textbf{$T_{eff}$ [MeV]}  & \textbf{Deuterons} & \textbf{Anti-deuterons} \\
\hline
Minimum Bias&519 $\pm$ 27	&512 $\pm$ 32\\
0-20\%	  &536 $\pm$ 32	&562 $\pm$ 51\\
20-92\%  &475 $\pm$ 29	&456 $\pm$ 35\\
\hline
\textbf{$dN/dy$}  &	  &	    \\
\hline
Minimum Bias&0.0250 $\pm^{0.0006(stat.)}_{0.005(sys.)}$	
&0.0117 $\pm^{0.0003(stat.)}_{0.002(sys.)}$\\
0-20\% &0.0727 $\pm^{0.0022(stat.)}_{0.0141(sys.)}$
&0.0336 $\pm^{0.0013(stat.)}_{0.0057(sys.)}$\\
20-92\% &0.0133 $\pm^{0.0004(stat.)}_{0.0029(sys.)}$	
&0.0066 $\pm^{0.0002(stat.)}_{0.0015(sys.)}$\\
\hline
\textbf{$\langle p_T \rangle$ [GeV/$c$]}&	  &	    \\
\hline
Minimum Bias&1.54 $\pm^{0.04(stat.)}_{0.13(sys.)}$
&1.52 $\pm^{0.05(stat.)}_{0.12(sys.)}$\\
0-20\%&1.58 $\pm^{0.05(stat.)}_{0.13(sys.)}$	
&1.62 $\pm^{0.07(stat.)}_{0.1(sys.)}$\\
20-92\%&1.45 $\pm^{0.05(stat.)}_{0.15(sys.)}$	
&1.41 $\pm^{0.06(stat.)}_{0.15(sys.)}$\\
\hline
\end{tabular}
\end{table}

 The extrapolated yields constitute $\approx$ 42\% of our total yields. 
$<p_T>$ was obtained in a similar manner, by taking the ratio of integrals:
\be
<p_T> = \frac{\int_0^\infty p_T^2 f(p_T)dp_T}{\int_0^\infty p_T f(p_T)dp_T}
\ee

Systematic uncertainties on $dN/dy$ and $\langle p_T \rangle$ are estimated 
by using an 
exponential in $p_T$ (instead of in $m_T$) and a ``truncated'' Boltzman 
distribution (assumed flat for $p_T < 1.1$ GeV/$c$) for alternative 
extrapolations. Systematic uncertaintiess on $dN/dy$ and $<p_T>$ were 
estimated by using two different functional forms:

\benn

\item $p_T$ exponential fit, of the form:
\be
\frac{d^2n}{2\pi m_tdm_tdy} = Ae^{-p_t/T_{const.}}
\ee
The $p_T$ exponential fits are shown in 
Figures~\ref{fig:ppg020_pt2},~\ref{fig:ppg020_pt0},~\ref{fig:ppg020_pt1}.
\begin{figure}[h!]
\begin{center}
\epsfig{file = 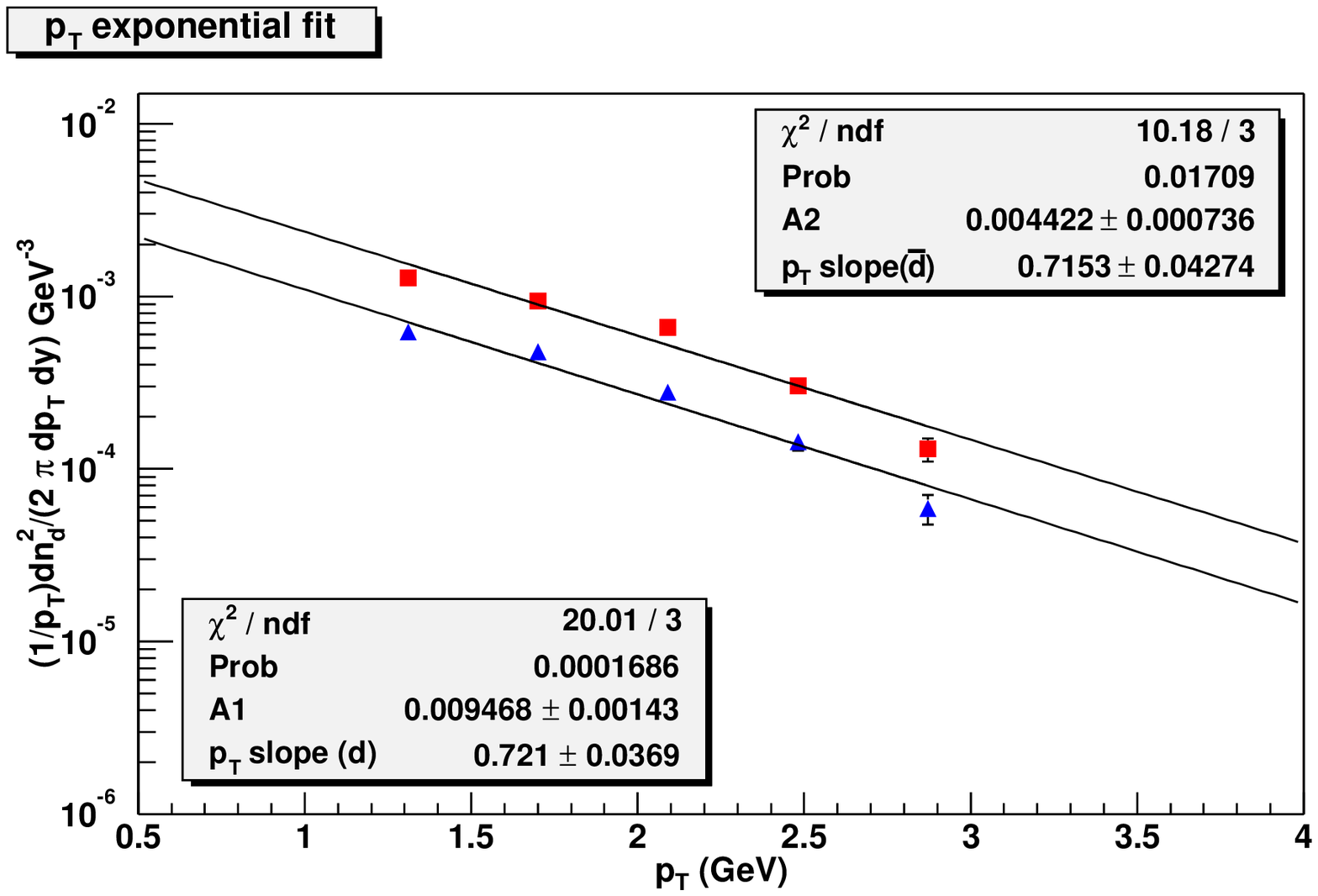, width = 12cm, clip = }
\end{center}
\caption{Fits using a $p_T$ exponential for min. bias data.}
\label{fig:ppg020_pt2}
\end{figure}

\begin{figure}[h!]
\begin{center}
\epsfig{file = 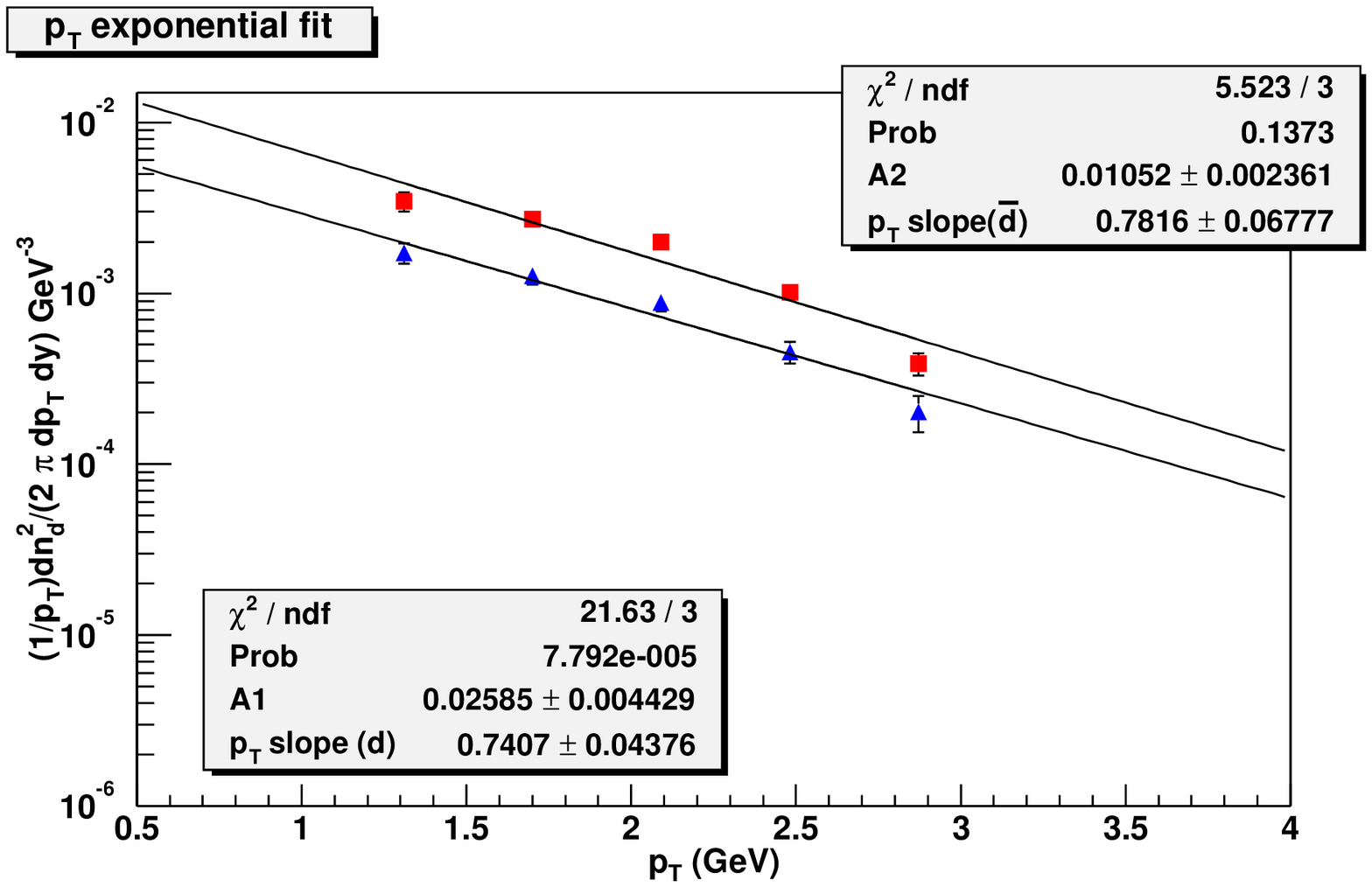, width = 12cm, clip = }
\end{center}
\caption{Fits using a $p_T$ exponential for 0-20\% centrality.}
\label{fig:ppg020_pt0}
\end{figure}

\begin{figure}[h!]
\begin{center}
\epsfig{file = 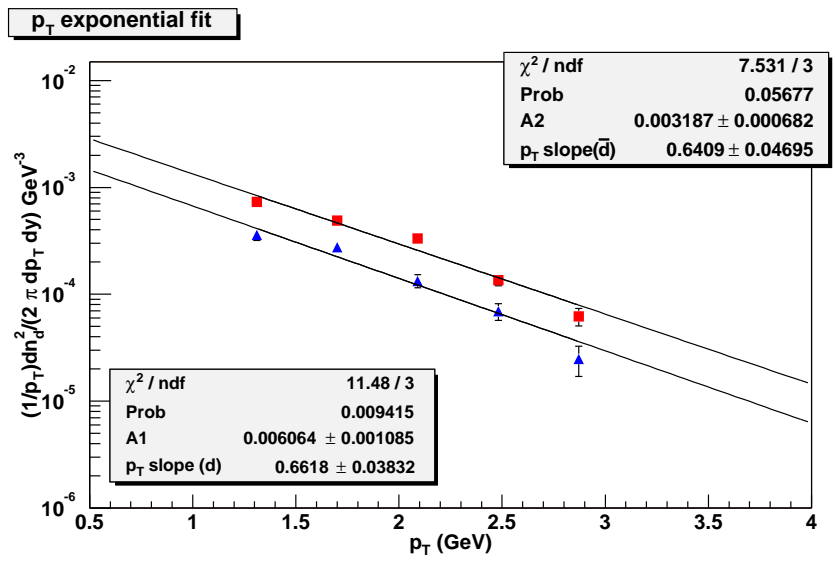, width = 12cm, clip = }
\end{center}
\caption{Fits using a $p_T$ exponential for 20-92\% centrality.}
\label{fig:ppg020_pt1}
\end{figure}
\clearpage
\item Truncated Boltzmann fit, in which we assume a flat distribution for
$p_T<1.1$ GeV (see Figures~\ref{fig:ppg020_tb2},~\ref{fig:ppg020_tb0},
~\ref{fig:ppg020_tb1}):
\begin{figure}[h!]
\begin{center}
\epsfig{file = 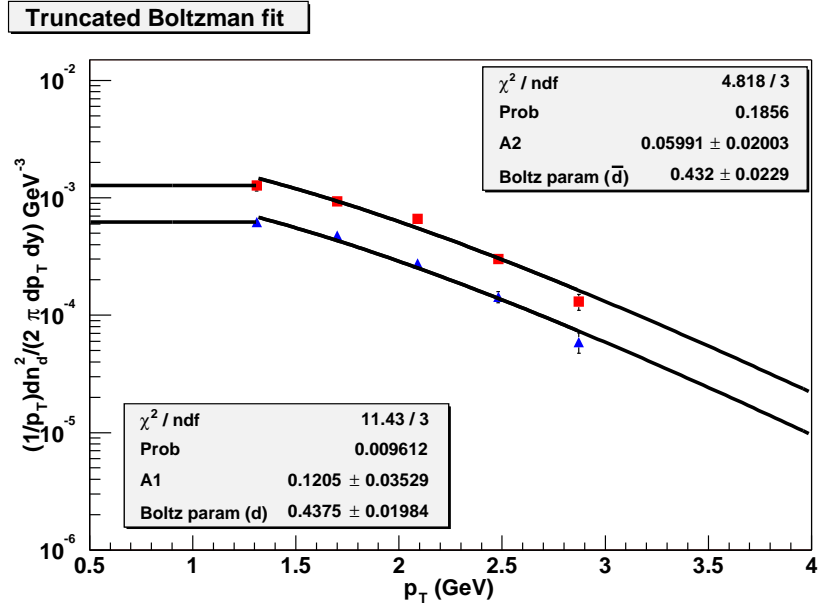, width = 12cm, clip = }
\end{center}
\caption{Fits using a Truncated Boltzmann Distribution for min. bias data.}
\label{fig:ppg020_tb2}
\end{figure}

\begin{figure}[h!]
\begin{center}
\epsfig{file = 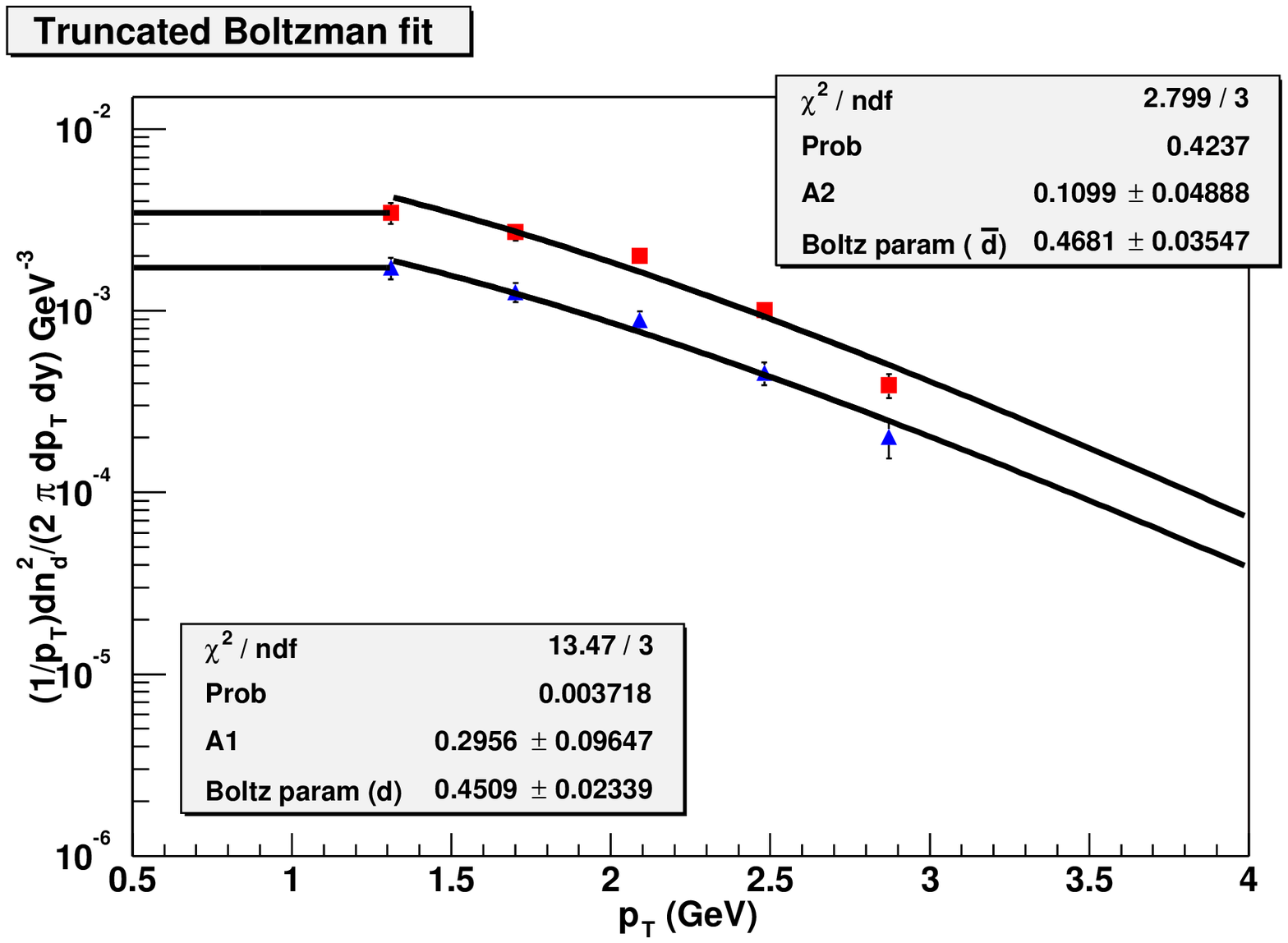, width = 12cm, clip = }
\end{center}
\caption{Fits using a Truncated Boltzmann Distribution for 0-20\% centrality.}
\label{fig:ppg020_tb0}
\end{figure}

\begin{figure}[h!]
\begin{center}
\epsfig{file = 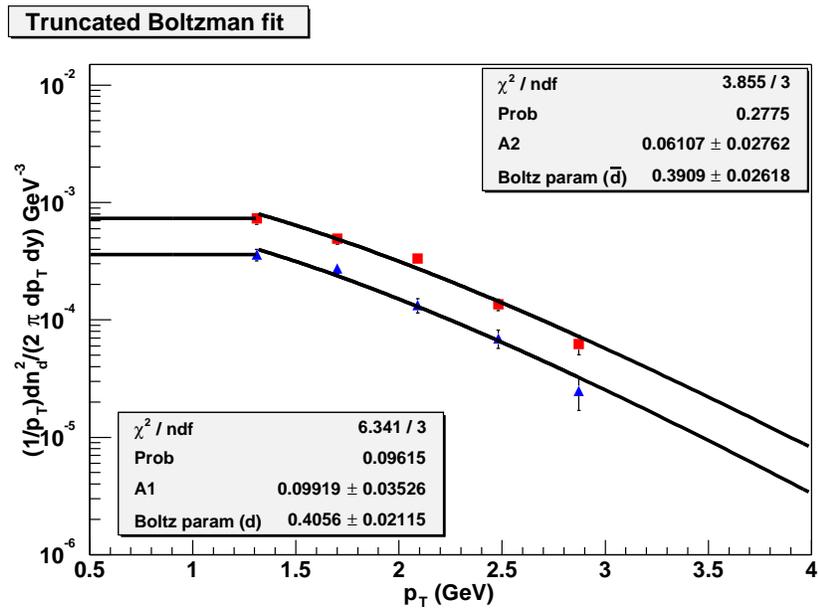, width = 12cm, clip = }
\end{center}
\caption{Fits using a Truncated Boltzmann Distribution for 20-92\% centrality.}
\label{fig:ppg020_tb1}
\end{figure}
\eenn  

 The inverse slope parameter $T_{eff}$, rapidity distributions $dN/dy$, and 
the mean transverse momenta $\langle p_T \rangle$, are compiled 
in Table~\ref{tab:teff} for three different centrality bins.

\clearpage

\section{Coalescence parameter $B_2$}

With a binding energy of 2.24 MeV, the deuteron is a very loosely bound
state. Thus, it is formed only at a later stage in the collision, most
likely after elastic hadronic interactions have ceased; the proton and
neutron must be close in space and tightly correlated in velocity to
coalesce. As a result, $d$ and $\bar{d}$ yields are a sensitive measure of
correlations in phase space and can provide information about the
space-time evolution of the system. If deuterons are formed by coalescence
of protons and neutrons, the invariant deuteron yield can be
related~\cite{butler} to the primordial nucleon yields by:
\begin{equation}
E_d\frac{d^3N_d}{d^3p_d}\biggr|_{p_d=2p_p} = B_2\left(E_p\frac{d^3N_p}{d^3p_p}\right)^2
\label{eq:coal}
\end{equation}
where $B_2$ is the coalescence parameter, with the subscript implying that
two nucleons are involved in the coalescence. The above equation includes
an implicit assumption that the ratio of neutrons to protons is unity. The
proton and antiproton spectra~\cite{ppg026} are corrected for feed-down
from $\Lambda$ and $\bar{\Lambda}$ decays by using a MC simulation tuned
to reproduce the particle ratios: ($\Lambda/p$ and
$\bar{\Lambda}/\bar{p}$) measured by PHENIX at 130 GeV~\cite{lambda130}.

We calculated $B_2$ by taking the scaled ratio of the deuteron and
anti-deuteron  spectra with the square of the spectra of the proton and 
anti-proton spectra, for each $p_T$ bin in comparable centralities. The data
for the proton and anti-proton spectra was taken from published PHENIX
papers~\cite{ppg026}. Systematical uncertainties are mostly same as for the
$d$, $\bar{d}$ spectra, except that the $p$, $\bar{p}$ spectra have an 
additional uncertainty due to feeding down from $\Lambda$, $\bar{\Lambda}$
decays, which leads to an uncertainty of 10.2\% in $B_2$.

\begin{figure}[thb]
\includegraphics[width=1.0\linewidth]{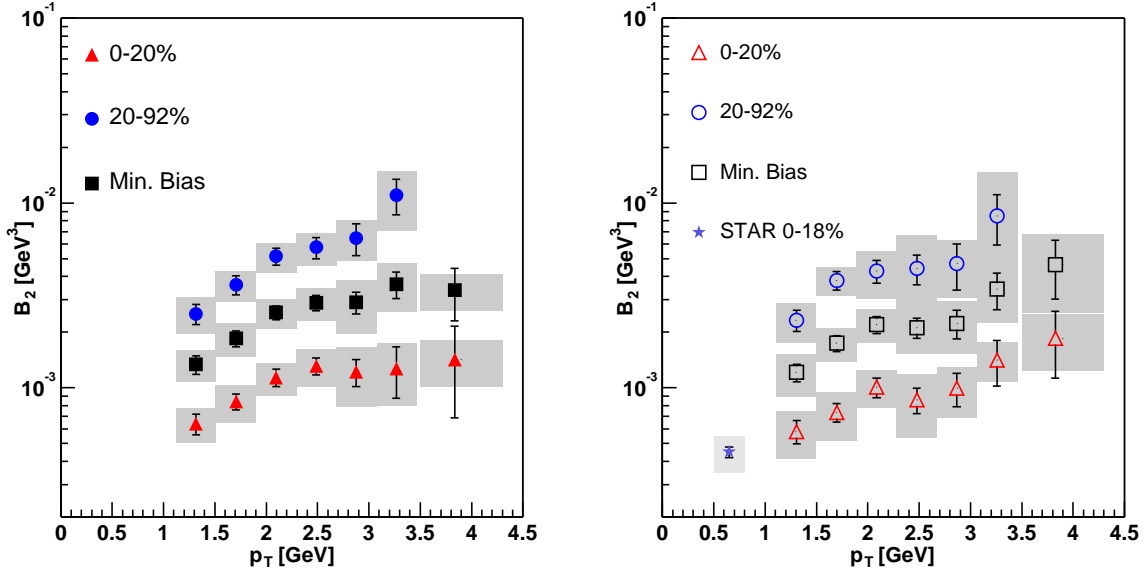}  
\caption{\label{fig:ppg020_fig3}
Coalescence parameter $B_2$ vs $p_T$ for deuterons (left panel) and anti-deuterons (right panel).
Grey bands indicate the 
systematic errors. Values are plotted at the ``true'' mean value of $p_T$ of 
each bin, the extent of which is indicated by the width of the grey bars along
x-axis.}
\end{figure}

Figure~\ref{fig:ppg020_fig3} displays the coalescence parameter $B_2$ as a
function of $p_T$ for different centralities (the values
are given in Table~\ref{tab:b2param}). We notice some important trends:

\begin{table}[h!]
\caption{Coalescence parameter $B_2$ for different centralities} 
\bcc
\begin{tabular}[]{|l|l|l|l|}
\hline
Centrality & $p_T$ [GeV]&$B_2$ [GeV$^2$/c$^3$] (deuterons) &$\bar{B_2}$ [GeV$^2$/c$^3$] (anti-deuterons) \\
\hline
	&	1.31193	&	0.000639055	$\pm$	7.50E-05	&	0.000581546	$\pm$	8.32E-05	\\
	&	1.70166	&	0.000843443	$\pm$	8.91E-05	&	0.000736467	$\pm$	8.50E-05	\\
	&	2.09137	&	0.00113993	$\pm$	9.86E-05	&	0.00100924	$\pm$	0.000121727	\\
0-20\% 	&	2.48116	&	0.00131118	$\pm$	0.000132695	&	0.000862111	$\pm$	0.000137338	\\
	&	2.87113	&	0.00122166	$\pm$	0.000192384	&	0.00099636	$\pm$	0.000203503	\\
	&	3.26136	&	0.00127308	$\pm$	0.000279434	&	0.00141501	$\pm$	0.000393006	\\
	&	3.83075	&	0.00141949	$\pm$	0.000521348	&	0.00185868	$\pm$	0.000731906	\\

\hline
	&	1.31193	&	0.00251278	$\pm$	0.000315446	&	0.00232644	$\pm$	0.000304454	\\
	&	1.70166	&	0.00361427	$\pm$	0.000426224	&	0.00381079	$\pm$	0.00044324	\\
	&	2.09137	&	0.00514855	$\pm$	0.000552146	&	0.00427978	$\pm$	0.000604291	\\
20-90\%	&	2.48116	&	0.00574697	$\pm$	0.000756338	&	0.00442243	$\pm$	0.000807512	\\
	&	2.87113	&	0.00644265	$\pm$	0.00124428	&	0.00469217	$\pm$	0.0013238	\\
	&	3.26136	&	0.0109891	$\pm$	0.00239219	&	0.00849642	$\pm$	0.002594	\\
	&	3.83075	&	0.00950758	$\pm$	0.00417112	&	0.00858541	$\pm$	0.00581748	\\

\hline
	&	1.31193	&	0.00133957	$\pm$	0.000153907	&	0.00121209	$\pm$	0.000132676	\\
	&	1.70166	&	0.00185072	$\pm$	0.000187566	&	0.0017466	$\pm$	0.000174225	\\
	&	2.09137	&	0.00255682	$\pm$	0.000216065	&	0.00219478	$\pm$	0.00023407	\\
Min. Bias.	&	2.48116	&	0.00288005	$\pm$	0.000275283	&	0.00211784	$\pm$	0.000267098	\\
	&	2.87113	&	0.00290036	$\pm$	0.000386863	&	0.00223415	$\pm$	0.000397725	\\
	&	3.26136	&	0.00364134	$\pm$	0.000592348	&	0.00341505	$\pm$	0.000760932	\\
	&	3.83075	&	0.00336927	$\pm$	0.00105975	&	0.00463725	$\pm$	0.00162501	\\
\hline
\end{tabular}
\ecc
\label{tab:b2param}
\end{table}

\benn
\item The decreased $B_2$ in more
central collisions implies that in larger sources, the average relative
separation between nucleons increases, thus decreasing the probability of
formation of deuterons.  
\item We observe that $B_2$ increases with $p_T$. This
is consistent with an expanding source because position-momentum correlations
lead to a higher coalescence probability at larger $p_T$.  The $p_T$-dependence
of $B_2$ can also provide information about the density profile of the source as
well as the expansion velocity distribution. It has been
shown~\cite{heinz_prc99,polleri} that generally a Gaussian source density profile leads
to a constant $B_2$ with $p_T$ as it gives greater weight to the center of the
system, where radial expansion is weakest.  This is not supported by our data,
which shows a rise in $B_2$ with $p_T$.

\begin{figure}[thb]
\includegraphics[width=1.0\linewidth]{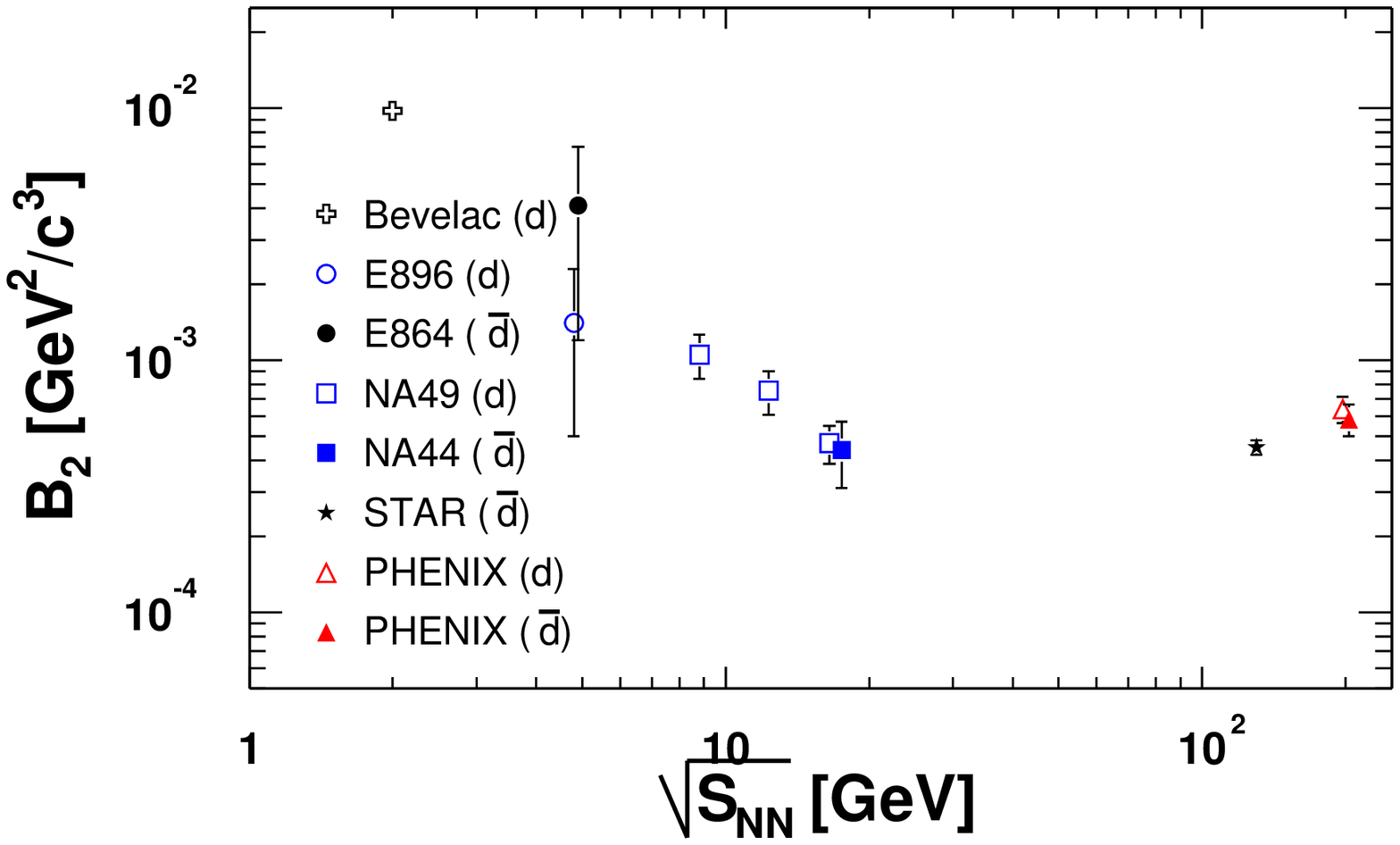}  
\caption{\label{fig:ppg020_sqsplot} (color online)
Comparison of the coalescence parameter for deuterons
and anti-deuterons ($p_T$ = 1.3 GeV/$c$) with other experiments at different 
values of $\sqrt{s}$.}
\end{figure}

\item Figure~\ref{fig:ppg020_sqsplot} compares $B_2$ for most central 
collisions to results at lower 
$\sqrt{s}$~\cite{eos,e896,e864,na49prc,na44prl,star}. 
Note that $B_2$ is nearly independent of $\sqrt{s}$, 
indicating that the source volume does not change appreciably with 
center-of-mass energy (with the caveat that $B_2$ 
varies as a function of $p_T$, centrality and rapidity). Similar behavior is
seen for $B_2$ for deuterons~\cite{na49prc} as a function of $\sqrt{s}$. 
This observation is consistent with what has been observed in Bose-Einstein 
correlation Hanbury-Brown Twiss (HBT) analysis at RHIC~\cite{hbt130,starhbt130}
for identified particles. The coalescence parameter $B_2$ for $d$ and 
$\bar{d}$, is equal within errors, indicating that nucleons and 
antinucleons have the same temperature, flow and freeze-out 
density distributions. 

\item  Thermodynamic models~\cite{mekjian} 
predict that $B_2$ scales with the inverse of the effective volume $V_{eff}$
($B_2 \propto 1/V_{eff}$). 
$d$ and $\bar{d}$ spectra are affected by radial flow, which 
concentrates the coalescing protons and neutrons, affecting
phase space correlations, thereby limiting the applicability of 
a simple thermodynamical model to determine an effective source size.
$B_2$ can also be used to obtain a 
source radius, analogous to a two particle Bose-Einstein
correlation~\cite{heinz_prc99} measurement. Although the 
``correct'' physical interpretation of $B_2$ is still sometimes
debated, thermodynamic models can be used to extract the radius
of the source from $B_2$, albeit in a model 
dependent way. For a fireball model in thermal and chemical 
equilibrium~\cite{csernai,mekjian}, the following 
relation holds:

\begin{equation}
R^3 = \alpha R_{np}(\hbar c)^3\frac{m_d}{m_p^2}B_2^{-1}
\label{eq:rcoal}
\end{equation}

where $\alpha = (3/4)\pi^{3/2}$ for a gaussian source and and $\alpha = 
(9/2)\pi^{2}$ for a hard sphere and $R_{np}$ is the ratio of neutrons to 
protons (assumed to be unity here). Assuming a gaussian distributed particle 
source, we find $R = 4.9 \pm 0.2$ fm for the $0-20\%$ most central 
collisons for deuterons at $p_T$ = 1.3 GeV (equivalent to proton momentum of
$p_p$ = 0.65 GeV). It should 
be noted that deuteron spectra are affected by radial flow, which 
concentrates the coalescing protons and neutrons, affecting
phase space correlations and limiting the applicability of the simple 
Eq.~\ref{eq:rcoal}.

\item The coalescence parameter $B_2$ for $d$ and 
$\bar{d}$, is equal within errors, indicating that nucleons and 
antinucleons have the same temperature, flow and freeze-out 
density distributions. This, alongwith the values of $\bar{d}/d$ ratios 
indicate that, within errors, $\mu_n \geq \mu_p$. This is 
expected since the entrance Au+Au channel has larger 
net neutron density than net proton density. 
\eenn

\chapter{Nuclear modification factor and the initial conditions}

In the previous chapters, by looking at the yields of deuterons and 
anti-deuterons, formed from the dense system of particles from the collision
zone, as it expanded and cooled, we were able study the final state effects
in Ultra Relativistic Heavy Ion Collisions of Au+Au nuclei at 
$\sqrt{s_{NN}}=200$ GeV. In this chapter we shift to the second part of this 
thesis: studying the initial state effects using hadronic probes. 

\section{Parton Distribution Functions}

In the 1960s, scattering  experiments were conducted at
the Stanford Linear Accelerator Center (SLAC) in which very high energy 
electron beams were fired on protons. This experiment was similar to 
Rutherford's classic experiment of bombarding a thin gold foil with 
$\alpha$-particles, which revealed
that the atom consisted of a small massive nucleus, which had most of the 
atom's positive charge, and was surrounded by a cloud of electrons. The SLAC 
experiments found that more electrons were scattered with large momentum 
transfers
than expected. This indicated the presence of discrete scattering centers 
inside the proton, very much the same way large scattering angles of the 
$\alpha$-particles indicated the existence of a small and massive nucleus. 
Moreover, the distribution of the scattering data showed `scale-invariance', 
which indicated that these scattering
centers were `point-like' i.e., they did not have any substructure (at least at
the energy scale they were being probed). These were given the generic name: 
partons. This lead to the developement of the parton model by 
Feynman~\cite{hadron} (and also by Bjorken and Paschos~\cite{bjork69}), in 
which the nucleon was envisaged to consist of essentially free point-like 
constituents, the ``partons'', from which the electron scatters incoherently. 
At a more quantitative level 
they also introduced the concept of parton distributions $q_i(x)$, which 
measure the probability of finding a parton of type $i$ in the proton, with a 
momentum fraction $x$ of the proton momentum. In the parton model, these 
parton distributions are independent of the energy scale at which the proton 
is being probed. The data also indicated that the charged scattering centers 
were spin $\frac{1}{2}$ fermions. Subsequently, they were identified with the 
quarks proposed by Murray Gell-Man~\cite{gell64}. Later experiments with 
neutrinos also supported this view. Parton distributions (PDFs) are essential 
for a detailed understanding of the nucleon structure as well as experiments 
involving hadronic initial states, and they evolve as one goes from one 
momentum scale (or equivalently length scale) to another. From a given initial
distribution, it is possible to calculate the evolution of PDFs using the 
framework of the DGLAP evolution equations~\cite{dglap}. The partons in the nucleon 
are either:
\bitm
\item Valence quarks: $u,d$-quarks usually with large momentum fractions. 
Valence parton distribution functions (PDFs) peak around 1/3 and go to zero at
momentum fraction of $x$ of 0. A typical PDF for valence quarks at $Q^2=100$ 
GeV shown in Figure~\ref{fig:cmpval}. 
\item Sea partons consisting of gluons, and quark-antiquark pairs. These 
usually have a softer distribution and diverge at small-$x$. A typical PDF for
gluons at $Q^2=100$ GeV shown in Figure~\ref{fig:cmpgl}. This increase in gluon density
has been observed at HERA~\cite{hera} for high $Q^2$ (momentum transfer) and small-$x$.
\eitm

\begin{figure}[h!]
\bcc
\includegraphics[width=0.75\linewidth]{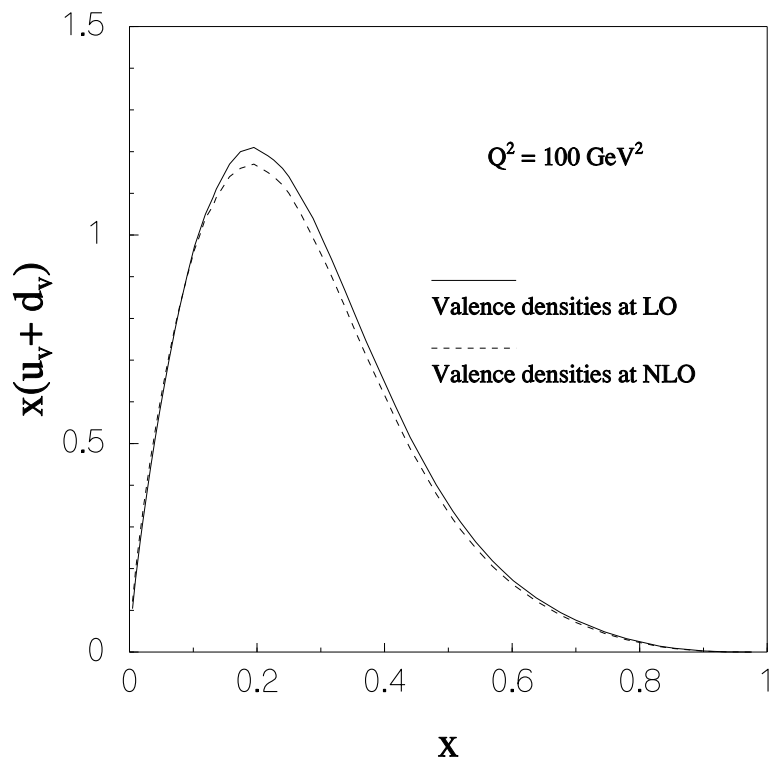}  
\ecc
\caption{A typical parton distribution function (PDF) for valence quarks at 
$Q^2=100$ GeV.}
\label{fig:cmpval}
\end{figure}

\begin{figure}[h!]
\bcc
\includegraphics[width=0.75\linewidth]{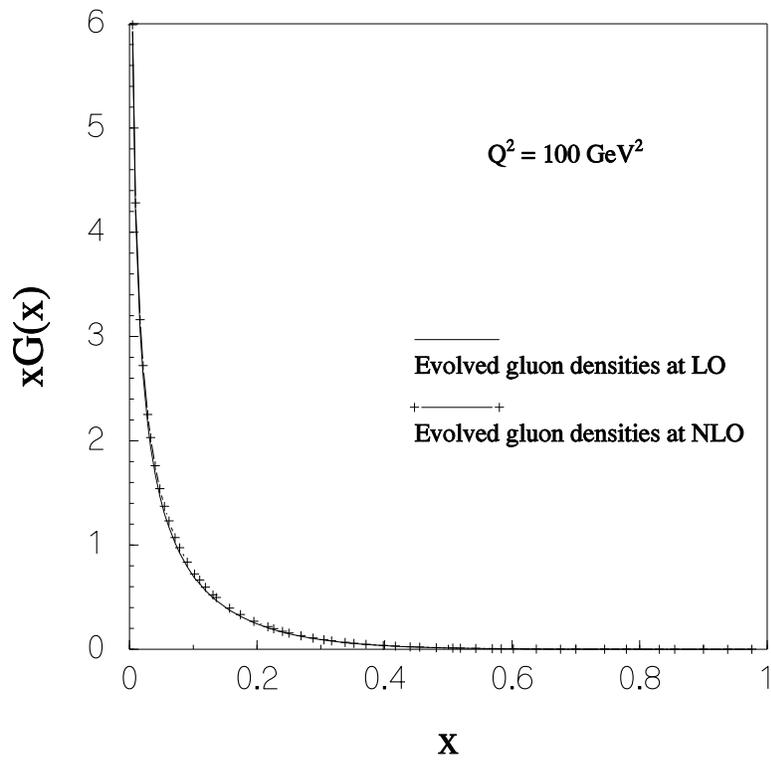}  
\ecc
\caption{A typical PDF for gluons at $Q^2=100$ GeV.}
\label{fig:cmpgl}
\end{figure}
\clearpage
\section{Hard Scattering}

In high energy collisions hard scatterings can produce jets of particles. A
schematic depiction of jet production via hard scattering is shown in
Figure~\ref{fig:hardscattering}. The production of energetic high $p_T$ 
particles depends on the distribution of the scattering centers i.e., quarks
(valence and sea) and gluons. As a result study of jet production can shed
light into PDFs.
\begin{figure}[h!]
\bcc
\includegraphics[width=0.5\linewidth]{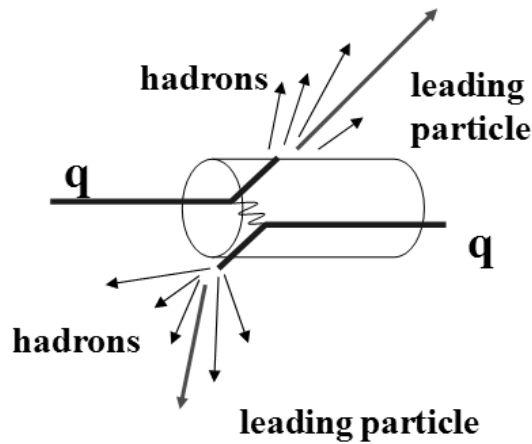}  
\ecc
\caption{A schematic depiction of jet production through hard scattering.}
\label{fig:hardscattering}
\end{figure}

In previous section we saw how the gluon (and sea quark) PDFs diverge at 
small-$x$. However, this seems to be contrary to needs of unitarity principle.
The non-Abelian nature of QCD and the fact that gluons themselves carry color
``charge'' can lead to gluon fusion processes of type $g+g\rightarrow g$. 
This can lead to the depletion of the small-$x$ partons in a nucleus compared 
to those in a nucleon. This phenomenon is known as shadowing~\cite{shadowing}.
Naively shadowing can be understood from the uncertainty principle: small-$x$
quarks and gluons can spread over a distance comparable to the nucleon-nucleon
separation, leading to a spatial overlap and fusion. The depletes the number
of partons in small-$x$ region (shadowing) and increasing the density of 
high momentum partons (anti-shadowing). The increasing gluon fusion processes 
at small-$x$ can lead the gluon PDFs to stop increasing (gluon saturation). 
The Color Glass Condensate theory~\cite{cgc_klm} gives a universal QCD 
explanation for the low-$x$ shadowing. 

\section{Nuclear modification factor: $R_{cp}$}

An experimentally simple and convenient way to study particle production is to
look at the ratio $R_{cp}$ of the particle yield in central collisions with 
the particle yield in peripheral collisions, each normalized by number of 
nucleon nucleon inelastic scatterings ($N_{coll}$). Also known as the nuclear 
modification factor, $R_{cp}$ is defined as:
\begin{equation}
R_{cp} = \frac{\left( \frac{dN}{d\sigma dp_T} \right)^{Central}/N_{coll}^{Central}}{\left(\frac{dN}{d\sigma dp_T}\right)^{Peripheral}/N_{coll}^{Peripheral}}
\label{eq:rcp}
\end{equation}
A Glauber model~\cite{glauber} and BBC simulation was used to obtain 
$N_{coll}$. $R_{cp}$ is what we get divided by what we expect. If particle 
production in Heavy Ion Collisions was
simply an incoherent sum of p+p collisions, then $R_{cp}$ would be unity.
Any deviation from unity would indicate a different kind of physics. Thus, we 
assume that peripheral collisions are similar to p+p collisions, allowing us
to normalise $R_{cp}$ independent of the p+p reference spectrum. As a result 
many systematics related to detector efficiencies cancel, leading to a 
relatively clean measurement with minimal systematics. 

Measurement of the $R_{cp}$ variable, has yielded some of the most interesting 
results at RHIC. At mid-rapidity, $R_{cp}$ was observed to be suppressed at 
Au + Au collisions at $\sqrt{s_{NN}} = 200$ 
GeV~\cite{phenix_130supre,phenix_130cent,star_130supre,phenix_200pi}. This 
could be explained in either in terms of:
\benn
\item Jet Suppression: energy loss of energetic particles due to dense 
partonic matter formed as a result of deconfinement. This energy loss 
$\simeq$ GeV/fm, and is mostly due to gluon bremmstrahlung processes. This 
results in a decrease in the yield of high energy particles or jet 
suppression~\cite{jetsup_wang,jetsup_vitev,gyulassy_jetsup}. This would be 
mean that
the observed suppression in Au+Au collisions at RHIC is a final state effect.
\item Color Glass Condensate (CGC)~\cite{cgc_klm,gluonsat1,gluonsat2}: at 
sufficiently high energies and low $x$-values (momentum fraction of nucleon 
carried by the parton) gluon fusion processes can deplete the number of 
scattering centers. This can lead to lower particle multiplicities and hence 
suppression in $R_{cp}$. This would mean that the observed suppression is an 
initial state effect.
\eenn

In order to determine whether the observed suppression in $R_{cp}$ was due
to initial state effects (CGC or gluon saturation) or due to final state
effects (deconfinement leading to jet suppression) a control experiment was 
performed at RHIC, using deuteron on gold collisions at the same energy 
$\sqrt{s}=200$ GeV. By comparing results from d+Au collisions with those from 
Au+Au collisions we can attempt to distinguish between effects that could 
potentially be due to deconfinement, versus effects of cold nuclear matter. 
No suppression in $R_{cp}$ was observed in the d+Au collisions at 
mid-rapidity~\cite{phenix_nosup}, instead an enhancement was observed. This 
seems to indicate that that the observed suppression at mid-rapidity in Au+Au
collisions was likely due to final state effects. This enhancement is referred 
to as Cronin effect and is theorized to be a result of initial state 
multiple scattering of the partons.

\section{$R_{cp}$ and rapidity}

Although the d+Au results seem to be
inconsistent with the CGC hypothesis, there is a possibility that the 
saturation effects might be observable at forward and backward rapidities at
RHIC. It turns out that a new regime of parton physics at small-$x$ can be 
reached by going to large rapidities. By looking at $R_{cp}$ at 
forward (and backward) rapidities we can probe momentum fraction $x$, which
can be related to the rapidity and tranvese energy of the particle by the 
following relations:

\be
x=\frac{M_T}{\sqrt{s}}e^{-y}
\label{eq:x_au}
\ee

\be
x=\frac{M_T}{\sqrt{s}}e^{y}
\label{eq:x_d}
\ee

Due to their forward and backward rapidity coverages, the PHENIX Muon Arms 
(described in more detail in the next chapter) can be used to probe regimes
of both small and large-$x$. The North Arm (d going direction) probes low-$x$ 
partons from the Au nucleus, allowing us to probe the saturation/shadowing 
region, while the South Arm (Au going direction) probes high-$x$ partons from 
Au, allowing us to study the anti-shadowing/Cronin regime. 

Our present analysis seeks to
extend the $R_{cp}$ measurement for charged hadrons to forward rapidity
(approximate pseudorapidity range $1.2<\eta<2.0$) using the Phenix Muon Arms.
Any variation of $R_{cp}$ from Deuteron going side to the gold going side
from the nominal value of unity would have important implications. Naively 
one would expect $R_{cp}$ at forward rapidities to behave not very differently
from mid-rapidity and show a Cronin enhancement for both Gold going direction 
and for the Deuteron going direction with a slightly greater increase in the 
Deuteron going direction. However, shadowing effects at small-$x$ (where
$x$ is the fraction of the nucleon momentum carried by the parton) could 
lead to a suppression on the Deuteron going side. CGC hypothesis too, would 
lead to a decrease in the number of scattering centers causing suppression.
Hence, in order to get a deeper understanding of the
perennial interplay between Cronin-like enhancements due to multiple 
scattering and depletion of scattering centers due to shadowing-like effects,
we need to measure $R_{cp}$ at forward and backward rapidities.

\chapter{Hadron identification using Muon Arms}

\section{PHENIX Muon Arms}
The PHENIX detector has two muon arms: North and South. South Muon Arm 
acceptance covers from -2.2 to -1.2 in pseudo-rapidity $\eta$ and full 
azimuth, whereas the North Muon Arm acceptance covers from
1.2 to 2.4 in $\eta$ and full azimuthal acceptance. Each arm is designed to
track and identify muons, while rejecting pions and kaons of the order of
$\approx 10^{-3}$. Each muon arm is a radial magnetic field spectrometer with
a  muon tracker (MuTr) and a muon identifier (MuID) consisting of layers of 
absorber and tracking. In addition, the pole tips of the Central Magnet (CM)
made of steel and brass, in the apertures of the muon north and south arms, 
act as hadron absorbers to reject pions and kaons. The MuTr consists of three 
stations  of cathode-strip readout tracking chambers mounted inside 
conical-shaped muon magnets. For an illustrative sketch see 
Figure~\ref{fig:mms_sketch}. 

\begin{figure}[h]
\includegraphics[width=1.0\linewidth]{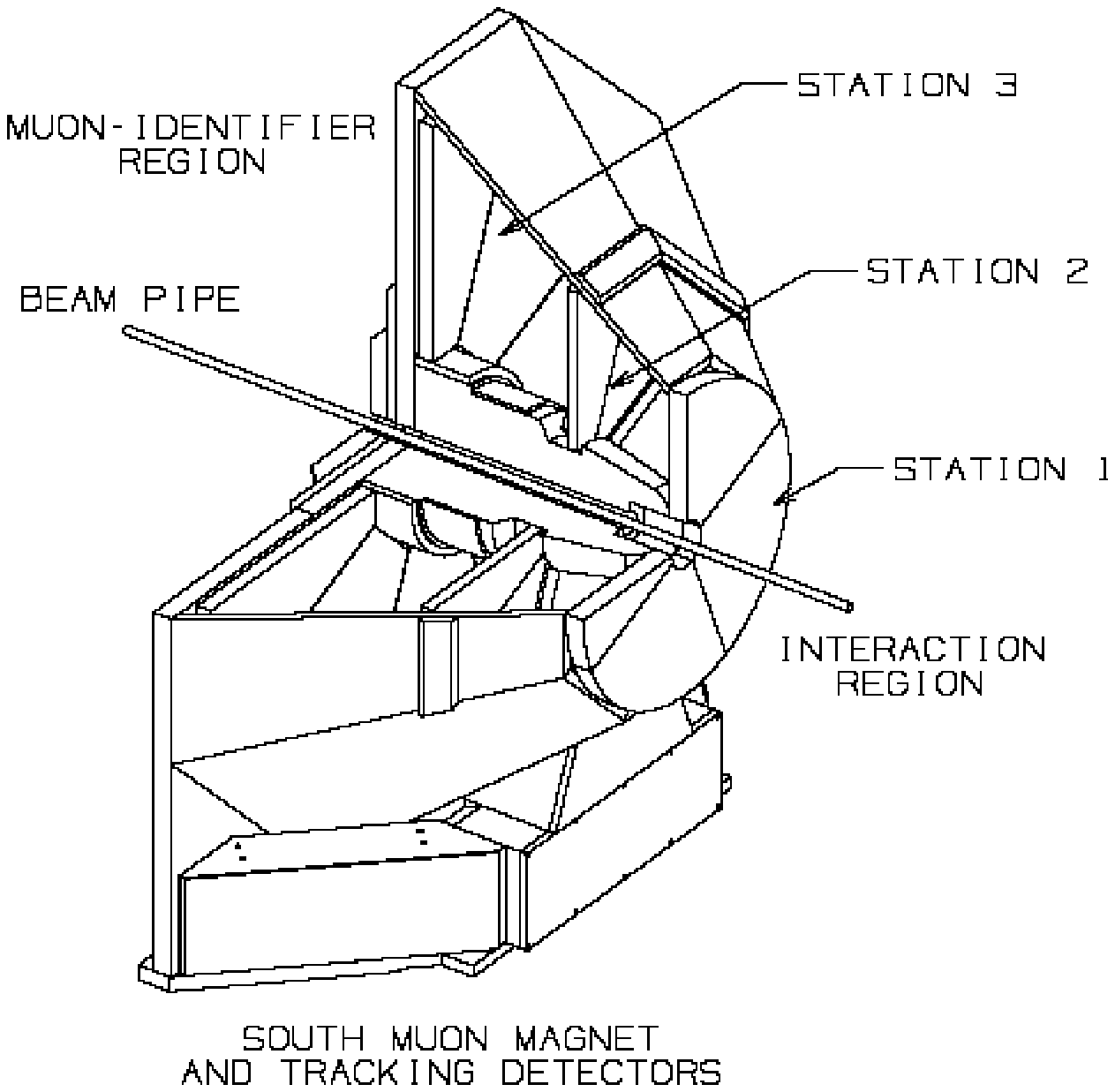}  
\caption{The Muon South Arm: muons from the collision point, travel into the 
Station 1 and so on, eventually reaching the MuID plates behind(not shown in
this figure).}
\label{fig:mms_sketch}
\end{figure}

The Station 1 tracking chambers are located closest to the interaction region
and therefore are the smallest (approximately 1.25 m from inside radius to
outside radius), and have the highest occupancy per strip. This also leads to
a stringent requirement of a minimum of 95\% active area within the acceptance.
Station 3 is the last chamber before the MuID and is 2.4 m long and 2.4 m wide.
The chamber uses a gas mixture comprising of $50\% Ar + 30\% CO_2 + 20\% CF_4$ 
and operates a typical high voltage (HV) of 1850 V, with a gain of 
$\approx 2\times 10^4$. A minimum ionizing particle (MIP) is assumed to 
deposit 100 
electrons, which leads to a total cathode charge of 80 fC. In order to 
maintain a good momentum resolution down to 1.5 GeV/$c$, the thickness at the 
Station 2 detector was required to be $\leq 0.1\%$ of a radiation length. This
was done by making the Station 2 cathodes (in octant shape) of etched 25 
micron copper coated mylar foils. 

A track measured in MuTr is identified as a muon if it hits the MuID. A side
view of the Muon Arms is shown in Figure~\ref{fig:muonarm_side}. To
reject pions of upto 4 GeV/$c$, steel of depth 90 cm (5.4 hadronic interaction
lengths) is required. Since the muon magnet backplate is 30 cm, we need an 
additional 60 cm in the MuID. This is implemented by segmenting the absorber
into 4 layers of thicknesses 10, 10, 20 and 20 cm. This type of segmentation
can improve the measurement of trajectories in the MuID. The 5 gaps between
the steel segments are instrumented with Iarocci tubes, which are planar drift
tubes consisting of 100 $\mu$m gold-coated CuBe anode wires at the center of 
long channels of a graphite-coated plastic cathode. Groups of Iarocci tubes are
arrayed in $x$ and $y$ directions. The MuID for the South Arm is same as the
North Arm, except that the Muon Magnet backplate is 20 cm instead of 30 cm
in the North Arm.

\begin{figure}[h!]
\includegraphics[width=1.0\linewidth]{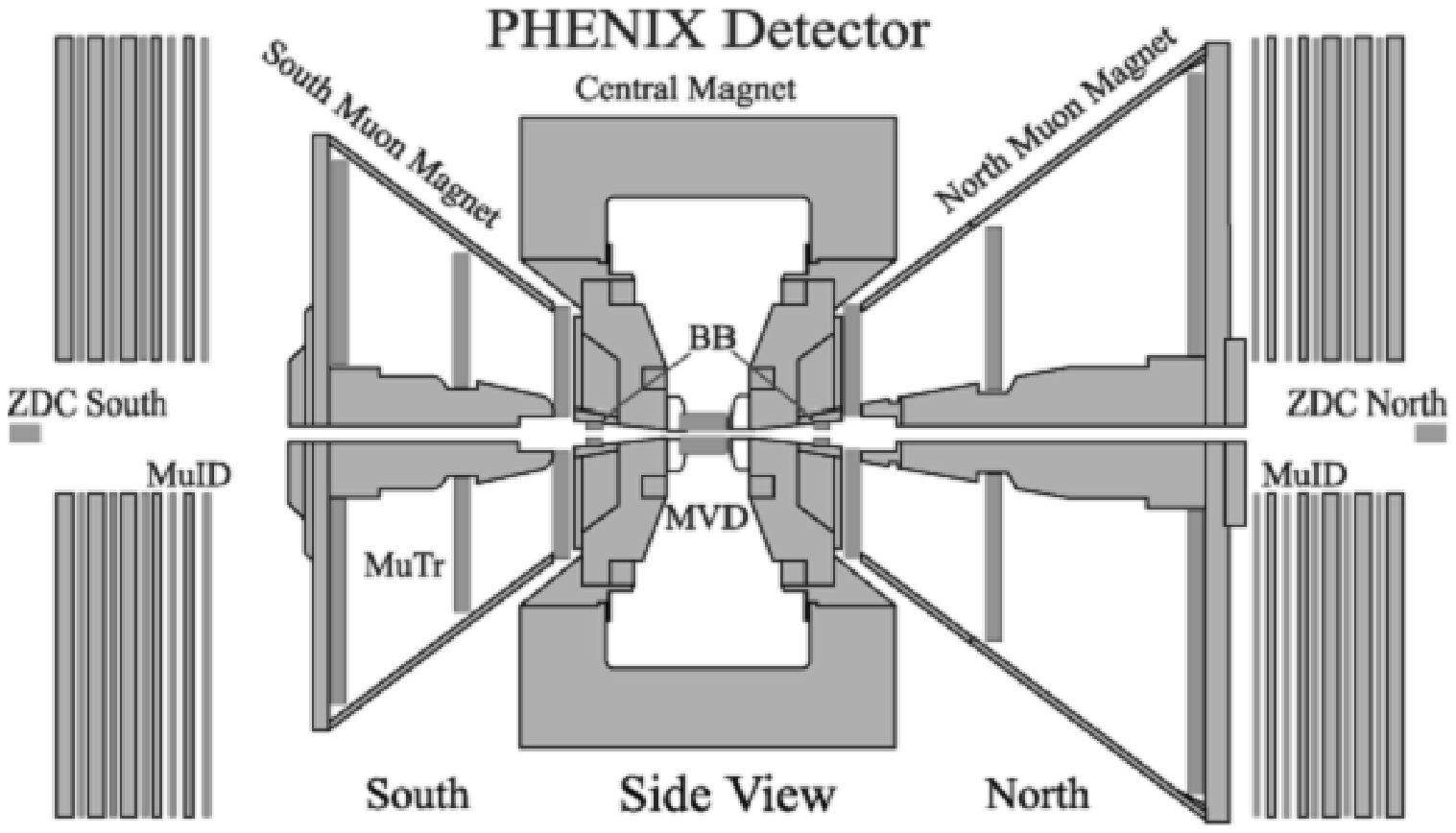}  
\caption{A side view of the muon arms shows the muon tracker (MuTr) and the
muon identifier (MuID). Particles travel from the collision point, through the
muon magnet absorber into the MuTr and finally into the MuID.}
\label{fig:muonarm_side}
\end{figure}
\clearpage

\section{Selection of hadrons}
Although the PHENIX Muon Arm Spectrometers~\cite{phenixnim,muonnim} were 
designed to detect muons, we use them in this analysis to measure charged 
hadrons by rejecting muons: ``One man's meat (hadrons) is another man's 
poison (muons)''. Although both muons and hadrons lose their energy when 
passing through materials via Bethe-Bloch ionization energy 
loss~\cite{bethe-bloch}, hadrons can lose a much larger portion of their 
energy due to hadronic interactions. As mentioned earlier, the PHENIX Muon
Arms consist of a tracker (MuTr) with a steel copper absorber in front 
of the tracker to reduce hadronic background and get better a muon signal (See
Figures~\ref{fig:mms_sketch} and~\ref{fig:muonarm_side}). This
is followed by layers of the identifier (MuID), which are interleaved with 
steel plates again to reduce hadronic background. This means that if a 
particle penetrates to the deepest MuID gap, without getting absorbed in 
between it was probably a muon, whereas if it stops in the shallow gaps, it 
is more likely to be a hadron. If we look at the distribution of the total 
momentum $p_{tot}$ as measured at Station 1~\footnote{Track momentum cannot 
be measured directly at the vertex due to presence of absorber in between, 
instead it is calculated by propagating particle tracks through the absorber. 
Momentum at Station 1 is a more direct experimental measure.} (behind the 
absorber) stopped at given shallow MuID gaps, (Fig.~\ref{fig:pgaphist}) we 
see that there is a peak of mostly slow muons around 1.3 GeV, and a longer 
hadronic tail. This is similar to what is seen in Monte Carlo (MC) 
simulations~\cite{chunswork}. Thus, by applying appropriate cuts in $p_{tot}$ 
and MuID depth, we can identify hadrons.

\begin{figure}
\includegraphics[width=1.0\linewidth]{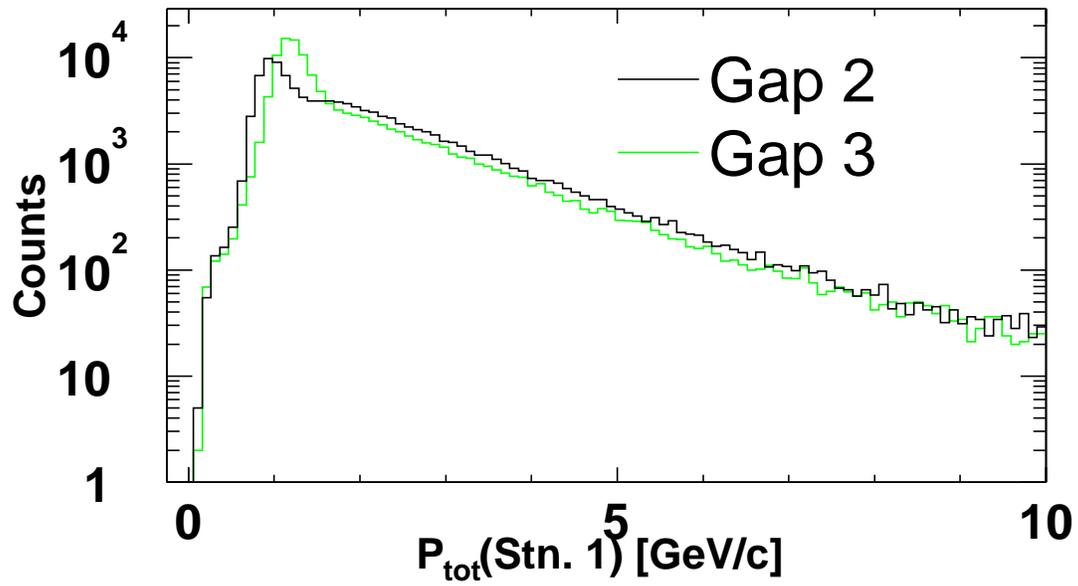}  
\caption{Histogram of momentum measured at station 1 of MuTr for tracks 
stopping at different MuID gaps from d+Au min. bias data. Around 1.3 GeV
there is a peak of mostly stopped muons, followed by a long hadronic tail.}
\label{fig:pgaphist}
\end{figure}

As described previously, the raw data from a run is reconstructed and then
saved in easily digestible format, based on detector subsystems or physics
type. For this analysis, we used PHENIX Muon nanodsts (MwDSTs), which contain
information from the PHENIX Muon Arms. As in previous run, the Zero Degree 
Calorimeters (ZDCs) and the Beam Beam Counters (BBCs) were used to determine 
the event centrality (or impact parameter) by looking at multiplicities. For 
this analysis, we used Minimum Bias (MB) events, which are essentially 
minimally triggered events in which the BBC and the ZDC fired and are as 
unbiased as can be made. The event vertex was restricted to $|z|<35$ cm of the 
collision vertex, primarily for reasons to do with detector acceptances.

After excluding events rejected by our global cuts, we analysed 67 million 
events. In order to select hadrons we fitted a Gaussian to the peak on 
Fig.~\ref{fig:pgaphist} and obtained a width of 200 MeV/$c$, centered at 
1.3 GeV/$c$. In addition, tracks were selected for optimal signal to 
background ratio, by using the following cuts:
\begin{itemize}
\item Number of hits in the tracker MuTr: $nhits >= 12$.
\item Track stopping at MuID gap = 2 or 3, exclusively. Since particles that 
penetrate deeper into the MuID, without stopping in the intervening steel are
more likely to be muons, we select only those which stop exclusively in the
shallow gaps.
\item Total momentum $p_{tot}$(Station 1) $> 1.3+3\times0.2$ = 1.9 GeV/$c$. 
Since the slow muons (stopped at the shallow gaps) peak around $p_{tot}$ = 
1.3 GeV/$c$, we can reject them by excluding particles in that range. 
\item Track $\chi^2< 10$ (helps in rejection of ghost tracks).
\item $-2.0<\eta<-1.3$ for South Arm and $2.2>\eta>1.3$ for 
North Arm. These acceptance cuts are applied to reduce the small angle 
background, mainly from beam gas interactions.  
\end{itemize}

\section{Sources of contamination}
We have two main sources of contamination:
\benn
\item Muon contamination through prompt muons (which are produced at the 
collision point) and muons from meson decays.
\item Hadronic showers.
\eenn

Most muons are produced through meson decays, primarily pions and kaons.
Although most pions and kaons produced in the collision are stopped at the
absorber, some decay into muons along their flight path. As a result, the
further the collision vertex is from the absorber greater is the probability 
that the pion (or kaon) will decay into a muon. This leads to a characteristic
linear $z$-vertex dependence of the decay muons that can be used to cross 
check our muon contamination. Fig.~\ref{fig:zplotOO_nocut} shows the 
normalized BBC 
collision vertex distribution for the events where tracks are measured in the
South (top panels) and North Muon Arms (bottom panels), before making any cuts.
From left to right each panel shows the distributions at Gap 2, 3 and 4 of the 
MuID. Since Gap 4 is the last gap, it always has a
preponderance of muons, which can be used for comparision. Our hadronic
signal is mostly in Gaps 2 and 3, which are the shallow gaps. As expected the
distributions at all the gaps, show the characteristic slope seen in decay
muons. By comparing the slopes at Gaps 2 \& 3 with Gap 4, it is possible to 
estimate that decay muon contamination before making our 
hadron selection cuts is $65 \pm 4\%$. After we make the cuts (see 
Fig.~\ref{fig:zplotOO_allcut}), the $z$-vertex distributions at Gaps 2 \& 3 
become flat and the slopes are consistent with a decay muon contamination of 
$0\pm 5\%$. The $z$-vertex distribution at Gap 4 (extreme right panels at top 
and bottom) still shows a slope, indicating that we still have a lot of muons 
at the last Gap (Gap 4), as expected.

\begin{figure}[h]
\begin{center}
\includegraphics[width=1.0\linewidth]{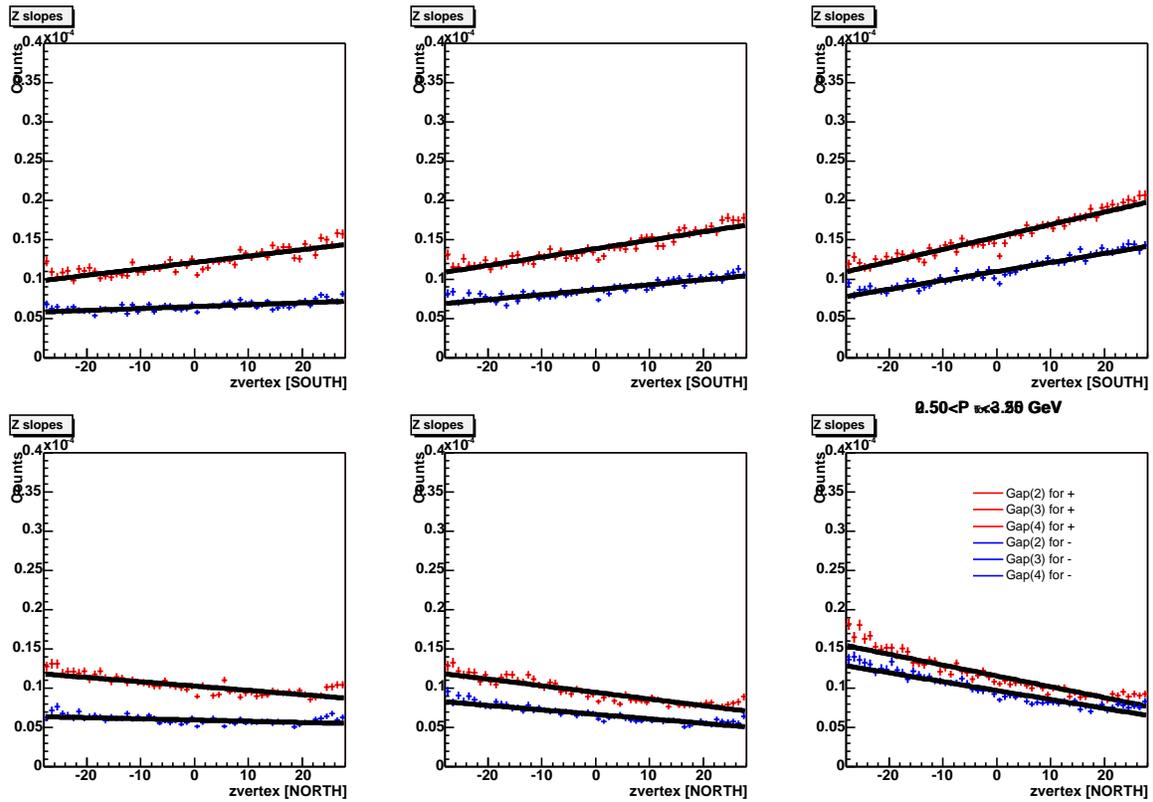}  
\end{center}
\caption{Normalized BBC vertex distributions at different gaps before cuts. 
Top panels are for South Arm, bottom panels are for North Arm and the Gaps 
go from 2 to 4 as we move left to right.}
\label{fig:zplotOO_nocut}
\end{figure}

\begin{figure}[h]
\begin{center}
\includegraphics[width=1.0\linewidth]{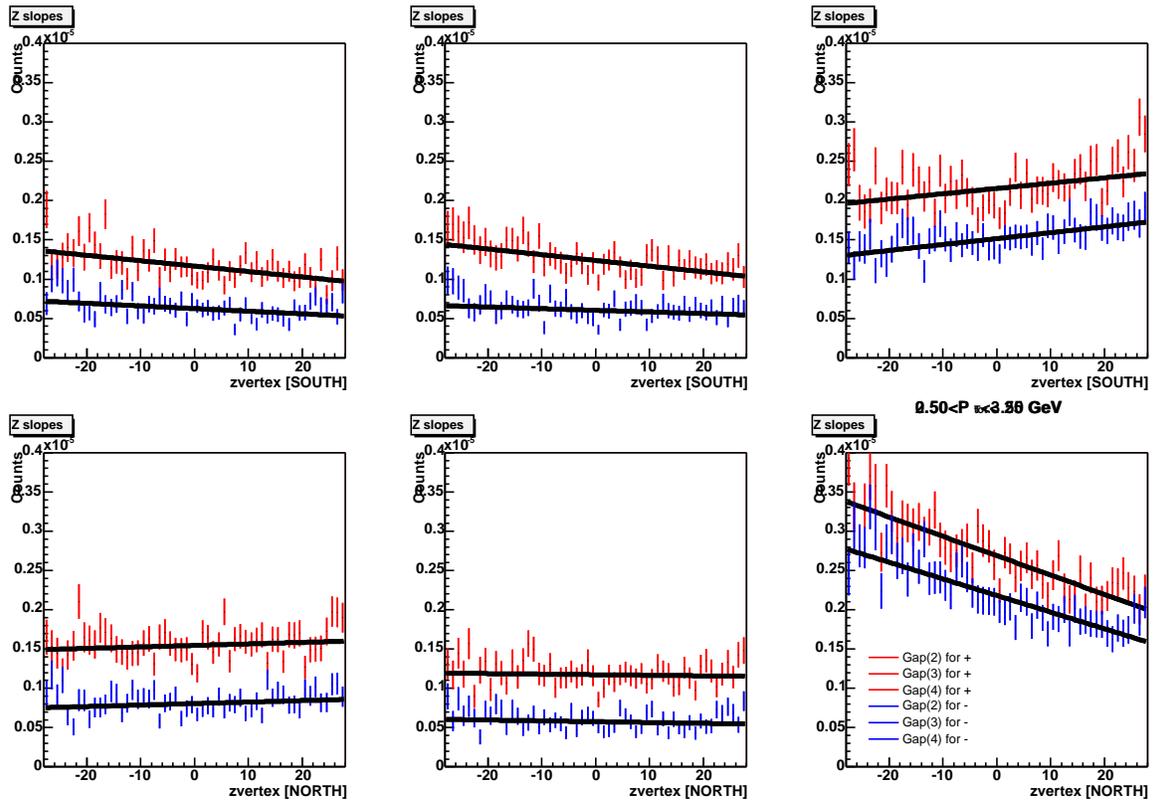}  
\end{center}
\caption{Normalized BBC vertex distributions at different gaps after cuts. 
Top panels are for South Arm, bottom panels are for North Arm and the Gaps 
go from 2 to 4 as we move left to right.}
\label{fig:zplotOO_allcut}
\end{figure}
\clearpage

After rejecting muons, we are mostly left with either primary hadrons that are
produced at the vertex or secondary hadrons produced by hadronic showers.
Although primary hadrons lose energy via Bethe-Bloch process~\cite{bethe-bloch}
as they traverse through the absorber, the energy loss $dE/dx$ does not vary
much in the momentum range of interest ($1-100$ GeV/$c$). This can lead to an 
uncertainty of about 5\% in our momentum measurement as the hadrons travel
through the detector and the absorbers.

In order to estimate and remove particles from hadronic showers, we look at
the angular distribution of particles from the vertex. Hadronic inelastic 
scattering can lead to larger angular spreads compared to regular multiple
scattering. We look the angular difference between the momentum $p_{tot}$ at 
Station 1 (angle $\theta_{p1}$), and the vector joining the vertex with the 
first measurement point in Station 1 (angle $\theta_{Vst1}$). This is shown in
a sketch in Figure~\ref{fig:station1}. We make a histogram of the variable
$\Delta\theta = \theta_{Vst1} - \theta_{p1}$ after using the appropriate 
Jacobian for weighting ($\frac{1}{\theta_{avg}}\frac{p}{\theta_{avg}}$), (see 
Fig.~\ref{fig:showerbkg_cut}) we notice that:
\benn
\item There is narrow multiple scattering peak and a wide shower tail and this 
changes as a function of $p_t$ (see Figs.~\ref{fig:showerbkg_5_1},
~\ref{fig:showerbkg_1_2},~\ref{fig:showerbkg_cut},~\ref{fig:showerbkg_3_4}),
as expected. MC simulations indicate similar behaviour~\cite{chunswork}.
\item After the cut $\Delta\theta < 0.030$ only about 15\% of the shower 
``background'' remains in our signal at $2<p_t<4$ GeV.
\item We also observe that our signal to background varies as a function of
$p_T$, starting out high at low $p_T$ (0.5 - 1.0 GeV/$c$), goes down to 15\% 
and then at high $p_T$ (3.0 - 4.0 GeV/$c$) it goes up a little to 23\%.
\eenn

\begin{figure}[h]
\begin{center}
\epsfig{file = 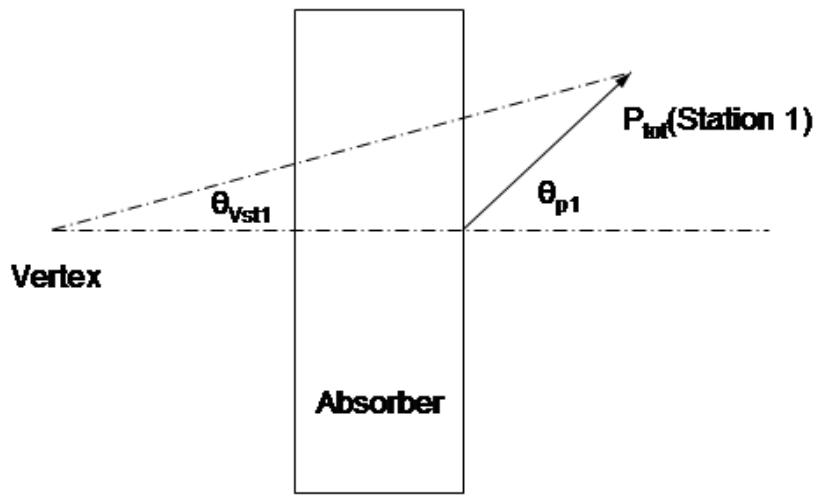, width = 12cm, clip = }
\end{center}
\caption{Sketch showing $\Delta\theta$ angular cut.}
\label{fig:station1}
\end{figure}

\begin{figure}[h]
\begin{center}
\epsfig{file = 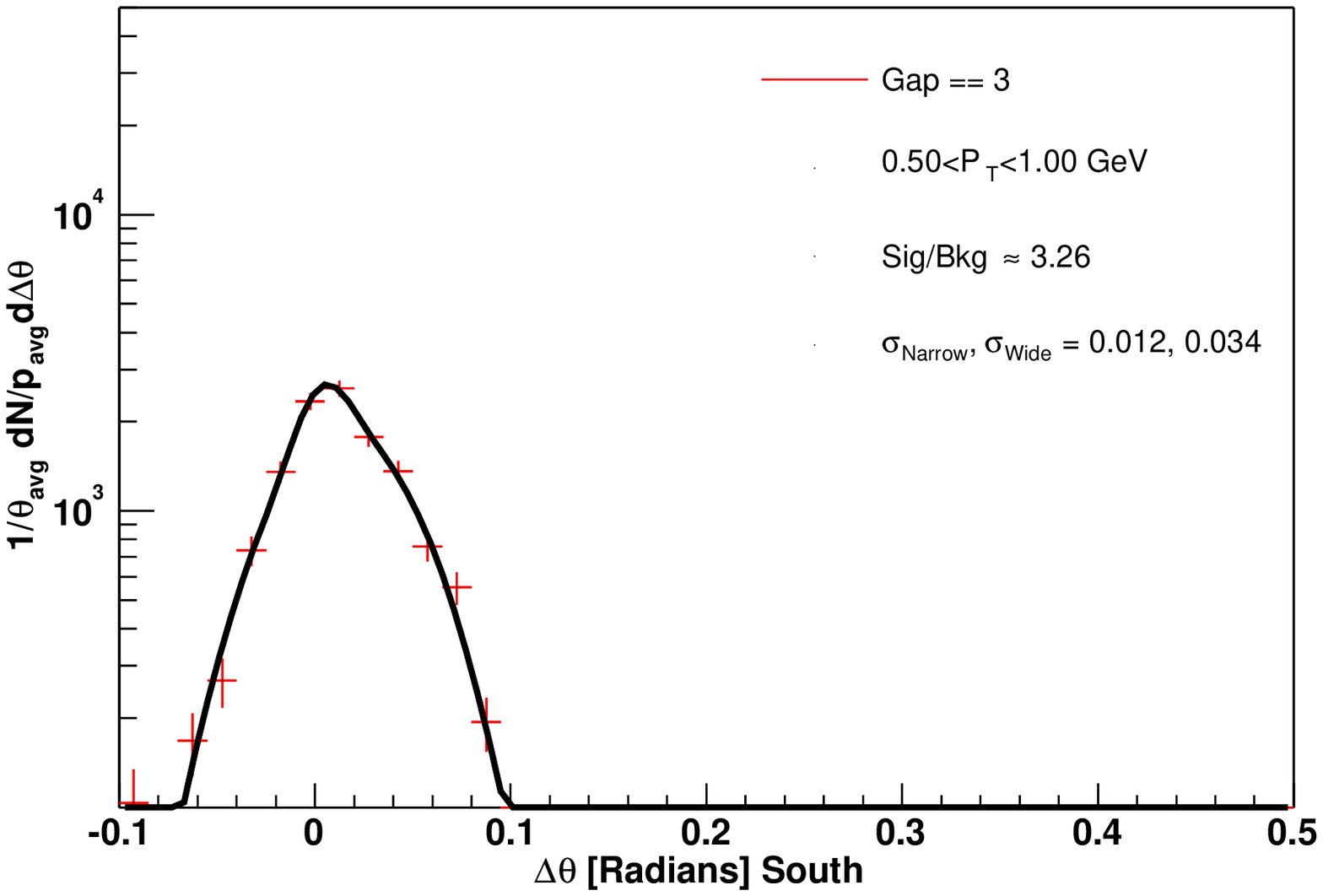, width = 12cm, clip = }
\end{center}
\caption{$\Delta\theta$ histogram for $0.5 < p_T < 1.0 $ GeV/$c$, showing
the wide peak of hadronic scattering and narrow peak of signal.}
\label{fig:showerbkg_5_1}
\end{figure}

\begin{figure}[h]
\begin{center}
\epsfig{file = 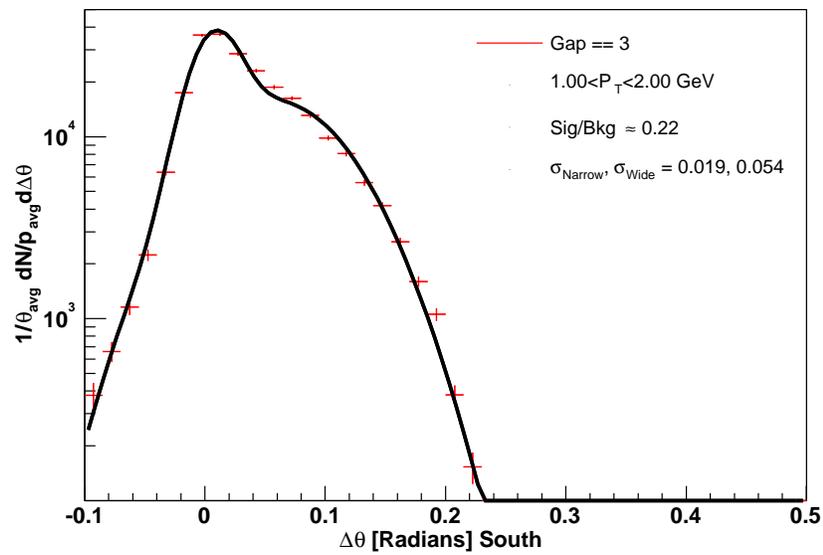, width = 12cm, clip = }
\end{center}
\caption{$\Delta\theta$ histogram for $1.0 < p_T < 2.0 $ GeV/$c$.}
\label{fig:showerbkg_1_2}
\end{figure}

\begin{figure}[h]
\begin{center}
\epsfig{file = 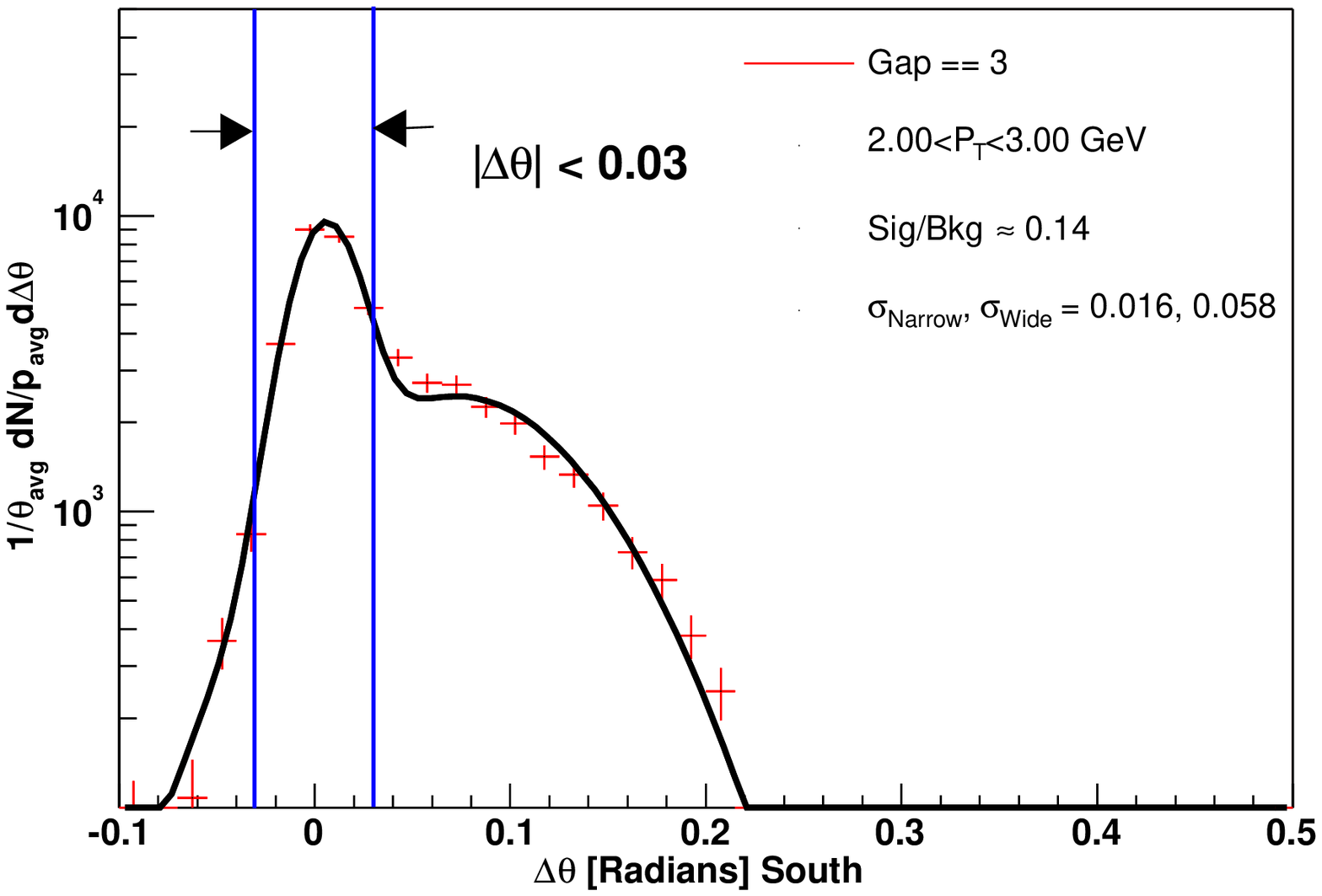, width = 12cm, clip = }
\end{center}
\caption{$\Delta\theta$ histogram for $2.0< p_T < 3.0 $ GeV/$c$, showing
the wide peak of hadronic scattering and narrow peak of signal.}
\label{fig:showerbkg_cut}
\end{figure}

\begin{figure}[h]
\begin{center}
\epsfig{file = 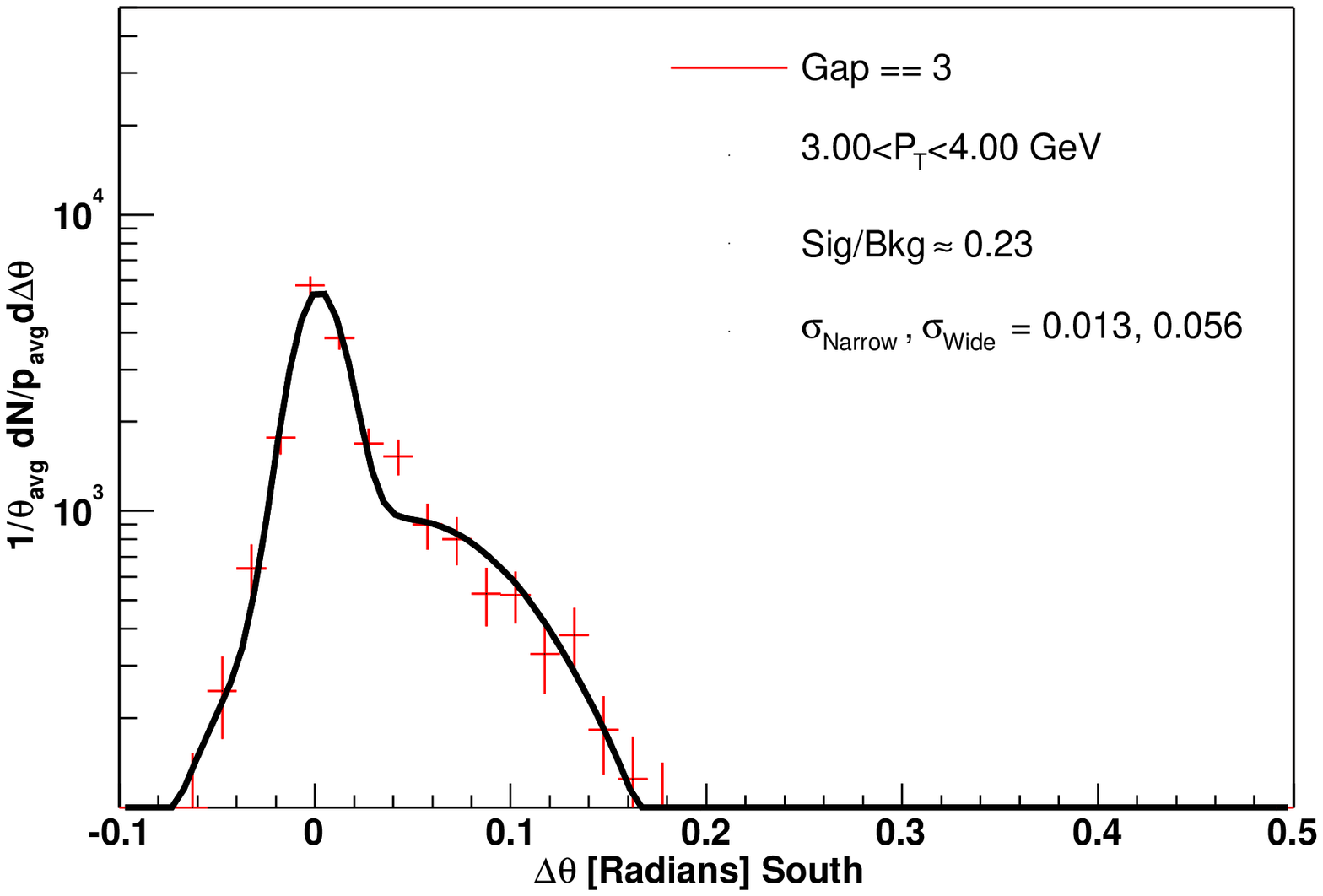, width = 12cm, clip = }
\end{center}
\caption{$\Delta\theta$ histogram for $3.0 < p_T < 4.0 $ GeV/$c$.}
\label{fig:showerbkg_3_4}
\end{figure}

For greater understanding we plot a 2D histogram of 
$\theta_1 - \theta_{Vst1}$ vs $\theta_{Vst1}$ and $p_{tot}$ 
(see Fig.~\ref{fig:thetascatter}). We immediately see
a large tail consistent with our expectation of showers. To cut this away 
without cutting too much signal, we apply a simple line cut 
$\Delta\theta<0.03$. 

\begin{figure}[h]
\begin{center}
\epsfig{file = 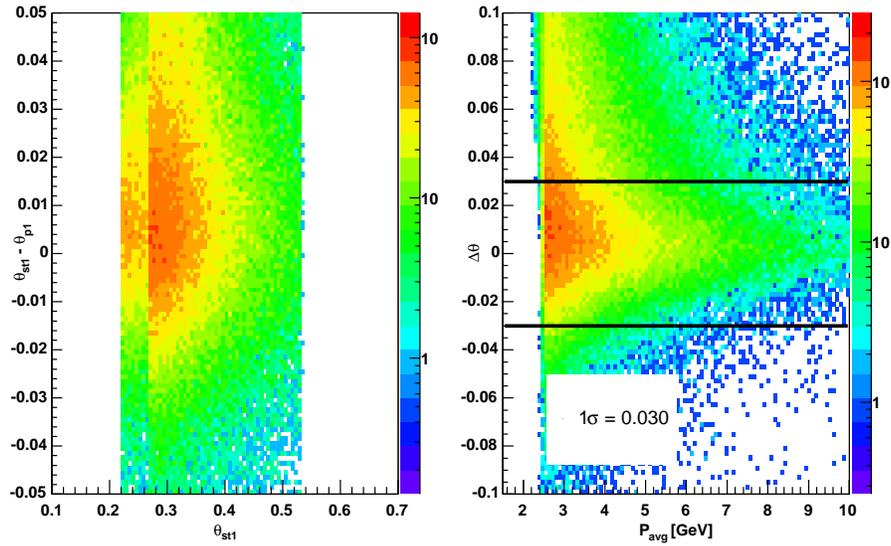, width = 12cm, clip = }
\end{center}
\caption{2D histogram of $\theta_1 - \theta_{Vecst1}$ vs $\theta_1$ and 
$p_{tot}$ ($0.5<p_T<3.5$ GeV) showing the cut.}
\label{fig:thetascatter}
\end{figure}

\clearpage

\section{Run QA}

	Runs used for this data analysis were selected after doing a Physics 
QA, by looking at several quantities like the average multiplicities. If the 
values were out of the average bands, the runs were rejected. This was done to
reduce uncertainties from variation in detector acceptances due to tripped or
dead channels, or beam background.
\benn
\item {\bf Beam Background}: Due to the location of the muon arms, background 
due to beam can potentially be large. There are several cuts which can 
minimize this:
\benn
\item It is expected that such a background will be coming from behind the 
muon arms and hence will hit the deepest MuID layers first. Since we are
vetoing on the LASTGap (Gap 4), i.e., rejecting all tracks that make it to the
LASTGap, this should considerably reduce this.
\item If we look at the run-by-run variation in the number of hits (see 
Fig.~\ref{fig:QA_npart}), then beam background should lead to large 
fluctuations in this. And the variation in this number should give us an upper 
limit on the beam background.
\item The $\Delta\theta$ angular cuts also reduce this because the angular
distribution of the background is expected to be different from regular tracks.
\eenn

\item {\bf MuID Efficiency}: We are selecting hadrons by vetoing on the 
LASTGap (Gap 4). If the Gap 4 is 
switched off, or else if there is some sort of inefficiency in that panel, 
then all tracks would stop at Gap 3 instead of Gap 4, leading to a muonic
contamination in our hadronic tracks. This contamination can be reduced by
run QA. We looking at the run-by-run variation in the average value of the
LASTGap (see Fig.~\ref{fig:QA_lastgap}). Any substantial inefficiency in 
Gap 4 (relative to our statistics), will lead to a drop in this value for 
that run.
\eenn

\begin{figure}[h]
\begin{center}
\epsfig{file = 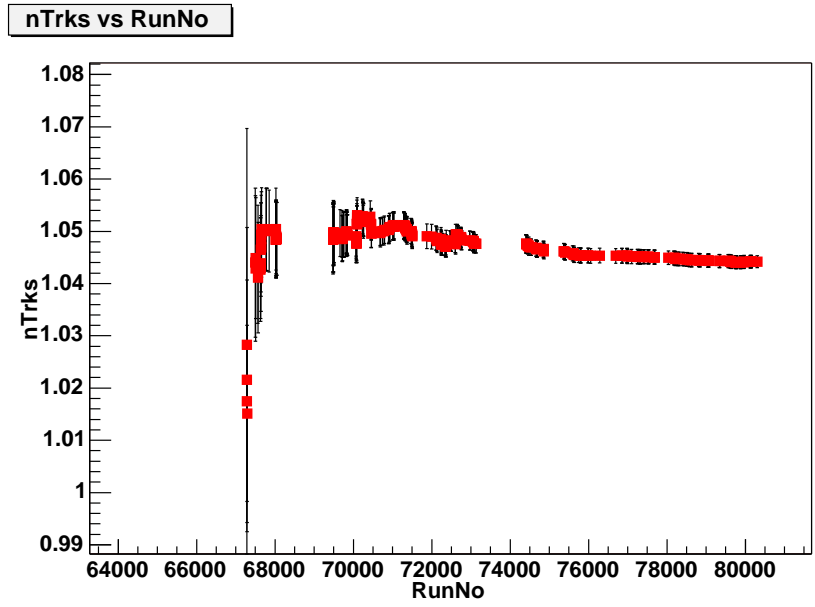, width = 12cm, clip = }
\end{center}
\caption{Run-by-run variation in npart.}
\label{fig:QA_npart}
\end{figure}

\begin{figure}[h]
\begin{center}
\epsfig{file = 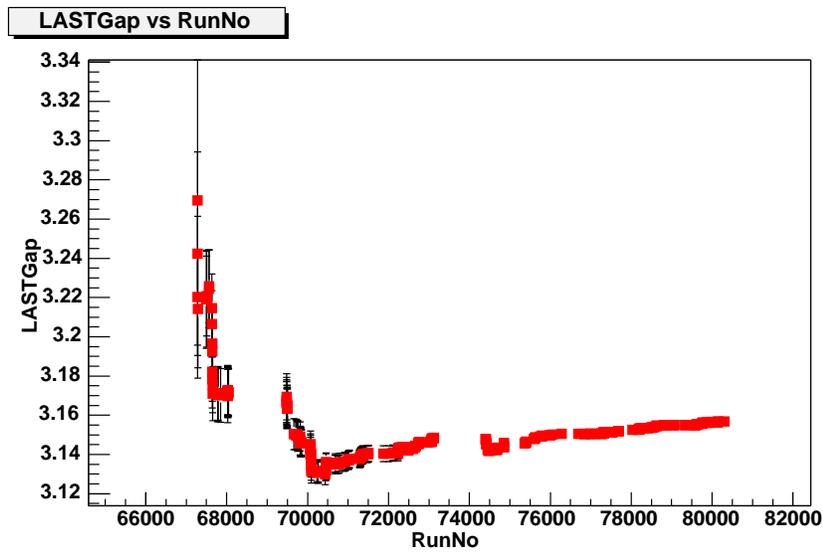, width = 12cm, clip = }
\end{center}
\caption{Run-by-run variation in LASTGap.}
\label{fig:QA_lastgap}
\end{figure}

\chapter{Hadron $R_{cp}$ measurement}
We look at the particle multiplicities in forward and backward directions 
for d+Au collisions at $\sqrt{s_{NN}}=200$ GeV, and examine the behaviour
as compared to mid-rapidity, as well their variation as a function of 
centrality. We can also compare results from d+Au collisions with those 
from Au+Au collisions in an attempt to distinguish between effects 
that could potentially be due to deconfinement, versus effects of cold 
nuclear matter. 

\section{$R_{cp}$ vs $p_T$}
After obtaining the hadronic multiplicities as described in the previous 
chapter, we obtained the nuclear modification factor $R_{cp}$ by taking the
ratios of multiplicities at a given centrality bin with the peripheral 
(60-92\%) multiplicity scaled with the number of collisions $N_{coll}$, as
defined previously in Eq.~\ref{eq:rcp}:
\begin{equation}
R_{cp} = \frac{\left( \frac{dN}{d\sigma dp_T} \right)^{Central}/N_{coll}^{Central}}{\left(\frac{dN}{d\sigma dp_T}\right)^{Peripheral}/N_{coll}^{Peripheral}}
\end{equation}

We plotted $R_{cp}$ both as a function of transverse momentum $p_T$ and 
pseudorapidity $\eta$, for different centralities. The momentum fraction
$x$ can be related to $p_T$ as: $x \approx p_T/\sqrt{s}$.
Thus, greater the energy, lower is the value of $x$ we can get for a given
$p_T$. It is important to stay at higher values of $p_T$ preferably $p_T>3$ 
GeV/$c$, so that we stay away from the soft regime, which is not very well
described by standard perturbative QCD. 

Figures~\ref{fig:rcp_pt2}, ~\ref{fig:rcp_pt1} and ~\ref{fig:rcp_pt0} show 
$R_{cp}$ vs $p_T$ for north and south arms for centralities 0-20\%, 20-40\% 
and 40-60\% respectively (for data points see Table~\ref{tab:rcppt}). We 
immediately notice:
\benn
\item $R_{cp}$ shows a suppression in the North Arm (d going direction) 
 and an enhancement in the South Arm (Au going direction).
\item The suppression is maximum for 0-20\% centrality and decreases for 
less central collisions. The enhancement too is maximum for 0-20\% centrality 
and decreases for less central collisions. 
\eenn

\begin{figure}[h]
\begin{center}
\includegraphics[width=1.0\linewidth]{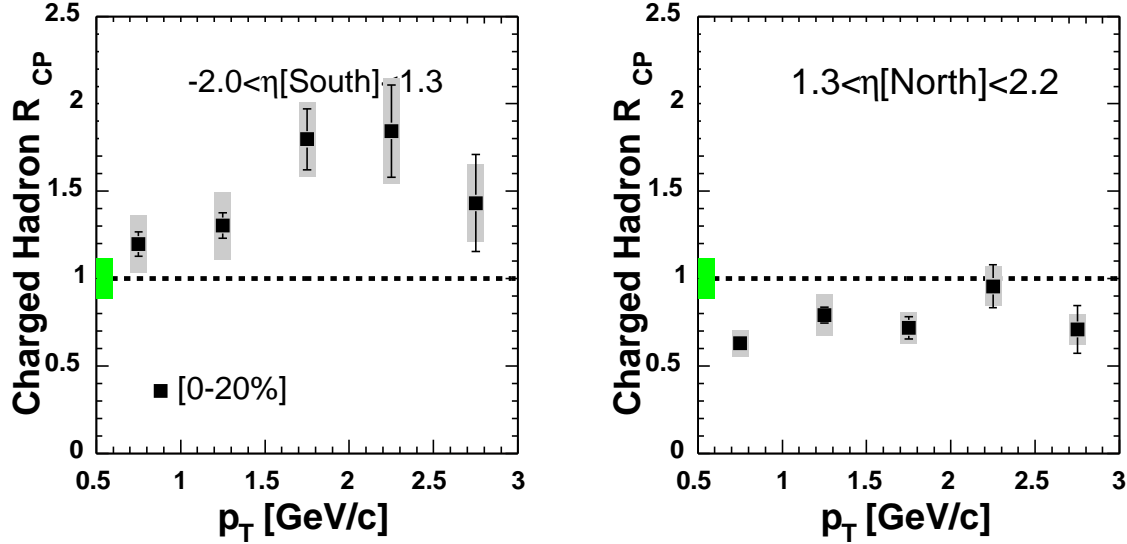}  
\end{center}
\caption{$R_{cp}$ vs $p_T$ for all charged hadrons for 0-20\% most central 
events. Left panel shows data from the South Arm ($-2.2<\eta<-1.2$), while the
right panel shows data from North Arm($1.2<\eta<2.4$). Systematic errors 
that vary point by point are shown by grey bars at
each data point, while errors are same for all points are depicted by the green
bar. {\bf (PHENIX Preliminary)}}
\label{fig:rcp_pt2}
\end{figure}

\begin{figure}[h]
\begin{center}
\includegraphics[width=1.0\linewidth]{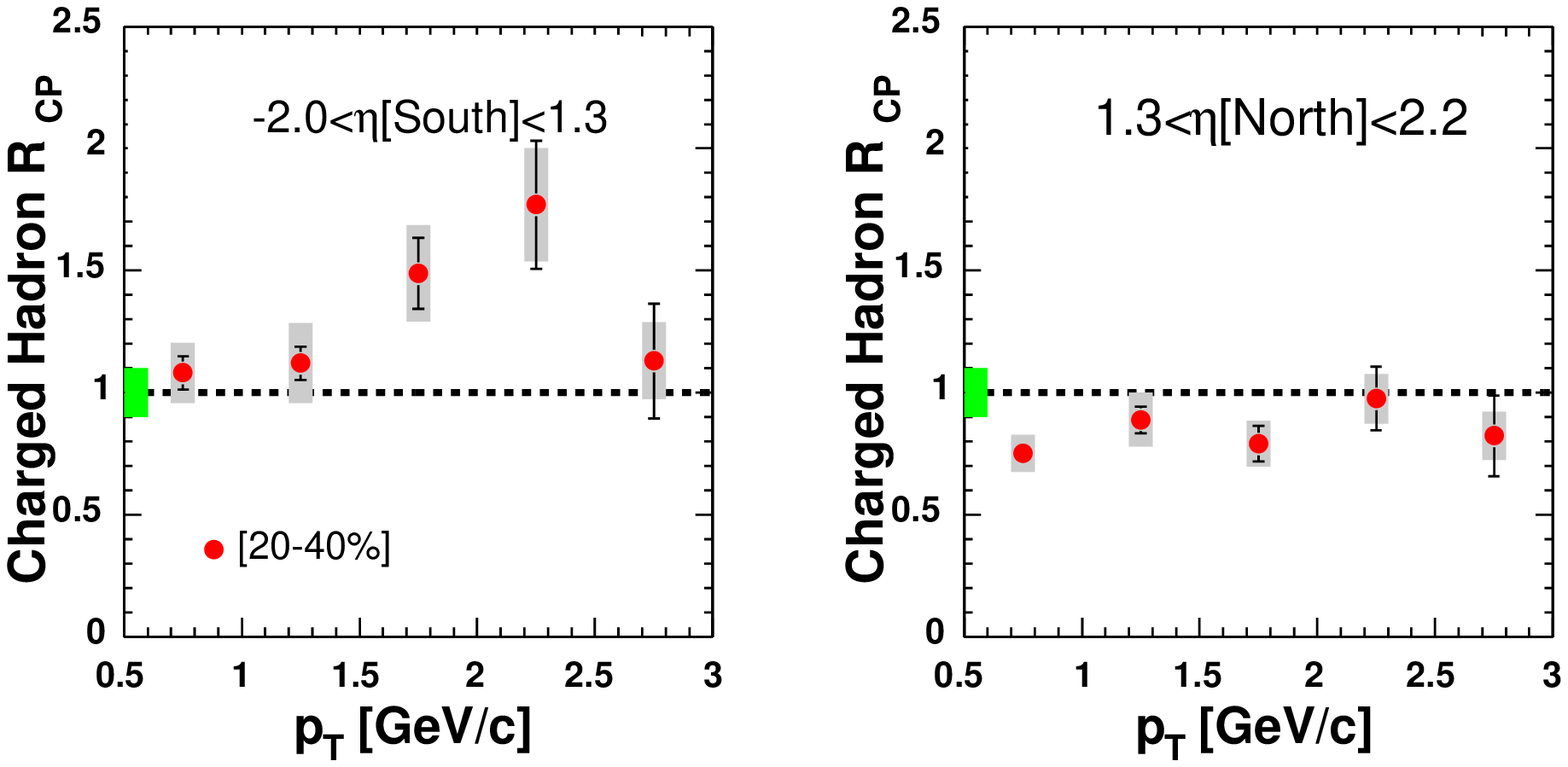}  
\end{center}
\caption{$R_{cp}$ vs $p_T$ for all charged hadrons for 20-40\% most central 
events. Left panel shows data from the South Arm ($-2.2<\eta<-1.2$), while the
right panel shows data from North Arm($1.2<\eta<2.4$). Systematic errors that 
vary point by point are shown by grey bars at
each data point, while errors are same for all points are depicted by the green
bar.{\bf (PHENIX Preliminary)}}
\label{fig:rcp_pt1}
\end{figure}

\begin{figure}[h]
\begin{center}
\includegraphics[width=1.0\linewidth]{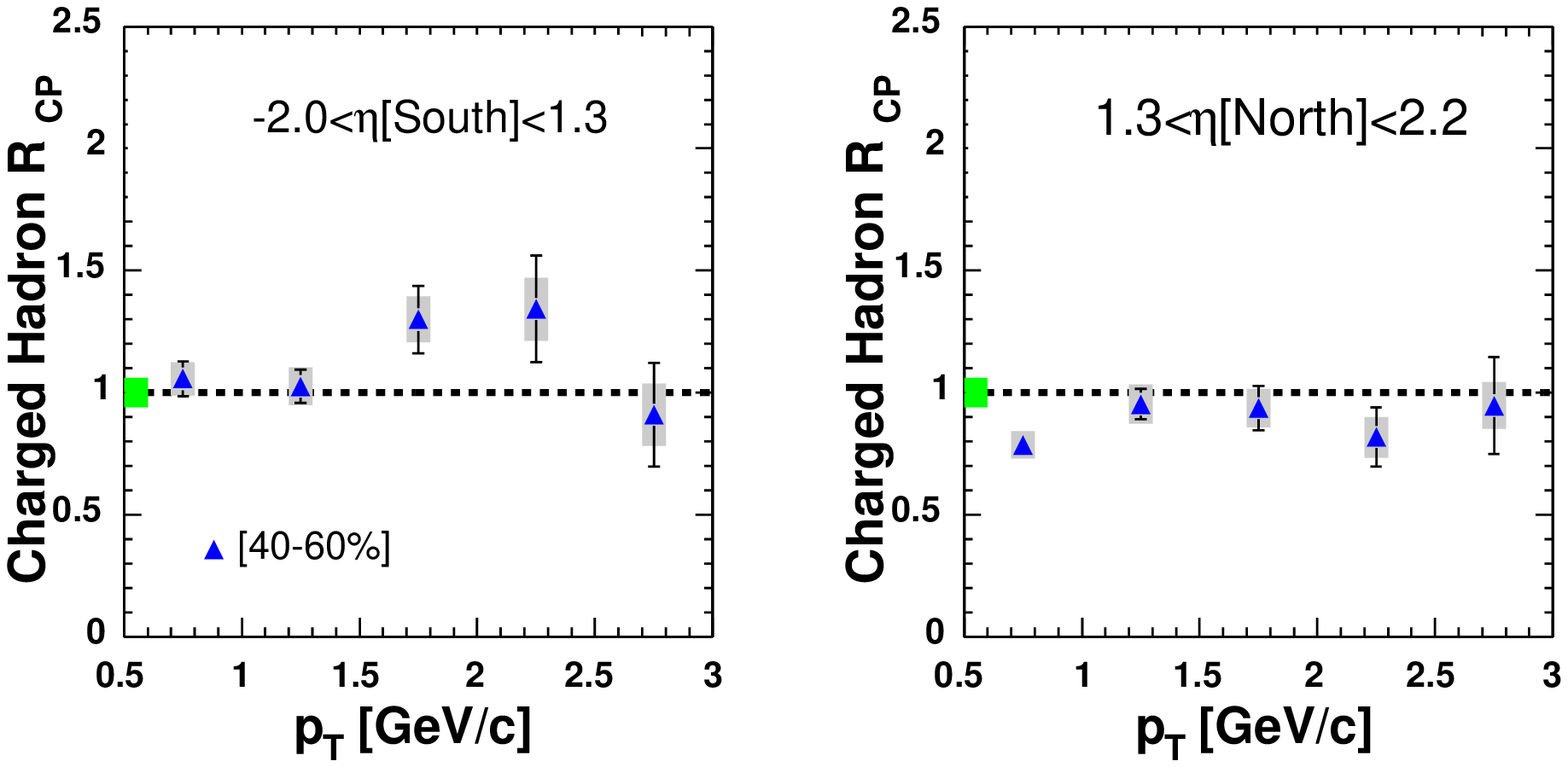}  
\end{center}
\caption{$R_{cp}$ vs $p_T$ for all charged hadrons GeV for 40-60\% most 
central events. Left panel shows data from the South Arm ($-2.2<\eta<-1.2$), 
while the right panel shows data from North Arm($1.2<\eta<2.4$). Systematic 
errors that vary point by point are shown by grey bars at
each data point, while errors are same for all points are depicted by the green
bar.{\bf (PHENIX Preliminary)}}
\label{fig:rcp_pt0}
\end{figure}

We classify our systematic uncertainties for $R_{cp}$ vs $p_T$ measurement
in two categories:
\benn

\item Those which vary point to point for each data point in $p_T$ (listed in
Tables~\ref{tab:rcppt_southsys} and ~\ref{tab:rcppt_northsys} for the South
and North Arms respectively):
\benn
\item Potential shower contamination: this was determined by varying 
the polar angle $\Delta\theta$ cut from 0.03 to 0.06.
\item Potential beam background: this can be estimated by varying the 
acceptance cut from $15^o$ to $12^o$.
\eenn

\item Those which are constant for each data point(listed in 
Table~\ref{tab:rcppt_norm}):
\benn
\item Uncertainty on number of binary collisions $N_{coll}$: 
$10.8\%$~\cite{phenix_nosup} for 0-20\% centrality. It varies as a function of
centrality. 
\item Tracking/road finding efficiency:$\approx 4\%$. This primarily occurs 
because the track reconstruction system in the muon arms can misidentify 
tracks.
\eenn
\eenn

\begin{table}[h!]
\caption{Data table for $R_{cp}$ vs $p_T$ for North and South Arms. All 
errors are statistical. Systematic uncertainties are given in 
Tables~\ref{tab:rcppt_southsys},~\ref{tab:rcppt_northsys} 
and~\ref{tab:rcppt_norm}.}
\bcc
\begin{tabular}{|l|l|l|l|}
\hline
Centrality&$p_T$ [GeV/$c$]&$R_{cp}$(South)&	$R_{cp}$(North)	\\
\hline
	&0.5 - 1.0 	&1.19735$\pm$0.0706	&0.629646$\pm$0.0321	\\
	&1.0 - 1.5	&1.30357$\pm$0.0743	&0.790778$\pm$0.0467	\\
0-20\%	&1.5 - 2.0	&1.79743$\pm$0.1744	&0.718265$\pm$0.0639	\\
	&2.0 - 2.5	&1.84448$\pm$0.2638	&0.955506$\pm$0.1223	\\
	&2.5 - 3.0 	&1.43172$\pm$0.2778	&0.708701$\pm$0.1361	\\
\hline
	&0.5 - 1.0 	&1.08141$\pm$0.0681	&0.751891$\pm$0.0399	\\
	&1.0 - 1.5	&1.12044$\pm$0.0683	&0.888502$\pm$0.0551	\\
20-40\%	&1.5 - 2.0	&1.48803$\pm$0.1458	&0.789919$\pm$0.0735	\\
	&2.0 - 2.5	&1.77024$\pm$0.262	&0.974458$\pm$0.1306	\\
	&2.5 - 3.0 	&1.13006$\pm$0.2351	&0.823329$\pm$0.1647	\\
\hline
	&0.5 - 1.0 	&1.05717$\pm$0.0719	&0.785904$\pm$0.0448	\\
	&1.0 - 1.5	&1.0257	$\pm$0.0677	&0.95276$\pm$0.0629	\\
40-60\%	&1.5 - 2.0	&1.29969$\pm$0.1378	&0.936201$\pm$0.0908	\\
	&2.0 - 2.5	&1.3429	$\pm$0.2175	&0.817881$\pm$0.1227	\\
	&2.5 - 3.0 	&0.909452$\pm$0.2119	&0.946805$\pm$0.1979	\\
\hline
\end{tabular}
\ecc
\label{tab:rcppt}
\end{table}

\begin{table}[h!]
\caption{Point to point systematic uncertainties for $R_{cp}$ in South Arm.}
\bcc
\begin{tabular}{|c|c|c|c|c|}
\hline									
Centrality	&$p_T$[GeV/$c$]	&$\Delta\theta$ cut&Acceptance
&Total $p_T$ dependent systematics\\		
\hline	
	&	0.5-1.0	&	2.90\%	&	6.80\%	&	7.39\%	\\
	&	1.0-1.5	&	5.70\%	&	7.40\%	&	9.34\%	\\
0-20\%	&	1.5-2.0	&	1.10\%	&	2.40\%	&	2.64\%	\\
	&	2.0-2.5	&	8.00\%	&	8.30\%	&	11.53\%	\\
	&	2.5-3.0	&	9.30\%	&	5.10\%	&	10.61\%	\\
\hline									
	&	0.5-1.0	&	1.50\%	&	5.10\%	&	5.32\%	\\
	&	1.0-1.5	&	8.40\%	&	6.30\%	&	10.50\%	\\
20-40\%	&	1.5-2.0	&	7.30\%	&	4.40\%	&	8.52\%	\\
	&	2.0-2.5	&	1.30\%	&	8.20\%	&	8.30\%	\\
	&	2.5-3.0	&	9.50\%	&	0\%	&	9.50\%	\\
\hline									
	&	0.5-1.0	&	1.80\%	&	0.50\%	&	1.87\%	\\
	&	1.0-1.5	&	3.20\%	&	3.00\%	&	4.39\%	\\
40-60\%	&	1.5-2.0	&	3.20\%	&	1.70\%	&	3.62\%	\\
	&	2.0-2.5	&	2.80\%	&	6.70\%	&	7.26\%	\\
	&	2.5-3.0	&	11.30\%	&	5.30\%	&	12.48\%	\\
\hline								
\end{tabular}
\ecc
\label{tab:rcppt_southsys}
\end{table}

\begin{table}[h!]
\caption{$p_T$ systematic uncertainties for $R_{cp}$ in North Arm.}
\bcc
\begin{tabular}{|c|c|c|c|c|}						
\hline						
Centrality&$p_T$[GeV/$c$]&$\Delta\theta$ cut&Acceptance
&Total $p_T$ dependent systematics\\
\hline	
	&	0.5-1.0	&	2.80\%	&	1.90\%	&	3.38\%	\\
	&	1.0-1.5	&	6.80\%	&	6.90\%	&	9.69\%	\\
0-20\%	&	1.5-2.0	&	2.20\%	&	4.30\%	&	4.83\%	\\
	&	2.0-2.5	&	0.30\%	&	3.10\%	&	3.11\%	\\
	&	2.5-3.0	&	1.30\%	&	4.30\%	&	4.49\%	\\
\hline								
	&	0.5-1.0	&	0.50\%	&	1.40\%	&	1.49\%	\\
	&	1.0-1.5	&	4.10\%	&	5.80\%	&	7.10\%	\\
20-40\%	&	1.5-2.0	&	5.50\%	&	2.90\%	&	6.22\%	\\
	&	2.0-2.5	&	1.50\%	&	1.10\%	&	1.86\%	\\
	&	2.5-3.0	&	3.80\%	&	5.00\%	&	6.28\%	\\
\hline									
	&	0.5-1.0	&	2.50\%	&	2.60\%	&	3.61\%	\\
	&	1.0-1.5	&	2.80\%	&	4.90\%	&	5.64\%	\\
40-60\%	&	1.5-2.0	&	4.30\%	&	3.50\%	&	5.54\%	\\
	&	2.0-2.5	&	8.00\%	&	0.40\%	&	8.01\%	\\
	&	2.5-3.0	&	5.20\%	&	6.00\%	&	7.94\%	\\
\hline	
\end{tabular}
\ecc
\label{tab:rcppt_northsys}
\end{table}

\begin{table}[h!]
\caption{$p_T$ independent systematics for $R_{cp}$.}
\bcc
\begin{tabular}{|c|c|c|c|c|}						
\hline						
Centrality	&$n_{binary}$&Tracking/roadfinding	\\
\hline	
0-20\%	&	10.80\%	&	4\%	\\
\hline	
20-40\%	&	9.30\%	&	4\%	\\
\hline	
40-60\%	&	4.80\%	&	4\%	\\
\hline	
\end{tabular}
\ecc
\label{tab:rcppt_norm}
\end{table}
\clearpage

\section{$R_{cp}$ vs $\eta$}
Using the same methodology described above, we calculated $R_{cp}$ vs $\eta$
by integrating our data in the range $1 < p_t< 3$ GeV/$c$, for each $\eta$
bin. The $\eta$ range goes from -2.2 to -1.2
for the South Arm to 1.2 to 2.4 for the North Arm. In addition there is also a
data point at midrapidity using previous results from the PHENIX Central 
Arm~\cite{phenix_nosup}. Figures~\ref{fig:rcp_eta2},
~\ref{fig:rcp_eta1} and ~\ref{fig:rcp_eta0} show plots of $R_{cp}$ vs $\eta$ 
for the centrality ranges 0-20\%, 20-40\% and 40-60\% respectively (for data 
points see Table~\ref{tab:rcpeta}). We notice two main trends:
\benn
\item $R_{cp}$ shows a suppression in the North Arm (d going direction) 
 and an enhancement in the South Arm (Au going direction) and this increase
is continuous and almost linear as we go from $\eta=-2.2$ in the South Arm
to $\eta=2.4$ in the North Arm.
\item The suppression (and enhancement) is maximum for 0-20\% centrality and 
decreases for less central collisions. 
\eenn

\begin{table}[h!]
\caption{$R_{cp}$ vs $\eta$ in the range $1 < p_t< 3$ GeV/$c$.}
\bcc
\begin{tabular}[ ]{|c|c|c|}
\hline
Centrality   &$\eta$ & $R_{cp}$ \\
\hline
	&	-1.88333	&	1.76019	$\pm$	0.13991	\\
	&	-1.65	&	1.64172	$\pm$	0.0631918	\\
	&	-1.41667	&	1.53932	$\pm$	0.0892065	\\
0-20\%	&	1.4125	&	1.00895	$\pm$	0.0741331	\\
	&	1.6375	&	0.922162	$\pm$	0.0442295	\\
	&	1.8625	&	0.795191	$\pm$	0.0279444	\\
	&	2.0875	&	0.781327	$\pm$	0.0771171	\\
\hline							
	&	-1.88333	&	1.40951	$\pm$	0.119334	\\
	&	-1.65	&	1.40715	$\pm$	0.0573594	\\
	&	-1.41667	&	1.36865	$\pm$	0.0838189	\\
20-40\%	&	1.4125	&	0.965126	$\pm$	0.0754662	\\
	&	1.6375	&	0.937888	$\pm$	0.0475359	\\
	&	1.8625	&	0.8784	$\pm$	0.0323008	\\
	&	2.0875	&	0.890123	$\pm$	0.0914602	\\
\hline							
	&	-1.88333	&	1.26621	$\pm$	0.115834	\\
	&	-1.65	&	1.21456	$\pm$	0.0539039	\\
	&	-1.41667	&	1.15746	$\pm$	0.0776094	\\
40-60\%	&	1.4125	&	1.09681	$\pm$	0.0899434	\\
	&	1.6375	&	0.953508	$\pm$	0.0519437	\\
	&	1.8625	&	0.919708	$\pm$	0.0362131	\\
	&	2.0875	&	0.962761	$\pm$	0.105112	\\
\hline
\end{tabular}
\ecc
\label{tab:rcpeta}
\end{table}

Systematics for $R_{cp}$ vs $\eta$ are again of two types: 
\benn
\item Point-to-point systematic uncertainties, which vary from one data point
to another (see Table~\ref{tab:rcpetasys}):

\benn
\item  Systematic uncertainty in the $p_{tot}$ cut (which is used to reject 
muons stopped in the shallow MuID Gaps) is calculated by varying our
$p_{tot}$ cut by $1\sigma$ from 1.9 GeV to 2.1 GeV.
\item  Systematic uncertainty in $p_{T}$ is calculated by varying our
$p_{T}$ range from 1.0---2.0 GeV, to 1.2---2.2 GeV.
\item  Since we integrate over a given $p_T$ range for each $\eta$ bin, we need
to determine the systematic uncertainty due to $p_{T}$. This is calculated by 
varying our $p_{T}$ range from 1.0--3.0 GeV, to 1.2--3.2 GeV.
\item  Systematic uncertainty in $\Delta\theta$ cut (which is used to reject 
hadronic showers) is calculated by varying 
it from $\Delta\theta < 0.030$ to $\Delta\theta < 0.060$.
\eenn
\item Then there are systematic uncertainties that are independent of  $\eta$.
These consist of uncertainty in $N_{coll}$, the number of collisions in a given
centrality bin and tracking/roadfinding efficiency. These were tabulated 
earlier in Table~\ref{tab:rcppt_norm}.
\eenn

\begin{table}[h!]
\caption{Point-to-point systematics for $R_{cp}$ vs $\eta$ in the range 
$1 < p_t< 3$ GeV/$c$.}
\bcc
\begin{tabular}[ ]{|c|c|c|c|c|c|}
\hline
Centrality&$\eta$ &$p_{tot}$&$p_{T}$& $\Delta\theta$& Total point-to-point systematics\\
\hline
	&	-1.88333&	0.42\%&	7\%&	4.36\%	&	8.26\%	\\
	&	-1.65	&	1.11\%&	1.1\%&	0.86\%	&	1.78\%	\\
	&	-1.41667&	0.47\%&	8.52\%&	0.88\%	&	8.58\%	\\
0-20\%	&	1.4125	&	2.48\%&	0.63\%&	0.41\%	&	2.59\%	\\
	&	1.6375	&	2.57\%&	3.57\%&	1.65\%	&	4.70\%	\\
	&	1.8625	&	0.08\%&	5.06	&1.94\%	&	5.42\%	\\
	&	2.0875	&	0\%&	4.7\%	&0.55\%	&	4.73\%	\\
\hline							
	&	-1.88333&	0.1\%&	1.44\%	&0.56\%	&	1.55\%	\\
	&	-1.65	&	1.32\%& 1.83\%	&0.48\%	&	2.31\%	\\
	&	-1.41667&	0.37\%	&2.95\%	&0.41\%	&	3.00\%	\\
20-40\%	&	1.4125	&	1.12\%	&2.84\%	&0.98\%	&	3.21\%	\\
	&	1.6375	&	2.62\%	&0.56\%	&0.44\%	&	2.72\%	\\
	&	1.8625	&	0.14\%	&3.56\%	&1.48\%	&	3.86\%	\\
	&	2.0875	&	0\%	&3.08\%	&1.76\%	&	3.55\%	\\
\hline							
	&	-1.88333&	0.19\%	&3.82\%	&5.35\%	&	6.58\%	\\
	&	-1.65	&	1.3\%	&1.93\%	&1.23\%	&	2.63\%	\\
	&	-1.41667&	2.97\%	&2.63\%	&0.09\%	&	3.97\%	\\
40-60\%	&	1.4125	&	3.15\%	&1.53\%	&0.07\%	&	3.50\%	\\
	&	1.6375	&	2.2\%	&4.16\%	&0.49\%	&	4.73\%	\\
	&	1.8625	&	0.07\%	&2.65\%	&2.64\%	&	3.74\%	\\
	&	2.0875	&	0\%	&3.44\%	&5.44\%	&	6.44\%	\\
\hline
\end{tabular}
\ecc
\label{tab:rcpetasys}
\end{table}

\begin{figure}[h]
\includegraphics[width=1.0\linewidth]{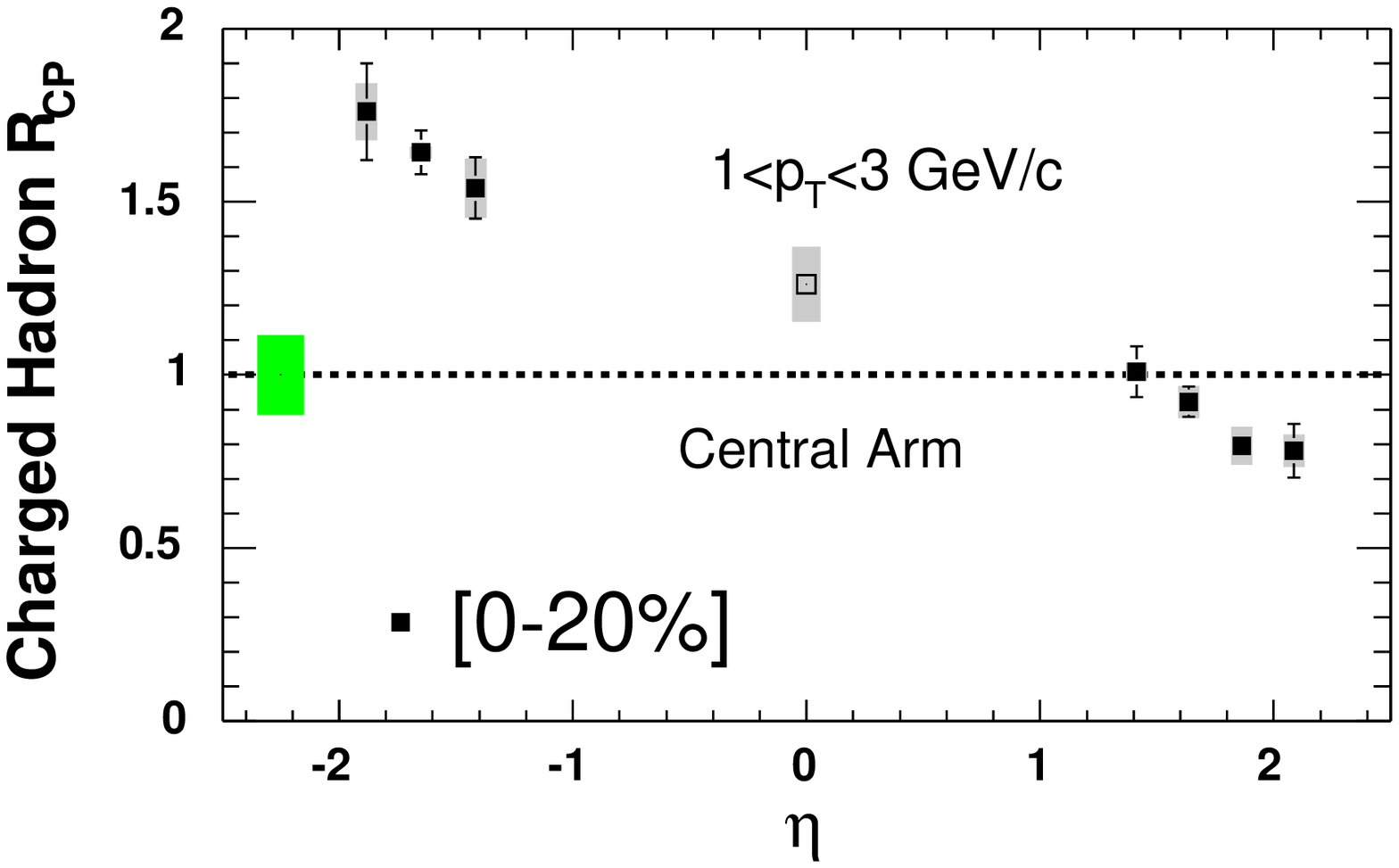}  
\caption{$R_{cp}$ vs $\eta$ for all charged hadrons in the range $1<p_t<3$ 
GeV/$c$ for 0-20\% most central events. Au going direction is along negative
$\eta$ (South Arm) and d going direction is along positive $\eta$ (North Arm).
Systematic errors that vary point by point are shown by grey bars at
each data point, while errors are same for all points are depicted by the green
bar.{\bf (PHENIX Preliminary)}}
\label{fig:rcp_eta2}
\end{figure}

\begin{figure}[h]
\includegraphics[width=1.0\linewidth]{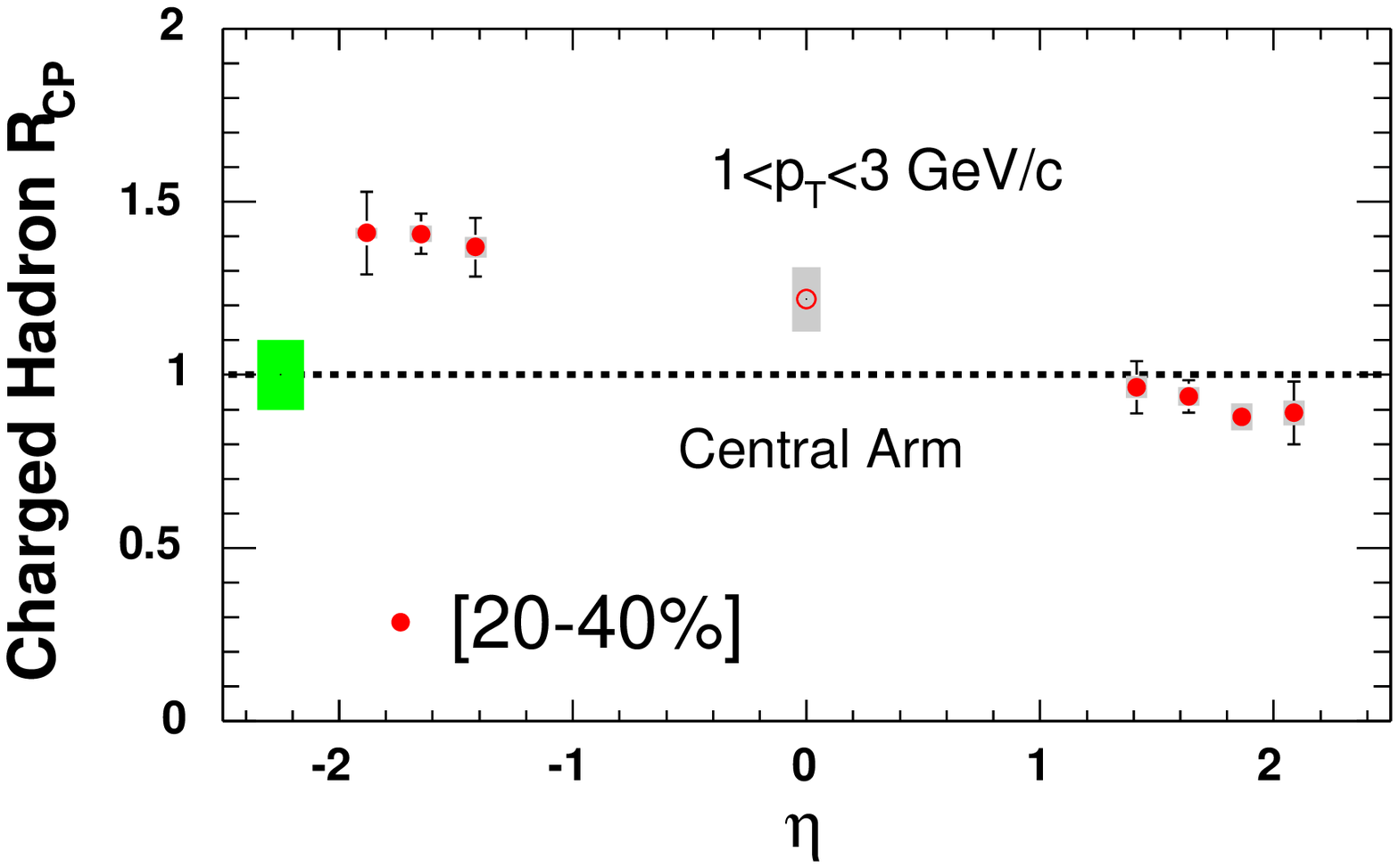}  
\caption{$R_{cp}$ vs $\eta$ for all charged hadrons in the range $1<p_t<3$ 
GeV/$c$ for 20-40\% most central events. Au going direction is along negative
$\eta$ (South Arm) and d going direction is along positive $\eta$ (North Arm).
Systematic errors that vary point by point are shown by grey bars at
each data point, while errors are same for all points are depicted by the green
bar.{\bf (PHENIX Preliminary)}}
\label{fig:rcp_eta1}
\end{figure}

\begin{figure}[h]
\includegraphics[width=1.0\linewidth]{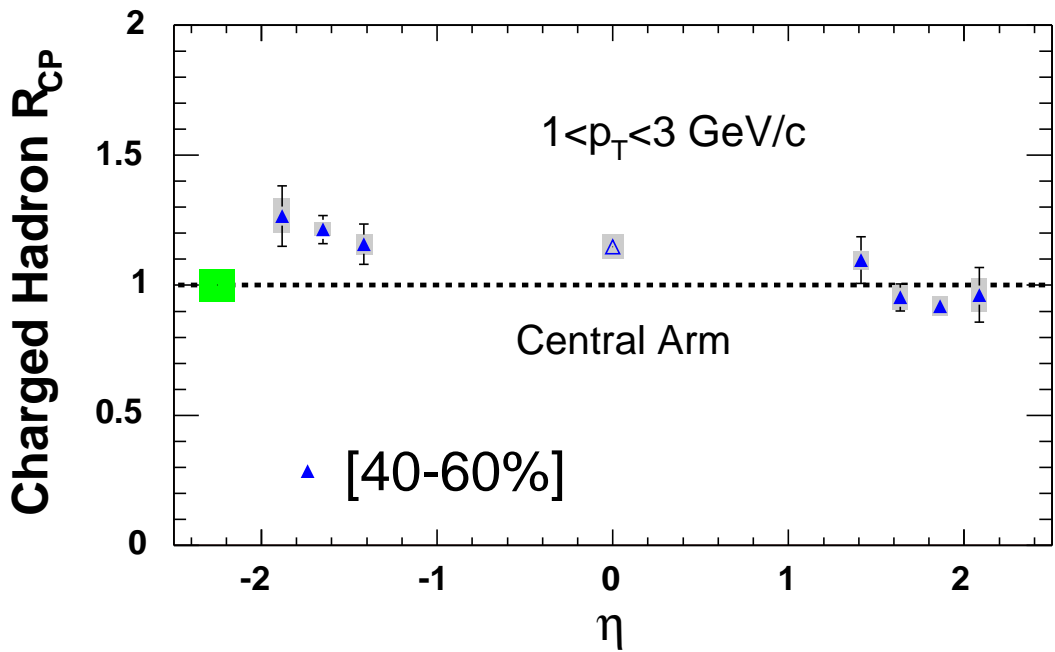}  
\caption{$R_{cp}$ vs $\eta$ for all charged hadrons in the range $1<p_t<3$ 
GeV/$c$ for 40-60\% most central events. Au going direction is along negative
$\eta$ (South Arm) and d going direction is along positive $\eta$ (North Arm).
Systematic errors that vary point by point are shown by grey bars at
each data point, while errors are same for all points are depicted by the green
bar.{\bf (PHENIX Preliminary)}}
\label{fig:rcp_eta0}
\end{figure}
\clearpage

The plots shown earlier in Figures~\ref{fig:rcp_eta2}, \ref{fig:rcp_eta1},
\ref{fig:rcp_eta0}, \ref{fig:rcp_pt2}, \ref{fig:rcp_pt1}, \ref{fig:rcp_pt0}
were for minimum bias data only. In order to improve statistics and increase
the momentum range, triggered data was used to obtain better 
plots~\cite{chunswork}, which can be seen in Figures~\ref{fig:ppg036_fig2}
and \ref{fig:ppg036_fig3}. Special thanks goes to Chun Zhang and Ming X. Liu
for Figures~\ref{fig:ppg036_fig2} and \ref{fig:ppg036_fig3}.

\begin{figure}[h!]
\begin{center}
\includegraphics[width=0.75\linewidth]{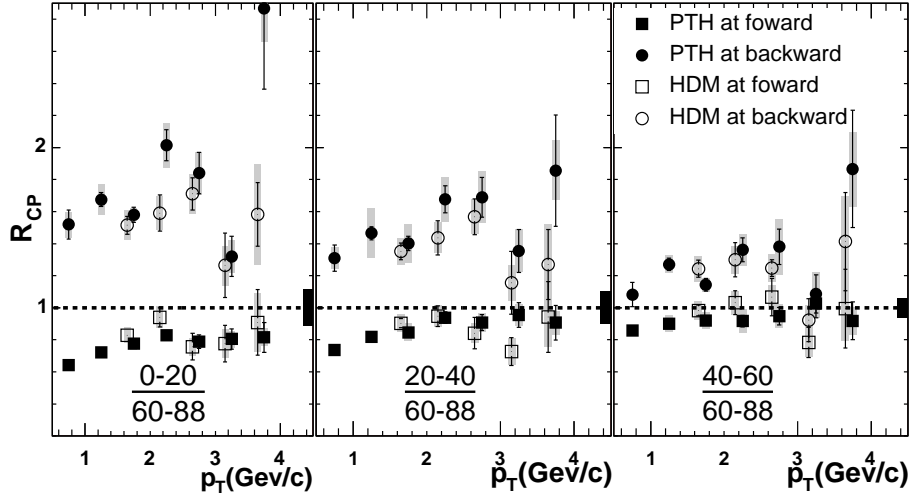}  
\end{center}
\caption{$R_{cp}$ vs $p_T$ for triggered data. PTH stands for punch through
hadrons, whereas HDM stands for data from hadron decay mesons.}
\label{fig:ppg036_fig2}
\end{figure}

\begin{figure}[h!]
\begin{center}
\includegraphics[width=0.75\linewidth]{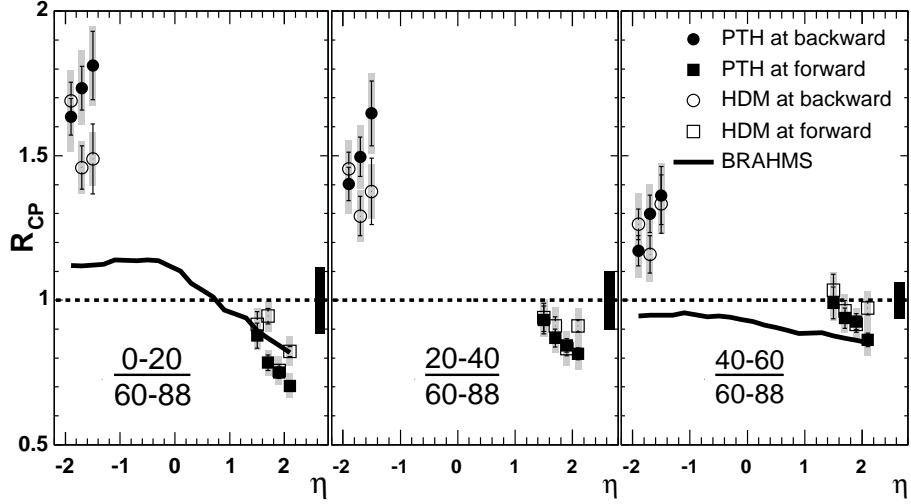}  
\end{center}
\caption{
$R_{cp}$ vs $\eta$ for triggered data. PTH stands for punch through
hadrons, whereas HDM stands for data from hadron decay mesons. The solid line
depicts a fit to the BRAHMS data~\cite{brahms_rcp} for comparison. 
{\bf (PHENIX Preliminary Data)}}
\label{fig:ppg036_fig3}
\end{figure}

\section{Comparisons}

Our data on hadron $R_{cp}$ indicates a suppression on the forward region 
(d going direction). Similar behaviour was also seen by BRAHMS experiment 
REFS? at RHIC. This is qualitatively consistent with shadowing/saturation 
type effects in the small-$x$ region being probed at forward rapidities. 
Moreover, the central to peripheral $R_{cp}$ (Figure~\ref{fig:rcp_eta2}) is 
more suppressed as compared to semi-central to peripheral $R_{cp}$
(Figure~\ref{fig:rcp_eta1}). 

As mentioned previously (Chapter 5), the Color Glass Condensate model gives 
a universal QCD explanation for low-$x$ shadowing and predicts a depletion in
scattering centers due to gluon processes. The small-$x$ regions can be 
probed by going to forward rapidities (see Eq.\ref{eq:x_au}). A toy model 
prediction~\cite{cgc_klm} is made regarding variation of $R_{cp}$ as a 
function of rapidity. In Figure~\ref{fig:cgc_pred} $R^{pA}$ (which is 
equivalent to $R_{cp}$ if we assume that peripheral collisions are analogous
to p+p collisions) is plotted vs $k/Q_s$, where $k$ is the transverse momentum
of the partons and $Q_s$ is the saturation scale, given by:
\be
Q_s(y) \approx Q_{s0}e^{2\bar{\alpha}y}
\label{eq:sat_scale}
\ee
As the rapidity (or equivalently energy) increases, the upper solid curve 
slowly turns into the lower solid curve. It is also predicted that as the
collision centrality increases $R^{pA}$ will decrease, consistent with our
observations. 

\begin{figure}[h!]
\bcc
\includegraphics[width=0.75\linewidth]{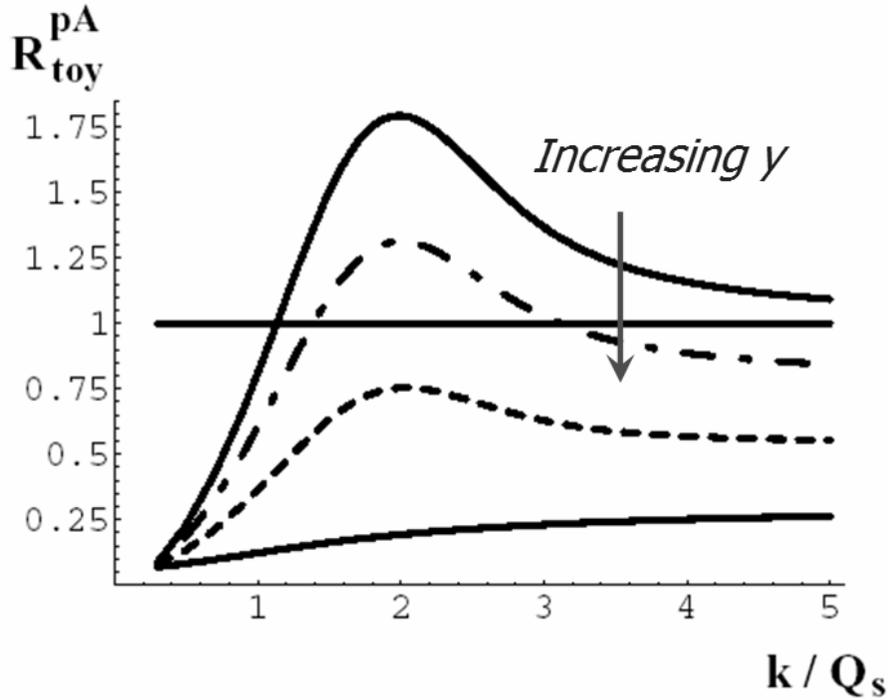}  
\ecc
\caption{$R^{pA}$ vs $k/Q_s$ as predicted by CGC model~\cite{cgc_klm}. The
nuclear modification factor decreases with rapidity.}
\label{fig:cgc_pred}
\end{figure}

On the other hand, the backward rapidity region (Au going direction) is very 
different. We see an enhancement, which is not satisfactorily explained by 
the CGC model, which predicts a flat distribution in the backward region. This
sort of enhancement has been seen at other experiments and is known as the 
Cronin effect, and is usually attributed to initial multiple scatterings at
the partonic level.  Another candidate is anti-shadowing, which occurs when
$g + g \rightarrow g$ processes deplete partons at small-$x$ partons but 
increase those at larger-$x$ (and momenta). Although, we should note that our momentum
range of 1--3 GeV/$c$ is not completely in the hard scattering regime. Hence we 
factorisation might not apply and we might be sensitive to soft physics. In such a 
scenario, the backward enhancement could be explained by rapidity exchange. Early 
studies~\cite{stopping_power}
have shown that in a p+Pb collision with p incident at 100 GeV/$c$ on Pb at rest, the
rapidity exchange $\simeq$ 2.5. However RHIC is at a much higher energy and so this 
rapidity exchange is expected to a small effect. Thus, at present, there is no single
satisfactory explaination for the enhancement in the backward direction.

In addition we can also include
results from the PHENIX $J/\psi$ measurements for comparison. It is 
interesting to look at the $J/\psi$ because due to its heavy mass, it is 
mostly produced by gluon gluon fusion processes (see Figure~\ref{fig:gg2jpsi}).
Hence $J/\psi$ production is an independent probe of gluon PDFs. In 
Figure~\ref{fig:brahms_rcpcomp} we plot hadrons from PHENIX and BRAHMS (in a 
different $\eta$ range) as well as the $J/\psi$ from PHENIX, we notice that 
that $R_{cp}$ (vs $\eta$) is almost a straight line. 
\begin{figure}[h!]
\bcc
\includegraphics[width=0.5\linewidth]{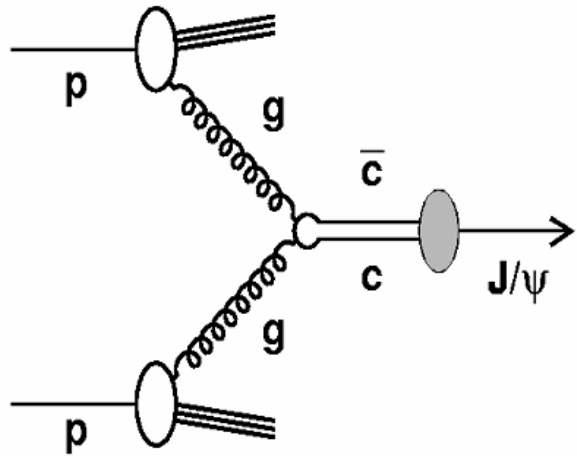}  
\ecc
\caption{Due to its heavy mass, at RHIC energies $J/\psi$ is produced mostly 
via $g + g \rightarrow J/\psi$.}
\label{fig:gg2jpsi}
\end{figure}

\begin{figure}[h!]
\begin{center}
\includegraphics[width=1.0\linewidth]{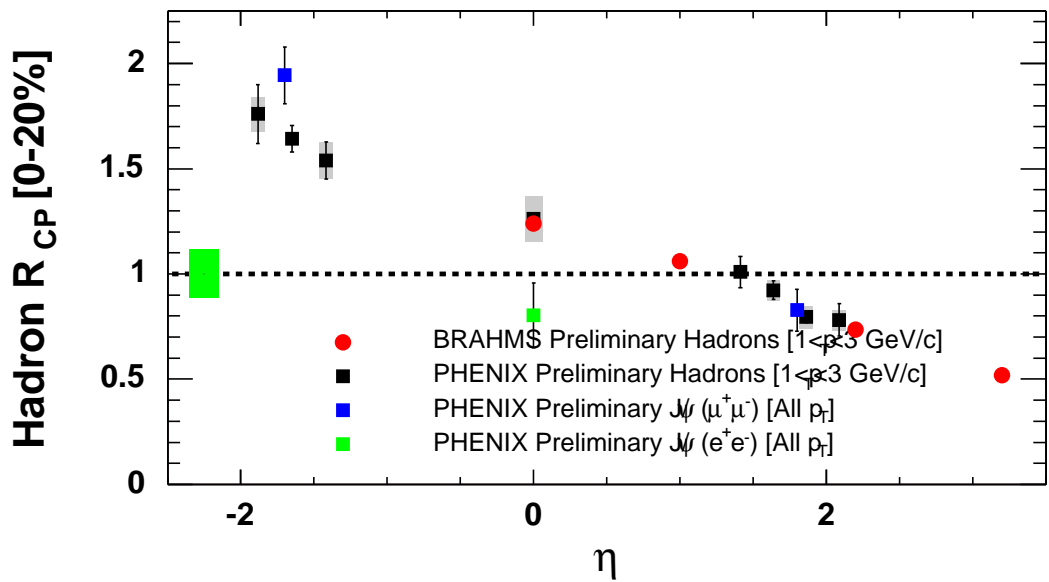}  
\end{center}
\caption{Comparison of PHENIX Preliminary hadron $R_{cp}$ with that of BRAHMS,
and PHENIX $J/\psi$.}
\label{fig:brahms_rcpcomp}
\end{figure}

There is no simple explanation why all this data falls in a straight line. 
The $J/\psi$ results are especially puzzling because of the differing 
production mechanisms. Other models like the recombination model~\cite{hwa}
make only make some qualitative predictions regarding the $\eta$ dependence.

\chapter{Summary}

We studied the final state effects of the produced matter from 
ultrarelativistic Au+Au collisions at $\sqrt{s_{NN}} = 200$ GeV, by looking at 
the production of the simplest nuclei: deuterons and anti-deuterons.  
The transverse momentum spectra of $d$ and $\bar{d}$ in the
range $1.1<p_T<4.3$~GeV/$c$ were measured at mid-rapidity and were found to 
be less steeply falling than proton (and antiproton) spectra. This seems to 
be consistent with a constant (flat) source density profile. The extracted 
coalescence parameter $B_2$ increases with $p_T$, which is indicative of an 
expanding source. $B_2$ decreases for more central collisions, consistent 
with an increasing source size with centrality.  The $B_2$ measured in 
nucleus-nucleus collisions is independent of $\sqrt{s_{NN}}$ above 12 GeV, 
consistent with Bose-Einstein correlation measurements of the radii of the
source. $B_2$ is equal within errors for both deuterons and
anti-deuterons. From the measurements, it is estimated that 
$\bar{n}/n$ = 0.64 $\pm$ 0.04.

We studied the effect of cold nuclear matter in d+Au collisions at 
$\sqrt{s_{NN}} = 200$ GeV, by looking at particle production in forward
and backward directions.  
$R_{cp}$ was observed to be suppressed at Au+Au collisions at 
$\sqrt{s_{NN}} = 200$ 
GeV~\cite{phenix_130supre,phenix_130cent,star_130supre,phenix_200pi}, 
at mid-rapidity. This was consistent either with prediction of jet suppression
due to energy loss in the dense partonic matter created in Au+Au collisions
~\cite{jetsup_wang,jetsup_vitev} or with the depletion of low-$x$ partons 
due to gluon saturation or Color-Glass Condensate (CGC) 
hypothesis~\cite{cgc_klm,gluonsat1,gluonsat2}. When the d+Au control experiment
was done to verify whether or not the suppression seen in Au+Au was due to 
final state effects (jet suppression) or due to initial state effects (changes
in the parton distribution functions) an enhancement was seen for $R_{cp}$ at 
mid-rapidity~\cite{phenix_nosup,star_nosup}. This means that CGC does not 
describe the mid-rapidity distributions and not is applicable in that regime.
	There is a different story at forward rapidity, where the position of
the Muon Arms enables us to probe small values of the momentum fraction $x$. 
We found hadron $R_{cp}$ to be suppressed at forward rapidities
(d going direction). This is qualitatively consistent with 
shadowing/saturation type effects in the small-$x$ region being probed at 
forward rapidities. $R_{cp}$ was enhanced at backward rapidities (Au going 
region). If we include other data: the PHENIX $J/\psi$ and charged hadron 
$R_{cp}$ from BRAHMS, and plot $R_{cp}$ vs $\eta$, we found that the data 
almost lies on a straight line. Although the suppression at forward rapidites
can be qualitatively understood using the CGC model, however the enhancement
at backward rapidities remains a mystery. Although, we should note that our 
momentum range of 1--3 GeV/$c$ is not completely in the hard scattering 
regime. Hence factorisation might not apply and we might be sensitive to soft 
physics. In such a scenario, the backward enhancement could be explained by 
rapidity exchange. Early studies~\cite{stopping_power} have shown that in a 
p+Pb collision with p incident at 100 GeV/$c$ on Pb at rest, the rapidity 
exchange $\simeq$ 2.5. However RHIC is at a much higher energy and so this 
rapidity exchange is expected to a small effect. Thus, at present, there is 
no single satisfactory explaination for the enhancement in the backward 
direction.

In the near future, we plan to try to measure $R_{cp}$ for the prompt muons, 
i.e., muons produced at the vertex mostly from charm decays. This could give 
us some insight on flavor dependence of saturation effects in d+Au collisions. 
For Au+Au, this could give us insight on jet suppression mechanisms, for 
instance gluon bremstrahlung type radiative suppression will depend on the mass
of the quark, and be lesser for charm and other heavy quarks. Another 
measurement would be doing d+Au collisions at an intermediate energy between 
AGS and current RHIC energies to fill in the gaps in previous experiments.
Further in the future, exciting new measurements can be made using the proposed
electron beam accelerator eRHIC at BNL, which will be capable of e+A (and 
polarized protons). Color Glass Condensate model makes very different 
predictions for the gluon PDFS as compared to conventional pQCD calculations.
eRHIC can help us to gain insight into the small-$x$ behaviour of gluons. 
Another interesting study would be the parton propagation through nuclei, 
which can lead to $p_T$ broadening and help us gain a better understanding of
the Cronin effect.

\appendix

\chapter{Comparison between Data and Monte Carlo}

In order to ensure that our matching cuts (or 
track residuals): $\sigma_{\phi}(TOF, PC3)$ and $\sigma_z(TOF, PC3)$ are well
matched between data and MC, we have plotted them against variables 
like $z$, centrality, $p_T$ and so on, for first for the data and then for
the Monte Carlo (MC). 

\section{Matching systematics for data}

\begin{figure}[h]
\begin{center}
\epsfig{file = 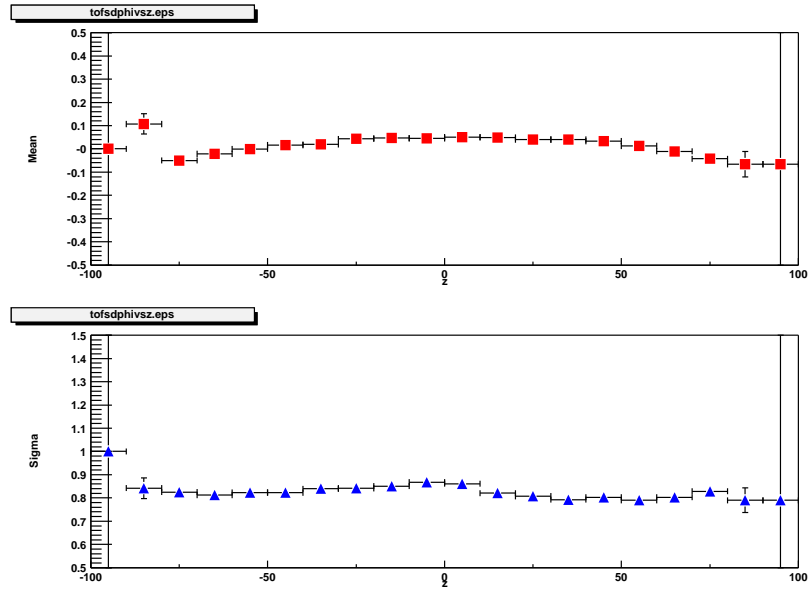, width = 12cm, clip = }
\end{center}
\caption{$\sigma_{\phi}(TOF)$ 
matching variable vs $z$ for data.}
\label{fig:tofsdphivsz}
\end{figure}

\begin{figure}[h]
\begin{center}
\epsfig{file = 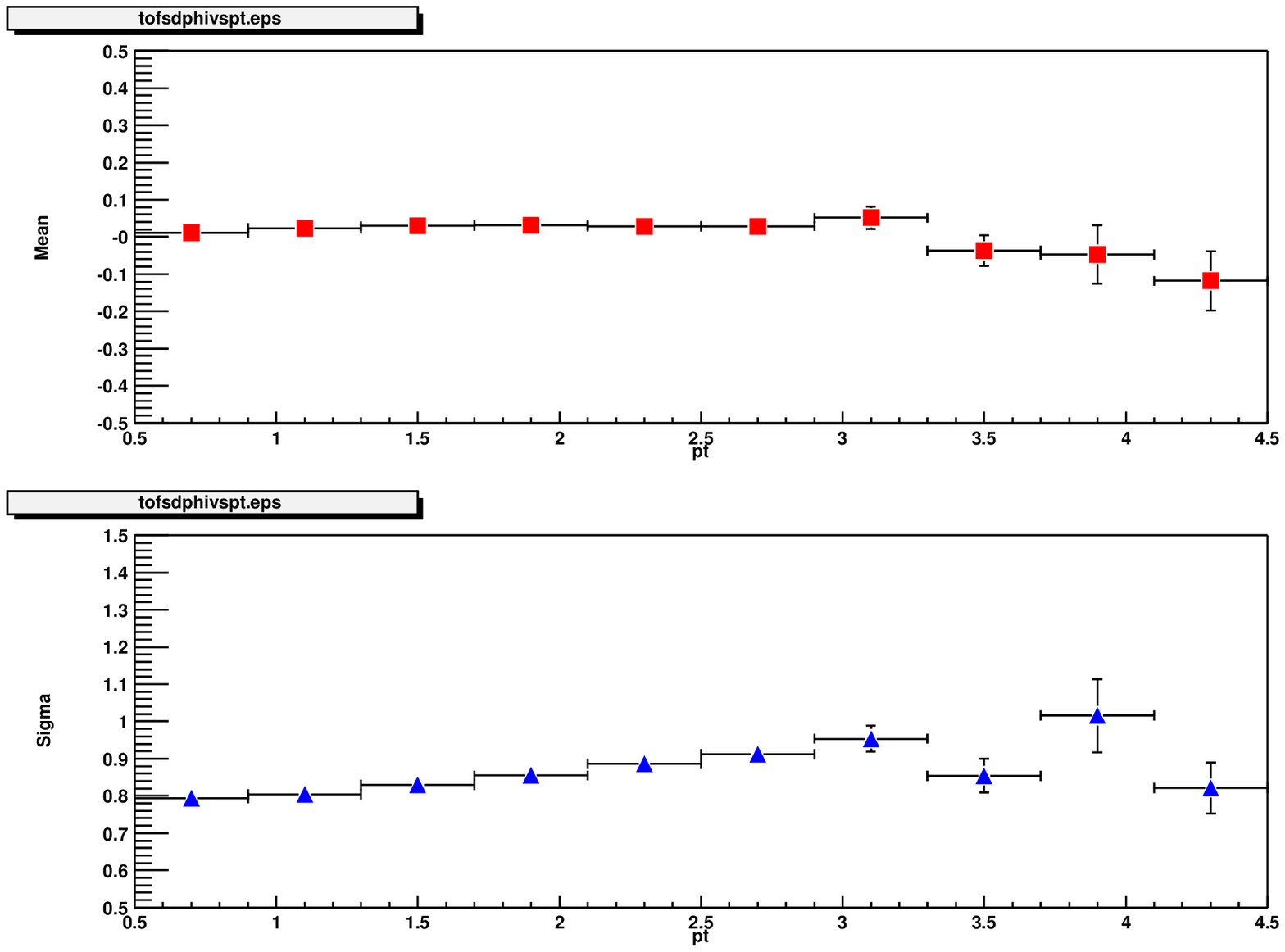, width = 12cm, clip = }
\end{center}
\caption{$\sigma_{\phi}(TOF)$ matching variable vs $p_T$ for data.}
\label{fig:tofsdphivspt}
\end{figure}

\begin{figure}
\begin{center}
\epsfig{file = 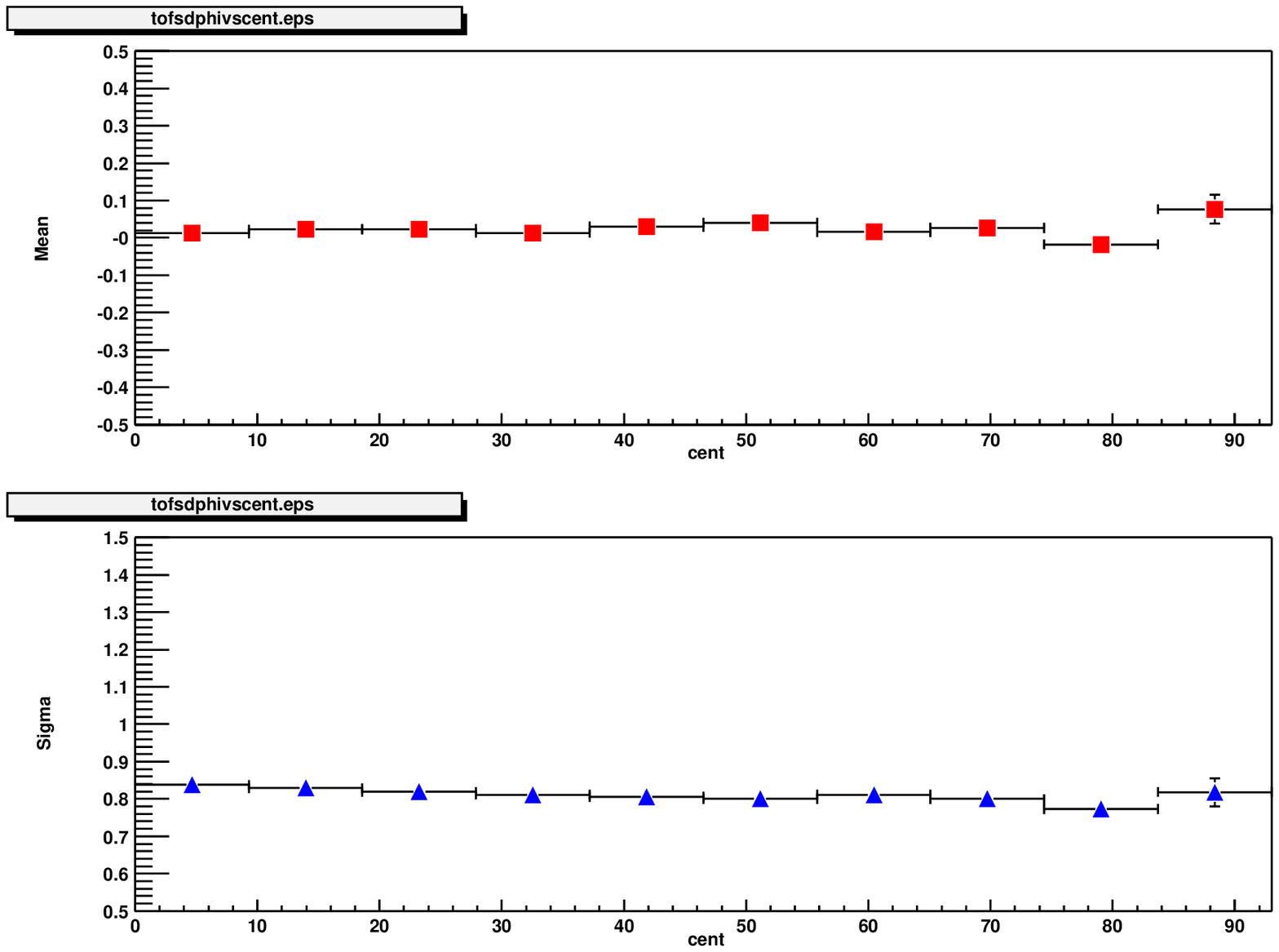, width = 12cm, clip = }
\end{center}
\caption{$\sigma_{\phi}(TOF)$ matching variable vs Centrality for data.}
\label{fig:tofsdphivscent}
\end{figure}

\begin{figure}
\begin{center}
\epsfig{file = 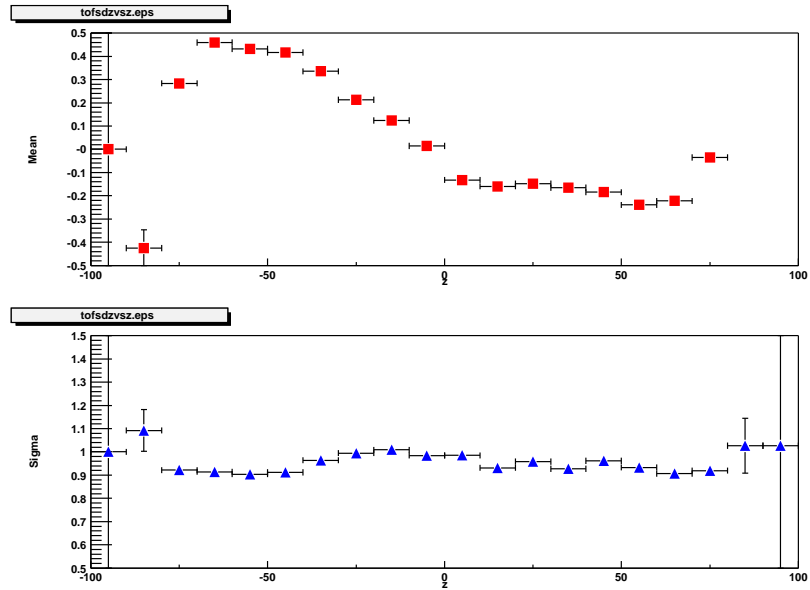, width = 12cm, clip = }
\end{center}
\caption{$\sigma_{z}(TOF)$ matching variable vs $z$ (as measured in DC)
 for data.}
\label{fig:tofsdzvsz}
\end{figure}

\begin{figure}
\begin{center}
\epsfig{file = 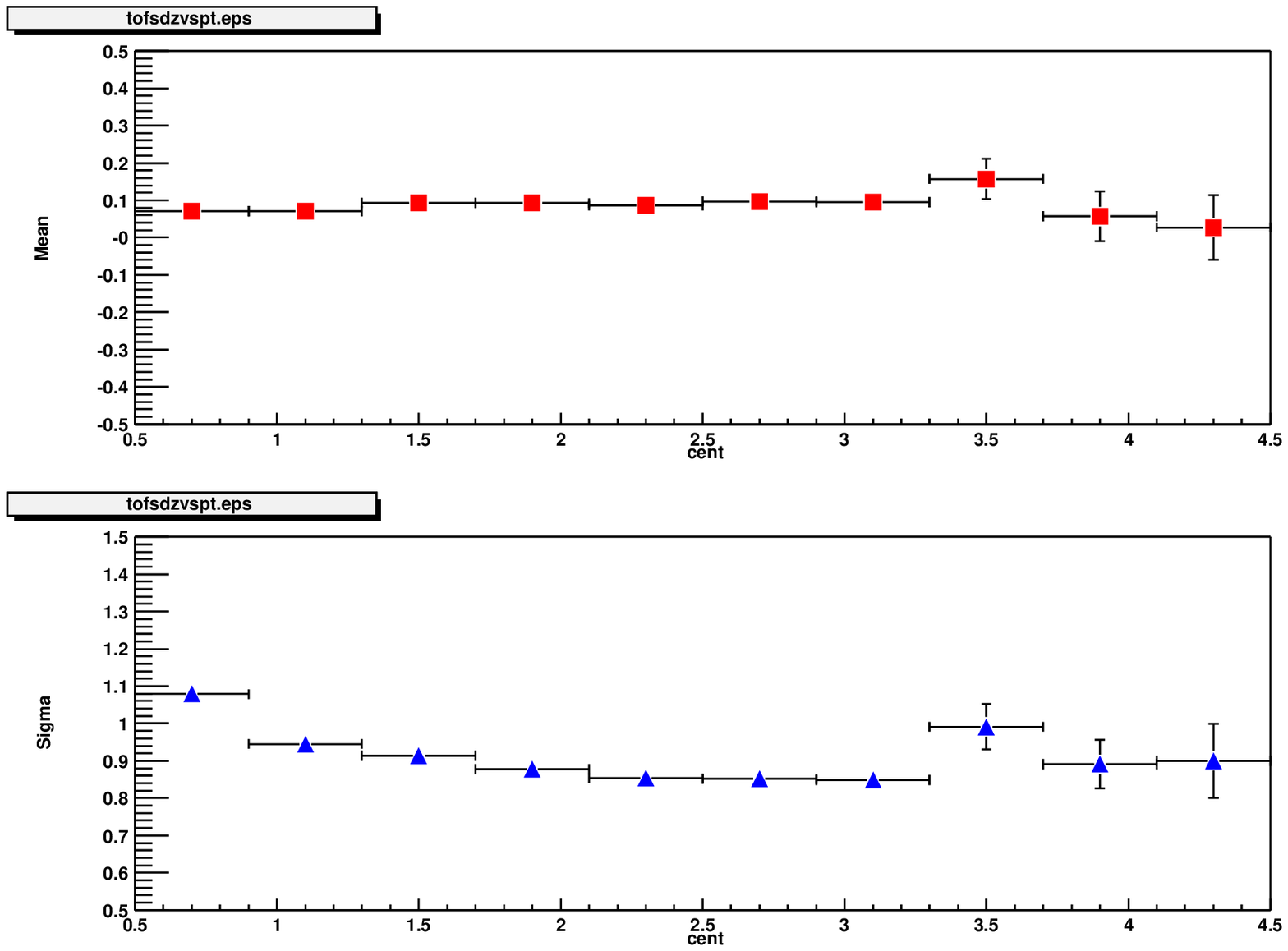, width = 12cm, clip = }
\end{center}
\caption{$\sigma_{z}(TOF)$ matching variable vs $p_T$ for data.}
\label{fig:tofsdzvspt}
\end{figure}

\begin{figure}
\begin{center}
\epsfig{file = 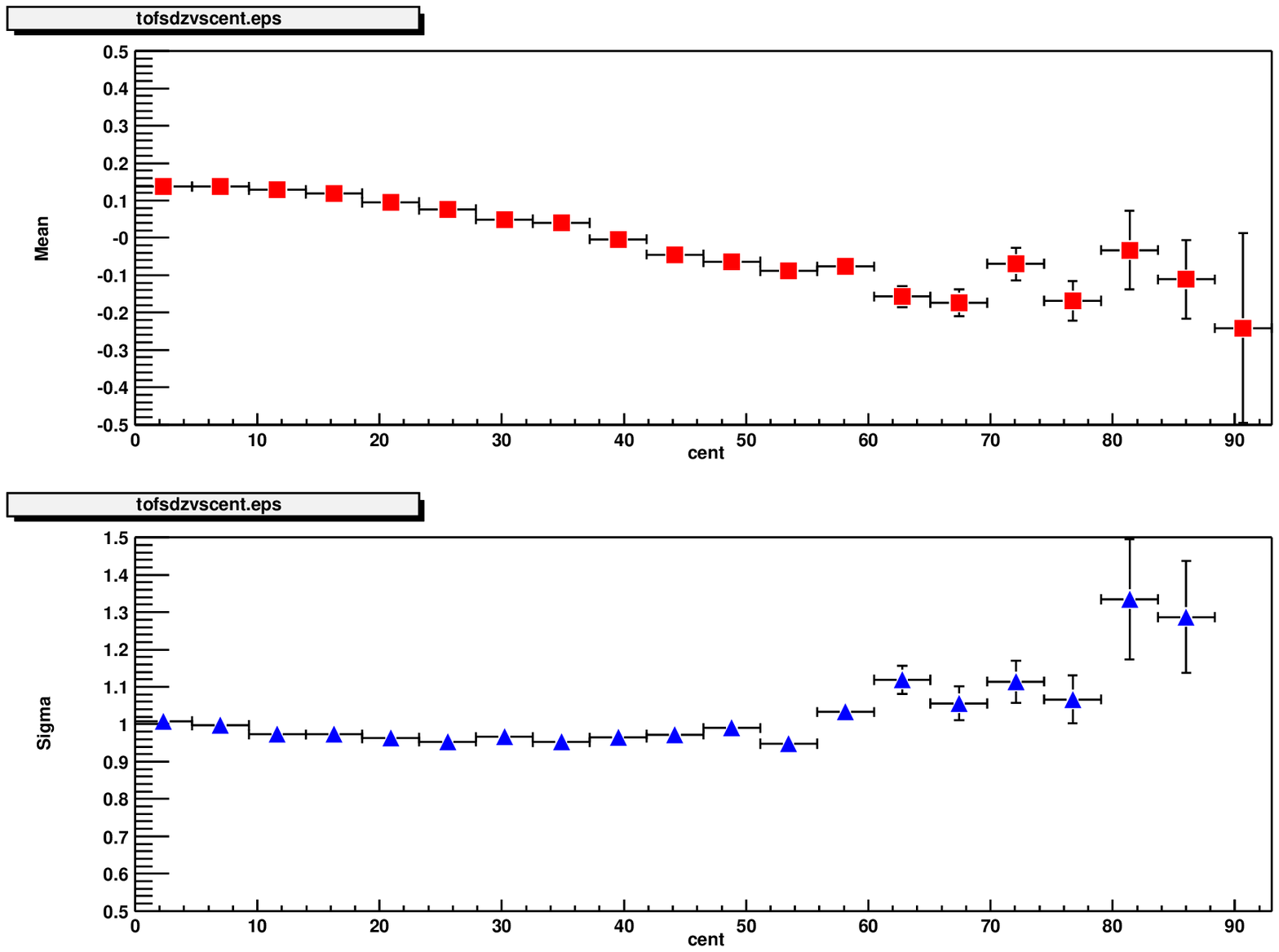, width = 12cm, clip = }
\end{center}
\caption{$\sigma_{z}(TOF)$ matching variable vs Centrality for data.}
\label{fig:tofsdzvscent}
\end{figure}

\begin{figure}
\begin{center}
\epsfig{file = 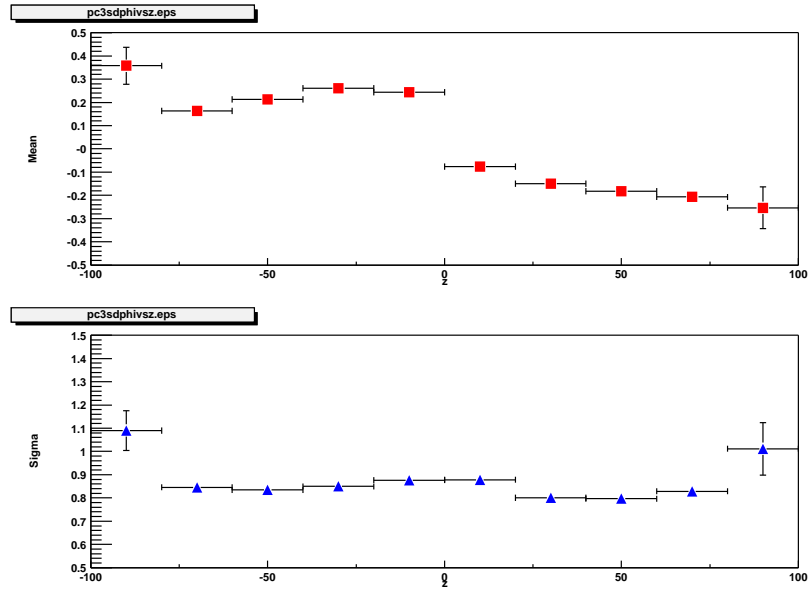, width = 12cm, clip = }
\end{center}
\caption{$\sigma_{\phi}(PC3)$ matching variable vs $z$ (as measured in DC) for data.}
\label{fig:pc3sdphivsz}
\end{figure}

\begin{figure}
\begin{center}
\epsfig{file = 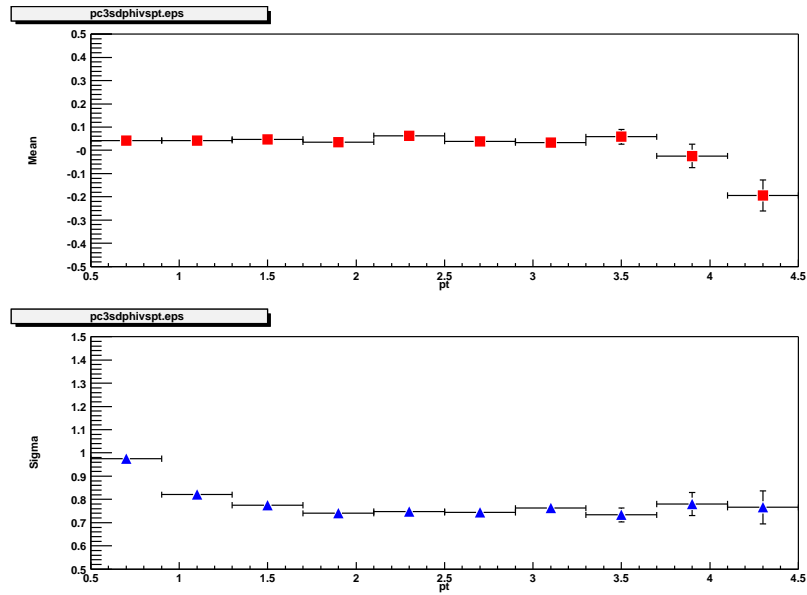, width = 12cm, clip = }
\end{center}
\caption{$\sigma_{\phi}(PC3)$ matching variable vs $p_T$ for data.}
\label{fig:pc3sdphivspt}
\end{figure}

\begin{figure}
\begin{center}
\epsfig{file = 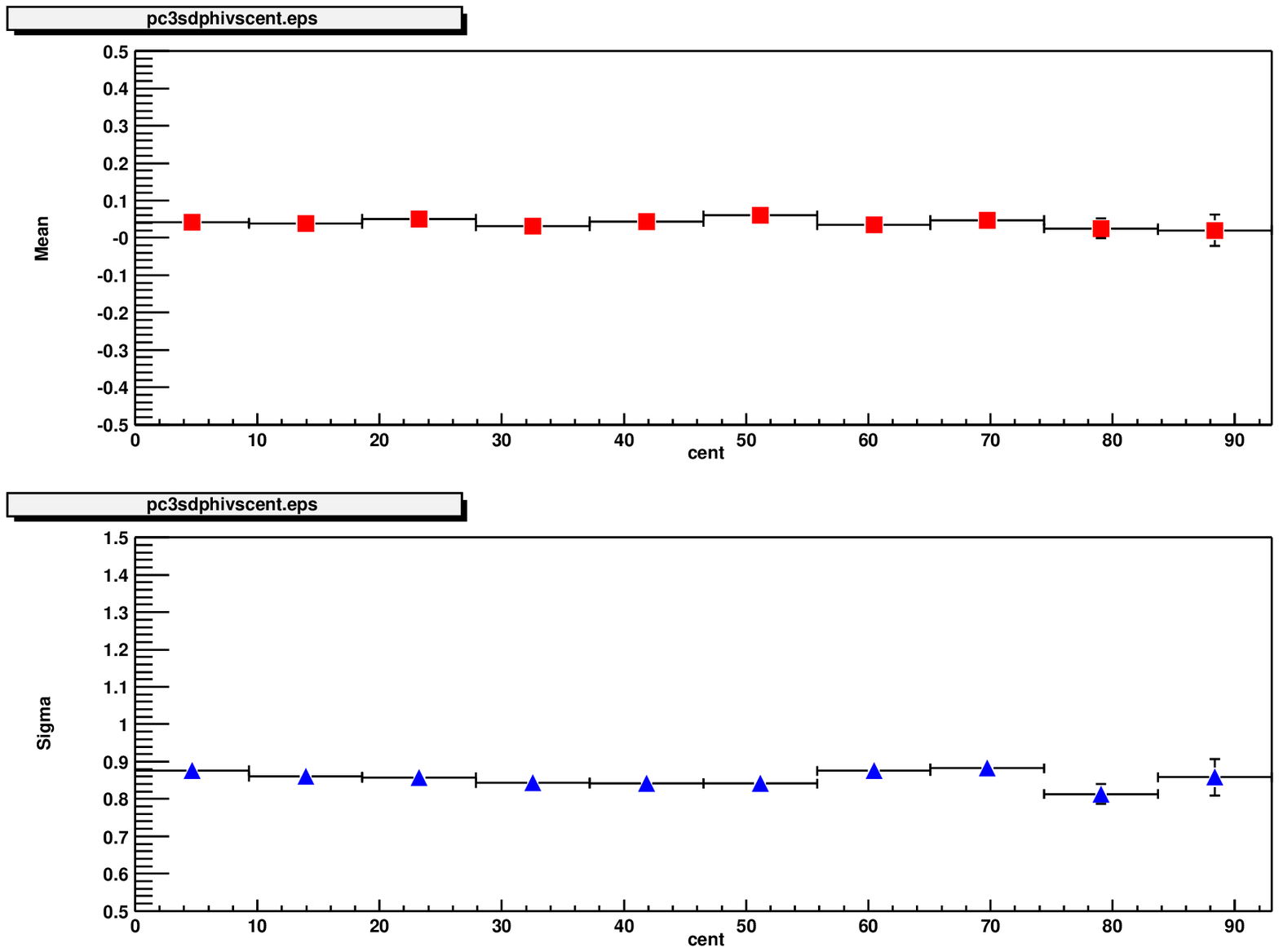, width = 12cm, clip = }
\end{center}
\caption{$\sigma_{\phi}(PC3)$ matching variable  vs centrality for data.}
\label{fig:pc3sdphivscent}
\end{figure}

\begin{figure}
\begin{center}
\epsfig{file = 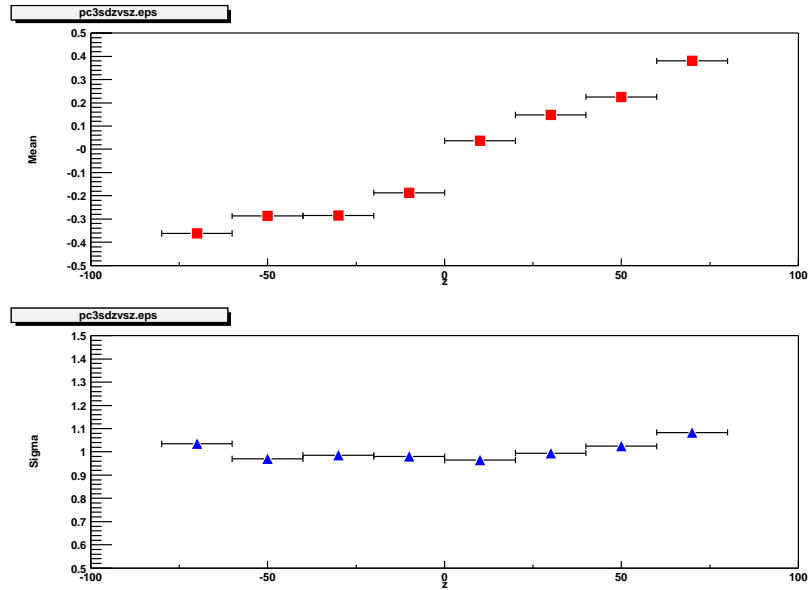, width = 12cm, clip = }
\end{center}
\caption{$\sigma_z(PC3)$  matching variable vs $z$ for data.}
\label{fig:pc3sdzvsz}
\end{figure}

\begin{figure}
\begin{center}
\epsfig{file = 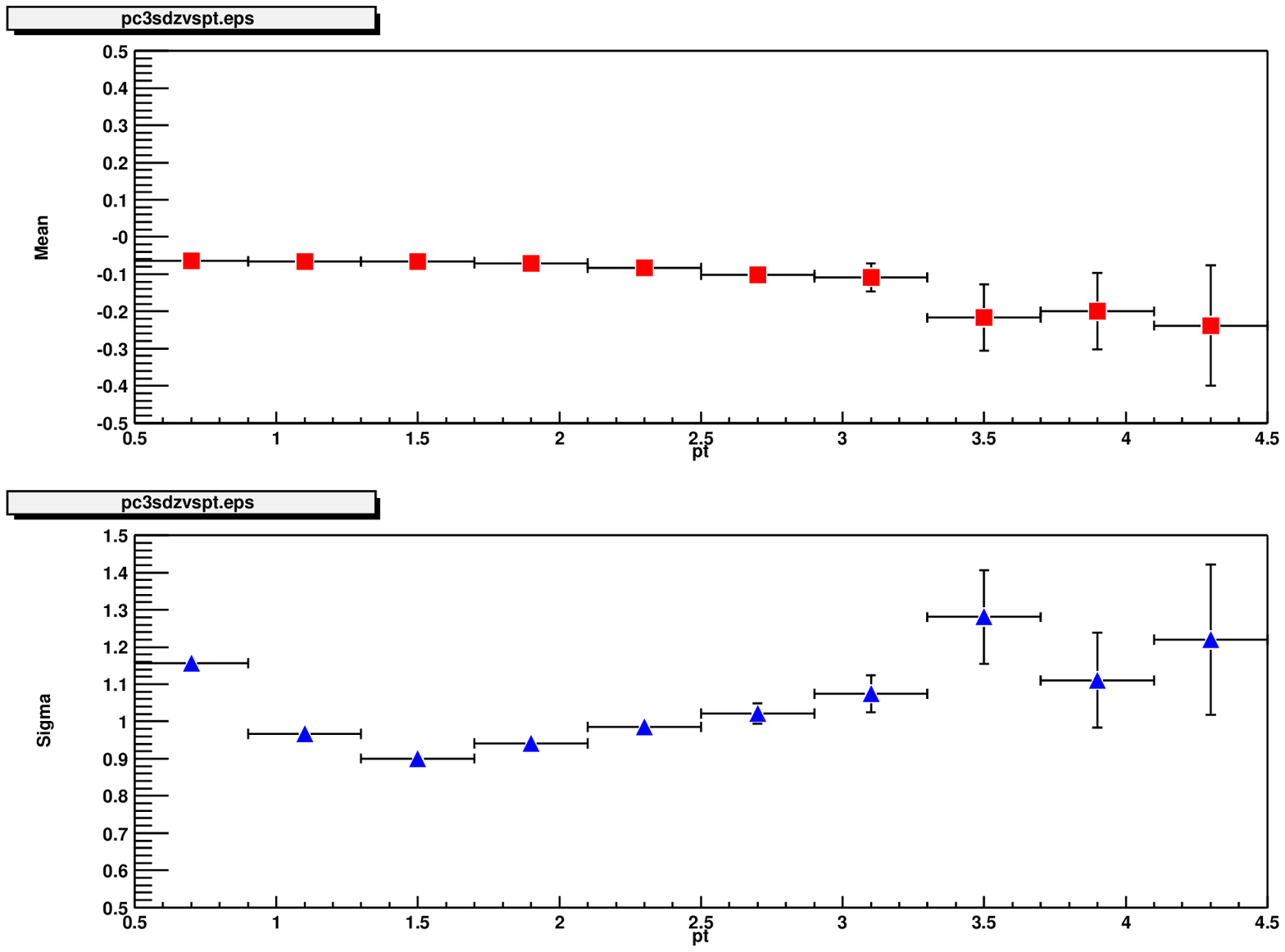, width = 12cm, clip = }
\end{center}
\caption{$\sigma_z(PC3)$  matching variable vs $p_T$ for data.}
\label{fig:pc3sdzvspt}
\end{figure}

\begin{figure}
\begin{center}
\epsfig{file = 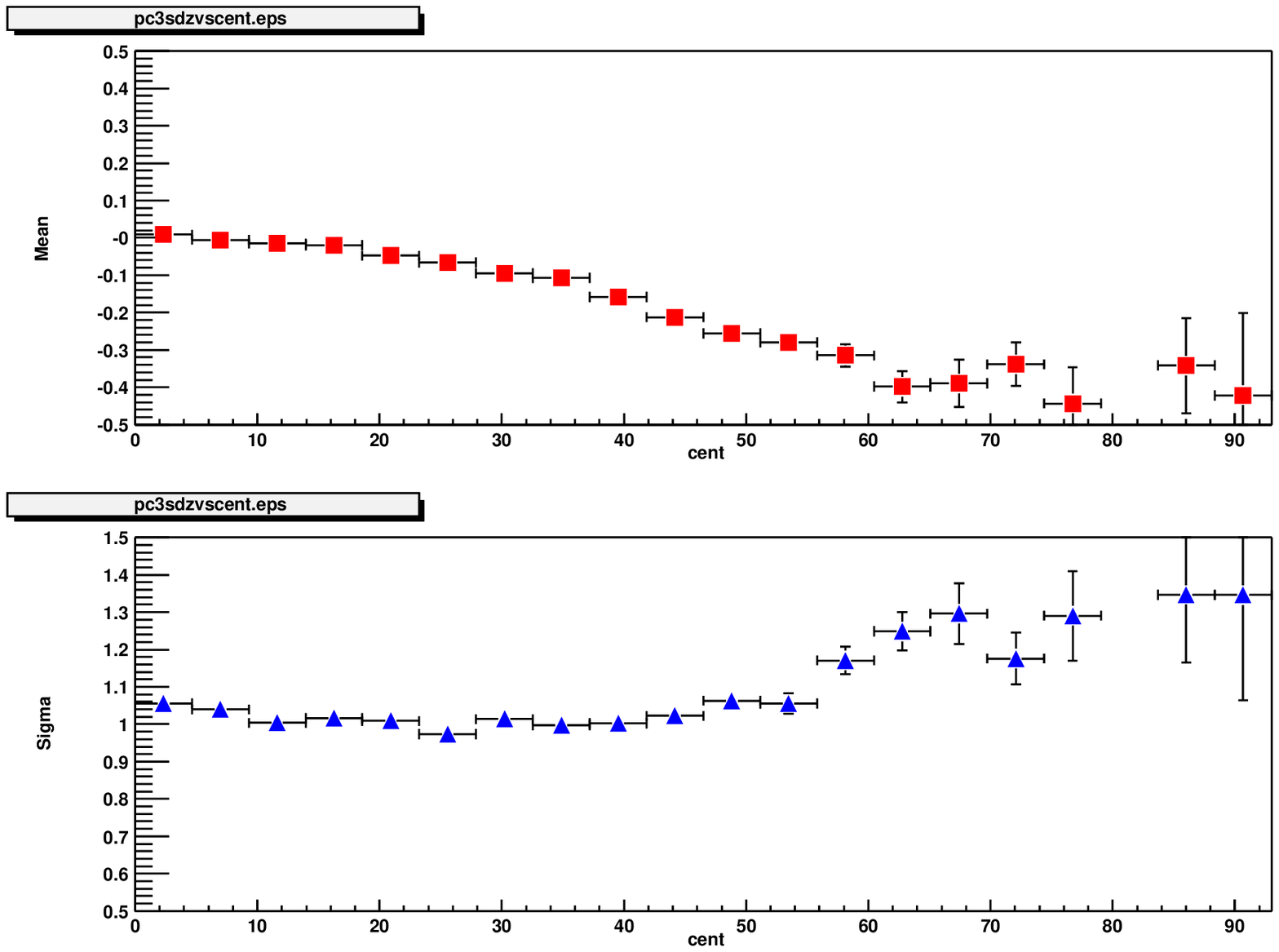, width = 12cm, clip = }
\end{center}
\caption{$\sigma_z(PC3)$  matching variable  vs centrality for data.}
\label{fig:pc3sdzvscent}
\end{figure}

\clearpage
\newpage
\section{Matching systematics in MC}

\begin{figure}[h]
\begin{center}
\epsfig{file = 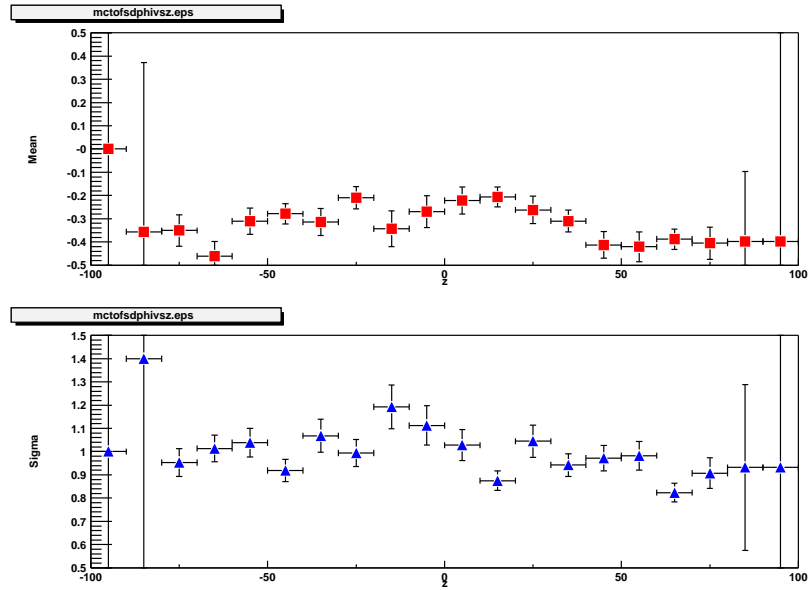, width = 12cm, clip = }
\end{center}
\caption{$\sigma_{\phi}(TOF)$ matching variable vs $z$ in MC}
\label{fig:mctofsdphivsz}
\end{figure}

\begin{figure}[h]
\begin{center}
\epsfig{file = 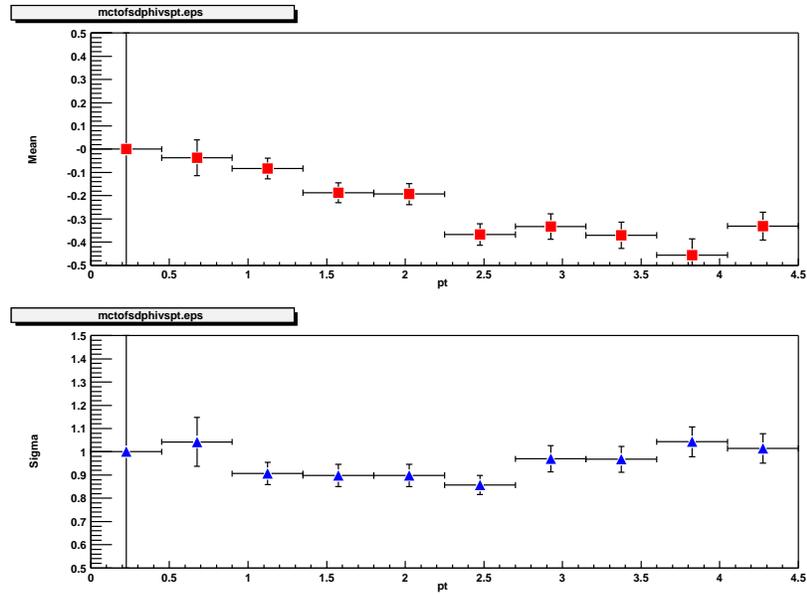, width = 12cm, clip = }
\end{center}
\caption{$\sigma_{\phi}(TOF)$ matching variable vs $p_T$ in MC}
\label{fig:mctofsdphivspt}
\end{figure}

\begin{figure}[h]
\begin{center}
\epsfig{file = 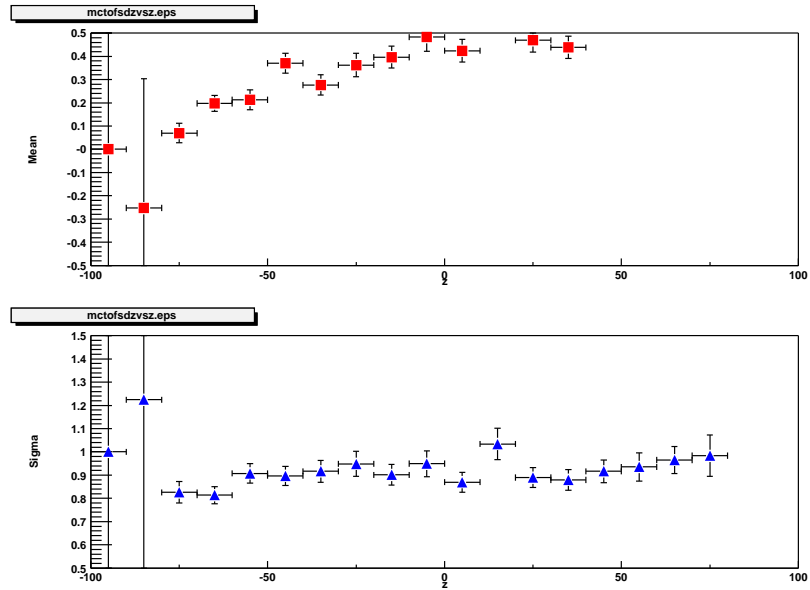, width = 12cm, clip = }
\end{center}
\caption{$\sigma_{z}(TOF)$ matching variable vs $z$ in MC}
\label{fig:mctofsdzvsz}
\end{figure}

\begin{figure}
\begin{center}
\epsfig{file = 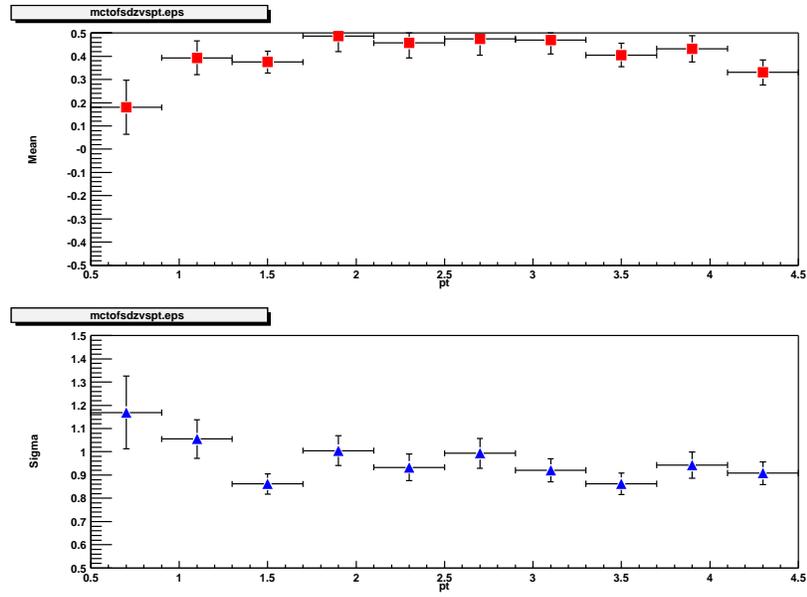, width = 12cm, clip = }
\end{center}
\caption{$\sigma_{z}(TOF)$ matching variable vs $p_T$ in MC}
\label{fig:mctofsdzvspt}
\end{figure}

\begin{figure}
\begin{center}
\epsfig{file = 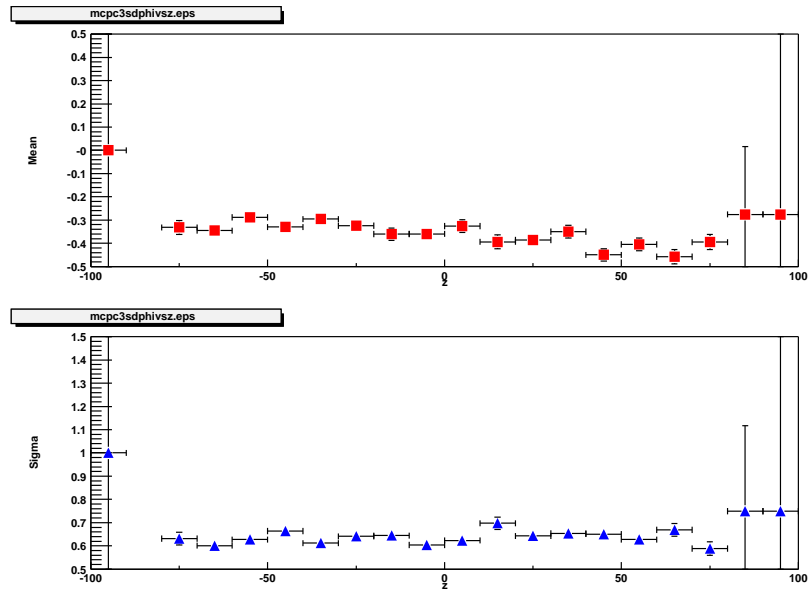, width = 12cm, clip = }
\end{center}
\caption{$\sigma_{\phi}(PC3)$ matching variable vs $z$ in MC}
\label{fig:mcpc3sdphivsz}
\end{figure}

\begin{figure}
\begin{center}
\epsfig{file = 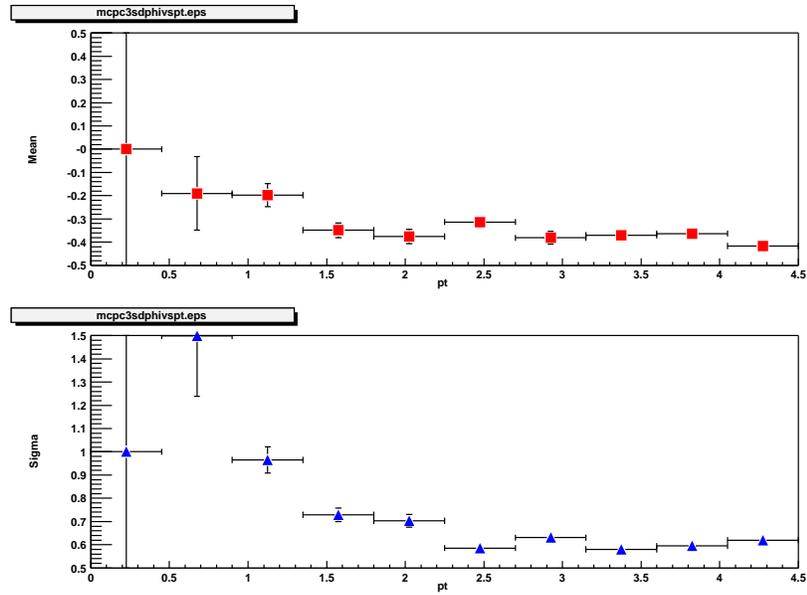, width = 12cm, clip = }
\end{center}
\caption{$\sigma_{\phi}(PC3)$ matching variable vs $p_T$ in MC}
\label{fig:mcpc3sdphivspt}
\end{figure}

\begin{figure}
\begin{center}
\epsfig{file = 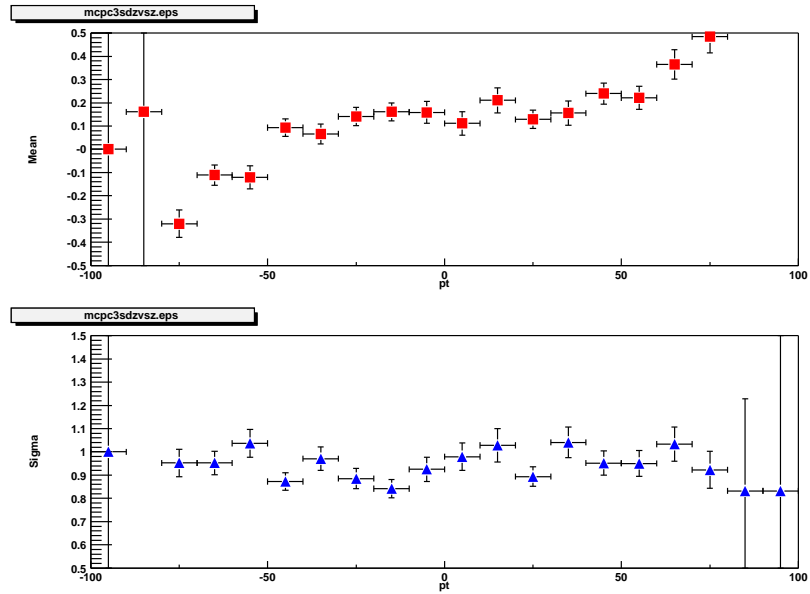, width = 12cm, clip = }
\end{center}
\caption{$\sigma_z(PC3)$  matching variable vs $z$ in MC}
\label{fig:mcpc3sdzvsz}
\end{figure}

\begin{figure}
\begin{center}
\epsfig{file = 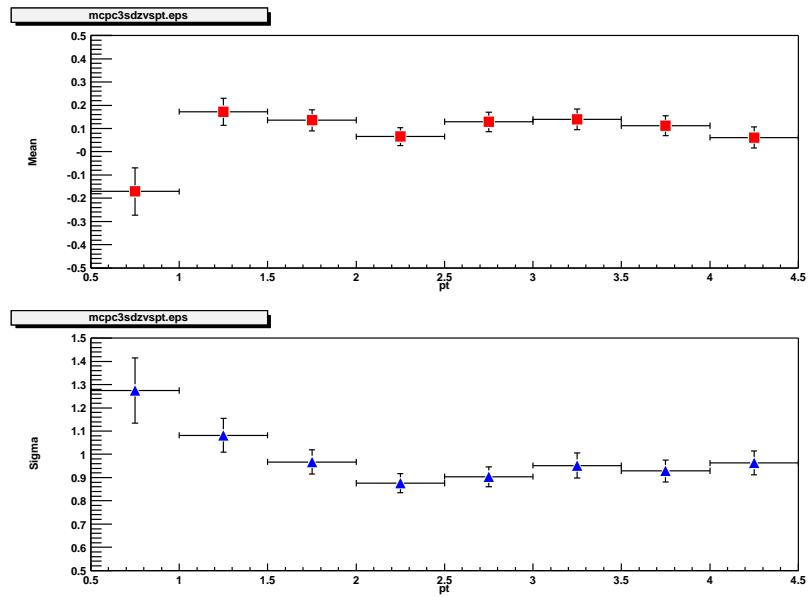, width = 12cm, clip = }
\end{center}
\caption{$\sigma_z(PC3)$  matching variable vs $p_T$ in MC}
\label{fig:mcpc3sdzvspt}
\end{figure}

\clearpage
\newpage
\section{Comparison of acceptance between data and MC}

We also need to check that our MC simulations match each detector's
characteristics as accurately as possible. To cross check this, we look at
track and hit distributions for acceptance both in MC and data.  

\begin{figure}
\begin{center}
\epsfig{file = 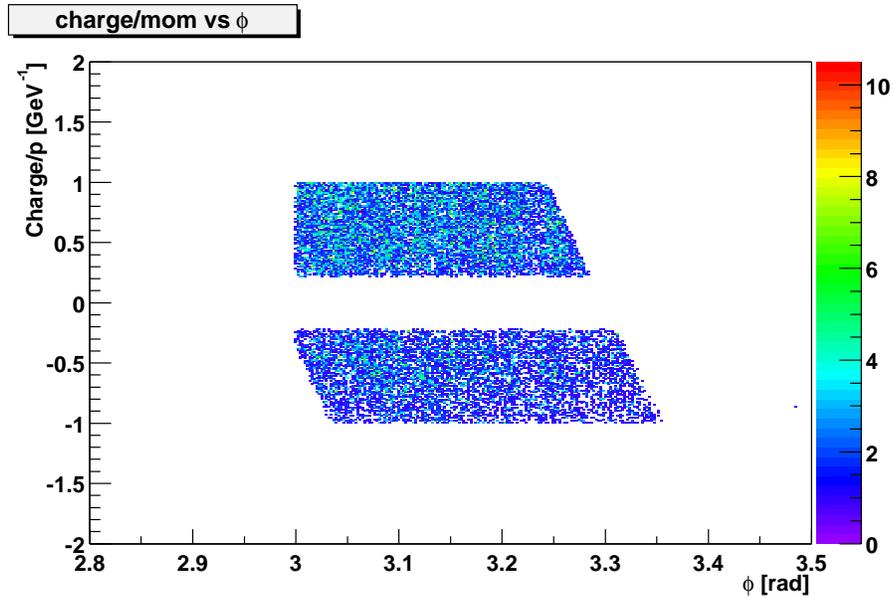, width = 12cm, clip = }
\end{center}
\caption{Charge/$p$ vs Dch $\phi$ in data.}
\label{fig:datafidu_momphi}
\end{figure}

\begin{figure}
\begin{center}
\epsfig{file = 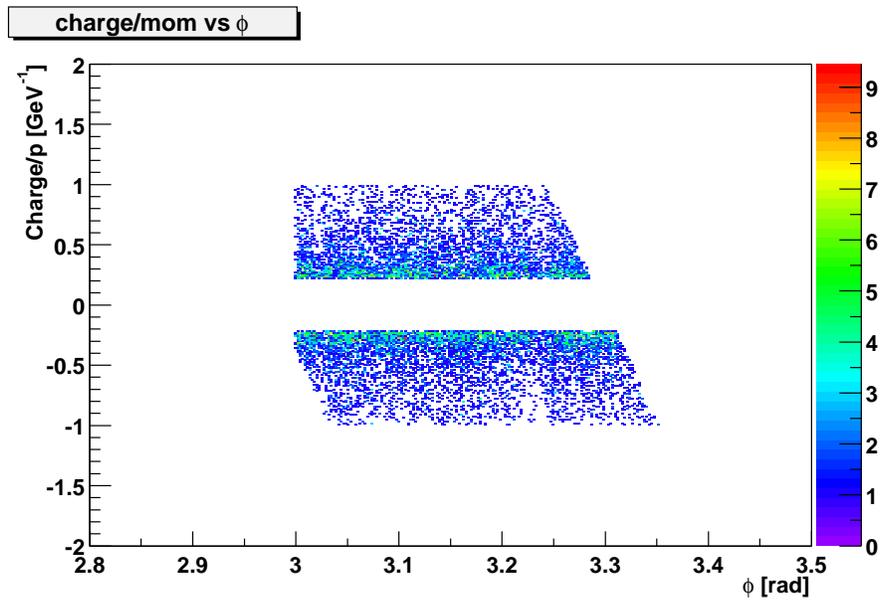, width = 12cm, clip = }
\end{center}
\caption{Charge/$p$ vs DC $\phi$ in MC.}
\label{fig:mcfidu_momphi}
\end{figure}

\begin{figure}
\begin{center}
\epsfig{file = 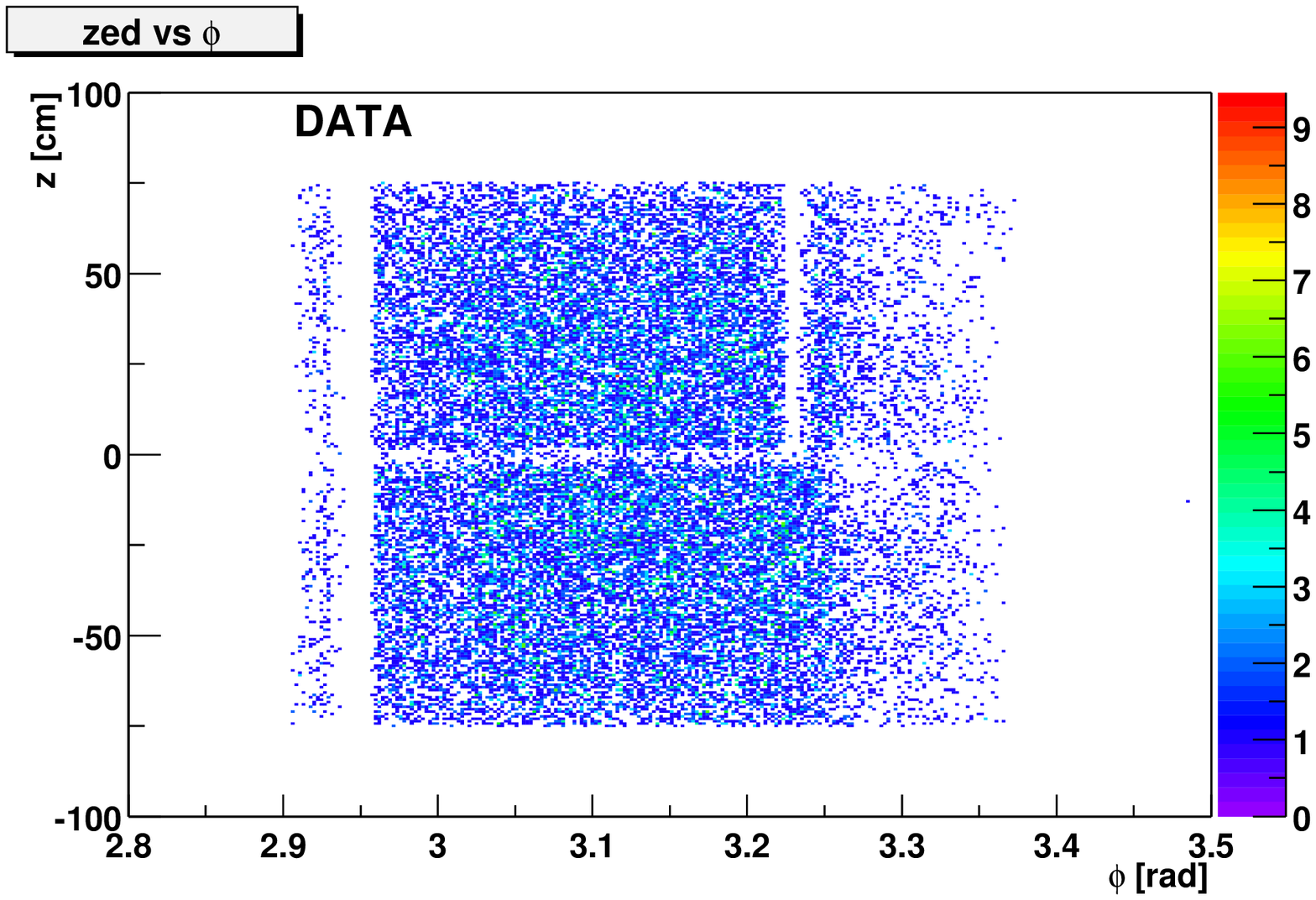, width = 12cm, clip = }
\end{center}
\caption{$z$ vs DC $\phi$ in data.}
\label{fig:datafidu_zphi}
\end{figure}

\begin{figure}
\begin{center}
\epsfig{file = 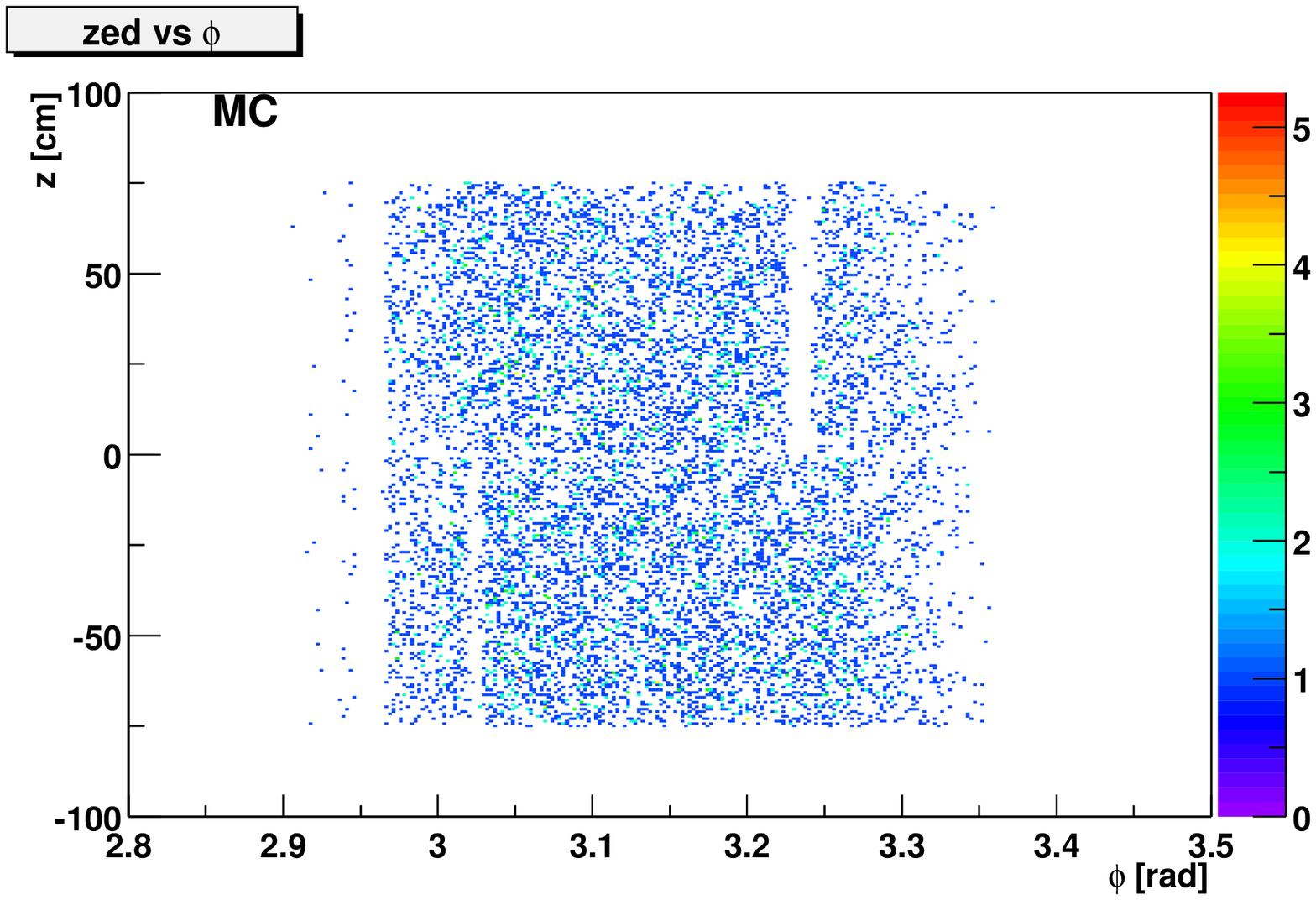, width = 12cm, clip = }
\end{center}
\caption{$z$ vs DC $\phi$ in MC.}
\label{fig:mcfidu_zphi}
\end{figure}

\addtocontents{toc}{\hspace{-2cm}\bf References  \hspace{14.0cm}}


\begin{thebibliography}{99}

\bibitem{microbb} J. D. Bjorken, \Journal{\PRD} {27}{140}{1983}

\bibitem{qcdprediction} For a review see: F. Karsch, 
\Journal{\NPA} {698}{199c}{2002}

\bibitem{nsac} A long range plan for nuclear science, DOE/NSF, Dec. 1983

\bibitem{ruuskanen} For a review of dilepton production see: 
P. V. Ruuskanen, \Journal{\NPA}{522}{255c}{1991} and P. V. Ruuskanen, 
\Journal{\NPA}{544}{169c}{1992}.

\bibitem{lichard} For a review of photon production see: 
J. Kapusta, P. Lichard and D. Siebert, \Journal{\NPA}{544}{485c}{1992} and 
P. V. Ruuskanen, \Journal{\NPA}{544}{169c}{1992}.

\bibitem{strangeness} J. Rafelski \Journal{\NPA}{544}{279c}{1992}; H. C. Eggers
and J. Rafelski, Int. J. Mod. Phys. {\bf A6}, 1067 (1991); P. Koch, 
B. M\"{u}ller and J. Rafelski, Phys. Rep. {\bf 142}, 167 (1986).

\bibitem{satz} T. Matsui and H. Satz, \Journal{\PLB}{178}{416}{1986}; 
T. Matsui, \Journal{\ZPC}{38}{245}{1988}.

\bibitem{hbt_original} R. Hanbury-Brown and R. Q. Twiss, Phil. Mag. 45, 633 
(1954).

\bibitem{gyulassy_jetsup} M. Gyulassy and M. Pl\"{u}mer, 
\Journal{\PLB}{243}{432}{1990}; X. N. Wang and M. Gyulassy, 
\Journal{\PRL}{68}{1480}{1992}; R. Baier {\it et al}, 
\Journal{\PLB}{345}{277}{1995}; 

\bibitem{mult1} K. Adcox {\it et al.} [PHENIX Collaboration], 
\Journal{\PRL} {86}{3500}{2001}. 

\bibitem{phobosback} B.~B.~Back {\it et al.}  [PHOBOS Collaboration],
\Journal{\PRL} {88}{022302}{2002}. 

\bibitem{ppg006} K. Adcox {\it et al.} [PHENIX Collaboration], 
\Journal{\PRL} {88}{242301}{2002}. 

\bibitem{phobospbar} B.~B.~Back {\it et al.}  [PHOBOS Collaboration],
\Journal{\PRC} {67}{021901}{2003}.

\bibitem{starpbar} C.~Adler {\it et al.} [STAR Collaboration],
\Journal{\PRL}{86}{4778}{2001}.

\bibitem{csernai} L. P. Csernai and J. I. Kapusta, 
Phys. Rep. {\bf 131}, 223 (1986).

\bibitem{mekjian} A. Z. Mekjian, 
\Journal{\PRC} {17}{1051}{1978}.

\bibitem{ppg009} K. Adcox {\it et al.} [PHENIX Collaboration], 
\Journal{\PRC} {69}{024904}{2004}.

\bibitem{ppg026} S. S. Adler {\it et al.} [PHENIX Collaboration], 
{\it nucl-ex/0307022}.

\bibitem{phenix_130supre} K. Adcox {\it et al.} [PHENIX Collaboration], 
\Journal{\PRL}{88}{022301}{2002}. 

\bibitem{phenix_130cent} K. Adcox {\it et al.} [PHENIX Collaboration], 
\Journal{\PLB}{561}{82}{2003}. 

\bibitem{star_130supre} C. Adler {\it et al.} [STAR Collaboration], 
\Journal{\PRL}{89}{202301}{2002}. 

\bibitem{phenix_200pi} S. S. Adler {\it et al.} [PHENIX Collaboration], 
\Journal{\PRL}{91}{241803}{2002}. 

\bibitem{phenix_nosup} S. S. Adler {\it et al.} [PHENIX Collaboration], 
\Journal{\PRL}{91}{07203}{2003}. 

\bibitem{star_nosup} J. Adams {\it et al.} [STAR Collaboration], 
\Journal{\PRL}{91}{072304}{2003}. 

\bibitem{cgc_klm} D. Kharzeev, E. Levin and L. McLerran,
\Journal{\PLB}{561}{93}{2003}; D. Kharzeev, Y. V. Kovchegov and K. Tuchin,  
\Journal{\PRD}{68}{094013}{2003}.

\bibitem{gluonsat1} R. Baier {\it et al.} \Journal{\PRD}{68}{054009}{2003}

\bibitem{gluonsat2} J. Jallilian-Marian, Y. Nara and R. Venugopalan,
\Journal{\PLB}{577}{54}{2003}; A. Dumitru and J. Jallilian-Marian, 
\Journal{\PRL}{89}{022301}{2002};

\bibitem{phenixnim} K. Adcox {\it et al.} [PHENIX Collaboration], 
\Journal{\NIMA}{499}{469}{2003} and references therein. 

\bibitem{ppgmb} S. S. Adler {\it et al.} [PHENIX Collaboration], 
\Journal{\PRL} {91}{072301}{2003}. 

\bibitem{bethe-bloch} http://pdg.lbl.gov/2002/passagerpp.pdf.

\bibitem{pdg} K. Hagiwara {\it et al.}, [Particle Data Group],
\Journal{\PRD}{66}{010001}{2002}.

\bibitem{geant} GEANT 3.21, CERN program library.

\bibitem{jiaembed} Private Communication, Jiangyong Jia.

\bibitem{lambda130} K. Adcox {\it et al.} [PHENIX Collaboration], 
\Journal{\PRL} {89}{092302}{2002}. 

\bibitem{Moiseev} A. A. Moiseev and J. F. Ormes in Astroparticle Physics 
6(1997) 379-386.

\bibitem{Jaros} J. Jaros et al., Phys. Rev. C18(1978)2273.

\bibitem{Abdurak} E. O. Abdurakhmanov et al., Z. Phys. C5(1980)1.

\bibitem{lundwork} Private Communication, Joakim Nystrand, Rickard du Rietz and 
E. Stenlund.

\bibitem{heinz_prc99} R. Scheibl, and U. Heinz, 
\Journal{\PRC}{59}{1585}{1999}. 

\bibitem{polleri} A. Polleri, J. P. Bondorf, and I. N. Mishustin, 
\Journal{\PLB} {419}{19}{1998}.

\bibitem{braun_thermal}  P. Braun-Munzinger, K. Redlich and J. Stachel,
Invited review for Quark Gluon Plasma 3, eds. R. C. Hwa and Xin-Nian Wang, 
World Scientific Publishing.

\bibitem{butler} S. T. Butler and C. A. Pearson, 
Phys. Rev. {\bf 129}, 836 (1963).

\bibitem{eos} S. Wang {\it et al.}, [EOS Collaboration], 
\Journal{\PRL} {74}{2646}{1995}. 

\bibitem{e896} S. Albergo {\it et al.}, [E896 Collaboration], 
\Journal{\PRC} {65}{034907}{2002}. 

\bibitem{e864} T. A. Armstrong {\it et al.} [E864 Collaboration], 
\Journal{\PRL} {85}{2685}{2000}. 

\bibitem{na49prc} T. Anticic {\it et al.} [NA49 Collaboration], 
\Journal{\PRC} {69}{024902}{2004}. 

\bibitem{na44prl} I. G. Bearden {\it et al.} [NA44 Collaboration], 
\Journal{\PRL} {85}{2681}{2000}.

\bibitem{star} C. Adler {\it et al.}, [STAR Collaboration],
\Journal{\PRL} {87}{262301}{2001}.

\bibitem{hbt130} K. Adcox {\it et al.} [PHENIX Collaboration], 
\Journal{\PRL} {88}{192302}{2002}. 

\bibitem{starhbt130} C. Adler {\it et al.} [STAR Collaboration], 
\Journal{\PRL} {87}{082301}{2001}. 

\bibitem{hadron} R. P. Feynman: {\it Photon-Hadron Interactions}, W. A. 
Benjamin Inc., 1972.

\bibitem{bjork69} J. D. Bjorken, E. A. Paschos: {\it Phys. Rev.}, {\bf 185} 
(1969) 1975.

\bibitem{gell64} M. Gell-Mann: {\it Phys. Lett.}, {\bf 8} (1964) 214.

\bibitem{dglap} G. Altarelli and G. Parisi, {\it Nucl. Phys.}, {\bf B126} 298 (1977); Y.L. Dokshitzer, {\it Sov. Phys. JETP},{\bf 46} 641 (1997); V.N. Gribov, and L.N. Lipatov, {\it Sov J. Nucl. Phys.} {\bf 15} 438; 675  (1972). 

\bibitem{hera} H. Abramowicz and A.C. Caldwell {\it Reviews of Modern Phys.}, Vol. 
{\bf 71}, No.{\bf 5} (1999).  

\bibitem{shadowing}  Arneodo, Phys. Rep. {\bf 240}, 301 (1994); Nicolaev \& 
Zakharov \Journal{\PLB}{55}{397}{1975}; Mueller \& Qiu, 
\Journal{\NPB}{268}{427}{1986}.

\bibitem{glauber} In the present work the Woods-Saxon nuclear density 
parameters: radius R = 6.38 fm, diffusivity $a$ = 0.54 fm and N-N cross 
section $\sigma_{NN}^{inel}$ = 42 mb were used. The deuteron is described
by a Hulth\'{e}n wave function (L. Hulth\'{e}n and M. Sagawara, Handbuch der
Physik {\bf 39} (1957)) with $\alpha$ = 0.228 fm$^{-1}$ and $\beta$ = 1.18 
fm$^{-1}$.

\bibitem{jetsup_wang} M. Gyulassy and X. N. Wang,
\Journal{\NPB}{420}{583}{1994}; X. N. Wang \Journal{\PRC}{58}{2321}{1998}.

\bibitem{jetsup_vitev} I. Vitev and M. Gyulassy,
\Journal{\PRL}{89}{252301}{2002}.

\bibitem{muonnim} K. Adcox {\it et al.} [PHENIX Collaboration], 
\Journal{\NIMA}{499}{469}{2003}. 

\bibitem{chunswork} Private Communication, Chun Zhang and Ming X. Liu.

\bibitem{brahms_rcp} I. Arsene {\it et al.}  [BRAHMS Collaboration] nucl-ex/0401025. 

\bibitem{stopping_power} W. Busza and A.S. Goldhaber {\it Phys. Lett.} B{\bf 139} 235 
(1984).

\bibitem{hwa} R. Hwa {\it et al.} nucl-th/0404066.


\end{thebibliography}
\end{document}